\documentclass[aps,pra,groupedaddress,superscriptaddress,nofootinbib,notitlepage,floatfix]{revtex4-2}
\usepackage{times,bm,bbm,amsmath,amssymb,amsfonts,graphics,graphicx,color,xcolor,braket,theorem,soul} 
\usepackage{mathtools}
\usepackage{setspace}

\usepackage{hyperref}
\hypersetup{colorlinks,linkcolor={magenta},citecolor={magenta},urlcolor={magenta}}
\newtheorem{theorem}{Theorem}
\newtheorem{definition}{Definition}
\newtheorem{element}{Element}

\newcommand{\ignore}[1]{}

\let\oldsqrt\sqrt
\def\sqrt{\mathpalette\DHLhksqrt}
\def\DHLhksqrt#1#2{%
\setbox0=\hbox{$#1\oldsqrt{#2\,}$}\dimen0=\ht0
\advance\dimen0-0.2\ht0
\setbox2=\hbox{\vrule height\ht0 depth -\dimen0}%
{\box0\lower0.4pt\box2}}
\DeclareFontFamily{OT1}{pzc}{}
\DeclareFontShape{OT1}{pzc}{m}{it}
              {<-> s * [1.25] pzcmi7t}{}
\DeclareMathAlphabet{\mathpzc}{OT1}{pzc}
                                 {m}{it}
\usepackage[T1]{fontenc}
\usepackage{newtxtext,newtxmath}
\begin{document} 

\title{Does Quantum Mechanics Breed Larger, More Intricate Quantum Theories? \\
The Case for Experience-Centric Quantum Theory and the Interactome of Quantum Theories} 

\author{Alireza Tavanfar}
\email{alireza.tavanfar.research@gmail.com}
\affiliation{Champalimaud Research, Champalimaud Center for the Unknown, 1400-038 Lisbon, Portugal} 
\affiliation{Institute of Neuroscience, University of Oregon, Eugene, OR 97403, United States}
\affiliation{Centre of eXplainable Artificial Intelligence, Technological University Dublin, Dublin, Ireland}

\author{S. Alipour}
\email{s.alipoor@gmail.com}
\affiliation{QTF Centre of Excellence, Department of Applied Physics, Aalto University, P. O. Box 11000, FI-00076 Aalto, Espoo, Finland}

\author{A. T. Rezakhani}
\email{rezakhani@sharif.edu}
\affiliation{Department of Physics, Sharif University of Technology, Tehran 14588, Iran}
%%%%%%%%%%%%%%%%%%%%%%%%%%%%%%%%%%%%%%%%%%%%%%%%%%%%%%%%%%%%%%%%

%%%%%%%%%%%%%%%%%%%%%%%%%%%%%%%%%%%%%%%%%%%%%%%%%%%%%%%%%%%%%%%%
\begin{abstract}
We pose and address the radical question of whether quantum mechanics, known for its firm internal structure and enormous empirical success, carries in itself the genomes of larger quantum theories that have higher internal intricacy and phenomenological versatility. In other words, we consider, at the basic level of closed quantum systems and regardless of interpretational aspects, whether standard quantum theory (SQT) harbors quantum theories with context-based deformed principles or structures, having definite predictive power within much broader scopes. We answer this question in the affirmative following complementary evidence and reasoning arising from quantum-computation-based quantum simulation and fundamental, general, and abstract rationales within the frameworks of information theory, fundamental or functional emergence, and participatory agency. In this light, as we show, one is led to the recently proposed \textit{experience-centric quantum theory} (ECQT), which is a larger and richer theory of quantum behaviors with drastically generalized quantum dynamics. ECQT allows the quantum information of the closed quantum system’s developed state history to continually contribute to defining and updating the many-body interactions, the Hamiltonians, and even the internal elements and ``particles'' of the total system. Hence, the unitary evolutions are continually impacted and become guidable by the agent system's \textit{experience}. The intrinsic interplay of unitarity and non-Markovianity in ECQT brings about a host of diverse behavioral phases, which concurrently infuse closed and open quantum system characteristics, and it even surpasses the theory of open systems in SQT. From a broader perspective, a focus of our investigation is the existence of the \textit{quantum interactome}---the interactive landscape of all coexisting, independent, \textit{context-based} quantum theories that emerge from inferential participatory agencies---and its predictive phenomenological utility.
\end{abstract}
\maketitle 
%%%%%%%%%%%%%%%%%%%%%%%%%%%%%%%%%%%%%%%%%%%%%%%%%%%%%%%%%%%%%%%%
\begin{spacing}{.935}
{\hypersetup{linkcolor={blue}}
{\small{\tableofcontents}}
}
\end{spacing}

%%%%%%%%%%%%%%%%%%%%%%%%%%%%%%%%%%%%%%%%%%%%%%%%%%%%%%%%%%%%%%%%
\section{A Top-Down Invitation to an Interactome of Independent Quantum Theories} 
\label{sec:intro}

Quantum physics is one of the most successful grand theories of nature, within which a broad range of theories, e.g., applying to elementary particles and condensed matter systems, are hosted. This unique nature, together with the fact that quantum physics is constructible in general information-theoretic frameworks, makes it potentially amenable to accommodating yet more broader phenomena in both natural and artificial domains. On the other hand, closer explorations of this expansive attribute of quantum physics can itself contribute to a deeper and clearer understanding of \textit{what quantum behaviors are} in their broadest spectrum from the elementary to highly emergent. Such deeper reconsiderations of quantum theory are especially welcomed by our long-standing searches for (i) a compelling treatment of the quantum aspects of gravitational and cosmological phenomena---such as black holes, cosmological horizons, or the pregeometric universe---(ii) (abstract or tangible) \textit{emergent complex systems}. It is known that states and properties of quantum systems are fundamentally different from classical systems, which can enable diverse and rich quantum advantages, e.g., in information processing and computation. In particular, a key role is played by \textit{quantum dynamics} itself. In standard quantum theory (SQT), namely quantum mechanics and its finite-dimensional reductions, the unitary evolution of a closed system is by postulate time-local and linear, such that the Hamiltonian is not influenceable or updatable by the state history of the system. As strict as this postulate has been maintained, it is intriguing and plausible to scrutinize diverse quantum behaviors to see \textit{whether} and \textit{how} in proper contexts and ways time-nonlocal features such as the developed state history of a closed quantum system---or put differently, its \textit{experience}---can influence the instantaneous Hamiltonians and internal interactions of the system. Moreover, one can ask whether such a drastic deformation of SQT may exist or emerge in nature, in particular in contexts broader than the traditional domains of quantum physics. \\

Conducting a multifaceted investigation, we address these questions in the intersection of four independent, complementary directions; quantum-computation-based simulation, comparative formal theorems, behavioral analysis, and abstract, fundamental rationales. First, we devise a novel quantum simulation technique to examine and bridge the SQT itself to a larger, more intricate quantum theory, namely, experience-centric quantum theory (ECQT) proposed in Ref. \cite{ECQT-1}. By definition, ECQT is any quantum theory which allows the instantaneous Hamiltonians, internal interactions, and even the degrees of freedom of a closed quantum system to be experience centric (EC), that is, reformable and updatable by the developed state history of the closed system. In our scheme of quantum simulation, we demonstrate that a phenomenologically promising subclass of EC evolutions can in principle be simulated within SQT, albeit at the expense of exceedingly large overhead resources and extra quantum degrees of freedom. However, arbitrary EC quantum dynamics still remains out of reach of our quantum simulation scheme by using finite resources. Going ahead, we elevate the comparison between ECQT and SQT to the formal, general level by proving theorems stating that SQT admits various EC representations, whereas (at least without exceedingly large reservoirs of extra degrees of freedom) SQT is only a measure-zero subset of ECQT. Moreover, it is demonstrated that, due to the inherent interplay of unitarity and non-Markovianity, ECQT features diverse exotic dynamical phases and behaviors, which are fundamentally distinct from SQT (even its open-system derivatives), and may offer novel applications in their relevant phenomenological contexts. Moreover, we show that even slight modifications of SQT by some specific EC perturbations can yield nonnegligible observable effects. \\

Finally, upon presenting rationales at abstract fundamental levels, we put forward a formulation and characterization of \textit{general quantum behaviors}, which in turn are branched out to independent and context-based categories. In light of this characterization, which is based on the participatory-agentive ``it-from-(qu)bit'' paradigm, we elevate quantum mechanics to identify \textit{quantum theory} with the grand theory of (fundamental or emergent) general quantum behaviors, and show how it can branch out into an interacting system of structurally distinct, context-based quantum theories. This way, following our analysis of SQT vs. ECQT, we are led to the existence and phenomenological utility of a \textit{quantum interactome} (QI) which hosts and links interactively all the consistent, context-based quantum theories---SQT and beyond. This \textit{interactive landscape of context-based quantum theories} can be explored mainly by investigating mutual relations and structural interactions between its constituent theories. Trespassing the limiting boundaries of SQT by harnessing such \textit{enriched quantumness} within the formalism of ECQT offers a \textit{systematic, rigorous way} of understanding, connecting, and synergizing seemingly remote concepts, and paves the way toward novel theoretical descriptions and empirical predictions. As we see and contend, our findings signify that QI and its major component ECQT are promising and may lead to \textit{new physics and horizons}.
   
%%%%%%%%%%%%%%%%%%%%%%%%%%%%%%%%%%%%%%%%%%%%%%%%%%%%%%%%%%%%%%%%
\section{Quantum Simulation of a Larger Theory} 
\label{sec:SQTsim}

Classical simulation of quantum systems has been practiced extensively in physics. Methods such as mean-field theory \cite{MFT}, density functional theory \cite{DFT}, density matrix renormalization group \cite{DMRG}, and several other approximative techniques all aim to make the quantum problems more tractable on a classical computer by reducing the degrees of freedom and hence classical resources needed for simulation.  It is also important to know that when a quantum dynamics can be simulated on a classical computer with reasonable amount of classical resources \cite{Keshari}. \\

Despite all these attempts, the main idea of quantum simulation was developed due to the intractability of simulation of large quantum systems on classical computers \cite{Feynman-qsim}. It has been shown that rather than classical systems, ``quantum computers can be programmed to simulate any local quantum system'' \cite{Lloyd-qsim}. This has led to the advancement of numerous powerful techniques for quantum simulation \cite{qsim-techniques:PRXQ}.\\

Here, however, we aim to use the idea of quantum simulation for a different, novel purpose: simulating a theory \textit{larger} than SQT itself. It is known from classical simulation of quantum dynamics that the \textit{larger} dynamical space of quantum systems can in principle be simulated, under some conditions, on classical computers when \textit{sufficient} (and potentially large) amounts of classical resources are provided. Based on this, one may speculate that in general it is possible to simulate a larger theory, in the sense of dynamical spaces, by a smaller one when sufficient resources are available. \\

Inspired by this, the main question that we are addressing here is that, having sufficient \textit{quantum resources}, whether it is possible \textit{to use quantum simulation (based on SQT) to simulate a theory larger than SQT}. To make the question more specific and tractable, we look for quantum simulation of a specific dynamical space which is larger than what is allowed in SQT. SQT is a linear theory in that the dynamics of the state of a \textit{closed} quantum system is governed by a unitary operation which is generated by a Hamiltonian. More importantly, based on an implicit principle, the Hamiltonian must be state/history independent and can depend at most on time and some external control parameters. That is, a state-history-dependent Hamiltonian for a closed system is forbidden in SQT. \\

In the following, we show that it is possible to simulate a dynamics governed by a specific family of state-history-dependent Hamiltonians within the framework of SQT. This means that we can access to a larger quantum dynamical model using resources that has been allowed within the postulates of SQT. \\

For now, we restrict ourselves to a specific family of ``experience-centric'' (EC) unitary quantum dynamics proposed in Ref. \cite{ECQT-1} which is generated by the following polynomial state-history-dependent Hamiltonians:
\begin{align}
\mathbbmss{H}_{t}=\textstyle{\sum}_{j=1}^{M} \lambda_{\Gamma_j} \varrho_{t-a_{j_{1}}} \varrho_{t-a_{j_{2}}} \cdots \varrho_{t-a_{j_N}}+ \mathrm{h.c.}
\label{state-dep-H}
\end{align}
where $\lambda_{\Gamma_{j}}\in\mathbb{C}$ and here (throughout this paper) we have adopted the notation $\varrho_{t_{k}}$ to represent the state of the system at time $t_{k}$. Here each term (or monomial) is constructed from products of the states of the system at different points in the past times as well as present time. We confine ourselves to $N$ past (and present) points of the time with specific distances $a_i$ from the current time $t$, i.e., the states from which the Hamiltonian is constructed are chosen from the given set of past (or present) states at times $\{ t-a_{1}, t-a_{2}, \ldots, t-a_{N}\}$, where $0 \leqslant a_{N} \leqslant  a_{N-1} \cdots \leqslant a_{1}$. Each term in the Hamiltonian can be multiplication of the states at $N$ different points of time chosen from the history set.  \\

Here, we assume that $\lambda_{\Gamma}$ is a complex number such that 
\begin{equation}
\lambda_{\Gamma} = \mathrm{Tr}[\mathsf{U}\Gamma],
\label{lambda-gamma}
\end{equation}
where $\mathsf{U}$ is a unitary operator and $\Gamma$ represents a set of past or present states on which $\lambda_{\Gamma}$ depends. We note, however, that this is only a particular dependence of the couplings on the history, but it will be necessary to adopt an existing quantum interferometric circuit \cite{Ekert-etal} for our protocol. For general dependence see Sec. \ref{sec:ECQT}. \\

It is evident that conventional Hamiltonian simulation techniques \cite{qHsim, Poulin-qsim-tH} do not work in this case where the terms in the Hamiltonian are state or history dependent (hence unknown \textit{a priori}). Conventional techniques can be used only if we do quantum tomography on the system state at each time, which is inefficient in this case. \\

To propose a simulation technique which does not require state tomography, we use the recently developed technique of density matrix exponentiation \cite{Lloyd-DME}, which has also been realized experimentally \cite{DME-experimental}. We start with an immediate example, where the memory distances is two such that the Hamiltonian has only two monomials, 
\begin{equation}
\mathbbmss{H}=\lambda_{\Gamma} \varrho \varrho' + \lambda_{\Gamma}^{*} \varrho' \varrho.
\label{q.sim-1}
\end{equation}

Before going through the details of the simulation protocol, it should be noted that the dynamics is not simulatable for the duration $0 \leqslant t \leqslant a_{1}$ due to the lack of information, i.e., the prehistory evolution of the system is needed to simulate this time interval. Hence in this interval the dynamics should be given to enable simulation in the next times. This is a feature of ``delay differential equations'' in which rather than a single initial point we need a \textit{time interval} as the initial condition \cite{delay-DE}.

%%%%%%%%%%%%%%%%%%%%%%%%%%%%%%%%%%%%%%%%%%%%%%%%%%%%%%%%%%%%%%%%
\begin{figure}[tp]
\includegraphics[scale=0.34]{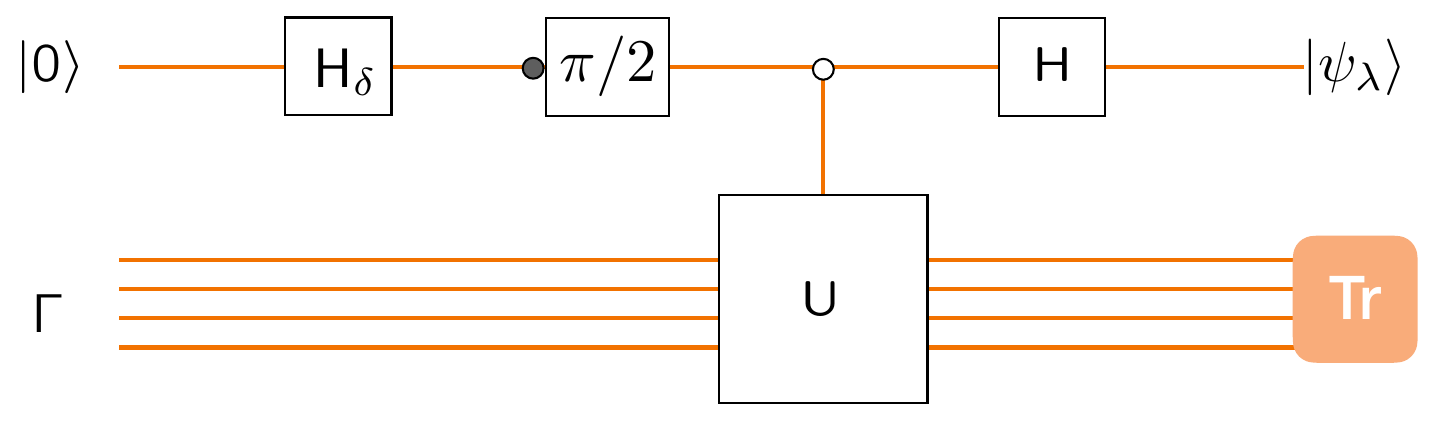}
\caption{Interferometric circuit for preparation of the $|\psi_{\lambda}\rangle$ state [Eq. \eqref{psilambda}]: By applying an unbalanced-Hadamard gate $\mathsf{H}_{\delta}$, controlled-\textsc{phase} gate with phase shift $\pi/2$, a controlled-$\mathsf{U}$ gate which is applied when the control system is in the state $|1\rangle$, and a final Hadamard gate $\mathsf{H}$ on the initial state $|0\rangle \langle 0| \otimes \Gamma$, we can prepare the state $|\psi_{\lambda}\rangle$ with $\lambda=\mathrm{Tr}[\mathsf{U} \Gamma]$. In this section, ``$\mathrm{Tr}$'' denotes tracing operation.}
\label{fig:int}
\end{figure}
%%%%%%%%%%%%%%%%%%%%%%%%%%%%%%%%%%%%%%%%%%%%%%%%%%%%%%%%%%%%%%%%

%%%%%%%%%%%%%%%%%%%%%%%%%%%%%%%%%%%%%%%%%%%%%%%%%%%%%%%%%%%%%%%%
\subsection{Quantum Simulation Protocol}

Our simulation protocol consists of different elements and is comprised of techniques for (i) estimation of nonlinear scalar functions of states using the proposed interferometric quantum circuit in Refs. \cite{Ekert-etal, Ericsson-thesis, Oi, OiAb, Pati-etal}, (ii) implementation of density matrix exponentiation by applying controlled-\textsc{swap} gates \cite{Marvian-Lloyd}, and for (iii) Trotter-Suzuki expansion. We emphasize that our protocol achieves significantly higher than simply simulating nonlinear scalar functions of quantum states and exponentiation density matrices; we can in principle simulate \textit{nonlinear} and \textit{non-Hermitian} quantum \textit{operations} beyond Refs. \cite{Ekert-etal, Lloyd-DME}. \\

\begin{figure}[tp]
\includegraphics[scale=0.3]{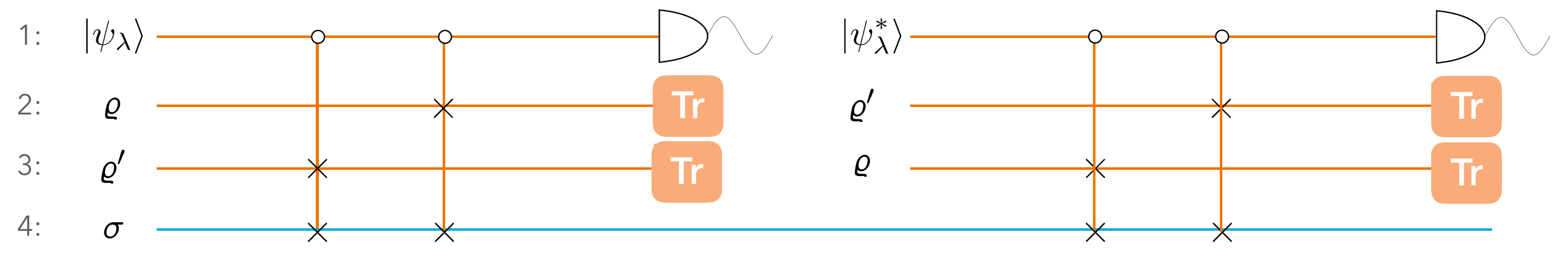}
\caption{Interferometry circuit for quantum simulation of a Hamiltonian with state-history-dependent couplings as in Eq. \eqref{q.sim-1}. Controlled-\textsc{swap} gates act if their control qubit is in $|0\rangle$. Measurements on the control qubits are on the $|0\rangle$ basis.}
\label{fig:int-new}
\end{figure}

We assume that the current state of the simulator is $\sigma$ and we aim to simulate the dynamics $e^{-i \delta\, \mathbbmss{H}}$ on it. Consider the following steps:
\begin{itemize}
\item (i) We first prepare a quantum state 
\begin{equation}
|\psi_{\lambda}\rangle= |-\rangle -i \delta \lambda_{\Gamma}\, |+\rangle,
\label{psilambda}
\end{equation}
where $\sigma^{1}|\pm\rangle = \pm |\pm\rangle$, with $\sigma_{x}$ being the first Pauli operator for spin-$1/2$ systems and $|\pm\rangle =(|0\rangle \pm |1\rangle)/\sqrt{2}$ in the basis of the eigenvectors of the third Pauli operator $\sigma^{3}$. Here $\delta$ is a sufficiently small real number ($\delta\ll1/|\lambda_{\Gamma}|$). To perform the preparation, we note that the value of $\lambda$ is not given. We are only given systems whose total state is what we call $\Gamma$, without even knowing (or needing to know) the state $\Gamma$ itself. We also know the relation between $\lambda$ and the unknown state $\Gamma$, which is given below Eq. \eqref{q.sim-1}. To prepare the state $|\psi_{\lambda}\rangle$ we use a modified ``interferometric'' quantum circuit, which has already been proposed to estimate nonlinear scalar functions of a state $\Gamma$, which obviated the need for tomography of that state \cite{Ekert-etal, Ericsson-thesis}---Fig. \ref{fig:int}.  

\item (ii) Inspired by the density matrix exponentiation technique \cite{Lloyd-DME, Marvian-Lloyd}, we apply suitable controlled-\textsc{swap} gates (i.e., Fredkin gates) on a combined system comprised of a system which is already prepared in $|\psi_{\lambda}\rangle$, the simulator, and the two quantum systems with (unknown) states $\varrho$ and $\varrho'$, hence the total system is in the state $|\psi_{\lambda}\rangle \langle \psi_{\lambda}| \otimes \varrho \otimes \varrho'\otimes \sigma$. The suitable controlled-\textsc{swap} gates use the prepared system in $|\psi_{\lambda}\rangle$ as the control system, and one of the systems in $\varrho$ or $\varrho'$ and the simulator as the target systems. The gates are applied when the control system is in the state $|+\rangle$. In the next step, we perform a selective measurement on the control system in the $\{|0\rangle,|1 \rangle$ basis and choose the output when it is in the state $|0 \rangle$. Finally, we trace out over all the ancillary systems, i.e., all the systems other than the simulator. Up to this step, the (unnormalized) state of the simulator becomes $\widetilde{\sigma}= e^{-i \delta \lambda_{\Gamma} \varrho \varrho'} \sigma \, e^{i \delta \lambda^{\ast}_{\Gamma} \varrho' \varrho}+ O(\delta^{2})$. Interestingly, this is equivalent to simulating a \textit{state-dependent} non-Hermitian Hamiltonian $\lambda_{\Gamma} \varrho \varrho'$.

\item (iii) To complete the simulation of the dynamics with the evolution $e^{i \delta \lambda^{\ast}_{\Gamma} \varrho' \varrho}\, \widetilde{\sigma}\, e^{-i \delta \lambda_{\Gamma} \varrho \varrho'}$, we simply need to repeat steps (i) and (ii) by preparing a new control system in 
\begin{equation}
|\psi_{\lambda}^{\ast}\rangle = |-\rangle - i \delta \lambda_{\Gamma}^{\ast}\, |+\rangle
\end{equation}
(rather than $|\psi_{\lambda}\rangle$) and exchanging $\varrho$ and $\varrho'$, such that the state of the total system becomes $|\psi_{\lambda}^{\ast}\rangle \langle \psi_{\lambda}^{\ast}| \otimes \varrho' \otimes \varrho \otimes \widetilde{\sigma}$.

Parts (ii) and (iii) can be performed through the quantum circuit of Fig. \ref{fig:int-new}.

\item (iv) Using the Trotter-Suzuki expansion to generalize simulation to the general case of Hamiltonians of Eq. \eqref{state-dep-H} by repeating steps (i) through (iii) for each term in this state-history-dependent Hamiltonian. 

\end{itemize}
%%%%%%%%%%%%%%%%%%%%%%%%%%%%%%%%%%%%%%%%%%%%%%%%%%%%%%%%%%%%%%%%
%%%%%%%%%%%%%%%%%%%%%%%%%%%%%%%%%%%%%%%%%%%%%%%%%%%%%%%%%%%%%%%%

We now explain in more detail our simulation protocol. Inspired by Ref. \cite{Ekert-etal}, we propose the circuit in Fig. \ref{fig:int}, where $\mathsf{H}$ is the Hadamard gate (or beam splitter in a quantum-optical sense) and $\mathsf{H}_{\delta}$ is an unbalanced Hadamard gate (unbalanced beam splitter in a quantum-optical sense) defined as
\begin{equation}
\mathsf{H}_\delta=\left(\begin{array}{cc}
\sin \delta & \cos \delta \\ 
\cos \delta & - \sin \delta \\ 
\end{array}\right) =\left(\begin{array}{cc}
\delta & 1 \\ 
1 & - \delta \\ 
\end{array}\right) + O(\delta^{2}). 
\end{equation}  
Next by applying the controlled-$\mathsf{U}$ gate (which acts when the control qubit is in the state $|+\rangle$) followed by the Hadamard gate, the (unnormalized) output state of the total system is 
\begin{align}
\tau \, = &\,  \Big(\delta^{2} |+\rangle\langle +|\otimes \mathsf{U} \Gamma \mathsf{U}^{\dag}+(1-\delta^{2}) |-\rangle\langle -|\otimes \Gamma +\delta  e^{-i\phi}|+\rangle\langle -|\otimes \mathsf{U} \Gamma +\delta e^{i\phi}|-\rangle\langle +|\otimes \Gamma \mathsf{U}^{\dag}\Big)+ O(\delta^{2}) \nonumber\\
\, = & \,\Big( |-\rangle\langle -|\otimes \Gamma +\delta  e^{-i\phi}|+\rangle\langle -|\otimes \mathsf{U} \Gamma +\delta e^{i\phi}|-\rangle\langle +|\otimes \Gamma \mathsf{U}^{\dag}\Big) + O(\delta^{2}).
\end{align}

Choosing $\phi=\pi/2$ and tracing out over the ancillary qubits, we have 
\begin{align}
|\psi_{\lambda}\rangle \langle \psi_{\lambda}| & = \big( |-\rangle\langle -| -i \delta \lambda_{\Gamma}|+\rangle\langle -| +i \delta \lambda^{\ast}_{\Gamma}|-\rangle\langle +|\big) + O(\delta^{2}) \nonumber\\
& = \big(|-\rangle -i \delta \lambda_{\Gamma} |+\rangle\big) \big(\langle -| +i \delta \lambda^{\ast}_{\Gamma} \langle +|\big) + O(\delta^{2}).
\end{align}

Now by putting the above state, within which the first qubit is used as the ``control'' qubit (``$\mathrm{c}$''), together with three other systems with states $\varrho$, $\varrho'$, and $\sigma$, which will be used as the target systems, we apply controlled-\textsc{swap} gates $\mathsf{S}_{\mathrm{c}}$ on the target systems if the control qubit is in the state $|0\rangle$. After applying the two gates $\mathsf{S}_{\mathrm{c},24} \mathsf{S}_{\mathrm{c},34}$ we obtain
\begin{align}
\mathsf{S}_{\mathrm{c},24} \mathsf{S}_{\mathrm{c},34} \big ( |\psi_{\lambda}\rangle &\langle \psi_{\lambda}| \otimes \varrho \otimes \varrho' \otimes \sigma \big) \mathsf{S}_{\mathrm{c},34} \mathsf{S}_{\mathrm{c},24}= \nonumber\\
 &\, \Big(|-\rangle\langle -| \otimes \varrho \otimes \varrho' \otimes \sigma 
  -i \delta \lambda_{\Gamma}\, |+\rangle\langle -| \otimes \big(\mathsf{S}_{24} \mathsf{S}_{34} \varrho \otimes \varrho' \otimes \sigma \big) 
 + i \delta \lambda_{\Gamma}^{\ast}\, |-\rangle\langle +| \otimes \big(\varrho \otimes \varrho' \otimes \sigma \mathsf{S}_{34} \mathsf{S}_{24}\big) \Big).
\end{align}
By performing a selective $|0 \rangle \langle 0|$ measurement on the control qubit, the (unnormalized) state of the total system reduces to
\begin{align}
|0\rangle\langle 0|\otimes \Big(&\varrho \otimes \varrho' \otimes \sigma -i \delta  \lambda_{\Gamma} \otimes 
\big(\mathsf{S}_{24} \mathsf{S}_{34} \varrho \otimes \varrho' \otimes \sigma\big) + i\delta \lambda_{\Gamma}^{\ast} \big(\varrho \otimes \varrho' \otimes \sigma \mathsf{S}_{34} \mathsf{S}_{24} \big)\Big).
\end{align}
After tracing over the other systems and using the \textsc{swap} identities
\begin{equation}
\begin{split}
&\mathrm{Tr}[\mathsf{S} (A\otimes B)] = \mathrm{Tr}[A B], \\
&\mathrm{Tr}_{1}[\mathsf{S} (A\otimes B)] = A B, \\
&\mathrm{Tr}_{1}[(A\otimes B)\mathsf{S}] = B A, 
\end{split}
\label{swapidentity}
\end{equation}
the (unnormalized) state of the simulator now becomes  
\begin{align}
\widetilde{\sigma} & = \big(\sigma -i \delta \lambda_{\Gamma} \varrho \varrho' \sigma + i \delta \lambda^{*}_{\Gamma} \sigma \varrho' \varrho \big) + O(\delta^{2}) \nonumber\\
& = \, e^{-i \delta \lambda_{\Gamma} \varrho \varrho'} \sigma \, e^{i \delta \lambda^{\ast}_{\Gamma} \varrho' \varrho}+ O(\delta^{2}). 
\end{align}
Here $\mathrm{Tr}_{1}$ (and similarly $\mathrm{Tr}_{2}$) denotes partial tracing over system $1$ ($2$). \\

To simulate the rest of the dynamics, we repeat the above steps by modifying the circuit of Fig. \ref{fig:int} as $\mathsf{U} \to \mathsf{U}^{\dag}$. These changes lead to the preparation of the state $|\psi^{\ast}_{\lambda}\rangle$. We also replace $\varrho \leftrightarrow \varrho'$. Putting these three ancillary systems besides the simulator which is now in the (unnormalized) state $\widetilde{\sigma}$, the whole system state is $|\psi^{\ast}_{\lambda}\rangle \langle \psi^{\ast}_{\lambda}| \otimes \Gamma\otimes \varrho' \otimes \varrho \otimes \widetilde{\sigma}$ (note the new order of $\varrho$ and $\varrho'$ compared to the previous step). We input this state to a quantum circuit similar to the one in the first step and do the final selective $|0\rangle\langle 0|$ measurement. Following a similar reasoning, the (normalized) final state of the simulator becomes  
\begin{align}
\mathrm{Tr}_{123} \big[\mathsf{S}_{\mathrm{c},24} \mathsf{S}_{\mathrm{c},34}\,(|\psi^{\ast}_{\lambda}\rangle \langle \psi^{\ast}_{\lambda}| \otimes \varrho' \otimes \varrho \otimes \widetilde{\sigma}) \, \mathsf{S}_{\mathrm{c},34} \mathsf{S}_{\mathrm{c},24}\big] & = \big(\widetilde{\sigma} - i\delta \lambda_{\Gamma}^{*}  \varrho' \varrho \,\widetilde{\sigma} +i \delta  \lambda_{\Gamma}\,\widetilde{\sigma} \varrho \varrho' \big) + O(\delta^{2})  \nonumber\\ 
 & = \big(\sigma -i \delta  \lambda_{\Gamma}\, \varrho \varrho' \sigma +i \delta \lambda_{\Gamma}^{*}\, \sigma \varrho' \varrho
-i \delta \lambda_{\Gamma}^{*}\, \varrho' \varrho\, \sigma +i \delta \lambda_{\Gamma}\, \sigma \varrho \varrho'  \big) + O(\delta^{2}) \nonumber\\
& = \sigma -i \delta [\lambda_{\Gamma}\, \varrho\,\varrho' +\lambda_{\Gamma}^{*}\, \varrho' \varrho, \sigma]+ O(\delta^{2}) 
\nonumber\\
& = e^{-i\delta(\lambda_{\Gamma}\, \varrho\varrho' +\lambda_{\Gamma}^{*}\, \varrho' \varrho)} \sigma e^{-i\delta(\lambda_{\Gamma}\, \varrho \varrho' +\lambda_{\Gamma}^{*}\, \varrho' \varrho)} + O(\delta^{2}).
\end{align}

The overall outcome of this protocol is to simulate the unitary dynamics generated by the state-history-dependent Hamiltonian (\ref{q.sim-1}) on the simulator system. As indicated before, this dynamics was in principle forbidden in SQT for a \textit{closed} quantum system, whereas our construction shows that it can in principle be simulated by SQT \textit{on a larger system}. In fact, by using multiple copies of a state, we can unleash the forbidden operations. 

%%%%%%%%%%%%%%%%%%%%%%%%%%%%%%%%%%%%%%%%%%%%%%%%%%%%%%%%%%%%
\subsection{Resource Analysis for Simulation}
\label{subsec:resourceanalysis}

In this part we analyze the resource complexity of our quantum simulation protocol. Three remarks are in order: \\

(i) If the state-dependent Hamiltonian of interest (\ref{state-dep-H}) has $M$ monomials, the above quantum circuit will have $M$ interferometry circuits, one per each term in accordance to step (iv) of the simulation protocol. \\

(ii) Assuming that $\lambda_{\Gamma}$'s depend on at most $N$ states from the state history (e.g., $\{\varrho_{t},\varrho_{t-a},\ldots,\varrho_{t-(N-1)a}\}$), each $|\psi_{\lambda}\rangle$ (and $|\psi_{\lambda}^{\ast}\rangle$) preparation part in the circuit of Fig. \ref{fig:int} requires $N$ past and present states. \\

(iii) The quantum circuit for simulating dynamics generated by a term of $l \leqslant L$ factors, e.g., $\lambda_{\Gamma}\, \varrho^{1}\ldots \varrho^{l}$, will be basically similar to the half of the circuit in Fig. \ref{fig:int-new}, but with $l+2$ lines (labeled ``$0,$'' \ldots``$l+1$'') and $l$ controlled-\textsc{swap} gates. \\

Thus, the number of the total system copies needed to implement each circuit includes $N+1$ systems for simulating $\lambda_{\Gamma}$, and at most $L+1$ copies of the system for the interaction part ($\varrho \varrho' \varrho'' \ldots \varrho'''$). That is, 
\begin{equation}
\# \, \mathrm{systems} = O\big(2M(N+L+1)\big).
\end{equation}

Thus, this means that at a given time $a_1$ we need $O\big(2M(N+L+1)\big)$ copies of the system state at present or past times. To simulate the dynamics for the interval $[a_1, t+a_1]$ we need to break down the dynamics in $m$ steps so that we can have (setting $\hbar \equiv 1$ throughout this paper)
\begin{align}
\mathbbmss{U}_{a_{1} \to a_{1}+t} =&\textbf{\textsf{T}} e^{-i\int_{a_{1}}^{a_{1}+t} ds\, \mathbbmss{H}_{s}}\\
=& e^{-i \delta \, \mathbbmss{H}_{t_{m}}} e^{-i\delta \, \mathbbmss{H}_{t_{m-1}}} \ldots e^{-i\delta \, \mathbbmss{H}_{t_{1}}} e^{-i\delta \, \mathbbmss{H}_{a_{1}}} + O(m\delta^{2}),\nonumber
\end{align}
where $\textbf{\textsf{T}}$ denotes the time-ordering operation and $t_{k}=a_{1}+k\delta $, with $k\in\{0,1,\ldots,m\}$ and $\delta=t/m$. Each term in this expansion is in turn expanded to $M$ separate terms corresponding to each monomial in $\mathbbmss{H}_{t_{k}}$ as
\begin{align}
e^{-i\delta \, \mathbbmss{H}_{t_{k}}} =e^{-i \delta \sum_{\mathrm{monomials}} (\lambda_{\Gamma} \varrho \varrho'\ldots \varrho''' + \mathrm{h.c.})} + O(M \delta^{2}).
\end{align}
Thus, the total error for simulation of $\mathbbmss{U}_{a_{1} \to a_{1}+t}$ by implementing dynamics of each monomial becomes $O(m M \delta^{2})=O(M t^{2}/m)$. \\

To simulate the dynamics for a period $[a_{1},a_{1}+t]$ within error $\epsilon$, we need 
\begin{equation}
m \geqslant O(M t^{2}/\epsilon)
\end{equation}
repetitions. However, it should be noted that to go one step forward in time in the simulation, we need to reconstruct all the elements (i.e., the past or the present system states) which are needed in $\mathbbmss{H}_{t_1}$. This is because the $O\big(2M(N+L)\big)$ copies of the state that we had at hand have already been used in the process of constructing $\mathbbmss{H}_{a_1}$. For constructing each of these elements we need $\alpha\equiv O\big(2M(N+L)\big)$ copies of the system. Hence the number of needed copies grows exponentially as $k(\epsilon,t)\equiv O(\alpha^{m(\epsilon,t)})$. This in turn can translate into the statement that to simulate EC dynamics of a quantum system of $O(d)$ Hilbert-space dimension in the time interval $[a_{1},a_{1}+t]$ within error $\epsilon$, our SQT simulation protocol requires a quantum system of $O(d^{k(\epsilon,t)})$ dimension. \\

\textit{Remark.---}Note that we have not proved optimality of our quantum simulation protocol. Thus, devising more efficient algorithms still remains a possibility.\\

It is important to note that simulating a generic EC Hamiltonian requires computational resources that grow super exponentially with simulation time. This exponential scaling is unavoidable, as it stems from the nonlinear nature of EC theory, while we are constrained to simulate it using standard linear quantum theory. Otherwise, if this were not the case---if we could efficiently simulate such systems using linear tools---it would be concluded that EC theory is essentially equivalent to quantum theory, and thus it could not be considered as a larger emergent theory.\\

As we elaborate in the forthcoming sections, EC Hamiltonians do not belong to the same complexity class as standard quantum dynamics. In fact, we shall later substantiate that ECQT is a closed theory by itself which can emerge out of sufficiently complex systems that can be microscopically quantum mechanical or even classical. Accordingly, the (super) exponential expense of simulation with quantum mechanics is a clear signature of genuine emergence characterized in the sense of complex system theories. Nevertheless, our simulation algorithm provides a systematic method for emulating the behavior of a class of such EC systems---provided sufficient resources are available. As a result, we have offered a principled approach to designing and testing EC systems within existing quantum simulation platforms.
 
%%%%%%%%%%%%%%%%%%%%%%%%%%%%%%%%%%%%%%%%%%%%%%%%%%%%%%%%%%%%
\subsection{Simulation Protocol for a General State-History-Dependent Hamiltonian: Tomography Needed}
 
A significant advantage of the quantum simulation protocol of the previous subsection is that it does not require any quantum state tomography. As stated earlier, this feature depends closely on the specific form of the coupling we have assumed---Eq. (\ref{lambda-gamma}). There is no guarantee that this protocol may work for arbitrary $\lambda$'s whose dependence on the state history is not in the simple, linear form of Eq. (\ref{lambda-gamma}). This, in turn, puts a strict condition on simulatability of ECQT with SQT. We shall return to this important point in Theorem \ref{thm:C} of Sec. \ref{sec:thms}. \\

In the following we show that how we can in principle simulate a general state-history-dependent Hamiltonian by using \textit{state tomography} and density-matrix exponentiation technique. We should, however, remark that the resources needed to simulate such general dynamics this way can grow significantly in a way that it may basically render this simulation intractable and inefficient. \\

Here we show that for a given Hamiltonian of the form \eqref{state-dep-H} with \textit{general} couplings $\{\lambda\}$, with arbitrary dependence on past and present states, one can in principle simulate the related dynamics within the SQT framework. Our proof works for even more general cases of any imaginable state-history dependence, but with the caveat we hinted above. \\
 
We first note that it is always possible to rewrite any given Hermitian operator $H$ as 
\begin{align}
H=H^{(+)}-H^{(-)},
\end{align}
where $ H^{(+)}=P^{(+)} H P^{(+)}$ and $ H^{(-)}=-P^{(-)} H P^{(-)}$ are two positive operators and $P^{(+)}$ ($P^{(-)}$) is the projector onto the subspace spanned by the eigenvectors of $H$ with positive (negative) eigenvalues. Note that $P^{(+)}P^{(-)}=0$ and hence $[P^{(+)},P^{(-)}]=0$ and $[ H^{(+)}, H^{(-)}]=0$. Commutation of $ H^{(+)}$ and $ H^{(-)}$ implies that if we can exponentiate $ H^{(+)}$ and $ H^{(-)}$ separately, it would be straightforward to simulate the unitary dynamics $e^{-i\Delta t\,H}$ from the concatenation of the subsequent unitary evolutions $e^{-i\Delta t\, H^{(\pm)}}$. \\

Let us first briefly recall the density matrix exponentiation algorithm \cite{Lloyd-DME}. This technique hinges on the relation
\begin{align}
\mathrm{Tr}_{1}[\mathsf{U}^{(\textsc{swap})}_{\delta} (\varrho \otimes \sigma) \mathsf{U}^{(\textsc{swap})\,\dag}_{\delta}]= \sigma - i \delta [\varrho, \sigma] + O(\delta^{2}) = e^{-i\delta \varrho} \sigma e^{i\delta \varrho} + O(\delta^{2}),
\end{align} 
where we have defined the unitary gate $\mathsf{U}^{(\textsc{swap})}_{\delta}=e^{-i \mathsf{S}\delta}$. Now after $n = O(t^{2}/\epsilon)$ times applications with $n$ copies of $\varrho$ allows simulation of $e^{-it \varrho} \sigma e^{it \varrho} $ to accuracy $\epsilon$, with $t=n\delta$. This way we simulate the dynamics $U=e^{-it \varrho}$ on the state $\sigma$. \\

Thus, in a similar vein, to simulate $U^{(+)}$ and $U^{(-)}$ by the density matrix exponentiation technique, we first need to construct valid density matrices out of $ H^{(\pm)}$, e.g., as $\varrho^{(\pm)}= H^{(\pm)}/\mathrm{Tr}[H^{(+)}]$; $H=\mathrm{Tr}[ H^{(+)}]\varrho^{(+)} - \mathrm{Tr}[ H^{(-)}]\varrho^{(-)}$. By setting $n^{(\pm)}$ such that 
\begin{equation}
\Delta t\, \mathrm{Tr}[ H^{(\pm)}]= n^{(\pm)} \delta^{(\pm)},
\label{cond:n_pm}
\end{equation}
we can see that using $n^{(+)}$ copies of $\varrho^{(+)}$
 \begin{align}
\mathrm{Tr}_{1}\big[\mathsf{U}^{(\textsc{swap})}_{\delta^{(+)}} \big(\varrho^{(+)} \otimes\cdots \mathrm{Tr}_{1}[\mathsf{U}^{(\textsc{swap})}_{\delta^{(+)}} \big(\varrho^{(+)} \otimes \varrho_{t} \big) \mathsf{U}^{(\textsc{swap})\,\dag}_{\delta^{(+)}} ]\ldots\big) \mathsf{U}^{(\textsc{swap})\,\dag}_{\delta^{(+)}} \big] \approx e^{-i \Delta t\, \mathrm{Tr}[ H^{(+)}] \varrho^{(+)}}~\varrho_{t} ~ e^{i \Delta t\, \mathrm{Tr}[ H^{(+)}] \varrho^{(+)}} =: \boldsymbol{\sigma}_{t}.
\end{align} 

Now by replacing $\mathsf{U}^{(\textsc{swap})}_{\delta^{(+)}}$ with $\mathsf{U}^{(-\textsc{swap})}_{\delta}= \mathsf{U}^{(\textsc{swap})\dag}_{\delta}$ in the exponentiation technique and applying the quantum circuit on the outcome of the above process we obtain 
\begin{align}
\mathrm{Tr}_{1}\big[\mathsf{U}^{(-\textsc{swap})}_{\delta^{(-)}} \big(  \varrho^{(-)} \otimes \ldots \mathrm{Tr}_{1}[\mathsf{U}^{(-\textsc{swap})}_{\delta^{(-)}}  \big(\varrho^{(-)} \otimes \boldsymbol{\sigma}_{t} \big) \mathsf{U}^{(-\textsc{swap})\,\dag}_{\delta^{(-)}} ]\ldots\big) \mathsf{U}^{(-\textsc{swap})\,\dag}_{\delta^{(-)}} \big] & \approx e^{i \Delta t\,\mathrm{Tr}[ H^{(-)}] \varrho^{(-)}} \boldsymbol{\sigma}_{t} \, e^{-i \Delta t\, \mathrm{Tr}[ H^{(-)}] \varrho^{(-)}} \nonumber \\
& =e^{-i \Delta t H} \,\varrho_{t}  \,e^{i \Delta t H} + O\big(\textstyle{\sum_{p=\pm}} n^{(p)} (\delta^{(p)})^{2}\big).
\end{align} 

To achieve an error $\epsilon$ in simulating the evolution $e^{-i \Delta t H}$ at time $t$, we need $n^{(+)} = O(\Delta t^{2} \, \mathrm{Tr}[ H^{(+)}]^{2}/\epsilon)$ copies of $\varrho^{(+)}$ and $n^{(-)} = O(\Delta t^{2} \, \mathrm{Tr}[ H^{(-)}]^{2}/\epsilon)$ copies of $\varrho^{(-)}$. \\

Three remarks are in order here. \\

(i) Note that $\Delta t$ is typically small. The above protocol has achieved the evolution $\varrho_{t}\to e^{-i \Delta t H_{t}} \,\varrho_{t}  \,e^{i \Delta t H_{t}}$. In fact, $H_{t}$ is basically time dependent through its dependence on the state history. Hence simulation of the evolution for longer times shall need the time-ordered exponential, 
\begin{align}
U_{0\to t} =\textbf{\textsf{T}} e^{-i\int_{0}^{t} ds\, H_{s}}= \lim_{m\to \infty} e^{-i\Delta t\, H_{t_{m}}} e^{-i\Delta t\, H_{t_{m-1}}} \ldots e^{-i\Delta t\, H_{t_{1}}} e^{-i\Delta t\, H_{0}},
\end{align}
where $t_{k}=k\,\Delta t$, with $k\in\{0,1,\ldots,m\}$ and $\Delta t=t/m$. This implies that a finite-time simulation of the true dynamics by a concatenation of $\ldots e^{-i\Delta t H_{t_{k+1}}} e^{-i\Delta t H_{t_k}}\ldots$ also introduces further errors, which should be taken into account properly. \\

(ii) Since $\Delta t$ is small, we also observe from Eq. (\ref{cond:n_pm}) that, for given $\delta^{(\pm)}$, $n^{(\pm)}$ do not need to be large. This is a favorable feature of this simulation that at any instant $t$ it does not require large number of copies of $\varrho^{(\pm)}$ to simulate $e^{-i \Delta t H_{t}} \,\varrho_{t}  \,e^{i \Delta t H_{t}}$. \\

(iii) The simulation protocol we described above implies that given \textit{sufficient} resources, one can in principle simulate general ECQT dynamics with the SQT framework. However, we should note that in the above protocol not only at any instant $t$ we need $n^{(+)}$ and $n^{(-)}$ copies of $\varrho^{(+)}$ and $\varrho^{(+)}$, respectively, but also we need to perform tomography on $\varrho_{t}$ at each time step. In fact, to prepare the relevant systems in $\varrho^{(\pm)}$, $\Omega(d^{2})$ copies of the state at each time is needed for state tomography, where $d$ is the dimension of the Hilbert space of the system. This implies that this protocol, indeed, can require a significant overhead resource for simulation, which hence may render the very simulation infeasible or even impossible. This point will be important later in Sec. \ref{sec:sqt-vs-ecqt} for our discussion about comparison of ECQT and SQT. 

%%%%%%%%%%%%%%%%%%%%%%%%%%%%%%%%%%%%%%%%%%%%%%%%%%%%%%%%%%%%%%%%
\section{The Rosetta Stone of Experience-Centric Quantum Theory}
\label{sec:ECQT}

We have seen in the previous section that, within the context of SQT, one can in principle simulate quantum evolutions which are vastly larger than linear time-local evolutions of SQT. In this section, following Ref. \cite{ECQT-1}, we lay out that ``larger'' theory on its own and explain a number of its distinctive properties compared to SQT. In particular, we show that this theory has a significantly richer and genuinely larger structure in the theory land and includes SQT. For detailed discussions on the conceptual foundations and motivations of ECQT as a part of a larger framework, referred to as the \textit{quantum interactome} (QI) $\mathpzc{M}_{Q}$, see Sec. \ref{sec:toward}. In the following, we provide essential defining elements and formalism of ECQT (at points with somewhat different notations and terminology or subtle refinements compared to Ref. \cite{ECQT-1}). Hereafter we make the convention that lower indexes indicate time, and when there is any risk of ambiguity about upper indexes and powers, we use parentheses to distinguish them. Exceptions will be stated clearly. 

%%%%%%%%%%%%%%%%%%%%%%%%%%%%%%%%%%%%%%%%%%%%%%%%%%%%%%%%%%%%%%%%
\begin{element}
(Quantum) State history or the experience of a closed quantum system
\label{def:1}
\end{element}

Assume that $\mathcal{S}$ is a \textit{closed} system with Hilbert space $\mathpzc{H}^{(\mathcal{S})}$ and its \textit{states} are represented by positive semidefinite density operators $\varrho^{(\mathcal{S})}_{t'}$, with time represented as subscript, which are known in a (continuous or discrete) interval $t'\in[t_{0},t]$, with $t_{0}$ and $t$ denoting the initial and present times, respectively. The time-ordered (countable or uncountable) collection of all states in the identified past-to-present interval is defined as the (quantum) \textit{state history} of $\mathcal{S}$,
\begin{equation}
\label{l1}
\mathpzc{P}_{[t_{0} , t]}^{ (\mathcal{S})} \equiv \{ \varrho^{(\mathcal{S})}_{t'},\, \forall t' \in [t_{0}, t]\} \subset \lim_{t \to \infty} \mathpzc{P}_{[t_{0} , t]}^{ (\mathcal{S})} \equiv \mathpzc{P}_{[t_{0} , \infty)}^{ (\mathcal{S})}.  
\end{equation}
The evolution of the states in a closed system is assumed to be \textit{unitary}. This implies that the trace $\mathrm{Tr}[\varrho_{t'}^{(\mathcal{S})}]$ and purity $\mathrm{Tr}[(\varrho^{(\mathcal{S})}_{t'})^{2}]$ are preserved. One can, hence, have an entire \textit{pure-state} history with $\varrho^{(\mathcal{S})}_{t'} = |\Psi^{(\mathcal{S})}_{t'} \rangle \langle \Psi^{(\mathcal{S})}_{t'} |$, upon choosing an initial pure state. In any such initialization, the history becomes equivalent to a set of Hilbert space rays, which we call a \textit{pure-state history},
\begin{equation}
\label{lll}
\mathpzc{P}_{[t_{0} , t]}^{ (\mathcal{S})} \cong \mathpzc{P}_{[t_{0} , t]}^{ (\mathcal{S} | \mathrm{purity})} \equiv \{ | \Psi^{(\mathcal{S})}_{t'} \rangle ,\, \forall t' \in [t_{0}, t]\} .  
\end{equation}
\\

%%%%%%%%%%%%%%%%%%%%%%%%%%%%%%%%%%%%%%%%%%%%%%%%%%%%%%%%%%%%%%%%
\begin{element}
Substate histories (experiences) and ``universal'' histories (experiences) of a closed quantum system
\label{def:2}
\end{element}

We now consider a partitioning of the closed system $\mathcal{S}$ into $N_{t}^{(\mathcal{S})}$ \textit{subsystems} $\mathcal{S}^{(i_{t})}$, where $\mathpzc{H}^{(\mathcal{S})} = \otimes_{j_{t}=1}^{N_{t}^{(\mathcal{S})}}\, \mathpzc{H}^{(j_{t})}  = \mathpzc{H}^{(i_{t})} \otimes \mathpzc{H}^{(\bar{i}_{t})}$, at every present moment $t$. We call this a degree-$N_{t}^{(\mathcal{S})}$ \textit{resolution} of $\mathcal{S}$. It is clear that the familiar case of time-invariant system partitionings are included as special cases, but dynamical Hilbert-space factorizations are allowed and can be even useful. The (unitary) time evolution of $\varrho_{t}^{(\mathcal{S})}$ projects itself into the (generically nonunitary) time evolutions of its $N_{t}^{(\mathcal{S})}$ \textit{substates} $\varrho^{(i_{t})}_{t}$ (in the Hilbert space $\mathpzc{H}^{(i_{t})}$) which are obtainable from a consistent and relevant dimensionality-reduction procedure in correspondence, for concreteness the standard partial tracing recipe $\varrho^{(i_{t'})}_{t'} \equiv \mathrm{Tr}_{\bar{i}_{t'}} [\varrho^{(\mathcal{S})}_{t'}]$. For each subsystem $\mathcal{S}^{(i_{t})}$, a substate history (experience) of $\mathcal{S}$ is associated, which is the time-ordered set of all instantaneous states of the subsystem $\mathcal{S}^{(i_{t})}$ during $[t_{0},t]$, 
\begin{equation}
\label{l2}
\mathpzc{P}_{[t_{0} , t]}^ {(i_{t})} \equiv \{ \varrho^{(i_{t})}_{t'},\, \forall t' \in [t_{0}, t]\}.  
\end{equation}  
Moreover, every degree-$N_{t}^{(\mathcal{S})}$ resolution of $\mathcal{S}$ introduces its degree-$N_{t}^{(\mathcal{S})}$ \textit{universal} history (experience). It is defined as the complete $N_{t}^{(\mathcal{S})}$-tuple of all the (resolution identified) individual histories, as
\begin{equation}
\label{lu}  
\mathpzc{P}_{[t_{0} , t]}^ {(\mathcal{S}; N_{t}^{(\mathcal{S})}\geqslant 1 | \{ \mathcal{S}^{(i_{t})} \})} \equiv \big(\mathpzc{P}_{[t_{0} , t]}^{(1)} , \cdots , \mathpzc{P}_{[t_{0} , t]}^ { (N_{t}^{(\mathcal{S})})}  \big). 
\end{equation}
One now defines at every present moment $t$ \textit{a chosen experience} of the system by taking some arbitrary choices of the very states $\varrho_{t' \in [t_{0},t]}^{(\mathcal{S})}$'s or $\varrho_{t' \in [t_{0},t]}^{(i_{t})}$'s constituting the histories $\mathpzc{P}_{[t_{0} , t]}^{ (\mathcal{S})}$ or $\mathpzc{P}_{[t_{0} , t]}^{(\mathcal{S}; N_{t}^{(\mathcal{S})}\geqslant 1 | \{ \mathcal{S}^{(i_{t})} \})}$---for a specific resolution $\mathpzc{F}_{t}$ (of degree $N_{t}^{(\mathcal{S})} \geqslant 1$), which itself can be rechosen moment by moment---as follows:
\begin{equation}
\mathpzc{P}^{\mathrm{chosen}}_{[t_{0} , t]|\mathpzc{F}_{t}} \equiv \big\{ \varrho^{(\mathcal{S})}_{t'_{\mathrm{chosen}}} \in \mathpzc{P}_{[t_{0} , t]}^{ (\mathcal{S})} \big\}_{|\{t'_{\mathrm{chosen}}\}\in[t_{0},t]}  \textstyle{\bigcup_{i_{t}=1}^{N_{t}^{(\mathcal{S})}\geqslant 1}} \big\{\varrho^{(i_{t})}_{t'_{\mathrm{chosen}}} \in \mathpzc{P}_{[t_{0} , t]}^{(\mathcal{S}, N_{t}^{(\mathcal{S})} \geqslant 1| \{ \mathcal{S}^{(i_{t})} \})} \big\}_{|\{t'_{\mathrm{chosen}}\}\in[t_{0},t]}.
\label{clh-two}
\end{equation} 
In some contexts and examples, it is appropriate that one considers a collection of distinct resolutions of the system (or all possible subsystems) at once $\{\mathpzc{F}^{\alpha}_{t}\}_{\alpha}$. In these cases, the above definition of the chosen experience is generalized as follows: 
\begin{equation}
\mathpzc{P}^{\mathrm{chosen}}_{[t_{0} , t]} \equiv \textstyle{\bigcup_{\{\alpha\}\subset\mathrm{I\!N} }} \,\mathpzc{P}^{\mathrm{chosen}}_{[t_{0} , t]|\mathpzc{F}^{\alpha}_{t}}.
\label{aaaaa}
\end{equation} 
We shall denote the system's chosen experience at every moment $t$ (irrespective of the underlying resolution) by $\mathpzc{P}^{\mathrm{chosen}}_{[t_{0} , t]}$, for notational brevity. \\

\textit{Remark.---}One can allow, in appropriate contexts or examples, to define sub-states (i.e., subsystem states) $\varrho^{(i)}_{t'}$ in alternative consistent ways other than the standard recipe of partial tracing---see, e.g., Refs. \cite{Alicki, noTr1, Carroll-coarse, Subra, Kaber} as possible alternative approaches and related investigations.

%%%%%%%%%%%%%%%%%%%%%%%%%%%%%%%%%%%%%%%%%%%%%%%%%%%%%%%%%%%%%%%%
\begin{element}
Experience-centric Hamiltonians
\label{def:3}
\end{element}

An \textit{EC Hamiltonian} $\mathbbmss{H}_{t}$ at every present moment $t$ for a closed quantum system $\mathcal{S}$ is a Hermitian operator $O_{t}$ which has any form of \textit{structural dependence} on $\mathpzc{P}^{\mathrm{chosen}}_{[t_{0} , t]}$, that is, 
\begin{equation}
\begin{gathered}
\mathpzc{P}^{\mathrm{chosen}}_{[t_{0} , t]} \,\xlongrightarrow{\mathrm{Experience\,Centricity}} \, \mathbbmss{H}_{t},\\
\mathbbmss{H}_{t} =  O_{t} (\mathpzc{P}^{\mathrm{chosen}}_{[t_{0} , t]} ) = \mathbbmss{H}_{t}^{\dag}.
\end{gathered}
\label{clh}
\end{equation}
In particular, the above general formulas generate \textit{three major classes} of the EC Hamiltonians $\mathbbmss{H}_{t}$ which we rank in the increasing order of generality as follows: 

(i) the density operators in $\mathpzc{P}^{\mathrm{chosen}}_{[t_{0} , t]}$ contribute in the construction of $\mathbbmss{H}_{t}$ as part of its building blocks; 

(ii) $\mathbbmss{H}_{t}$ takes as part of its building blocks some partial time-labeled information which are contained within the density operators constituting $\mathpzc{P}^{\mathrm{chosen}}_{[t_{0} , t]}$; 

(iii) the structural composition of $\mathbbmss{H}_{t}$ has any form of conditional dependence on some time-labeled information within $\mathpzc{P}^{\mathrm{chosen}}_{[t_{0} , t]}$. \\

\textit{Remark 1.---}The defining formulation of EC Hamiltonians is based on quantum states in the form of density operators (rather than state vectors), as Eqs. (\ref{clh-two}) and (\ref{clh}) manifest. There are two properties which are automatically satisfied by this formulation: (i) it secures the physical irrelevance of the global phase of the pure states of a closed quantum system in the absence of any background field. In other words, the fact that rays live in complex projective Hilbert space is incorporated by construction. (ii) It establishes the covariance (form invariance) of the EC Hamiltonians upon changing from pure to mixed states. This property is particularly useful in EC Hamiltonians which depend on a combination of pure and mixed states, for example when $\mathpzc{P}^{\mathrm{chosen}}_{[t_{0} , t]} $ has elements of the total system as well as some of its subsystems. \\

\textit{Remark 2.}---The general definitions of EC Hamiltonians presented in Eqs. (\ref{clh-two}) and (\ref{clh}) allow the use of arbitrary parts of the total information contained in the (sub-)system's density operators as the building blocks of the instantaneous EC Hamiltonians. It it understood that such information contents are to be extracted from within the corresponding instantaneous states by means of operatorial actions with well-defined information-theoretic meanings. For example, the selected partial information can correspond to one's favored probabilities of specific measurement outcomes of arbitrarily chosen observables at arbitrary present or past moments. In these cases, the extractions are, e.g., implementable via applying the projection operators on the states.\\

\textit{Remark 3.}---Let $f^{(m)}_{t}$ be a set of state-history-dependent constraints with $A^{(\tilde{m})}$'s being a set of state-history-independent operators. A \textit{general expression} of the third major class of the EC Hamiltonians is given by
\begin{equation}
\mathbbmss{H}_{t} = O_{t} ,\,\, \mathrm{s.t.}\,\, f^{(m)}_{t} \Big(\{\varrho_{t'}\}\subset \mathpzc{P} ^{\mathrm{chosen}}_{[t_{0} , t]},  \{A^{(\tilde{m})}\} , O_{t} \Big) =0;\,\,\{m\},\{\tilde{m}\} \subset \mathrm{I\!N}, \,\forall m,\,\forall t.
\label{cond-const}
\end{equation}
It is important to stress that in a typical case in the third major class of EC Hamiltonians, which are expressed in Eq. (\ref{cond-const}), the operator $O_{t}$ that identifies the instantaneous EC Hamiltonian by solving the above system of constraint equations cannot be explicitly written down by the density operators comprising $\mathpzc{P}^{(\mathrm{chosen})}_{[t_{0},t]}$ or their information. Instead, $\mathbbmss{H}_{t} = O_{t}$ is to be obtained by solving moment by moment the coupled system of equations numerically. A pedagogical example of these EC Hamiltonians is given in Example 3 of Element \ref{def:7}. \\

It is important to highlight that the chosen histories which contribute to making the EC Hamiltonians of a closed quantum system at two (finitely distant or even immediately close) moments $t_{<}$ and $t_{>}$, $\mathpzc{P}^{\mathrm{chosen}}_{[t_{0} , t_{<}]}$ and $\mathpzc{P}^{\mathrm{chosen}}_{[t_{0} , t_{>}]}$, can themselves consist of completely different choices of the system or subsystem states. Here, generally speaking, even the tensorial factorization of the system to subsystems, $\mathpzc{F}_{t}$ or $\{\mathpzc{F}_{t}^{\alpha}\}$, can be updated at every present moment $t$. Furthermore, one has the freedom to let the operators defining the instantaneous EC Hamiltonians be different functions of the quantum information in the chosen histories. In this sense, the instantaneous EC Hamiltonians can be alternatively thought as \textit{experiential re-creations}.

%%%%%%%%%%%%%%%%%%%%%%%%%%%%%%%%%%%%%%%%%%%%%%%%%%%%%%%%%%%%%%%%
\begin{element}
Time evolution unitarity, isometry, and generalizations
\label{def:5-}
\end{element}

The time evolutions of closed system states in ECQT are generated by EC Hamiltonians \eqref{clh}. In fact, unitarity of such evolutions is guaranteed by the very Hermiticity of the state-history-dependent operators $O_{t} = O_{t}^{\dag}$, postulated in Element \ref{def:3}. Finite time evolutions of closed system states, $\varrho_{t_{1}}^{(\mathcal{S})} \to \varrho_{t_{2}}^{(\mathcal{S})}$, are realized by operators $\mathbbmss{U}_{t_{1} \to t_{2}}$ which are state-history dependent and unitary, $\mathbbmss{U}_{t_{1} \to t_{2}} \mathbbmss{U}_{t_{1} \to t_{2}}^{\dag} = \mathbbmss{U}_{t_{1} \to t_{2}}^{\dag} \mathbbmss{U}_{t_{1} \to t_{2}} =  \openone$, for all $t_{2} \geqslant t_{1} \geqslant t_{0} $. The general construction of the evolution operators is straightforward in discrete (where moments $\in\mathbb{Z}$ and $\geqslant k_{0}$) and continuous times (where moments $\in\mathrm{I\!R}$ and $\geqslant t_{0}$), respectively, as
\begin{equation}
\label{aiuvb} 
\begin{split}
& \mathbbmss{U}_{s \to s + m} = \textstyle{\prod_{l = 0}^{m-1}} \, \mathbbmss{U}_{s+ l \to s + l +1} \equiv \mathbbmss{U}_{s+m-1 \to s+m} \ldots \mathbbmss{U}_{s+1 \to s+2}  \mathbbmss{U}_{s \to s+1};\,\, \mathbbmss{U}_{k \to k +1} =  e^{- i \mathbbmss{H}_{k}  (\mathpzc{P}^{\mathrm{chosen}}_{[k_{0} , k ]} )}  \\
& \mathbbmss{U}_{t_{1} \to t_{2}} = \lim_{m \to \infty} \textstyle{\prod_{l = 0}^{m}} e^{- i\frac{t_{2}-t_{1}}{m} \mathbbmss{H}_{t_{1} + \frac{t_{2}-t_{1}}{m} l}} = \textbf{\textsf{T}}e^{-i\int_{t_{1}}^{t_{2}}dt'\, \mathbbmss{H}_{t'} (\mathpzc{P}^{\mathrm{chosen}}_{[t_{0} , t']})} .
\end{split}
\end{equation}
The time evolution of the system state during a time interval $t_{1}$ to $t_{2}$ is accordingly given by their unitary transformation on the Hilbert space,
\begin{equation}
\label{uts} 
\varrho_{t_{2}}^{(\mathcal{S})} = \mathbbmss{U}_{t_{1} \to t_{2}} \varrho_{t_{1}}^{(\mathcal{S})} \mathbbmss{U}_{t_{1} \to t_{2}}^{\dag}, \, \, \forall\, t_{0} \leqslant t_{1} \leqslant t_{2}.
\end{equation}

\textit{Remark 1.---}We emphasize again that the very \textit{unitarity} of EC Hamiltonian evolution directly follows from two facts. First, the Hermiticity of the state-history-dependent operator $O_{t}$ guarantees that the EC Hamiltonian of the moment enlarges the quantum history one step forward. Second, there is nothing special about the moment $t$; it can be pushed all the way back to the initial moment $t_{0}$. At the moment of initiation, the history consists of the initial state of the closed system with its possible resolution counterparts $\{\varrho_{0}^{(\mathcal{S})},\varrho_{0}^{(i)}\}_{i}$. The EC Hamiltonian at the initial moment, $\mathbbmss{H}_{0}$, can be any Hermitian operator of these state elements as well as state-history-dependent operators, provided that it generates a nontrivial evolution, $[\mathbbmss{H}_{0},\varrho_{0}^{(\mathcal{S})}] \neq 0$. The examples given in Eq. \eqref{ech:entanglement} such EC initiations can be observed in all terms individually. In conclusion, EC Hamiltonians \eqref{clh} generate unitary state histories across the moduli space of complete consistent solutions of the EC Schr\"{o}dinger and von Neumann's equations, that is, Eqs. (\ref{ecvovn}) -- (\ref{ecvodte}) in the next element. \\

We now turn to the \textit{isometric} generalization of the EC evolutions of closed quantum systems. In proper contexts or examples (see, e.g., Ref. \cite{cs1}), it is appropriate to replace the unitarity condition with the weaker condition of \textit{isometry}, that is, $\Vert \Psi_{t} \Vert = 1,\,\forall t$. In these cases, the defining formulation of EC Hamiltonians (\ref{clh}) takes a straightforward generalization where the only difference being that now $\mathbbmss{H}_{t}$'s can be non-Hermitian, state-history-dependent operators which by construction satisfy the dynamical constraint 
\begin{equation}
\label{cond-isometry}
\mathrm{Tr}[\varrho_{t}( \mathbbmss{H}_{t} - \mathbbmss{H}_{t}^\dagger)] = 0, \,\,\forall t.
\end{equation}
It is evident that Eq. (\ref{cond-isometry}) is a particular example of the Hamiltonian's experience centricity in the third major class as given in Eqs. (\ref{cond-const}), albeit now generalized to the case of isometry.\\

\textit{Remark 2.---}It is possible to further extend EC evolution operators to the alternative evolution operators in generalized ``finite quantum mechanics'' \cite{Kornyak1, Kornyak2, Banks-21}. However, the implementation of this point is beyond the interest of this paper. 

%%%%%%%%%%%%%%%%%%%%%%%%%%%%%%%%%%%%%%%%%%%%%%%%%%%%%%%%%%%%%%%%
\begin{element}
EC von Neumann and Schr\"{o}dinger equations
\label{def:5}
\end{element}

Infinitesimal temporal flows of closed system states under EC unitary transformations---Eqs. (\ref{aiuvb}) and (\ref{uts})---between moments $t_{1} = t$ and $t_{2} = t + dt$, leads to the differential equation which governs state-history-dependent unitary evolutions in ECQT. We refer to this equation as the \textit{EC von Neumann equation},
\begin{equation}
\label{ecvovn} 
i \dot{\varrho}^{(\mathcal{S})}_{t} =  [ \mathbbmss{H}_{t},\varrho_{t}^{(\mathcal{S})} ] =  [O_{t} (\mathpzc{P}^{\mathrm{chosen}}_{[t_{0} , t]}) , \varrho_{t}^{(\mathcal{S})}], 
\end{equation}
where dot denotes $d/dt$. If the EC dynamics begins with a pure state $\varrho_{0}=|\Psi_{0} \rangle \langle \Psi_{0}|$ and by noting that the unitarity of the dynamics preserves quantum state norm and purity, such that $\varrho_{t}=|\Psi_{t}\rangle \langle \Psi_{t}|\,\forall t$, the EC von Neumann equation (\ref{ecvovn}) reduces to the following \textit{EC Schr\"{o}dinger equation}:
\begin{equation}
\label{ecvos} 
i |\dot{\Psi}^{(\mathcal{S})}_{t}\rangle =  \mathbbmss{H}_{t} |\Psi_{t}^{(\mathcal{S})}\rangle =  O_{t}(\mathpzc{P}^{\mathrm{chosen}}_{[t_{0} , t]}) |\Psi_{t}^{(\mathcal{S})}\rangle.   
\end{equation} 

In ECQT defined on discrete time spans, Eqs. (\ref{ecvovn}) and (\ref{ecvos}) are replaced with the dynamical system of one-step forward EC evolutions. From Eqs. (\ref{aiuvb}) and (\ref{uts}) we have
\begin{equation}
\label{ecvodte}  
\begin{split}
\varrho_{k+1}^{(\mathcal{S})} &= e^{-i\mathbbmss{H}_{k}  (\mathpzc{P}^{\mathrm{chosen}}_{[k_{0} , k ]})} \varrho_{k}^{(\mathcal{S})} e^{i \mathbbmss{H}_{k}  (\mathpzc{P}^{\mathrm{chosen}}_{[k_{0} , k ]})},\\
|\Psi_{k+1}^{(\mathcal{S})}\rangle  &=  e^{-i\mathbbmss{H}_{k}  (\mathpzc{P}^{\mathrm{chosen}}_{[k_{0} , k ]})} |\Psi_{k}^{(\mathcal{S})}\rangle.
\end{split}
\end{equation} 

\textit{Remark 1.---}The crucial point to be highlighted about the behaviors described by the EC von Neumann and Schr\"{o}dinger equations (\ref{ecvovn}) and (\ref{ecvodte}) is as follows. The fundamental distinctions of ECQT (compared to SQT) is its in-built characteristic of dynamics: generic EC evolutions of closed systems realize an inseparable merge of (conventional open-system) non-Markovianity and (conventional closed-system) unitarity. ECQT by construction has no fundamental reason for lifting unitarity to benefit from system's memory effects, and especially, for system's learning from its experience. It allows unitarity and memory effects to join forces together in an intrinsic manner. The intrinsic cooperation between the evolution unitarity and recurrent experiential updates enables ECQT to naturally develop a broad spectrum of unprecedented coexisting phases and phase transitions of quantum information behaviors, which are remarkable for their high complexity, multiscale intricacy, and phenomenological or operational richness. We shall return to this in Sec. \ref{sec:phase2-perturbation}. \\

\textit{Remark 2.---}The general forms of the dynamical equations---Eqs. (\ref{ecvovn}) and (\ref{ecvos})---in ECQT reveal an important (physical and mathematical) characteristic of them: given their inherent structural dependence on the system's state history, typical EC Schr\"{o}dinger and von Neumann equations are simultaneously \textit{nonlinear} and \textit{time nonlocal}. Moreover, these time nonlocalities and nonlinearities are strongly intertwined such that they are typically \textit{inseparable}. In particular, we highlight that mathematical forms and physical behaviors of (the solutions to) these differential equations are essentially incommensurable with their counterparts known in the literature of \textit{nonlinear quantum mechanics}. Analysis of these equations and their solutions, as presented in Ref. \cite{ECQT-1} and in Sec. \ref{sec:phase2-perturbation} of this paper, manifest these points evidently. 

%%%%%%%%%%%%%%%%%%%%%%%%%%%%%%%%%%%%%%%%%%%%%%%%%%%%%%%%%%%%%%%%
\begin{element}
Polynomial EC Hamiltonians of the kind $[\hskip-.8mm[N_{t},L_{t}]\hskip-.8mm]$ (finite and infinite $N_{t}$ and $L_{t}$)
\label{def:6}
\end{element}

A natural, infinitely large and rich family of EC Hamiltonians \eqref{clh} are the ``$(N,L)$ EC Hamiltonians'' originally introduced in Ref. \cite{ECQT-1}. At every present moment a closed quantum system chooses $N$ density operators of its history $\mathpzc{P}^{\mathrm{chosen}}_{[t_{0} , t]}$, forms a Hermitian \textit{polynomial} of degree $L$ out of them, and employs it as the momentary Hamiltonian of its unitary evolution. The polynomials can (and typically do) have structural dependence on state-history-independent operators as well. These are the very EC Hamiltonians we shall focus on. \\

Consider a system $\mathcal{S}$ with its corresponding $d$-dimensional Hilbert space $\mathpzc{H}$. The complete input to define an $[\hskip-.8mm[N_{t},L_{t}]\hskip-.8mm]$ instantaneous EC Hamiltonian on $\mathpzc{H}$ is a triplet $\Xi \equiv [\hskip-1mm[N_{t},L_{t},\vec{a}_{t}]\hskip-1mm] \in \mathrm{I\!N} \times \mathrm{I\!N} \times {\rm I\!R}_{+}^{N_{t}}$, where $\vec{a}_{t} = (a_{1} , \ldots , a_{N_{t}})$ encodes a choice of $N_{t}$ memory distances, being measured backward from the present moment and ordered reversely. An EC Hamiltonian $\mathbbmss{H}_{t}^{\Xi }$ is defined based on $\Xi $ as a Hermitian linear combination of a chosen set of $d_{t}^{-}$ anti-Hermitian and $d_{t}^{+}$ Hermitian monomials ($\{\mathbbmss{h}^{j-}_{t}, \mathbbmss{h}^{k+}_{t}\}$), which are (partially or entirely) made out of the past-to-present density operators of the system $\varrho_{t'}^{(\mathcal{S})} \in \mathpzc{P}^{\mathrm{chosen}}_{[t_{0} , t]}$. That is, employing any set of real-valued couplings which themselves can be state-history-dependent, one has
\begin{equation}
\begin{gathered} 
\label{ech:g} 
\mathbbmss{H}_{t}^{\Xi } \equiv i \textstyle{\sum_{j = 1}^{d_{t}^{-}}} \lambda^{j-}_{t} \,\mathbbmss{h}^{j-}_{t}  +  \textstyle{\sum_{k = 1}^{d_{t}^{+}}}\lambda^{k+}_{t} \,\mathbbmss{h}^{k+}_{t}, \\  
\lambda^{\ell\pm}_{t}  = \lambda^{\ell\pm}_{t} (\mathpzc{P}^{\mathrm{chosen}}_{[t_{0} , t]} )  \in \mathrm{I\!R},\;\;\;\mathbbmss{h}^{\ell\pm}_{t} = \mathbbmss{h}^{\ell\pm}_{t} (\mathpzc{P}^{\mathrm{chosen}}_{[t_{0} , t]} ).
\end{gathered} 
\end{equation}
Now, selecting an arbitrary collection of (trivial or nontrivial) state-history-independent operators, \textit{collectively denoted by $A$'s}, and assuming purity of the whole state history \eqref{lll} (which is possible due to the unitarity of the evolution), the state-history monomials can be generally formulated as 
\begin{equation}
\begin{gathered}
\label{monomials} 
\mathbbmss{h}^{k+}_{t}  =  \textstyle{\prod_{s=1}^{L_{t}^{k}}} A_{k_{s}}  \varrho_{t-a_{k_{s}}}^{(\mathcal{S})}  A_{k_{s+1}} + \big(\prod_{s=1}^{L_{t}^{k}} A_{k_{s}}  \varrho_{t-a_{k_{s}}}^{(\mathcal{S})}  A_{k_{s+1}} \big)^{\dag} , \\
\mathbbmss{h}^{j-}_{t}  = \textstyle{\prod_{r=1}^{L_{t}^{j}}} A_{j_{r}}  \varrho_{t-a_{j_{r}}}^{(\mathcal{S})}  A_{j_{r+1}} - \big(\prod_{r=1}^{L_{t}^{j}} A_{j_{r}}  \varrho_{t-a_{j_{r}}}^{(\mathcal{S})}  A_{j_{r+1}} \big)^{\dag}, 
\end{gathered} 
\end{equation}
for $1 \leqslant k \leqslant d_{t}^{+}$ and all choices of $\{ k_{1}, \ldots, k_{L_{t}^{k}}\} \in \{1, \ldots, N_{t} \}^{L_{t}^{k}}$, and similarly for $1 \leqslant j \leqslant d_{t}^{-}$ and all choices of $\{ j_{1}, \ldots, j_{L_{t}^{j}}\} \in \{1, \ldots, N_{t} \}^{L_{t}^{j}}$. For extensions to mixed states, see Element \ref{def:7}. \\

Thus far we have formulated the case of polynomial EC Hamiltonians with finite $N_{t}$ and $L_{t}$. However, the case of (countable or uncountable) infinite $N_{t}$ or $L_{t}$ is not only allowed but also can be relevant in some contexts. For example, an infinitely large family of $[\hskip-.8mm[N_{t},\infty]\hskip-.8mm]$ EC Hamiltonians can be constructed by forming a typical (nonpolynomial) function of $[\hskip-.8mm[N_{t},L_{t}]\hskip-.8mm]$ EC Hamiltonians with finite $L_{t}$. More interestingly, an infinitely large and more diverse family of $[\hskip-.8mm[\infty,L_{t}]\hskip-.8mm]$ EC Hamiltonians can be formed by choosing the largest quantum memory distance $a_{t}<\infty$ as the width of an arbitrary time window and integrating one's chosen quantum information of the state-history of the system over the entire window, weighed by temporal functions which control the significance of the contribution of the instantaneous quantum information. See Example 4 as a concrete illustration. \\

\textit{Remark 1.---}We denote a special class of EC Hamiltonians as \textit{hybrid} when they are comprised of an SQT Hamiltonian deformed \textit{additively} by an EC Hermitian operator. Moreover, an important type of EC Hamiltonians are those whose defining state-history $\mathbbmss{h}^{\ell\pm}$ monomials involve no nontrivial state-history-independent operators, i.e., all $A$'s in Eq. (\ref{monomials}) are $\openone$. We refer to this type as \textit{primitive} EC Hamiltonians. \\

\textit{Remark 2.---}Hereafter we shorten the notation by dropping $t$ in $N_{t}$, $L_{t}$, and $\vec{a}_{t}$, and denote $\mathbbmss{H}_{t}^{\Xi}$ simply by $\mathbbmss{H}_{t}^{[\hskip-.8mm[N,L]\hskip-.8mm]}$. However, it should be kept in mind that the triplet $[\hskip-.8mm[N,L,\vec{a}]\hskip-.8mm]$ can be re-chosen moment by moment. \\

\textit{Remark 3.---}Following lessons from disordered manybody systems, an interesting family of EC Hamiltonians can be formed by assuming that some $\lambda$ couplings are randomly chosen from given probability distributions. \\

\textit{Example 1.---}To illustrate the $[\hskip-.8mm[N,L]\hskip-.8mm]$ EC Hamiltonians as defined above, consider the primitive EC Hamiltonians of the kind $[\hskip-.8mm[1,1]\hskip-.8mm]$ and $[\hskip-.8mm[2,2]\hskip-.8mm]$ with respective constant quantum memory distances $\vec{a}_{t}=a>0$ and $\vec{a}_{t} = (a,0)$. The general forms of these EC Hamiltonians read as
\begin{equation} 
\begin{split}
\mathbbmss{H}^{[\hskip-.8mm[1,1]\hskip-.8mm]}_{t}  &= \lambda_{t-a}\, \varrho_{t-a},   \\
\mathbbmss{H}^{[\hskip-.8mm[2,2]\hskip-.8mm]}_{t}  & = \lambda_{t} \varrho_{t} + \lambda_{t-a} \varrho_{t-a} + \lambda_{t-a,t}^{\mathrm{R}}\{ \varrho_{t-a} ,\varrho_{t}  \} + i \lambda_{t-a,t}^{\mathrm{I}} [\varrho_{t-a} , \varrho_{t} ], 
\end{split} 
\label{11,22} 
\end{equation}
where $(\lambda_{t-a}, \lambda_{t}, \lambda_{t-a,t}^{\mathrm{R}},\lambda_{t-a,t}^{\mathrm{I}}) \equiv \Lambda_{t}( [ \mathpzc{P}^{\mathrm{chosen}}_{[t_{0} , t]} ]) \in \mathrm{I\!R}^{4}$ can \textit{themselves} be state-history dependent. The first term in $\mathbbmss{H}^{[\hskip-.8mm[2,2]\hskip-.8mm]}_{t}$ is included only as a matter of formal definition, but it is dynamically irrelevant (hence one can drop it). Two other specific examples are special primitive EC Hamiltonians of the kind $[\hskip-.8mm[2,3]\hskip-.8mm]$ and $[\hskip-.8mm[3,3]\hskip-.8mm]$ with respective constant quantum memory distances $\vec{a}_{t} = (a,0)$ and $\vec{a}_{t} = (a,b,0)$ defined as
\begin{equation} 
\begin{split}
\mathbbmss{H}^{[\hskip-.8mm[2,3]\hskip-.8mm]}_{t}  & = i \lambda_{t-a,t}^{\mathrm{I}} [\varrho_{t-a} , \varrho_{t} ] + \zeta_{t-a,t}^{\mathrm{R}} \varrho_{t-a}\varrho_{t}\varrho_{t-a}, \\
\mathbbmss{H}^{[\hskip-.8mm[3,3]\hskip-.8mm]}_{t}  & = \kappa_{t-a,t-b,t} \varrho_{t-a} \varrho_{t-b} \varrho_{t} + \kappa^{*}_{t-a,t-b,t} \varrho_{t} \varrho_{t-b} \varrho_{t-a},
\end{split} 
\label{23,233} 
\end{equation}
where $(\lambda_{t-a,t}^{\mathrm{I}}, \zeta_{t-a,t}^{\mathrm{R}}, \kappa^{\mathrm{R}}_{t-a,t-b,t}, \kappa^{\mathrm{I}}_{t-a,t-b,t}) \equiv \Lambda_{t}([\mathpzc{P}^{\mathrm{chosen}}_{[t_{0} , t]}])  \in \mathrm{I\!R}^{4}$. \\

\textit{Example 2.}---Consider a one-qubit closed system $\mathcal{S}$ and a single quantum memory distance $a> 0$. An example of a hybrid, nonprimitive $[\hskip-.8mm[1,1]\hskip-.8mm]$ Hamiltonian is give by 
\begin{equation}
\label{ech:entanglement} 
\mathbbmss{H}_{t}^{(\mathcal{S})} =  \sigma^{1} + \big(1 - \mathrm{Tr}[\varrho_{t - a}^{(\mathcal{S})} \varrho_{t}^{(\mathcal{S})} ]\big) \varrho_{t - a}^{(\mathcal{S})} + 5  \{ \varrho_{t}^{(\mathcal{S})} , \sigma^{1}\} + \sigma^{3} \varrho_{t}^{(\mathcal{S})} \sigma^{3}.   
\end{equation}

\textit{Example 3.---}The Gross-Pitaevskii equation is the nonlinear Schr\"{o}dinger equation with the potential $V^{\mathrm{GP}}_{t}\propto(\Vert\Psi_{t}\Vert)^{2}$. It is an extensively analyzed nonlinear equation, appearing for example in Bose-Einstein condensations \cite{book:GP}. It is, however, essential to note that this nonlinear equation arises in SQT as an \textit{effective} dynamics. Now, however, we show that the GP Hamiltonian can be recast as a \textit{genuinely-EC} Hamiltonian of the nonprimitive $[\hskip-.8mm[1,1]\hskip-.8mm]$ kind as
\begin{align}
\mathbbmss{H}^{\mathrm{GP}}_{t} = \textstyle{\sum_{k}} (\vert\Psi_{t}^{k}\vert)^{2} \, |k \rangle \langle k|  = \textstyle{\sum_{k}} \mathbbmss{h}^{k+} \varrho_{t} \mathbbmss{h}^{k+},
\label{hgp}
\end{align}
where $\varrho_{t}=|\Psi_{t}\rangle\langle \Psi_{t}|$ and $\Psi_{t}^{k}=\langle k|\Psi_{t}\rangle$, with $\mathbbmss{h}^{k+} = |k \rangle \langle k|$ representing projections on a complete orthonormal basis. On a similar note, one can also see that Weinberg's nonlinear quantum mechanics \cite{Weinberg-testQM, Weinberg-testQM3} can be contained in ECQT. \\

\textit{Example 4.---}An example of polynomial EC Hamiltonians using an entire time window of state-history information is given by
\begin{equation}
\mathbbmss{H}_{t}^{[\hskip-.8mm[\infty,2]\hskip-.8mm]} =  \textstyle{\int_{t - a}^t} d t'\, f_{t' t} \big[\mu_{t'} (\mathpzc{P}^{\mathrm{chosen}}_{[t-a , t]}) \big( (\varrho_{t'})^{2}  - \varrho_{t}\big) + i  \lambda^{\mathrm{I}}_{t' t}(\mathpzc{P}^{\mathrm{chosen}}_{[t-a , t]} ) [\varrho_{ t'} , \varrho_{t}] + \lambda_{t' t}^{\mathrm{R}} (\mathpzc{P}^{\mathrm{chosen}}_{[t-a , t]}) \{ \varrho_{t'},\varrho_{t}\}\big],
\end{equation}
where $f_{tt'}$ are arbitrarily chosen weight functions for different past-state contributions. 

%%%%%%%%%%%%%%%%%%%%%%%%%%%%%%%%%%%%%%%%%%%%%%%%%%%%%%%%%%%%%%%%
\begin{element}
Generalizations of $[\hskip-.8mm[N,L]\hskip-.8mm]$ EC Hamiltonians: Mixed-State Histories, Subsystem Resolutions, and Discriminative Type
\label{def:7}
\end{element}

\begin{center}
(a) \textit{Mixed-State Histories and Resolution-Refined $[\hskip-0.8mm[N,L]\hskip-0.8mm]$ EC Hamiltonians}
\end{center}

As stated in Element \ref{def:3}, straightforward generalizations of $[\hskip-.8mm[N,L]\hskip-.8mm]$ Hamiltonians [Eqs. (\ref{ech:g}) and (\ref{monomials})] are obtained by dropping the state-history purity assumption, and independently using a chosen resolution of $\mathcal{S}$ with $N_{t}^{(\mathcal{S})}$ subsystems. This generalization follows from the general definition \eqref{clh-two} on the basis of (state vs. universal) histories introduced in Eqs. (\ref{l1}), (\ref{l2}), and (\ref{lu}). Resolution-based generalizations of $[\hskip-.8mm[N,L]\hskip-.8mm]$ EC Hamiltonians are formed if one alters some of their state-history monomials \eqref{monomials}, $\mathbbmss{h}^{\ell\pm}_{t} \to \mathbbmss{g}^{\ell\pm}_{t}$'s, defined by replacing the states at arbitrarily chosen $a_{k_{s}}$'s with arbitrary powers of arbitrarily chosen substates: $\varrho_{t-a_{k_{s}}}^{(\mathcal{S})} \to \big(\varrho_{t-a_{k_{s}}}^{(i)}\big)^{n_{k_{s}}}$, and correspondingly $A_{k_{s}} \to A^{(i)}_{k_{s}}\otimes A^{(\bar{i})}_{k_{s}}$. Mathematically, these alterations introduce a vast extension of $[\hskip-.8mm[N,L]\hskip-.8mm]$ Hamiltonians (\ref{ech:g}) and (\ref{monomials}). \\

\begin{center}
(b) \textit{Discrimination-induced $[\hskip-0.8mm[N,L]\hskip-0.8mm]$ EC Hamiltonians}
\end{center}

A further generalization of the resolution-refined $[\hskip-0.8mm[N,L]\hskip-0.8mm]$ family is given by EC Hamiltonians where some chosen eigenstates of some observables play a discriminative role as its building blocks. For example, consider an observables $\mathbbmss{A}^{(i)}$ and one of its eigenprojections $P^{(i)m}$ corresponding to a favored outcome of its measurement and identify $A^{(i)}=P^{(i)m}$ in Eq. (\ref{monomials}). See Example 2 below. \\

\textit{Example 1.---}Consider a degree-$2$ resolution of a $4$-dimensional closed quantum system. This is equivalent to a two-qubit closed system  $\mathcal{S}$, with the Hilbert space $\mathpzc{H}^{(\mathcal{S})} = \mathpzc{H}^{(1)} \otimes \mathpzc{H}^{(2)}$. An example of a resolution-based, hybrid, nonprimitive $[\hskip-.8mm[2,2]\hskip-.8mm]$ Hamiltonian for this system is given by 
\begin{equation}
\label{ech:entanglement-} 
\mathbbmss{H}_{t}^{(\mathcal{S})} = \big\Vert \varrho_{t}^{(\mathcal{S})}-\varrho^{(1)}_{t} \otimes \varrho^{(2)}_{t} \big\Vert \,\sigma^{2} \otimes \sigma^{1} + \Big(\openone \otimes \big(\varrho^{(2)}_{t-a}\big)^{2}  +  \big(\varrho^{(1)}_{t-a}\big)^{2} \otimes \openone \Big) - i[\varrho_{t}^{(\mathcal{S})},\varrho^{(1)}_{t} \otimes \varrho^{(2)}_{t}] + \sigma^{3} \otimes \sigma^{3} \varrho_{t}^{(\mathcal{S})} \sigma^{3} \otimes \sigma^{3},
\end{equation}
where $\Vert \cdot\Vert$ is the standard operator norm. \\

\textit{Example 2.---}Consider the same subsystem resolution as in Example 1 and assume a discriminative scheme where the $+1$ eigenvalue of $\sigma^{1}$ on the first subsystem is favored as the eigenprojection $P^{(1)\,+} = |+ \rangle \langle +|$ plays a special role in the Hamiltonian construction as follows:
\begin{equation}
\label{ech:entanglement--} 
\begin{gathered}
\mathbbmss{H}_{t}^{(\mathcal{S})} = (1+\mathrm{Tr}[P^{(1)+} \varrho_{t}]) \, \sigma^{3} \otimes \sigma^{3} +\chi^{(\mathcal{S}|12)}_{t} - \big\langle P^{(1)+} \varrho^{(1)}_{t'} P^{(1)+} \big\rangle_{a,t, \{f_{tt'}\}} \otimes \openone , \\
\chi^{(\mathcal{S}|12)}_{t}  \equiv \varrho_{t}^{(\mathcal{S})}-\varrho^{(1)}_{t} \otimes \varrho^{(2)}_{t}, \\
\big\langle P^{(1)+} \varrho^{(1)}_{t'} P^{(1)+} \big\rangle_{a,t, \{f_{tt'}\}}  \equiv \textstyle{\int_{t-a}^{t}} dt' \, f_{tt'} P^{(1)+} \varrho^{(1)}_{t'} P^{(1) +} /(\textstyle{\int}_{t-a}^{t}dt'\, f_{tt'}),
\end{gathered}
\end{equation}
where $\chi^{(\mathcal{S}|12)}_{t}$ is the \textit{correlation operator} of the system $\mathcal{S}$, encompassing the entanglement and classical correlation data between the subsystem $1$ and $2$. \\

\textit{Example 3.---}Consider the same subsystem resolution as in Example 1 and 2 and the EC Hamiltonian $\mathbbmss{H}^{(\mathcal{S})}_{t} = \sum_{\alpha,\beta} h_{t,\alpha \beta} D^{\alpha \beta}$ for the closed system $\mathcal{S}$, with $D^{\alpha \beta}$ being the Dirac operators. Assume by a proper shift we have set $\mathrm{Tr}[\mathbbmss{H}^{(\mathcal{S})}_{t}] = h_{t,00} = 0$. We now formulate, as an example, an EC Hamiltonian $\mathbbmss{H}^{(\mathcal{S})}_{t}$ in the third major class. This Hamiltonian is EC as all $h_{t,\alpha\beta}$'s depend on the information contained in the chosen universal histories $\mathpzc{P}_{[t_{0} , t]}^{ (\mathcal{S})} \cup \mathpzc{P}_{[t_{0} , t]}^ {(\mathcal{S}; 2 | \{ 1,2\})}$ by solving for $\{h_{t,\alpha\beta}\}$ (at every present moment $t$) the EC system of $15$ algebraic equations as follows: 
\begin{equation}
\begin{gathered}
\Vert \mathbbmss{H}^{(\mathcal{S})}_{t} \Vert   = \Vert \chi^{(\mathcal{S}|12)}_{t}\Vert, \,\,\,\, \Vert \mathbbmss{H}_{t}^{(\mathcal{S})} -  \chi^{(\mathcal{S}|12)}_{t-a^{(\mathcal{S})}_{i}} \Vert = c^{(\mathcal{S})}_{i}, \,\,\,\, \det[\mathbbmss{H}^{(\mathcal{S})}_{t}] = \langle P^{(\mathcal{S})+} \varrho^{(\mathcal{S})}_{t'} P^{(\mathcal{S})+} \rangle_{\max\{a^{(\mathcal{S})}_{i}\},t,\{f^{(\mathcal{S})}_{t't}\}}, \\
\Vert K_{t}^{(1)} \Vert  = \mathrm{Tr}[(\varrho^{(1)}_{t})^{2}] , \,\,\,\, \Vert  K^{(1)}_{t} - \varrho^{(1)}_{t-a^{(1)}_{i}} \Vert  =  c^{(1)}_{i}, \,\,\,\, \det [K^{(1)}_{t}] = \langle P^{(1)+} \varrho^{(1)}_{t'} P^{(1)+} \rangle_{\max\{a^{(1)}_{i}\},t,\{f^{(1)}_{t't'}\}},\\
\Vert K_{t}^{(2)} \Vert  = \mathrm{Tr}[(\varrho^{(2)}_{t})^{2}] , \,\,\,\, \Vert  K^{(2)}_{t} - \varrho^{(2)}_{t-a^{(2)}_{i}} \Vert  =  c^{(2)}_{i}, \,\,\,\, \det [K^{(2)}_{t}] = \langle P^{(2)+} \varrho^{(2)}_{t'} P^{(2)+} \rangle_{\max\{a^{(2)}_{i}\},t,\{f^{(2)}_{t't'}\}}.
\end{gathered}
\label{eq:conditions}
\end{equation}
The character specification in the above equations are as follows. The EC subsystem operators $K^{(1)}_{t}$ and $K^{(2)}_{t}$ are defined by partial tracing over $\mathbbmss{H}^{(\mathcal{S})}_{t}$, $K^{(i)}_{t}= \mathrm{Tr}_{\bar{i}}[\mathbbmss{H}_{t}^{(\mathcal{S})}]$, for $i\in\{1,2\}$. The parameters $(\{a_{i}^{(1,2,\mathcal{S})}\}_{i=1}^{3},\{c_{i}^{(1,2,\mathcal{S})}\}_{i=1}^{3})$ are, respectively, chosen positive numbers identifying the corresponding quantum memory distances and norm-based distances. Moreover, $P^{(\mathcal{S})+}$ denotes the projection operator of the system's favored eigenstate of the Dirac operator $D^{11}$. The instantaneous operators $\chi^{(\mathcal{S}|12)}_{t}$, as defined in Eq. (\ref{ech:entanglement--}), are momentary correlation operators between the two subsystems. The other characters, parameters, and especially all the averaging operations are defined as in Example 2. \\

\textit{Remark.---}As a utility example, resolution-based EC Hamiltonians of the kinds similar to the expression in Eqs. (\ref{ech:entanglement-}), (\ref{ech:entanglement--}), and (\ref{eq:conditions}) may become relevant for addressing (artificial or natural) intelligent behaviors---see Sec. \ref{sec:toward} and Ref. \cite{ECQT-1}. 

%%%%%%%%%%%%%%%%%%%%%%%%%%%%%%%%%%%%%%%%%%%%%%%%%%%%%%%%%%%%%%%%
\begin{element}
EC two-point functions and observables---and their descendants
\label{def:xx}
\end{element}

The EC Hamiltonians together with the EC von Neumann and Schr\"{o}dinger equations [Eqs. \eqref{ecvovn} and \eqref{ecvos}] can be naturally expressed in terms of the system's state-history \textit{two-point} functions (\textit{autocorrelations}) and their $n$-point descendants \cite{ECQT-1}. In the case of pure-state histories the corresponding EC Hamiltonians (\ref{ech:g}) and (\ref{monomials}), these state-history two-point functions $m_{t_{1}, t_{2}}$ together with their corresponding fidelities and phase variables $(w_{t_{1}, t_{2}},\alpha_{t_{1}, t_{2}})$ and their descendant $n$-point correlations $m_{t_{1},\ldots,t_{n}}$ are defined as follows:
 \begin{equation}
 \label{fam}
 \begin{split}
 m_{t_{1}, t_{2}} &\equiv  \langle\Psi_{t_{1}} | \Psi_{t_{2}}\rangle \equiv w_{t_{1}, t_{2}} \,e^{ i \alpha_{t_{1}, t_{2}}} = m_{t_{2}, t_{1}}^{*} ,\\
 w_{t_{1}, t_{2}} &= \sqrt{\mathrm{Tr}[ \varrho_{t_{1}} \varrho_{t_{2}}]}, \\  
 m_{t_{1}, \ldots, t_{n+1}} & \equiv \textstyle{\prod_{r=1}^{n}} m_{t_{r}, t_{r+1}} \equiv w_{t_{1}, \ldots t_{n}} \,e^{i \alpha_{t_{1}, \ldots, t_{n}}} = m^{*}_{t_{n+1}, \ldots t_{1}}.
\end{split} 
\end{equation}

\textit{Remark 1.---}The extension to mixed states is straightforward. \\

\textit{Remark 2.}---The state-history $n$-point functions as defined in Eq. (\ref{fam}), and their extensions for mixed states, shall be particularly useful in harnessing the applications of the EC evolutions in domains and disciplines pointed out later in Sec. \ref{sec:outlook0}. \\

Assuming the purity of the state history of the closed system (the case of our main interest in this paper), the $[\hskip-.8mm[1,1]\hskip-.8mm]$, $[\hskip-.8mm[2,2]\hskip-.8mm]$, and $[\hskip-.8mm[3,3]\hskip-.8mm]$ EC Hamiltonians (\ref{11,22}) and (\ref{23,233}) adopt the simpler expressions as
\begin{equation} 
\label{11,22:pure}
\begin{split} 
\mathbbmss{H}^{[\hskip-.8mm[1,1]\hskip-.8mm]}_{t}   = & \lambda_{t-a}\; |\Psi_{t-a}\rangle  \langle\Psi_{t-a}|,  \\
\mathbbmss{H}^{[\hskip-.8mm[2,2]\hskip-.8mm]}_{t}  = &  \lambda_{t-a} |\Psi_{t-a}\rangle  \langle\Psi_{t-a}| + \lambda_{t-a,t} m_{t-a,t} |\Psi_{t-a}\rangle  \langle \Psi_{t}| + \lambda_{t-a,t}^{*} m_{t-a,t}^{*} |\Psi_{t}\rangle  \langle\Psi_{t-a}|, \\
\mathbbmss{H}^{[\hskip-.8mm[3,3]\hskip-.8mm]}_{t}  = &  \kappa_{t-a,t-b,t} m_{t-a,t-b,t}|\Psi_{t-a}\rangle  \langle\Psi_{t}| + \kappa^{*}_{t-a,t-b,t} m^{*}_{t-a,t-b,t}|\Psi_{t}\rangle  \langle\Psi_{t-a}| . 
\end{split}
\end{equation}
Thus, the EC Schr\"{o}dinger equations (\ref{ecvos}) read as follows:
\begin{equation} 
\label{ecso118}
\begin{split}
& [\hskip-.8mm[1,1]\hskip-.8mm]:\, i |\dot{\Psi}_{t}\rangle  = \lambda_{t-a}\, m_{t-a,t} |\Psi_{t-a}\rangle, \\ 
& [\hskip-.8mm[2,2]\hskip-.8mm]:\, i |\dot{\Psi}_{t}\rangle  = (\lambda_{t-a} + \lambda_{t-a,t} )m_{t-a,t} |\Psi_{t-a}\rangle + \lambda_{t-a,t}^{*} (w_{t-a,t})^{2} |\Psi_{t}\rangle,\\
& [\hskip-.8mm[3,3]\hskip-.8mm]:\, i |\dot{\Psi}_{t}\rangle  = \kappa_{t-a,t-b,t} m_{t-a,t-b,t} |\Psi_{t-a}\rangle + \kappa^{*}_{t-a,t-b,t} m_{t,t-b,t-a,t}|\Psi_{t}\rangle.
\end{split}   
\end{equation}

%%%%%%%%%%%%%%%%%%%%%%%%%%%%%%%%%%%%%%%%%%%%%%%%%%%%%%%%%%%%%%%%
\begin{definition}
Experience-Centric Quantum Theory (ECQT)
\label{def:4}
\end{definition}

ECQT is established based on the conceptual and technical explanations especially the six prerequisite defining elements of general quantum behavior in $\mathpzc{M}_{Q}$ presented in Sec. \ref{sec:toward}, together with the definition of EC Hamiltonians. \textit{By definition}, ECQT is any quantum theory in $\mathpzc{M}_{Q}$ whose defining context and other specifications allow the total Hamiltonians, the internal interactions, and even the partition-induced degrees of freedom (``particles'') of a (fundamental or high-level emergent) closed quantum system to be continually EC according to Elements \ref{def:1} -- \ref{def:3}. \\

\textit{Remark 1.}---As the formulations in Elements \ref{def:1} -- \ref{def:3} manifest, in ECQT the information units and modules within the developed state history of the closed system and its subsystems can serve as (some of) the building blocks of the Hamiltonians and (according to subsystem resolution-refined versions formulated in Element \ref{def:2}) the manybody interactions. Moreover, ECQT allows that these state-history information units and modules to continually guide the (flows of) system's Hilbert-space factorizations (Element \ref{def:2}), hence experiential formations of the internal elements and particles of the system. \\

\textit{Remark 2.}---It is important to appreciate that the experience centricity in an ECQT need not be confined only to the quantum dynamics; it can have deformational projections---induced by context-based requirements or consistencies---on observables, their measurements, and even inner structures of the Hilbert space. 

%%%%%%%%%%%%%%%%%%%%%%%%%%%%%%%%%%%%%%%%%%%%%%%%%%%%%%%%%%%%%%%%
\begin{element}
Nature of quantum states and experience in ECQT
\label{def:8}
\end{element}

The preferred and primary way in ECQT to understand density operators and their roles is based on \textit{agentive Bayesian probabilities}. One considers a single quantum system and an agent who examines the system by measurement actions. The momentary density operator is an encapsulation of all measurement outcome probabilities quantifying the last updated version of the agent's expectations about the system's responses to her future actions. This statement applies to both pure and mixed density operators, where, unlike mixed states, pure states show maximally informative cases. \\

However, the above Bayesian primacy does not impose any fundamental incompatibility with frequentist probabilistic implementations of density operators in ECQT. That is, it remains generally possible to employ suitable representations of density operators as statistical ensembles of pure-state quantum systems under classical probability distributions. Such mixture-based representations can be used, for example, for the initial-state preparation of the system in ECQT. But the unitary evolution of the ensembles develops interesting and distinctive features compared to their counterparts in SQT, due to the state-history-dependent nature of the EC Hamiltonians. In fact, by mixture representations of density operators in ECQT one can use the pure-state density operators corresponding to some constituent elements of the ensemble as the building blocks of the total EC Hamiltonians. Such recipes resemble with the subsystem-based construction of EC Hamiltonians, which indeed fits in the general description of EC Hamiltonians given in Eqs. (\ref{clh-two}) and (\ref{clh}). \\

Finally, it is crucial to note that in the Bayesian paradigm of ECQT the state-history as defined in Eq. (\ref{l1}) and the corresponding ``experience'' are defined with respect to \textit{a definite pair of the closed quantum system and (at least) an agent}---and not for the system alone. However, bearing this in mind, for brevity throughout this paper we simply refer to the experience as that of the system.

%%%%%%%%%%%%%%%%%%%%%%%%%%%%%%%%%%%%%%%%%%%%%%%%%%%%%%%%%%%%%%%%
\section{SQT versus ECQT}
\label{sec:sqt-vs-ecqt}

In the first part of this section we explicitly show how a general SQT Hamiltonian $H$ can be represented in various EC forms. In particular, we develop two independent, complementary recipes, one of which is \textit{basis-dependent} and the other one is \textit{basis independent}. In the second part, we prove a number of inclusion theorems which show that reformulation of ECQT for a system with a given Hilbert space is generically impossible by SQT on the same Hilbert space.

%%%%%%%%%%%%%%%%%%%%%%%%%%%%%%%%%%%%%%%%%%%%%%%%%%%%%%%%%%%%%%%%
\subsection{Reformulation of SQT within ECQT: Basis-Dependent Recipe}
\label{subsec:bd}

In SQT it is often preferred to use basis states which are simple, convenient, and time-independent. For example, we usually choose a user-friendly basis consisting of the orthonormal eigenstates of a time-independent operator in order to furnish $\mathpzc{H}$ and to represent the observables and their evolutions. However, the conceptual nature of ECQT makes it natural to recast the Hilbert-space observables in terms of the system's experience (state history) itself. Therefore, one chooses a dynamical basis $\mathpzc{C}^{\mathrm{EC}}_{t}$, identified with sufficiently many developed states of the system, to expand all observables and their momentary evolutions. These basis states are some orthonormalization of sufficiently many past-to-present states of either the whole quantum system or its constituting subsystems. Consider a $d$-dimensional closed quantum system whose unitarily evolving states under an SQT Hamiltonian $H$ are $\varrho_{t'}=|\Psi_{t'}\rangle \langle\Psi_{t'}|$. One chooses an EC basis which at each moment $t$ is defined as follows: 
\begin{equation}
\begin{gathered}
\label{ecb}
\mathpzc{C}^{\mathrm{EC}}_{t}\subset\mathpzc{P}^{\mathrm{chosen}}_{[t_{0} , t]} \subset \mathpzc{P}_{[t_{0} , t]}, \\ 
\mathpzc{C}^{\mathrm{EC}}_{t}= \{\ket{\Psi_{t-a_{t}^{1}}},\ldots,\ket{\Psi_{t-a_{t}^{d}}}\} \;\;\;;\;\;\; a_{t}^{i=1,\ldots,d} \geqslant 0;\, \forall t, \\
w_{t-a_{t}^{l},t-a_{t}^{k}} \equiv  |m_{t-a_{t}^{l},t-a_{t}^{k}}|  < 1 \;\;;\;\; m_{t-a_{t}^{l},t-a_{t}^{k}} \equiv \langle\Psi_{t-a_{t}^{l}}|\Psi_{t-a_{t}^{k}}\rangle,\, \forall t,(l,k).
\end{gathered}  
\end{equation}
Upon this choice, all operators and all observables in the system Hilbert space are expandable in terms of the system's experience states. For example, considering a time-independent observable $O$, its expansion is expressible in terms of a rank-$d$ matrix of experience-dependent coefficients $O^{ij}(\{m_{t-a_{t}^{l},t-a_{t}^{k}}\})$ as  
\begin{equation} 
O = \textstyle{\sum_{i,j=1}^{d}}O^{ij}(\{m_{t-a_{t}^{l},t-a_{t}^{k}}\}) |\Psi_{t-a_{t}^{i}}\rangle \langle\Psi_{t-a_{t}^{j}}|.
\end{equation}
In particular, any SQT Hamiltonian $H$ of the closed quantum system (even time-independent or experience-indifferent) \emph{enjoys its EC manifestation in the form of a special $[\hskip-.8mm[d,2]\hskip-.8mm]$ EC Hamiltonian} as   
\begin{equation} 
H =  \mathbbmss{H}_{t} =\mathrm{poly}^{[\hskip-.8mm[d,2]\hskip-.8mm]} \big(\{ \varrho_{t-a_{t}^{\mathrm{I}}} \},\{ \lambda_{lk} (\{m_{t-a_{t}^{l},t-a_{t}^{k}}\} ) \}\big).
\label{cond:d-2}
\end{equation}

%%%%%%%%%%%%%%%%%%%%%%%%%%%%%%%%%%%%%%%%%%%%%%%%%%%%%%%%%%%%%%%%
\subsubsection{EC Reformulation of a General One-Qubit System}
\label{subsec: et}

We assume that $H$ is a Hamiltonian on the Hilbert space $\mathbb{C}^{2}$ of the closed system of a single qubit ($d=2$). This Hamiltonian can be arbitrary; it can be a state-independent operator (SQT); it can have arbitrary time dependence; it can have any dependence on the present-moment state or the past states of the system; or even it can have any sort of alternative features. Our aim is to find a $[\hskip-.8mm[2,2]\hskip-.8mm]$ EC representation for this Hamiltonian.  \\

The derivation we lay out here is straightforward and general. Although we can expand any input $H$ in terms of the three Pauli operators and the identity operator, our derivation is independent of such expansion. Now rather than the canonical time-independent basis $\mathpzc{C} = \{ |1\rangle , |2\rangle \}$ (e.g., the eigenvectors of the $\sigma^{3}$ Pauli operator, corresponding to $\pm1$ eigenvalues), we choose a state-history basis as $\mathpzc{C}^{\mathrm{EC}}_{t}  =  \{\ket{\Psi_{t-a_{t}}},|\Psi_{t}\rangle \,|\, a > 0,\, w_{t-a_{t},a} < 1,\forall t\}$. After orthonormalization and a particular phase choice, we choose
\begin{equation}
\hat{\mathpzc{C}}^{\mathrm{EC}}_{t} \equiv\{\ket{\hat{\Psi}_{t-a}},|\Psi_{t}\rangle \},
\label{orthonormal basis} 
\end{equation} 
where
\begin{equation}
\begin{split}
\ket{\hat{\Psi}_{t-a}} &= (1/\gamma)\big( e^{i \alpha_{t-a,t}} \ket{ \Psi_{t-a} } - w_{t-a,t} |\Psi_{t}\rangle  \big),\\
m_{t-a,t} &\equiv \langle\Psi_{t-a}|\Psi_{t}\rangle \equiv w_{t-a,t} e^{i \alpha_{t-a,t}},\,\,\ \gamma \equiv \textstyle{\sqrt{1 - (w_{t-a,t})^{2}}}.
\end{split} 
\end{equation}
Note that here for brevity, we have shortened $a_{t}$ to $a$. The Hamiltonian expansion in the basis \eqref{orthonormal basis} reads as
\begin{equation}
\begin{split}
H &= \hat{\lambda}_{t-a} |\hat{\Psi}_{t-a}\rangle\langle\hat{\Psi}_{t-a}| + \hat{\lambda}_{t} | \Psi_{t} \rangle\langle\Psi_{t} |  + \hat{\lambda}_{t-a,t} |\hat{\Psi}_{t-a} \rangle\langle \Psi_{t}| + \hat{\lambda}_{t-a,t}^{*} |\Psi_{t}\rangle \langle \hat{\Psi}_{t-a}|.
\label{lte}
\end{split}
\end{equation} 
By rewriting all operators and couplings in terms of density operators, we note 
\begin{equation}
\begin{split}
|\Psi_{t} \rangle  \langle\Psi_{t}| & = \varrho_{t}, \\
|\hat{\Psi}_{t-a}\rangle\langle\Psi_{t}| & = (1/\gamma) \big[(1/w_{t-a,a}) \varrho_{t-a} \varrho_{t}  -  w_{t-a,t} \varrho_{t} \big],  \\ 
|\hat{\Psi}_{t-a}\rangle\langle\hat{\Psi}_{t-a}| & = \big(\varrho_{t-a}  -  \{\varrho_{t-a},\varrho_{t}\}   +  (w_{t-a,t})^{2} \varrho_{t} \big)/(\gamma)^{2}, \\
\hat{\lambda}_{t} &= \mathrm{Tr}[\varrho_{t}H], \\
\hat{\lambda}_{t-a,t} &= (1/\gamma w_{t-a,t}) \mathrm{Tr}[\varrho_{t-a} H \varrho_{t}] - (1/\gamma) w_{t-a,t}\,\mathrm{Tr}[\varrho_{t} H],  \\ 
\hat{\lambda}_{t-a} &= \big(\mathrm{Tr}[\varrho_{t-a}H] - 2\mathrm{Re}\big(\mathrm{Tr}[\varrho_{t-a}H\varrho_{t}] \big) + (w_{t-a,t})^{2}\,\mathrm{Tr}[\varrho_{t} H] \big)/(\gamma)^{2} . 
\label{opb}
\end{split}
\end{equation}
The EC form of $H$ as a $[\hskip-.8mm[2,2]\hskip-.8mm]$ Hamiltonian as
\begin{equation}
H  =  \mathbbmss{H}^{[\hskip-.8mm[2,2]\hskip-.8mm]}_{t}=  \lambda_{t-a}\varrho_{t-a} + \lambda_{t}\varrho_{t}  + i \lambda_{t-a,t}^{\mathrm{I}} [\varrho_{t-a},\varrho_{t} ]  +  \lambda_{t-a,t}^{\mathrm{R}} \{\varrho_{t-a},\varrho_{t}\} 
\label{gfutl}
\end{equation}
implies that  
\begin{equation}
\begin{split}
\lambda_{t-a} &= \hat{\lambda}_{t-a}/(\gamma)^{2}, \\
\lambda_{t} &= \hat{\lambda}_{t} - (2/\gamma) w_{t-a,t} \hat{\lambda}_{t-a,t}^{\mathrm{R}}  + (w_{t-a,t}/\gamma)^{2}  \hat{\lambda}_{t-a},   \\ 
\lambda_{t-a,t}^{\mathrm{I}} &= (1/\gamma w_{t-a,t})\hat{\lambda}_{t-a,t}^{\mathrm{I}}, \\
\lambda_{t-a,t}^{\mathrm{R}} &= (1/\gamma w_{t-a,t})\hat{\lambda}_{t-a,t}^{\mathrm{R}}  - \hat{\lambda}_{t-a}/(\gamma)^{2}.
\label{lcvvlt}
\end{split}
\end{equation} 
Finally, combining Eqs. (\ref{opb}) and (\ref{lcvvlt}) gives the $\lambda$ couplings as
\begin{equation}
\begin{split}
\lambda_{t} &= \frac{1}{(\gamma)^{4}}\Big[\mathrm{Tr}[\varrho_{t}H]  +  (w_{t-a,t})^{2}\mathrm{Tr}[\varrho_{t-a}H]  - 2\,\mathrm{Re}\big(\mathrm{Tr}[\varrho_{t-a}H\varrho_{t}]\big) \Big],  \\ 
\lambda_{t-a} &= \frac{1}{(\gamma)^{4}}\Big[\mathrm{Tr}[\varrho_{t-a} H]  +  (w_{t-a,t})^{2} \mathrm{Tr}[\varrho_{t} H]  - 2\,\mathrm{Re}\big(\mathrm{Tr}[\varrho_{t-a} H \varrho_{t}]\big)\Big],  \\
\lambda_{t-a,t}^{\mathrm{R}} &= \frac{1}{(\gamma)^{4}} \Big[\big(1 + 1/(w_{t-a,t})^{2}\big)\, \mathrm{Re}\big(\mathrm{Tr}[\varrho_{t-a} H \varrho_{t}] \big)  - \big(\mathrm{Tr}[\varrho_{t-a} H]  +  \mathrm{Tr}[\varrho_{t} H] \big)\Big], \\ 
\lambda_{t-a,t}^{\mathrm{I}} &= \frac{-1}{(\gamma)^{4}}\Big[\big(1-1/(w_{t-a,t})^{2}\big)\,\mathrm{Im}\big(\mathrm{Tr}[\varrho_{t-a}H\varrho_{t}]\big)\Big] = \frac{i}{2 (\gamma w_{t-a,t})^{2}} \mathrm{Tr}\big[ H[\varrho_{t-a},\varrho_{t}] \big].    
\end{split}
\label{lco}
\end{equation}

We stress again that nothing other than the dimensionality and Hermiticity of $H$ were fed into the above derivation, hence the final result holds for any arbitrary one-qubit $H$. Interestingly, we note that the above result even applies to one-qubit EC Hamiltonians, that is, for every $H = \mathbbmss{H}_{t}^{[\hskip-.8mm[N,L]\hskip-.8mm]}$, regardless of $[\hskip-.8mm[N,L]\hskip-.8mm]$. As a result, one's favorite choice of a one-qubit EC Hamiltonian with $N \gg 2$ is identically reformulatable as a $[\hskip-.8mm[2,2]\hskip-.8mm]$ EC Hamiltonian. Here the effects of the other $N-2$ quantum memories are squeezed into the couplings in Eq. \eqref{lco}.  

%%%%%%%%%%%%%%%%%%%%%%%%%%%%%%%%%%%%%%%%%%%%%%%%%%%%%%%%%%%%%%%%
\subsubsection{EC Reformulation of Time-Independent One-Qubit SQT Hamiltonians}
\label{subsec: st}

As a special case, let us require that the Hamiltonian $H$ be time independent. In the case of an SQT Hamiltonian, it must also be state-history independent. We now want to find how the result of Eqs. (\ref{gfutl}) and (\ref{lco}) reformulates SQT as a structurally fine-tuned subset of ECQT. \\

Imposing the time independence of $H$ leads to the following conditions:
\begin{equation}
\begin{split}
\label{ihsotsqm}
\mathrm{Tr}[\varrho_{t}H] & =  \mathrm{Tr}[\varrho_{t-a}H ]  =  \mathrm{Tr}[\varrho_{0}H]  \equiv \nu=\mathrm{const.}, \\ 
i\frac{\partial  m_{t-a,t}}{\partial a} & =  \langle\Psi_{t-a} | H | \Psi_{t}\rangle,\\
\mathrm{Tr}[\varrho_{t-a}H\varrho_{t} ] & =  m^{*}_{t-a,t} \langle\Psi_{t-a} | H | \Psi_{t} \rangle =  i m^{*}_{t-a,t}\frac{\partial  m_{t-a,t}}{\partial a} , \\
\mathrm{Re}([\mathrm{Tr}[\varrho_{t-a}H\varrho_{t}]\big) & = -(w_{t-a,t})^{2}\frac{\partial  \alpha_{t-a,t}}{\partial a}, \\
\mathrm{Im}\big(\mathrm{Tr}[\varrho_{t-a} H \varrho_{t}] \big) & =  \frac{1}{2} \frac{\partial  (w_{t-a,t})^{2}}{\partial a}, \\ 
m_{t-a,t} & =  m_{0,a}  =  w_{0,a} e^{i \alpha_{0,a}}, \,\,\, \gamma   =  \textstyle{\sqrt{1 - (w_{0,a})^{2}}}.  
\end{split}
\end{equation}
Hence the $\lambda$ couplings in Eq. \eqref{lco} are simplified as  
\begin{equation}
\begin{split}
\label{lcoisqm}
\lambda_{t-a}  &=  \frac{1}{(\gamma)^{4}} \Big[\big(1 + (w_{0,a})^{2}\big) \nu  + 2(w_{0,a})^{2} \frac{\partial  \alpha_{0,a}}{\partial a}\Big]  =  \lambda_{t}, \\ 
\lambda_{t-a,t}^{\mathrm{R}}  &= -\frac{1}{(\gamma)^{4}} \Big[2\nu  +  \big(1 + (w_{0,a})^{2}\big) \frac{\partial  \alpha_{0,a}}{\partial a}\Big],  \\ 
\lambda_{t-a,t}^{\mathrm{I}}  &=  \frac{1}{(\gamma)^{2}} \frac{\partial \ln(w_{0,a})}{\partial a}.
\end{split}
\end{equation}
The physics in Eqs. \eqref{lcoisqm} becomes clearer by writing the right-hand side (RHS) in terms of the energy spectrum of $H$. Using $H = E_{1} |1\rangle\langle1| + E_{2} |2\rangle \langle 2|$ (where $\ket{1}$ and $\ket{2}$ are ground and excited states, respectively) and $|\Psi_{0}\rangle = c_{0,1} \ket{1} + c_{0,2} \ket{2}$ gives
\begin{equation}
m_{t-a,t} = m_{0,a} = \frac{1}{2} \big( e^{ - i a E_{1} } + e^{ - i a E_{2} } \big) +  \frac{s_{0}}{2} \big( e^{ - i a E_{2} } - e^{ - i a E_{1} } \big), 
\end{equation}
where $s_{0} = ( | c_{0,2} |)^{2} - (| c_{0,1}|)^{2}$ and $-1 \leqslant s_{0} \leqslant 1$. The state-history fidelity and the state-history phase factor (the argument of the autocorrelation) are accordingly obtained as 
\begin{equation}
\begin{split}
F_{t-a,t} &\equiv (w_{t-a,t})^{2} = (w_{0,a})^{2} = \frac{1}{2} \big[ 1 + \big(1-(s_{0})^{2}\big) \cos( a \,\Delta E) \big],\\
\alpha_{t-a,t} &= \alpha_{0,a} = - \arctan \Big[ \frac{(1 + s_{0}) \sin(a E_{2}) - (1 - s_{0}) \sin(a E_{1})} { (1 + s_{0}) \cos(a E_{2}) + (1 - s_{0}) \cos(a E_{1})} \Big], 
\label{u1}
\end{split}
\end{equation} 
where $\Delta E =  E_{2}  -  E_{1}$ is the energy gap. From Eq. \eqref{lcoisqm} the EC $\lambda$ couplings depend not only on $w_{0,a}$ but also on the derivates of the state-history fidelity and the phase factor with respect to the quantum memory distance and also on the initial-moment expectation of the energy, 
\begin{equation}
\label{u2}
\begin{split}
\frac{\partial F_{0,a}}{ \partial a}  &= - \frac{\big(1-(s_{0})^{2}\big) \Delta E}{2} \sin(a \, \Delta E), \\
\frac{\partial \alpha_{0,a}}{ \partial a} & = - \frac{(s_{0})^{2} (E_{1} + E_{2}) \sin^{2}(a \, \Delta E/2) + s_{0} \Delta E + 1 } {  1 + \big(1-(s_{0})^{2}\big) \cos(a \, \Delta E)}, \\
\nu & = \big( E_{1} + E_{2} + s_{0} \,\Delta E \big)/2.
\end{split}
\end{equation} 

It is straightforward to show that out of the four $(\lambda_{t} , \lambda_{t-a}, \lambda_{t-a,t}^{\mathrm{R}}, \lambda_{t-a,t}^{\mathrm{I}})$ couplings of a $[\hskip-.8mm[2,2]\hskip-.8mm]$ EC Hamiltonian, there are only two physically relevant ones, $\bar{\lambda}_{t-a,t}^{\mathrm{R}} \equiv \lambda_{t-a} + \lambda_{t-a,t}^{\mathrm{R}}$ and $\lambda_{t-a,t}^{\mathrm{I}}$ \cite{ECQT-1}. Every time-independent one-qubit SQT Hamiltonian $H$ is identically expressible within ECQT in the form of a $[\hskip-.8mm[2, 2]\hskip-.8mm]$ EC Hamiltonian $\mathbbmss{H}_{t}$ given in Eq. \eqref{gfutl}. The physically relevant couplings of the equivalent $\mathbbmss{H}_{t}$ are determined in terms of the energy spectrum of $H$ and the initial state of the qubit as follows: 
\begin{equation}
\begin{split}
\label{crh}
\bar{\lambda}_{t-a,t}^{\mathrm{R}} &= \frac {  2 - \big(1-(s_{0})^{2}\big) (E_{1} + E_{2}) + s_{0} \Delta\, E - [ (E_{1} + E_{2}) + s_{0} \big( 1 - (s_{0})^{2}\big) \Delta E  ] \cos ( a\, \Delta E )    }{ 1 - \big( 1 -(s_{0})^{2} \big)^{2} \cos^{2} ( a\, \Delta E)} ,   \\
\lambda_{t-a,t}^{\mathrm{I}} &= - \frac {\big( 1 -(s_{0})^{2} \big) \Delta E \sin(a\, \Delta E )}{ 1 - \big( 1 -(s_{0})^{2} \big)^{2} \cos^{2} ( a\, \Delta E)}.
\end{split}
\end{equation} 
For the two extreme choices of the initial conditions, $s_{0} = 0$, for the maximally symmetric initial state, and $s_{0} = \pm 1$, for which the initial state coincides with the ground or excited eigenstate of the SQT Hamiltonian, the above $\lambda$ couplings become
\begin{align}
&s_{0} = 0 \;:\;\;\;\;\;  \bar{\lambda}_{t-a,t}^{\mathrm{R}} = 2\; \frac{ 1 - (E_{1} + E_{2}) \cos^{2} (a\, \Delta E /2)}{ \sin^{2} ( a\, \Delta E)  }  \;\;\;,\;\;\; \lambda_{t-a,t}^{\mathrm{I}} = - \; \frac { \Delta E  }{  \sin( a \,\Delta E)} \\
&s_{0} = \pm 1 \;:\;\;\;\;\; \bar{\lambda}_{t-a,t}^{\mathrm{R}} = 2  \pm \Delta E - (E_{1} + E_{2})  \cos ( a\, \Delta E )      \;\;\;,\;\;\; \lambda_{t-a,t}^{\mathrm{I}} = 0.
\label{crhws}
\end{align} 

As an example, consider $H = \sigma^{3}$. Applying the choice to the result \eqref{ihsotsqm} and upon using 
\begin{align*}
e^{i a \sigma^{3}}  &=  \cos(a) \openone  + i \sin(a) \,\sigma^{3},\\
m_{t-a,t}  &=  \cos(a)  - i \nu\sin(a), \\
\nu  &=  \langle\Psi_{0} | \sigma^{3} | \Psi_{0}\rangle \\
\gamma^{2} &= 1 - (w_{0,a})^{2} = [1-(v)^{2}] \sin^{2}(a),
\label{pzr}
\end{align*}
we obtain $H = \sigma^{3} = \mathbbmss{H}^{[\hskip-.8mm[2,2]\hskip-.8mm]}_{t} $ of the form \eqref{gfutl} with the following couplings:
\begin{equation}
\begin{split}
\lambda_{t-a}  =&  \lambda_{t} =  -\frac{\nu}{1 - (\nu)^{2}}\csc^{2}(a) = - \frac{\nu}{(\gamma)^{2}}  ,\\ 
\lambda^{\mathrm{R}}_{t-a,t}  =& \nu/(\gamma w_{0,a})^{2}, \\ 
\lambda^{\mathrm{I}}_{t-a,t} =&  -\frac{\sin(2 a)}{2(\gamma w_{0,a})^{2}}[1 - (\nu)^{2}] = - \frac{\cot(a)}{(w_{0,a})^{2}} .   
\end{split}
\label{couplingsohs3}
\end{equation} 

It is interesting to note that the EC form for even this simple one-qubit SQT Hamiltonian $H = \sigma^{3}$ is highly nontrivial. The reason, however, is significant, manifesting that the EC couplings of the SQT Hamiltonians should have such \textit{fine-tuned} structure that (as a net effect) wash out all state-history dependences. In other words, this mirrors the statement that SQT is a highly constrained low-dimensional reduction of ECQT. We will return to this important point in Sec. \ref{sec:thms}.

%%%%%%%%%%%%%%%%%%%%%%%%%%%%%%%%%%%%%%%%%%%%%%%%%%%%%%%%%%%%%%%%
\subsection{Reformulation of SQT by ECQT: Basis- and Decomposition-Independent Recipe} 
\label{sec:sqt-reformulation}

Now we develop a second recipe to obtain representations of SQT Hamiltonians within ECQT with the following properties: (a) it applies indiscriminately to finite-dimensional closed quantum systems; (b) it is independent of choosing any particular basis states; (c) it is independent of any tensor-product decomposition; and (d) it is largely flexible in the state-history resources the system needs to exploit for making its EC Hamiltonian. As a result, this recipe is a flexible general algorithm. \\

A convenient way to present this reformulation algorithm is based on a straightforward \textit{identity} which is satisfied by state-history couplings of any EC Hamiltonian (\ref{ech:g}). In other words, this identity manifests Eq. (\ref{ech:g}) in an alternative way. For simplicity and specificity of the discussion, we only focus on the \textit{primitive} EC Hamiltonians. The details of this identity is relegated to Appendix \ref{sec:identity}. This identity basically establishes a relation between the couplings $\lambda_{t}^{s\pm}$, the operators $\mathbbmss{h}_{t}^{s\pm}$, and the EC Hamiltonian $\mathbbmss{H}^{\Xi}_{t}$, as defined in Eq. (\ref{ech:g}). \\

Let us assume that the states of the closed system of interest are in a Hilbert space $\mathpzc{H}$ of dimension $d$, and these states evolve unitarily under an SQT Hamiltonian $H$. This Hamiltonian must be independent of the state-history of the system, including the state at the present moment state and the initial state. For further simplicity and clarity, we consider a time-independent $H$---extension to time-dependent Hamiltonians in SQT is straightforward. Furthermore, as before we assume the purity of the state history. This does not impact the special physical context of our study. \\

We now want to reformulate the evolution of the above SQT system in the form of a pure-state-history $[\hskip-.8mm[N,L]\hskip-.8mm]$ EC Hamiltonians [Eqs. \eqref{ech:g} and \eqref{monomials}] subject to the the dimensionality-matching constraint $d^{-} + d^{+} = (d)^{2}$. The key is to obtain all the matching $\lambda_{t}$ couplings by directly imposing $\mathbbmss{H}_{t}^{\Xi} = H$. Following the terminology of Appendix \ref{sec:identity}, here we determine the elements of $\boldsymbol{h}_{t}\equiv \big( h_{t}^{1-},\ldots,h_{t}^{d^{-}-},h_{t}^{1+},\ldots,h_{t}^{d^{+}+}\big)^{T}$, where $h_{t}^{s\pm} \equiv \mathrm{Tr}\big[i\mathbbmss{h}_{t}^{s\pm} \mathbbmss{H}_{t}^{\Xi} \big]$, by imposing the condition $\mathbbmss{H}_{t}^{\Xi} = H$ and by using the spectral decomposition $H = \sum_{n=1}^{d} E_{n}|n \rangle \langle n|$ and specifying the initial state as $\varrho_{0} = |\Psi_{0}\rangle \langle \Psi_{0}|$, with $|\Psi_{0}\rangle = \textstyle{\sum_{n=1}^{d}} c_{0,n} |n\rangle$,
\begin{equation}
\label{scp:hpmisqm} 
\begin{split}
& h_{t}^{k+} =  2w [k]\,\textstyle{\sum_{n=1}^{d}} (|c_{0,n}|)^{2} E_{n} \cos\big[ E_{n} ( t_{k_{1}} - t_{ k_{{L_{k}}}} ) - \alpha[k]  \big],  \\
& h_{t}^{j-} =  2w [j]\,\textstyle{\sum_{n=1}^{d}} (|c_{0,n}|)^{2} E_{n} \sin \big[ E_{n} ( t_{j_{1}} - t_{ j_{{L_{j}}}}) - \alpha[j]  \big].
\end{split} 
\end{equation} 
The combined state-history fidelities $w[s]$ and argument factors $\alpha[s]$ can be determined using the following result for the state-history two-point functions for any time-independent $H$:
\begin{equation}
\label{scp:m} 
\begin{split}
 m_{t' t''} & = \langle \Psi_{t'} | \Psi_{t''} \rangle = \langle\Psi_{0} |e^{-i(t''-t')H} |\Psi_{0} \rangle = \textstyle{\sum_{n = 1}^{d}} (|c_{0,n}|)^{2} e^{- i E_{n} (t''-t')} = m_{0 , t''- t'} = w_{0 , t''- t'}  e^{i \alpha_{0 , t''- t'}},\\
 \alpha_{0 , t''- t'} & = \mathrm{arg} \Big(\textstyle{\sum_{n = 1}^{d}} (|c_{0,n}|)^{2}e^{- i E_{n} (t''-t')}\Big),\\
w_{0 , t''- t'} & = \left|\textstyle{\sum_{n = 1}^{d}} (|c_{0,n}|)^{2}\, e^{- i E_{n} (t''-t')}\right|,\,\forall t',t'' \in [t_0,t].
\end{split}
\end{equation}
Hence
\begin{equation}
\label{scp:mcm}
\begin{split}
w [s] &=  \textstyle{\prod_{r=1}^{L_{s}-1}}  w_{0 , ( t_{s_{r+1}} - t_{s_r})} = \alpha_{0 , ( t_{s_{r+1}} - t_{s_r})} = \textstyle{\prod_{r=1}^{L_{s}-1}}  \mathrm{arg} \Big(\textstyle{\sum_{n = 1}^{d}} (|c_{0,n}|)^{2} e^{- i E_{n} ( t_{s_{r+1}} - t_{s_r} )}\Big),\\
\alpha [s] & =  \textstyle{\sum_{r=1}^{L_{s}-1}} \alpha_{0 , ( t_{s_{r+1}} - t_{s_r})} = \textstyle{\sum_{r=1}^{L_{s}-1}} \big| \textstyle{\sum_{n = 1}^{d}} (|c_{0,n}|)^{2} e^{- i E_{n} ( t_{s_{r+1}} - t_{s_r} )} \big|.
\end{split}
\end{equation}
To sum up, what we presented above is the solution of the equation $\mathbbmss{H}_{t}^{\Xi} = H$ for any time-independent Hamiltonian of any (finite) Hilbert space dimension by replacing the RHS of Eq. (\ref{wgi:ech}) by the parameters computed above. It is straightforward to check that this solution reproduces all the results of the one-qubit closed system in SQT as derived in Secs. \ref{subsec: et} and \ref{subsec: st}. \\

We can also remark on the spectrum of the $[\hskip-.8mm[N,L]\hskip-.8mm]$ choices such that the corresponding EC Hamiltonian can represent SQT Hamiltonians. If for a moment we relax the assumption of the state-history purity, then a general $[\hskip-.8mm[N,L]\hskip-.8mm]$ EC Hamiltonian has $\sum_{k=1}^{L} (N)^{k}=N\big((N)^{L}-1\big)/(N-1)$ independent real parameters which are its couplings. Hence to find a representation of an SQT Hamiltonian defined on a Hilbert space of dimension $d$ as an $[\hskip-.8mm[N,L]\hskip-.8mm]$ EC Hamiltonian, we need to choose $N$ and $L$ such that $\sum_{k=1}^{L} (N)^{x} \geqslant (d)^{2}$---or simply $(N)^{L} \geqslant (d)^{2}$. This parameter counting is consistent with the dimensionality constraint $d^{+}+d^{-}=(d)^{2}$, and it shows that when we take $N=d$ then $L$ can be $\geqslant 2$. This should be contrasted with the basis-dependent recipe [Eq. (\ref{cond:d-2})] the universal choice of $N = \dim(\mathpzc{H})=d$ and $L=2$ is dictated by the very recipe. However, the basis-independent recipe has this flexibility that any SQT Hamiltonian here admits infinitely many equivalent $[\hskip-.8mm[N,L]\hskip-.8mm]$ EC representations. In this recipe the favorite choice of $[\hskip-.8mm[N,L]\hskip-.8mm]$ is given by measures such as context-dependent efficiencies---not by the recipe. 

%%%%%%%%%%%%%%%%%%%%%%%%%%%%%%%%%%%%%%%%%%%%%%%%%%%%%%%%%%%%%%%%
\subsection{Inclusion Theorems} 
\label{sec:thms}

A Hamiltonian belongs to (or is \textit{admissible} in) SQT if it has a representation either as a strictly time-independent Hermitian operator or a time-dependent Hermitian operator whose time dependence is independent of all the state-history of the system, namely, its dynamics is independent of both initial state and state trajectory of the system. \\

Let us consider a quantum system with a Hilbert space $\mathpzc{H}$. In the following we provide three theorems which compare dynamics of this system in SQT and ECQT. In particular, to make the comparison on an equal footing, below we presume that to obtain an SQT-admissible representation for an EC Hamiltonian we do not lift it to a larger Hilbert space. This caveat is important because as we demonstrated in Sec. \ref{sec:SQTsim}, the dynamical action of some subclasses of EC Hamiltonians can in principle be simulated within the SQT framework if we allow using a higher Hilbert space and multiple copies of the system. \\

%%%%%%%%%%%%%%%%%%%%%%%%%%%%%%%%%%%%%%%%%%%%%%%%%%%%%%%%%%%%%%%%
\begin{theorem}
There exist infinitely many $\mathbbmss{H}^{[\hskip-.8mm[1,1]\hskip-.8mm]}_{t}$'s without any representation as time-independent or time-dependent SQT Hamiltonians $H_{t}$.
\label{thm:A}
\end{theorem}

\textit{Proof}: Consider the family of $[\hskip-.8mm[1,1]\hskip-.8mm]$ Hamiltonians made of pure states and state-history-independent couplings, 
\begin{equation}
\mathbbmss{H}_{t}^{[\hskip-.8mm[1,1]\hskip-.8mm]} = \lambda_{t-a}\varrho_{t-a},
\label{11a}
\end{equation}
where $\lambda_{t-a}$ is assumed independent of $\mathpzc{P}_{[t_{0} , t]}$. \\

\textbf{\textsf{Part (a)}}: Nonexistence of \textit{time-independent} SQT-admissible representations \\

First, it is straightforward to prove that in general Eq. \eqref{11a} does not admit any time-independent operator representation. This EC Hamiltonian has $d$ eigenvalues, where $d = \dim(\mathpzc{H})$; one eigenvalue equals $\lambda_{t-a}$, with all others being equal to $0$. Correspondingly, up to linear combinations, it has one eigenstate associated to the nonzero eigenvalue and $d-1$ degenerate eigenstates as follows. $|\Psi_{t-a}\rangle$, the quantum state of the system at the moment $t-a$, is the eigenstate with eigenvalue $\lambda_{t-a}$. The basis states of the subspace orthogonal to $|\Psi_{t-a}\rangle$, namely $|\Psi_{t-a}^{\perp i}\rangle\, (i = 1, \ldots, d-1)$ are its degenerate eigenstates. The only point which we use in the proof is that $\mathbbmss{H}_{t}^{[\hskip-.8mm[1,1]\hskip-.8mm]}$ has a dynamical eigenspace; its eigenstates change in time. That is, unless special cases where the entire trajectory is trivial, time-independent states cannot be its eigenstates. Now assume that there is a representation of this Hamiltonian in the form of a time-independent operator $H=\mathbbmss{H}_{t}^{[\hskip-.8mm[1,1]\hskip-.8mm]}$. As a result, this would also imply the existence of a time-independent eigenstate with eigenvalue $\lambda_{t-a}$. This, however, does not comply with the eigenspace of $\mathbbmss{H}_{t}^{[\hskip-.8mm[1,1]\hskip-.8mm]}$. Thus, we conclude that $\mathbbmss{H}_{t}^{[\hskip-.8mm[1,1]\hskip-.8mm]}$'s does not have any time-independent SQT-admissible representation. \\

\textbf{\textsf{Part (b)}}: Nonexistence of \textit{time-dependent} SQT-admissible representations \\

Assume that $H_{t} = \mathbbmss{H}_{t}^{[\hskip-.8mm[1,1]\hskip-.8mm]}$ is a distinct representation of $\mathbbmss{H}_{t}^{[\hskip-.8mm[1,1]\hskip-.8mm]}$ in the form of a time-dependent Hamiltonian admissible in SQT. Thus, the unitary time-evolution operator of the system finds a representation as follows:
\begin{equation}
U_{t' \to t''} = \lim_{m \to \infty} \textstyle{\prod_{l = 0}^{m}} e^{- i\frac{t''-t'}{m} H_{t' + \frac{t''-t'}{m} l}}.
\label{aiu} 
\end{equation}
Note that, similar to $H_{t}$, $U_{t' \to t''}$ is also independent of the past-to-present states of the system, i.e., independent of all $\varrho_{t_{<}}$'s, $\forall t_{<} \in [t_{0},t]$, including all possible choices for the $\varrho_{0}$ initialization of the state of the system. Hence for every $\varrho_{0}$, $H_{t}$ must produce the exact same trajectory as $\mathbbmss{H}^{[\hskip-.8mm[1,1]\hskip-.8mm]}_{t}$ does. Combining Eqs. \eqref{11a} and \eqref{aiu}, we obtain 
\begin{equation}
H_{t} =\lambda_{t-a} \, U_{t_{0} \to t-a} \,\varrho_{0} \,U_{t_{0} \to t-a}^{\dag}.
\label{impie}
\end{equation}
This is an identity that state-history-independent $H_{t}$ should satisfy for any $\varrho_{0}$. However, it is evident that Eq. \eqref{impie}, across the defining space of the initial state, cannot be satisfied identically. Based on our assumption, the left-hand side (LHS) is state-history independent, hence it does not change with $\varrho_{0}$. However, because neither $\lambda_{t-a}$ nor the SQT-admissible $U_{t_{0}\to t}$ depend on the history of the system, the RHS unavoidably depends on $\varrho_{0}$ and varies with it. The only way for Eq. \eqref{11a} to satisfy Eq. \eqref{impie} is to enforce $U_{t_{0} \to t}$---hence $H_{t_{<}}$'s too---to depend on $\varrho_{0}$. This, however, violates SQT. Thus, we conclude that Eq. \eqref{11a} does not have any time-dependent SQT-admissible representation. \\

\textbf{\textsf{Part (c)}}: \\

We can extend the above proof to $\mathbbmss{H}_{t}^{[\hskip-.8mm[1,1]\hskip-.8mm]}$'s where now $\lambda_{t-a}$ has a generic arbitrary dependence on $\mathpzc{P}_{[t_{0} , t]}$. Part (a) of the proof, the absence of time-independent operator representations, goes identically the same. For Part (b) we reconstruct Eqs. \eqref{aiu} and \eqref{impie} exactly as before. The impossibility argument of Eq. \eqref{impie} holding for all possible $\varrho_{0}$'s as an identity goes similarly. The LHS is of course independent of $\varrho_{0}$, and for generic dependence of $\lambda_{t-a}$ on the state history, dependence of the RHS on $\varrho_{0}$ cannot be cancelled out obviously. Hence the impossibility holds also for generic $\mathbbmss{H}_{t}^{[\hskip-.8mm[1,1]\hskip-.8mm]}$'s as presented in Eq. \eqref{11a}.

\hfill$\blacksquare$
\\

\textit{Remark.---}The above arguments can be identically repeated for any other $\varrho_{t_{<}}$ with $t_{<}$ ranging from $t_{0}$ to $t$. That is, if we consider versions of $\mathbbmss{H}^{[\hskip-.8mm[1,1]\hskip-.8mm]}_{t}$ with $(t_{0},\varrho_{0})$ replaced with any $(t_{<},\varrho_{t_{<}})$, the same argument carries over, which implies that the impossibility is limited only to an enforced dependence on the initial state.

%%%%%%%%%%%%%%%%%%%%%%%%%%%%%%%%%%%%%%%%%%%%%%%%%%%%%%%%%%%%%%%%
\begin{theorem}
There exist infinitely many $\mathbbmss{H}_{t}^{[\hskip-.8mm[N,L]\hskip-.8mm]}$'s with $N,L > 1$ without any representation as time-independent or dynamical SQT Hamiltonians $H_{t}$.
\label{thm:B}
\end{theorem}

\textit{Proof}: Argument goes structurally similar to the proof of Theorem \ref{thm:A}. For specificity, let us consider the family of $\mathbbmss{H}_{t}^{[\hskip-.8mm[2,2]\hskip-.8mm]}$'s as
\begin{align}
\mathbbmss{H}_{t}^{[\hskip-.8mm[2,2]\hskip-.8mm]} = \lambda_{t-a,t}^{\mathrm{R}}\{ \varrho_{t-a} , \varrho_{t}\} + i \lambda_{t-a,t}^{\mathrm{I}}[ \varrho_{t-a} , \varrho_{t}], 
\label{22a}
\end{align}
where $\lambda_{t-a,t}^{\mathrm{I},\mathrm{R}}$ depend on $\mathpzc{P}_{[t_{0} , t]}$ generically---or in special cases may be independent of it. \\

\textbf{\textsf{Part (a)}}: \\

The eigensystem of this \textit{generic} EC Hamiltonian shall be time-dependent. Its nontrivial eigenvalues depend on $(w_{t-a,a} ; \lambda_{t-a,t}^{\mathrm{I}},\lambda_{t-a,t}^{\mathrm{R}})$, where now the nontrivial eigenstates are given by certain linear combinations of $|\Psi_{t}\rangle $ and $|\Psi_{t-a}\rangle$, with the defining coefficients depending on $(m_{t-a,a} ; \lambda_{t-a,t}^{\mathrm{I}},\lambda_{t-a,t}^{\mathrm{R}})$. Because the eigensystem of this $\mathbbmss{H}_{t}^{[\hskip-.8mm[2,2]\hskip-.8mm]}$ is time dependent in a generic sense, it is impossible to always find a \textit{time-independent} operator representation $H$ for it. \\

\textbf{\textsf{Part (b)}}: \\

If $\mathbbmss{H}_{t}^{[\hskip-.8mm[2,2]\hskip-.8mm]}$ can be identically represented as a \textit{time-dependent} $H_{t} = \mathbbmss{H}_{t}^{[\hskip-.8mm[2,2]\hskip-.8mm]}$, then the following equation should hold as an identity across the choice space of $\varrho_{0}$'s:
\begin{equation}
H_{t} = \lambda_{t-a,t}^{\mathrm{R}}\{ U_{t_{0} \to t-a} \varrho_{0}\; U_{t_{0}, t-a}^{\dag} , U_{t_{0} \to t}  \varrho_{0} U_{t_{0} \to t}^{\dag}\} + i \lambda^{\mathrm{I}}_{t-a,t} [ U_{t_{0}\to t-a} \varrho_{0} U_{t_{0} \to t-a}^{\dag},\,U_{t_{0}\to t} \varrho_{0} U_{t_{0}\to t}^{\dag}]. 
\label{impieet}
\end{equation}
The LHS is again independent of $\varrho_{0}$. All the $U$'s on the RHS are independent of $\varrho_{0}$ too. This equation becomes an identity for all $\varrho_{0}$ only if the total dependence of the RHS on $\varrho_{0}$ cancels out. The cancelation imposes, however, fine-tuned dependencies of the $\lambda^{\mathrm{I}}_{t-a,a}$ and $\lambda^{\mathrm{R}}_{t-a,a}$ couplings on the state history. Hence we conclude the nonexistence of an SQT-admissible $H_{t}$ for this $\mathbbmss{H}_{t}^{[\hskip-.8mm[2,2]\hskip-.8mm]}$ if $\lambda$'s are state-history independent or have \textit{generic} arbitrary dependence on the state history of the system. With appropriate adjustments, this proof carries over obviously for higher values of $N$ and $L$ and more general $[\hskip-.8mm[N,L]\hskip-.8mm]$ EC Hamiltonians.

\hfill$\blacksquare$
\\

%%%%%%%%%%%%%%%%%%%%%%%%%%%%%%%%%%%%%%%%%%%%%%%%%%%%%%%%%%%%%%%%
\begin{theorem}
The set of SQT Hamiltonians $S_{\mathrm{SQT}}$ is a measure-zero subset of the set of all $\mathbbmss{H}_{t}^{[\hskip-.8mm[N,L]\hskip-.8mm]}$ ECQT Hamiltonians $S^{[\hskip-.8mm[N,L]\hskip-.8mm]}_{\mathrm{ECQT}}$; i.e., for the cardinalities (or measures) of these sets we have $|S_{\mathrm{SQT}}\vert \ll \vert S^{[\hskip-.8mm[N,L]\hskip-.8mm]}_{\mathrm{ECQT}}\vert$. In addition, we also have $\bigcup_{N,L} S^{[\hskip-.8mm[N,L]\hskip-.8mm]}_{\mathrm{ECQT}}\subset S_{\mathrm{ECQT}}$ and $|\bigcup_{N,L} S^{[\hskip-.8mm[N,L]\hskip-.8mm]}_{\mathrm{ECQT}}\vert \ll \vert S_{\mathrm{ECQT}}\vert$.
\label{thm:C}
\end{theorem}

\textit{Proof}: We have shown in Sec. \ref{sec:sqt-vs-ecqt} that SQT is a subset of ECQT in the sense that any SQT-admissible Hamiltonian $H_{t}$ can also be represented as some EC Hamiltonian $\mathbbmss{H}_{t}$ (i.e., such that $\mathbbmss{H}_{t}=H_{t}$). Whereas, as proved in the above theorems, there exist \textit{infinitely} many $\mathbbmss{H}^{[\hskip-.8mm[N,L]\hskip-.8mm]}_{t}$'s that cannot admit any representation as a time-independent or time-dependent Hamiltonian in SQT. Moreover, in the proof of Theorem \ref{thm:B} we showed that $[\hskip-.8mm[N,L]\hskip-.8mm]$ EC Hamiltonians which are not SQT-admissible are in fact \emph{typical}. Hence $S_{\mathrm{SQT}} \subset \bigcup_{N,L} S^{[\hskip-.8mm[N,L]\hskip-.8mm]}_{\mathrm{ECQT}}$ and $S^{[\hskip-.8mm[N,L]\hskip-.8mm]}_{\mathrm{ECQT}} \nsubseteq S_{\mathrm{SQT}}$. We note also that by construction (Sec. \ref{sec:ECQT}) $S_{\mathrm{ECQT}}^{[\hskip-.8mm[N,L]\hskip-.8mm]}$ is itself a measure-zero subset of all EC Hamiltonians $S_{\mathrm{ECQT}}$, because polynomials are themselves measure-zero subset of all functions.   

\hfill$\blacksquare$
\\

The above argument can be illustrated explicitly in the case of one-qubit SQT Hamiltonians $H$. First we consider time-independent Hamiltonians worked out in Sec. \ref{subsec: st}, following Eqs. \eqref{lcoisqm} or equivalently \eqref{crh}. The state-history independence of the EC Hamiltonian imposes the following \emph{constraints} on its EC form:
\begin{equation}
\label{hyp:srf} 
\begin{split}
\mathrm{Tr}[H] & =  \lambda_{t} + \lambda_{t-a}  +  2 (w_{t-a,t})^{2} \lambda_{t-a,t}^{\mathrm{R}} = \text{(some state-history independent) const.}, \\
\mathrm{Tr}[H\varrho_{t}] & = \lambda_{t} + (w_{t-a,t})^{2} (\lambda_{t-a} + 2\lambda_{t-a,t}^{\mathrm{R}}) = \text{(some state-history independent) const.}
\end{split}   
\end{equation}
It is obvious from a geometric perspective that Eqs. (\ref{hyp:srf}) reduce to the generically $4$-dimensional coupling space of a $[\hskip-.8mm[2,2]\hskip-.8mm]$ EC Hamiltonian ($\mathrm{I\!R}^{4}$) to a codimension-$2$ hypersurface of $\mathrm{I\!R}^4$. This represents geometrically the \textit{measure-zero} inclusion of time-independent SQT Hamiltonians within ECQT. \\

Equations (\ref{hyp:srf}) reveal two physically interesting additional points. (i) For an ECQT Hamiltonian to be non-generically equivalent to to an SQT Hamiltonian, its EC couplings should necessarily be state-history dependent in fine-tuned manners. (ii) For such equivalence to occur, the EC Hamiltonian couplings should bear state-history dependences which are nonanalytic, e.g., of the form $1/(w_{t-a,t})^{2}$. \\

We now consider time-dependent one-qubit SQT Hamiltonians recast in an EC form. As we have derived in Sec. \ref{subsec: et}, the requirement $\mathbbmss{H}_{t}^{[\hskip-.8mm[2,2]\hskip-.8mm]} = H$ implies that, for example, the coupling $\lambda_{t-a,t}^{\mathrm{I}}$ must be in the form of Eq. \eqref{lco}, 
\begin{equation} 
\label{again}  
\lambda_{t-a,t}^{\mathrm{I}} =  \frac{i}{2 (\gamma w_{t-a,t})^{2}} \mathrm{Tr}\big[[\varrho_{t-a},\varrho_{t}]H\big]. 
\end{equation}
Moreover, in the same section we have proved that the $\lambda_{t-a,t}^{\mathrm{I}}$ coupling of an $\mathbbmss{H}^{[\hskip-.8mm[2,2]\hskip-.8mm]}$---with the form as in Eq. (\ref{gfutl})---satisfies the relation
\begin{equation}
\lambda_{t-a,t}^{\mathrm{I}} = \frac{i} {2 (\gamma w_{t-a,t})^{2}} \mathrm{Tr}\big[ [ \varrho_{t-a} , \varrho_{t} ] \mathbbmss{H}^{[\hskip-.8mm[2,2]\hskip-.8mm]}_{t}  \big].  
\label{identity} 
\end{equation}  
The point is that, although Eq. \eqref{identity} is an identity and always holds, Eq. \eqref{again} is a nontrivial equation, which does not always hold \textit{reversely}. That is, we do not always obtain a solution for Eq. \eqref{again} if we go the reverse way back from $\mathbbmss{H}_{t}$ and look for a corresponding SQT Hamiltonian. This single observation by itself is a clear demonstration of \textit{measure-zero} inclusion of SQT within ECQT. \\

As Ref. \cite{ECQT-1} has shown, the Schr\"{o}dinger or von Neumann equations which arise from EC Hamiltonians are not only nonlinear but also nonlocal in time. Indeed, in all the examples of the previous sections we have seen that for the equality of an SQT Hamiltonian and an $[\hskip-.8mm[N,L]\hskip-.8mm]$ EC Hamiltonian, the composite Hermitian operator which defines the EC Hamiltonian must be fine-tuned structurally. It is evident from where these structural fine-tunings come. They are necessary for canceling out the otherwise-inevitable nonlinearities and time nonlocalities of the deformed Schr\"{o}dinger equations which are resulted from the generic EC Hamiltonians. The statement holds for both EC Hamiltonians which are made entirely from the system's density operators and those which include fixed operators likewise. We can bring up the famous example of the Gross-Pitaevskii Hamiltonian which is a $[\hskip-.8mm[1,1]\hskip-.8mm]$ EC Hamiltonian, as shown in Eq. \eqref{hgp}. In addition, we can show that Weinberg's generalized quantum mechanics \cite{Weinberg-testQM, Weinberg-testQM3} can be reformulated as subclasses of EC Hamiltonians. These cases provides yet two other examples which highlight that EC Hamiltonians deviate from linear state-history-independent Hamiltonians of SQT, unless structural fine-tunings are imposed.

%%%%%%%%%%%%%%%%%%%%%%%%%%%%%%%%%%%%%%%%%%%%%%%%%%%%%%%%%%%%%%%%
\section{Behavioral Richness of EC Unitary Evolutions}
\label{sec:phase2-perturbation}

In this section we contrast EC unitary evolutions with those in quantum mechanics in order to elucidate the general lessons drawn from the formal inter-theoretical investigations in Sec. \ref{sec:sqt-vs-ecqt}. In doing so, in Sec. \ref{sec:phase1} and following Ref. \cite{ECQT-1}, we present a classification of one-qubit EC behavioral phases and demonstrate examples of dynamical phase transitions among them. Next we turn to a detailed study of the deformations and perturbations of SQT by means of additive EC Hamiltonians, which is done by means of time-local, near-Markovian, and deeply non-Markovian EC deformations throughout Secs. \ref{susbec:PRI-A} -- \ref{subsec:PRI-D}. Finally, we analyze further enriching behavioral impacts of state-history-dependent couplings in EC Hamiltonians. 

%%%%%%%%%%%%%%%%%%%%%%%%%%%%%%%%%%%%%%%%%%%%%%%%%%%%%%%%%%%%%%%%
\subsection{Dynamics and Behavioral Phases}
\label{sec:phase1}

Our behavioral analysis in this subsection concerns one-qubit EC unitary evolutions with $[\hskip-.8mm[N,L]\hskip-.8mm]$ EC Hamiltonians under the following specifications: (i) pure-state histories, (ii) with low-order $[\hskip-.8mm[N,L]\hskip-.8mm]$, especially $N ,L \leqslant 3$, with constant couplings. The $[\hskip-.8mm[1,1]\hskip-.8mm]$ and $[\hskip-.8mm[2,2]\hskip-.8mm]$ EC Hamiltonians are given in Eq. (\ref{11,22}), or equivalently Eq. (\ref{11,22:pure}) for pure-state histories. For concreteness we choose  the quantum memory distances to be, respectively, constants $a>0$ and $\vec{a}_{t} = (a_{1}, a_{2}) = (a,0)$. Choosing $\dot{a}_{1}=a_{2}=0$ is only for simplification while still leaving sufficient behavioral novelties in our analyses. The behavioral phase diagram of the one-qubit closed quantum systems under these EC Hamiltonians was investigated in Ref. \cite{ECQT-1}, upon numerically solving the EC Schr\"{o}dinger equation \eqref{ecso118}. The classification of behavioral phases, which we review now, mirrors the defining topological and geometrical characteristics of the EC dynamics of one-qubit pure states and their state-history two-point functions \eqref{fam} (or fidelities), formed at sufficiently late times, as \textit{behavioral dynamical attractors}.  \\

It is obtained that EC unitary evolutions of the closed one-qubit system under the above specifications develop \textit{five principal phases of behavioral attractors}, which are described as follows:

\begin{itemize}

\item Phase \textbf{\textsf{I}}: The wavefunction $|\Psi_{t}\rangle$ and its state-history fidelity $w_{t-a,t}$ develop fixed-point attractors. 
 
\item Phase \textbf{\textsf{II}}: The wavefunction develops large regular simple periodic oscillations such that the state-history fidelity has a fixed-point attractor, typically with tiny fluctuations. 
 
\item Phase \textbf{\textsf{III}}: The wavefunction and the state-history fidelity feature large irregular oscillations.
 
\item Phase \textbf{\textsf{IV}}: The wavefunction and the state-history fidelity develop large structural oscillations in the form of sequential repetitions of temporally invariant behavioral modules which are marked with robustly ordered patterns.
 
\item Phase \textbf{\textsf{V}}: The wavefunction features bistable attractor behavior of consecutively switching between two metastable states. These state swaps are in the form of sharp transitions. Accordingly and in a manner which is synchronized with the wavefunction swaps, the state-history fidelity features consecutive sharp minima which equally parse the plateau of $w_{t-a,t} = 1$.

\end{itemize} 

\begin{figure}[tp]
\includegraphics[width=\linewidth]{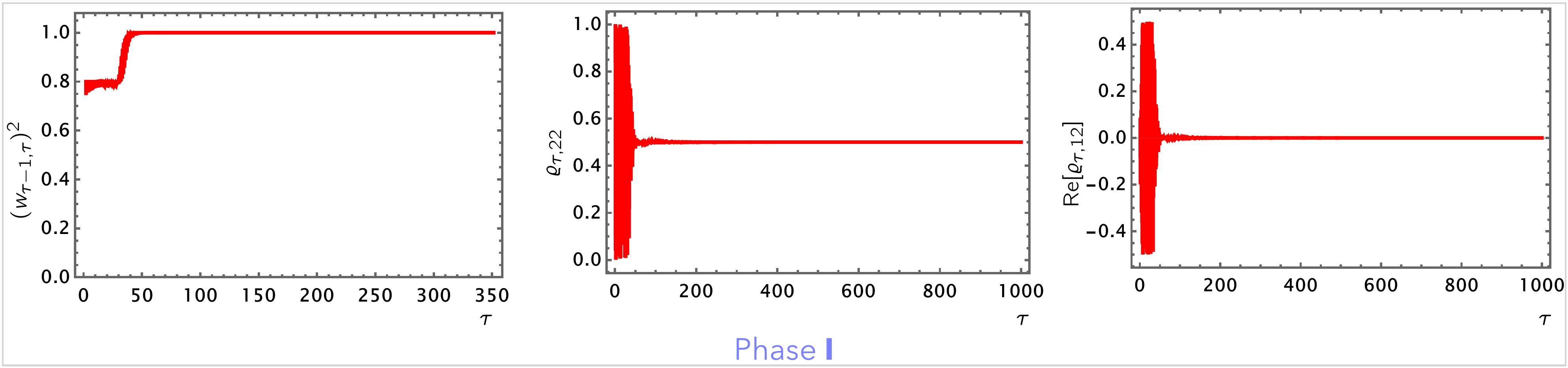}\\
\includegraphics[width=\linewidth]{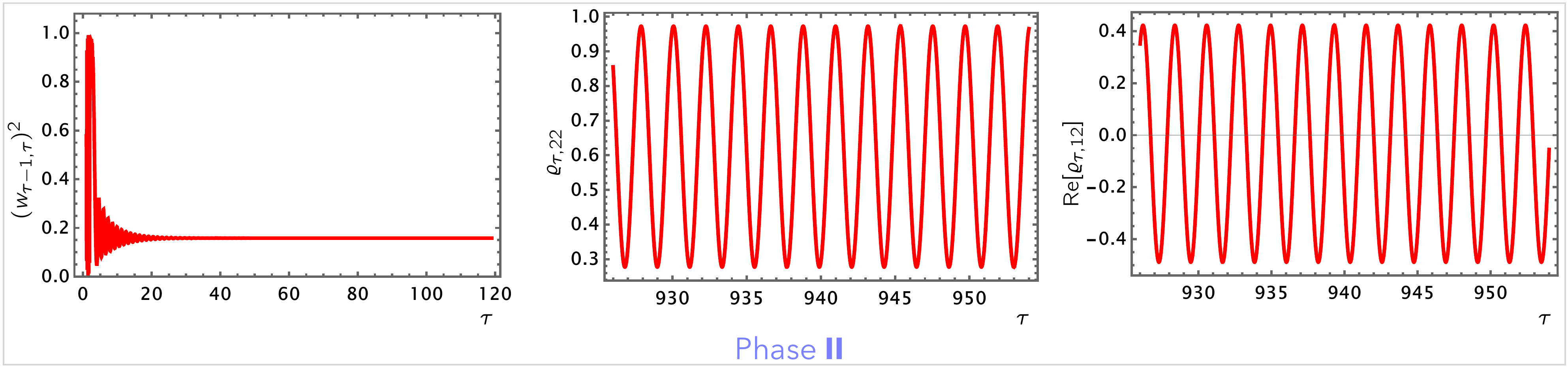}\\
\includegraphics[width=\linewidth]{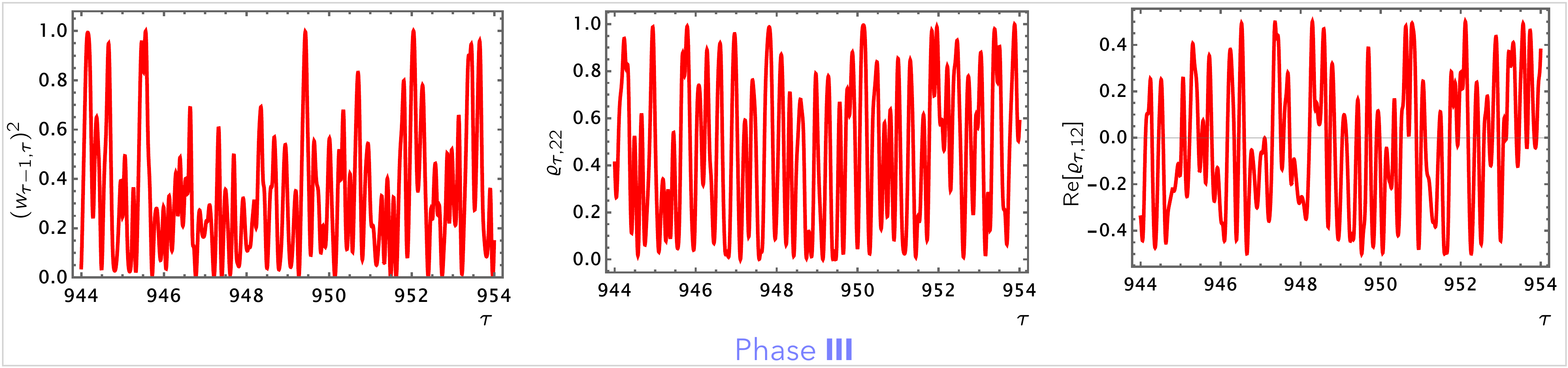}\\
\includegraphics[width=\linewidth]{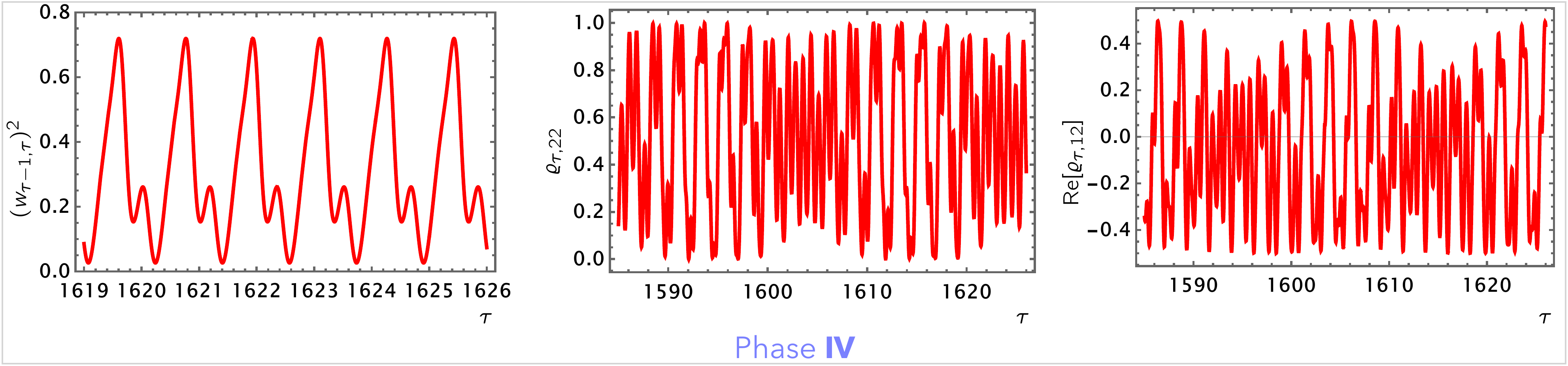}\\
\includegraphics[width=\linewidth]{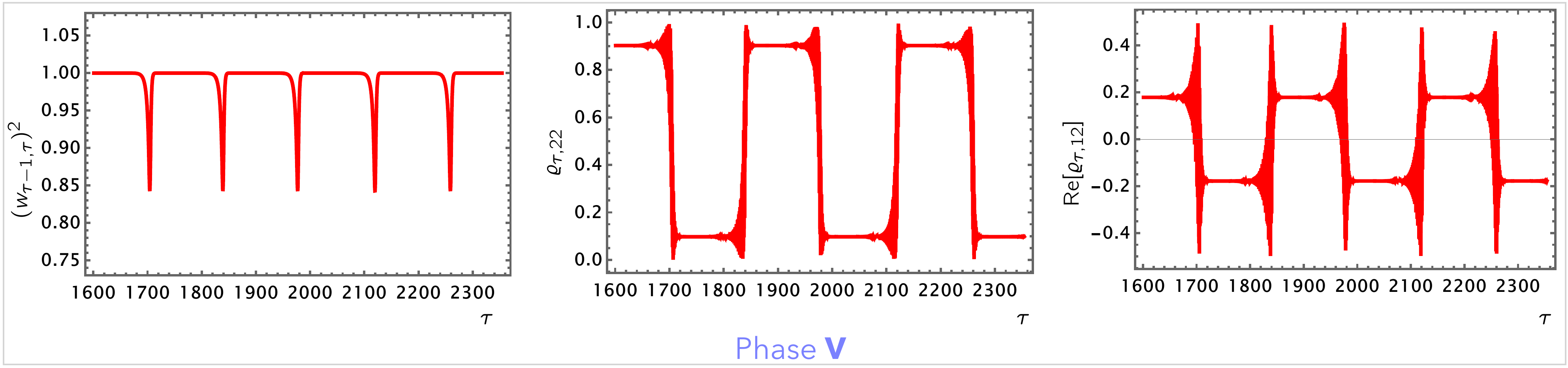}
\caption{The five behavioral phases of one-qubit EC unitary evolutions with $[\hskip-.8mm[2,2]\hskip-0.8mm]$ EC Hamiltonians given in Eq. (\ref{11,22}). The plots show the EC time evolutions of $\varrho_{t}$ the representative numerical solutions corresponding to the EC equation (\ref{ecvovn}), or equivalently, the EC Schr\"{o}dinger equation given in Eq. (\ref{ecso118}), with the initial history in the time interval $[0,a]$ developed by an SQT ``kicker'' Hamiltonian $H_{\mathrm{kicker}} = 5 \sigma^{2}$. We have assumed in all plots $a=3$, $\varrho_{0} = |+ \rangle \langle +| = (1/2)(\openone + \sigma^{1})$, $\lambda_{t-a}=0$, and we have dropped the dynamically irrelevant coupling $\lambda_{t}$. The $(\lambda^{\mathrm{I}}_{t-a,t}, \lambda^{\mathrm{R}}_{t-a,a})$ couplings for the five phases are as follows: Phase \textbf{\textsf{I}} $(12.49, 0)$; Phase \textbf{\textsf{II}} $(-0.4, 1.15)$; Phase \textbf{\textsf{III}} $(-2, 8.15)$; Phase \textbf{\textsf{IV}} $(-0.97, 1.98)$; Phase \textbf{\textsf{V}} $(2.19, 1.32)$. The EC dynamics in the plots are vs. the dimensionless variable $\tau\equiv t/a$, and we have used the basis of the eigenvectors $\{|1\rangle,|2\rangle\}$ of $\sigma^{3}$ to represent $\varrho_{\tau}$.}
\label{fig:phase1}
\end{figure}

Alongside the above five principal phases, one encounters their crossover deformations in the total behavioral phase diagram. These deformations exhibit a diverse variety of mixed behavioral phases which blend together some characteristics of five principal phases---see Ref. \cite{ECQT-1} for detailed explanations and demonstrations. A major lesson that one draws from such behavioral investigations of EC unitary evolutions is that the behavioral characteristics of SQT closed and open quantum systems are intrinsically interfused in ECQT. Indeed, EC behavioral phases feature significantly rich qualities which are unprecedented in SQT-based systems, as they can be generated by means of \textit{strongly synergetic interplays of non-Markovianity and unitarity}. This point is crucial and is further confirmed in the forthcoming toy models in the upcoming subsections of Sec. \ref{sec:phase2-perturbation}. Figure \ref{fig:phase1} shows some representative examples of the five principal behavioral phases of one-qubit wavefunctions.  \\

A remarkable characteristic of non-hybrid (primitive) EC unitary evolutions is a version of deep non-Markovianity which in Ref. \cite{ECQT-1} was referred to as ``robust non-Markovianity.'' It is the characteristic that such EC Hamiltonians generate nontrivial evolutions only when the corresponding largest quantum memory distances $a$ exceed (or are equal to) finite lower bounds set by the defining couplings of the EC Hamiltonians. In contrast, hybrid EC evolutions do not admit any lower bounds on the largest quantum memory distances---see, e.g., Secs. \ref{susbec:PRI-A} -- \ref{subsec:PRI-D}---such that they have nontrivial near-Markovian regimes.  \\
 
\begin{figure}[tp]
\includegraphics[width=\linewidth]{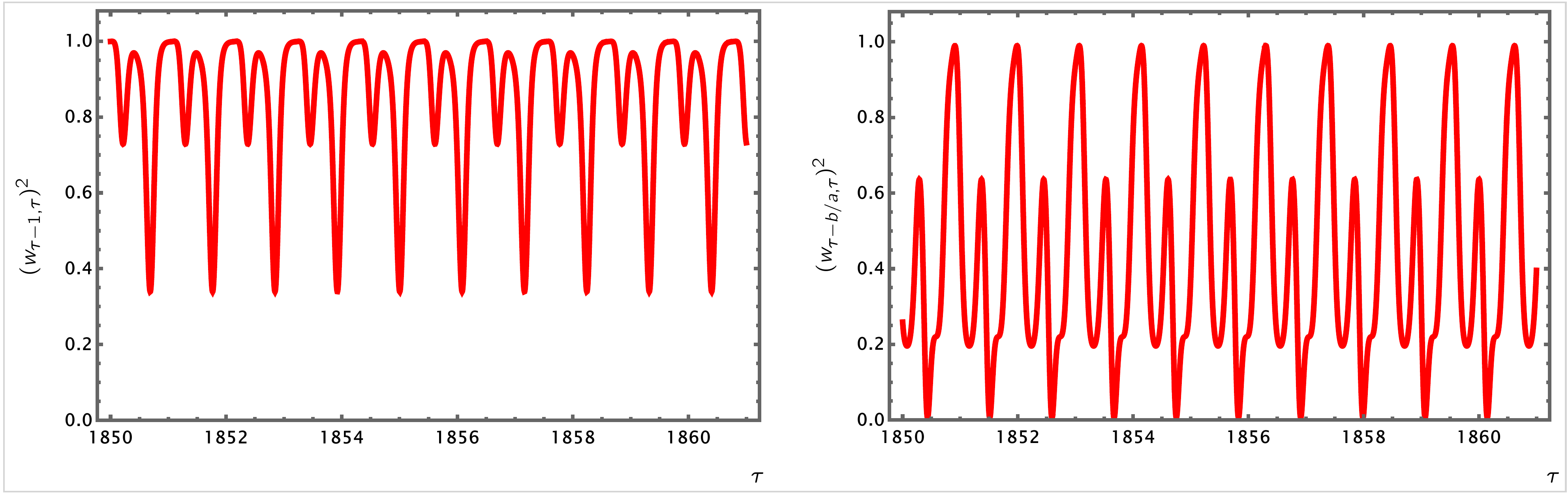}\\
\includegraphics[width=\linewidth]{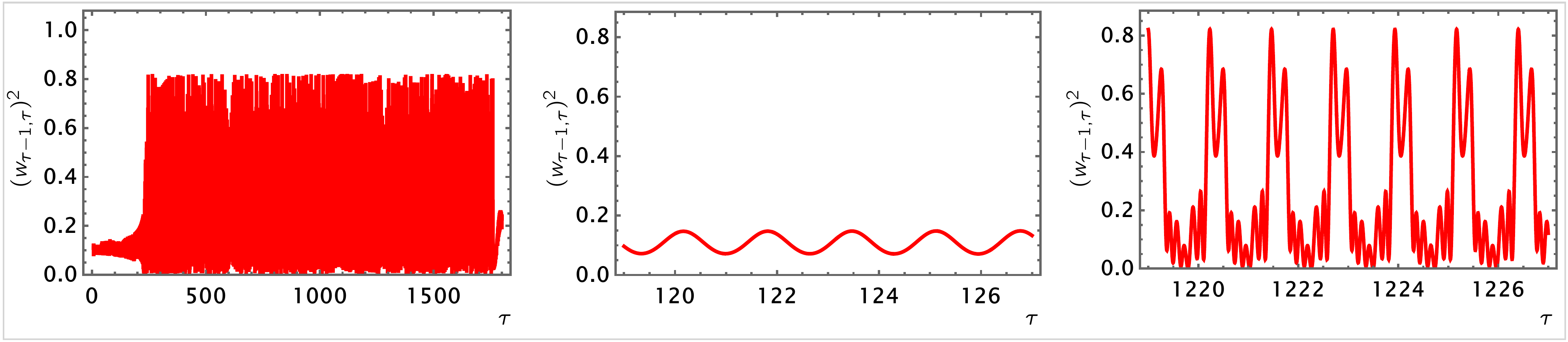}
\caption{Two highlighted examples of higher-order behavioral phases of a one-qubit closed system developed, respectively, by [top] $\mathbbmss{H}_{t}^{[\hskip-.8mm[3,3]\hskip-.8mm]}$ given in Eq. (\ref{23,233}) and [bottom] the same $\mathbbmss{H}_{t}^{[\hskip-.8mm[3,3]\hskip-.8mm]}$ additively deformed by a genuinely EC Hermitian operator which in the $\sigma^{3}$ basis is equal to $\nu(\varrho_{t})^{T}$. The behaviors correspond to the numerical solutions corresponding to the EC equation (\ref{ecvovn}), or equivalently (for the top one), the EC Schr\"{o}dinger equation given in Eq. (\ref{ecso118}), with the initial history in the time interval $[0,a]$ developed by an SQT ``kicker'' Hamiltonian $H_{\mathrm{kicker}} = 5 \sigma^{2}$. The specification data is given by: [top] $H_{\mathrm{kicker}} = 41 \sigma^{2}$, $\varrho_{0}=|+\rangle \langle +| = (1/2)(\openone + \sigma^{1})$, $a=3$, $a/b=1.35$, $\kappa^{\mathrm{R}}_{t-a,t-b,t}=3.855$, and $\kappa^{\mathrm{I}}_{t-a,t-b,t}=5.124$; [bottom] $H_{\mathrm{kicker}} = 41 \sigma^{2}$, $\varrho_{0}=|+\rangle \langle +| = (1/2)(\openone + \sigma^{1})$, $a=3$, $a/b=2$, $\nu=4.74$, $\kappa^{\mathrm{R}}_{t-a,t-b,t}=0$, and $\kappa^{\mathrm{I}}_{t-a,t-b,t}=-2$. The EC dynamics in the plots are vs. the dimensionless variable $\tau\equiv t/a$, and we have used the basis of the eigenvectors $\{|1\rangle,|2\rangle\}$ of $\sigma^{3}$ to represent $\varrho_{\tau}$.} 
\label{fig:higherphase}
\end{figure}

\begin{figure}[tp]
\includegraphics[width=\linewidth]{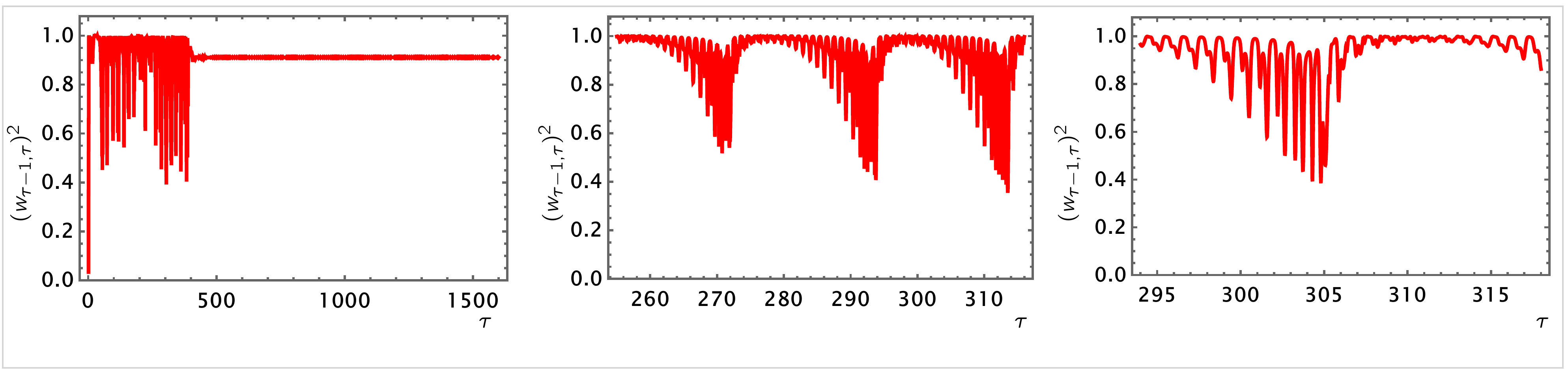}
\caption{A purely internal dynamical transition to the behavioral Phase \textbf{\textsf{II}} from the multiscale behavioral Phases (\textbf{\textsf{IV}},\textbf{\textsf{V}}), developed by a higher-order EC unitary evolution of the one-qubit wavefunction with the $[\hskip-.8mm[3,3]\hskip-.8mm]$ EC Hamiltonian (\ref{23,233}). The solution is obtained upon solving numerically the corresponding EC equation (\ref{ecvovn}), or equivalently, the EC Schr\"{o}dinger equation given in Eq. (\ref{ecso118}). The specification data is given by $\varrho_{0}=|+\rangle\langle +|=(1/2)(\openone + \sigma^{1})$, $H_{\mathrm{kicker}} = 800 \sigma^{2}$, $a=4$, $a/b\approx 1.73$, $\nu=0.05$, $\kappa^{\mathrm{R}}_{t-a,t-b,t}=3.25$, and $\kappa^{\mathrm{I}}_{t-a,t-b,t}=5$. The EC dynamics in the plots are vs. the dimensionless variable $\tau\equiv t/a$, and we have used the basis of the eigenvectors $\{|1\rangle,|2\rangle\}$ of $\sigma^{3}$ to represent $\varrho_{\tau}$.} 
\label{fig:higherphase2}
\end{figure}

We now turn to behavioral effects sourced by lowest \textit{higher-order interactions} among state-history information in EC unitary evolutions.  Specifically, we focus on one-qubit pure-state histories corresponding to the $[\hskip-.8mm[3,3]\hskip-.8mm]$ EC Hamiltonian given in Eq. (\ref{23,233}). Figures \ref{fig:higherphase} and \ref{fig:higherphase2} present three remarkable examples of such higher-order EC effects on the unitary evolution of the one-qubit closed quantum system. The two plots of the top row in Fig. \ref{fig:higherphase} present an $N=L=3$ EC evolution one-qubit wavefunction, with control parameters specified in its caption, where the evolution of the state-history fidelities $(w_{t-a,t})^{2}$ (left) and $(w_{t-b,t})^{2}$ (right) feature a highly ordered blending of the principal behavioral Phases \textbf{\textsf{IV}} and \textbf{\textsf{V}}. The three plots of the lower row in Fig. \ref{fig:higherphase} depict the EC evolution of the one-qubit state-history fidelity ($(w_{t-a,t})^{2}$) under an interestingly deformed $N=L=3$ EC Hamiltonian whose definition is given in the caption of the figure. As one observes, the autocorrelation $(w_{t-a,t})^{2}$ goes through an initial transient period of regular simple oscillations with small amplitude. The third-order interactions among state-history information, however, trigger a sharp dynamical transition to a behavioral Phase \textbf{\textsf{IV}} which is everlasting. In this post-transition phase, not only the oscillation amplitudes enlarge maximally (by about an order of magnitude), but also the defining structure of the constituting modules of Phase \textbf{\textsf{IV}} have become remarkably intricate. Moreover, one observes in Fig. \ref{fig:higherphase2} that the one-qubit closed quantum system makes a purely internal sharp dynamical transition---triggered by third-order interactions among state-history information---from a blend of behavioral Phases \textbf{\textsf{IV}} and \textbf{\textsf{V}} to the behavioral Phase \textbf{\textsf{II}}. \\

%%%%%%%%%%%%%%%%%%%%%%%%%%%%%%%%%%%%%%%%%%%%%%%%%%%%%%%%%%%%%%%%
%%%%%%%%%%%%%%%%%%%%%%%%%%%%%%%%%%%%%%%%%%%%%%%%%%%%%%%%%%%%%%%%
  
There are two points in order. (i) An important feature of higher-order EC evolutions is that the wavefunction and state-history autocorrelations can develop across different temporal scales distinct behavioral phases (particularly among the five principal phases) in manners which are highly orchestrated. Figure \ref{fig:higherphase2} serves as a clear example. (ii) Moreover, we highlight that one-qubit wavefunctions naturally develop a diverse variety of unprecedented purely-internal long-time dynamical transitions between these behavioral phases. These EC dynamical transitions are not only beyond the reach of usual quantum systems, but also show exotic physical qualities. As analyzed in Ref. \cite{ECQT-1}, the unitary evolutions of one-qubit wavefunctions under $[\hskip-.8mm[N,L]\hskip-.8mm]$ EC Hamiltonians with $N \geqslant 3$ or $L \geqslant 3$ can easily develop long-time-scale transitions between all the five principal behavioral phases. Note that these dynamical transitions of the behavior are not caused by any external control, nor are they effects of changing the couplings of the EC Hamiltonians. They are purely internal, caused by the inbuilt interactions between the quantum information in unitary state history of the closed system. A clear example of these dynamical transitions is shown in Fig. \ref{fig:higherphase2}, following the explanation given above. \\ 

Finally, we refer to Ref. \cite{ECQT-1} for the details of behavioral phase partitioning of these EC unitary evolutions in terms of the wide ranges of the defining couplings, along with analytic analyses of robust non-Markovianity, and the other relevant behavioral specificities.  
  
%%%%%%%%%%%%%%%%%%%%%%%%%%%%%%%%%%%%%%%%%%%%%%%%%%%%%%%%%%%%%%%%
\subsection{Deforming SQT Evolutions by Time-Local EC Hamiltonians} 
\label{susbec:PRI-A}

Consider a closed quantum system with a $d$-dimensional Hilbert space $\mathpzc{H}$ and pure states $\varrho_{t}  = |\Psi_{t}\rangle \langle\Psi_{t}|$, $\forall t$. We evolve this system unitarily under an EC Hamiltonian which has also an SQT part. This Hamiltonian represents deformation of a SQT Hamiltonian (denoted by $H$) by a Hermitian non-SQT Hamiltonian which we choose to be $\dot{\varrho}_{t}$. We stress that the deformation operator $\dot{\varrho}_{t}$ is indeed state-history dependent, because it depends on the system's initial state $\varrho_{0}$, and hence according to Element \ref{def:3} is itself an EC Hamiltonian. For simplicity we assume that $H$ and the deformation parameter $\xi $ are time independent. The hybrid EC Hamiltonian of the system reads as 
\begin{equation}
\label{wow} 
\mathbbmss{H}_{t} = H - \xi \dot{\varrho}_{t}.   
\end{equation} 

We are interested in the state-history-driven behavior of the closed system under Eq. \eqref{wow}, specifically how the wavefunction $|\Psi_{t}\rangle$ makes transitions between and across the eigenstates of the SQT Hamiltonian $H$ during its unitary evolution. We obtain the general exact solution, describing the dynamics of the system for general deformation parameter $\xi$. However, what specially interests us is the wavefunction behavior when the EC deformation is a small perturbation of SQT corresponding to infinitesimal $\xi $.
\\

%%%%%%%%%%%%%%%%%%%%%%%%%%%%%%%%%%%%%%%%%%%%%%%%%%%%%%%%%%%%%%%%
%%%%%%%%%%%%%%%%%%%%%%%%%%%%%%%%%%%%%%%%%%%%%%%%%%%%%%%%%%%%%%%%

The unitary evolution of the system is described by the EC von Neumann equation (\ref{ecvovn}). Considering an initial pure state $\varrho_{0}=|\Psi_{0} \rangle \langle \Psi_{0}|$, the dynamics is given by the EC Schr\"{o}dinger equation (\ref{ecvos}), or equivalently, 
\begin{equation} 
\label{ecsovn}   
(1 - i \xi ) \ket{\dot{\Psi}_{t}} = - i ( H - i\xi  \varepsilon_{t} \openone ) |\Psi_{t}\rangle,
\end{equation} 
where we have defined the system's ``energy function'' 
\begin{equation}
\label{etv} 
\varepsilon_{t} \equiv \mathrm{Tr}[\varrho_{t} \mathbbmss{H}_{t}].
\end{equation}
Note that $\varepsilon_{t}$ can also be rewritten as 
\begin{equation}
\label{eto} 
\varepsilon_{t} = \langle \Psi_{t} | \mathbbmss{H}_{t} | \Psi_{t}\rangle = -i \langle\dot{\Psi}_{t} | \Psi_{t}\rangle = \mathrm{Tr}[\varrho_{t} H] - \xi  \, \mathrm{Tr}[\varrho_{t}  \dot{\varrho}_{t}] = \mathrm{Tr}[\varrho_{t} H].   
\end{equation}
By using this function, interestingly the EC dynamics can be put in an equivalent form as
\begin{equation}
i |\dot{\Psi}_{t}\rangle = \frac{1}{1 - i \xi } (H - i \xi  \varepsilon_{t} \openone ) |\Psi_{t}\rangle  \equiv \mathbbmss{H}_{t}^{\mathrm{eff}} |\Psi_{t}\rangle,
\label{uii} 
\end{equation}
where
\begin{equation}
\mathbbmss{H}_{t}^{\mathrm{eff}} =  \frac{1}{1 + (\xi )^{2}} \big(H + (\xi )^{2} \varepsilon_{t} \openone\big) + i \frac{\xi }{1 + (\xi )^{2}} (H - \varepsilon_{t} \openone)
\end{equation}
is a non-Hermitian effective Hamiltonian. Note that when $H=\xi =0$, we have $|\dot{\Psi}_{t}\rangle=0$. \\

%%%%%%%%%%%%%%%%%%%%%%%%%%%%%%%%%%%%%%%%%%%%%%%%%%%%%%%%%%%%%%%%
%%%%%%%%%%%%%%%%%%%%%%%%%%%%%%%%%%%%%%%%%%%%%%%%%%%%%%%%%%%%%%%%

A comment is in order before we proceed to solve the dynamics of the system. What we observe in Eq. \eqref{uii} is that a non-Hermitian Hamiltonian is dynamically equivalent to the (original) Hermitian EC Hamiltonian \eqref{wow}. The physical equivalence of the Hamiltonian \eqref{wow}, whose Hermiticity guarantees both the time-evolution unitarity and the reality of the spectrum of the Hamiltonian, and a non-Hermitian Hamiltonian might sound peculiar. Nevertheless, this equivalence is mathematically consistent and represents one of the distinctive features of ECQT. We note that the non-Hermiticity of the Hamiltonian \eqref{uii} is harmless, respecting the unitarity of the evolution (hence $(\Vert\Psi_{t}\Vert)^{2}=1$) and the reality of the spectrum of the Hamiltonian. What underlies this equivalence is the following observation:
\begin{equation}
\label{harmless}
\begin{gathered}  
\mathrm{Tr} \big[ \varrho_{t}  \big( \mathbbmss{H}_{t}^{\mathrm{eff}} - (\mathbbmss{H}_{t}^{\mathrm{eff}})^{\dag} \big)  \big] = 2 i\, \mathrm{Im} \big(\langle\Psi_{t} |  \mathbbmss{H}_{t}^{\mathrm{eff}}|\Psi_{t}\rangle\big)  = 0,\, \forall t.
\end{gathered}
\end{equation} 
In SQT, Hamiltonians should be state independent, hence the dynamical constraint \eqref{harmless}, despite being weaker than Hermiticity, can be satisfied at all times only by Hermitian Hamiltonians. In the context of ECQT, unitary evolutions are generated by Hamiltonians which are typically state-history dependent, hence physically equivalent representations in terms of non-Hermitian Hamiltonians which nevertheless satisfy Eq. \eqref{uii} can be developed. \\

Now we proceed to obtain the EC behavior of the closed system. Equations (\ref{eto}) and (\ref{uii}) result in the following final form of the EC Schr\"{o}dinger equation \eqref{ecsovn}:
\begin{equation}
\label{uiiw8}  
\ket{\dot{\Psi}_{t} } = \frac{\xi  - i }{1 + (\xi )^{2}} \big( H - i \xi  \varepsilon_{t}\openone   \big) |\Psi_{t}\rangle .
\end{equation}

We now use the orthonormal basis defined by all the eigenstates of $H$ to work out the general solution to Eq. \eqref{uiiw8} and obtain the EC flow of the system's wavefunction across these $d $ energy levels which are assumed to have a nondegenerate ground state, named  $\ket{1}$, to a nondegenerate maximally excited stated, named $\ket{d}$. Hence one rewrites Eq. \eqref{uiiw8} in terms of the complete set of parameters $\{E_n\}$, the energy eigenvalues of $H$, and a complete set of paired variables $(p_{t,n},\varphi_{t,n})$ as  
\begin{equation}
\begin{gathered}
\label{hap} 
H = \textstyle{\sum_{n=1}^{d}} E_{n} |n \rangle \langle n|,\,\, E_{1} < E_{2}\leqslant \ldots <E_{d},\\
|\Psi_{t}\rangle  = \textstyle{\sum_{n=1}^{d}} c_{t,n} \ket{n} \equiv \sum_{n=1}^{d} \textstyle{\sqrt{p_{t,n}}} \,e^{i \varphi_{t,n}} |n\rangle, \\ 
p_{t,n} = \varrho_{t,nn} = (|c_{t,n}|)^{2}, \\
\varphi_{t,n} - \varphi_{t,m} = \mathrm{arg}\big( \varrho_{t,nm} \big).
\end{gathered}  
\end{equation}
The EC evolution of the variables $p_{n,t}$ (the probabilities we are interested in) and the arguments $\varphi_{t,n}$ are coupled, however, in a one-way manner. The dynamical system which describes their evolution reads
\begin{equation}
\label{lds}
\begin{gathered} 
\frac{\dot{p}_{t,n}}{p_{t,n}} = \frac{2 \xi }{1 + (\xi )^{2} } \big( E_{n} -  \varepsilon_{t} \big), \\ 
\dot{\varphi}_{t,n} = - \frac{1}{1 + (\xi )^{2} } \big( E_{n} +  (\xi )^{2} \varepsilon_{t} \big).  
\end{gathered}
\end{equation}
The energy function $\varepsilon_{t}$ which contributes to the RHS of the above equations is critical for restoring instantaneous unitarity of this closed-system EC dynamics. To play its role as explained, $\varepsilon_{t}$ must depend collectively on all momentary probabilities $p_{t,n}$'s in a specific unique form as
\begin{equation}
\label{urf} 
\varepsilon_{t} = \textstyle{\sum_{n=1}^{d}} p_{t,n} E_{n}.
\end{equation}

The central point of our study is obtained by the $\dot{p}_{t,n}$ equations in the dynamical system \eqref{lds}, without needing to know the specific form of $\varepsilon_{t}$ as a function of $\{p_{t,n}\}$. These equations imply that the evolving probabilities $p_{t,n}$ corresponding to the energy levels $n > 1$ and $n < d $ satisfy the following equations:
\begin{equation}
\label{ap8}
\begin{gathered} 
\frac{\dot{p}_{t,n}}{p_{t,n}} - \frac{\dot{p}_{t,1}}{p_{t,1}} = + \frac{2 (E_{n} - E_{1})}{1 + (\xi )^{2}} \xi,\, \forall n > 1  \\
\frac{\dot{p}_{t,n}}{p_{t,n}} - \frac{\dot{p}_{t,d}}{p_{t,d}} = - \frac{2 (E_{d} - E_n)}{1 + (\mu)^{2} }\xi,\, \forall n < d  
\end{gathered}
\end{equation}
from whence
\begin{equation}
\label{ap8ipr}
\begin{gathered} 
\frac{p_{t,n}}{p_{t,1}}  = \frac{p_{0,n}}{p_{0,1}} e^{2\xi \frac{E_{n} - E_{1}}{1 + (\xi )^{2} }t},  \, \forall n > 1 \\
\frac{p_{t,n}}{p_{t,d}}  = \frac{p_{0,n}}{p_{0,d}} e^{-2\xi \frac{E_{d} - E_{n} }{1 + (\xi )^{2} }t},  \, \forall n < d .
\end{gathered}
\end{equation}

It is observed that merely depending on the sign of the coupling $\mu$, all $p_{t,n > 1}/p_{t,1}$ or all $p_{t,n<d}/p_{t,d}$ monotonically decay to zero with exponentially fast rate in the unitary process. Following Eq. \eqref{ap8ipr}, we can also determine the reference probabilities $p_{t,1}$ and $p_{t,d}$. As mathematically implemented by the function $\varepsilon_{t}$ [Eq. \eqref{urf}] in Eqs. \eqref{lds}, time evolution unitarity of the EC system validates the constraint $\sum_{n=1}^{d} p_{t,n} = 1$ at all moments. Hence the probabilities are determined completely as 
\begin{equation}
\label{1Ns}
\begin{split} 
p_{t,1} &= \Big ( 1 + \textstyle{\sum_{n = 2}^{d}}  (p_{0,n}/p_{0,1}) \, e^{2 \xi  \frac{E_{n} - E_{1}}{1+(\xi )^{2}} t} \Big)^{-1}, \\  
p_{t,d} &=  \Big( 1 + \textstyle{\sum_{n = 1 }^{d  - 1}} (p_{0,n}/p_{0,d}) \, e^{-2 \xi  \frac{E_{d} - E_{n}}{1+(\xi )^{2}} t} \Big)^{-1}.  
\end{split}
\end{equation}
These two probabilities are indeed equal to the momentary fidelities between the system's state and the two edge eigenstates of the Hamiltonian $H$, $\ket{1}$ and $\ket{d}$, respectively. Hence for $\xi  > 0$ ($\xi  < 0$), the time derivate of the fidelity $F(\varrho_{t},|d \rangle \langle d|)$ ($F(\varrho_{t},|1\rangle \langle 1|)$) is positive definite, and the fidelity is continually drawn to maximum value $1$, as we see in Eqs. (\ref{ap8ipr}) and (\ref{1Ns}). Thus, either the edge eigenstate of the SQT Hamiltonian are the asymptotic fixed-point attractors of the unitary evolution generated by the hybrid EC Hamiltonian \eqref{wow}, regardless of the value of $\xi $ and solely depending on its sign. That is, 
\begin{equation}
\label{fdlw8}
\begin{gathered} 
\varrho_{\infty} \equiv \theta(-\xi ) |1 \rangle \langle 1| + \theta(\xi ) |d  \rangle \langle d|, \\ 
F(\varrho_{t} , \varrho_{\infty} ) \equiv \mathrm{Tr}[\varrho_{t}  \varrho_{\infty}];\, \frac{d}{dt} F(\varrho_{t},\varrho_{\infty} ) > 0,\, \forall t, \\ 
\lim_{t \to \infty} F(\varrho_{t},\varrho_{\infty} ) = 1,
\end{gathered} 
\end{equation} 
where $\theta(x)$ is the unit step function. Accordingly, combining Eqs. (\ref{ap8ipr}) and (\ref{1Ns}) with (\ref{lds}) and (\ref{urf}) the energy function $\varepsilon_{t}$ and the phase variables $\varphi_{t,n}$'s are obtained as
\begin{equation}
\label{urfvo}
\begin{gathered} 
 \varepsilon_{t} = E_{(\infty)} + O(e^{-z t} ),\,z>0, \\ 
 E_{(\infty)} \equiv \mathrm{Tr} [\varrho_{\infty} \, H ]  = \theta(-\mu)\, E_{1} + \theta(\xi ) \, E_{d},\\    
\theta(-\xi ) \big( (\varphi_{t,n} - \varphi_{t,1}) - (\varphi_{0,n} - \varphi_{0,1}) \big) +  \theta(\xi)\big(( \varphi_{t,n} - \varphi_{t,d}) - ( \varphi_{0,n} - \varphi_{0,d}) \big)  = - \frac{E_{n} - E_{(\infty)}}{1 + (\xi )^{2} }t,\, 1\leqslant n\leqslant d. 
\end{gathered}   
\end{equation}
It is interesting to note that the above behavior occurs irrespectively of the details of the Hamiltonian $H$ (e.g., its gap structure) or how strong $\xi $ is. In addition, from the above equations one can read a characteristic time for the evolution as
\begin{equation}
\tau(H,\xi ) \equiv \frac{1+(\xi )^{2}}{2|\xi |}\times\begin{cases} 1/(E_{2}-E_{1}); \,\,\,&\xi<0\\  1/(E_{d}-E_{d-1}),\,\,\, &\xi>0 \end{cases}.
\label{time-tau}
\end{equation}

%%%%%%%%%%%%%%%%%%%%%%%%%%%%%%%%%%%%%%%%%%%%%%%%%%%%%%%%%%%%%%%%
\begin{figure}[tp]
\includegraphics[width=\linewidth]{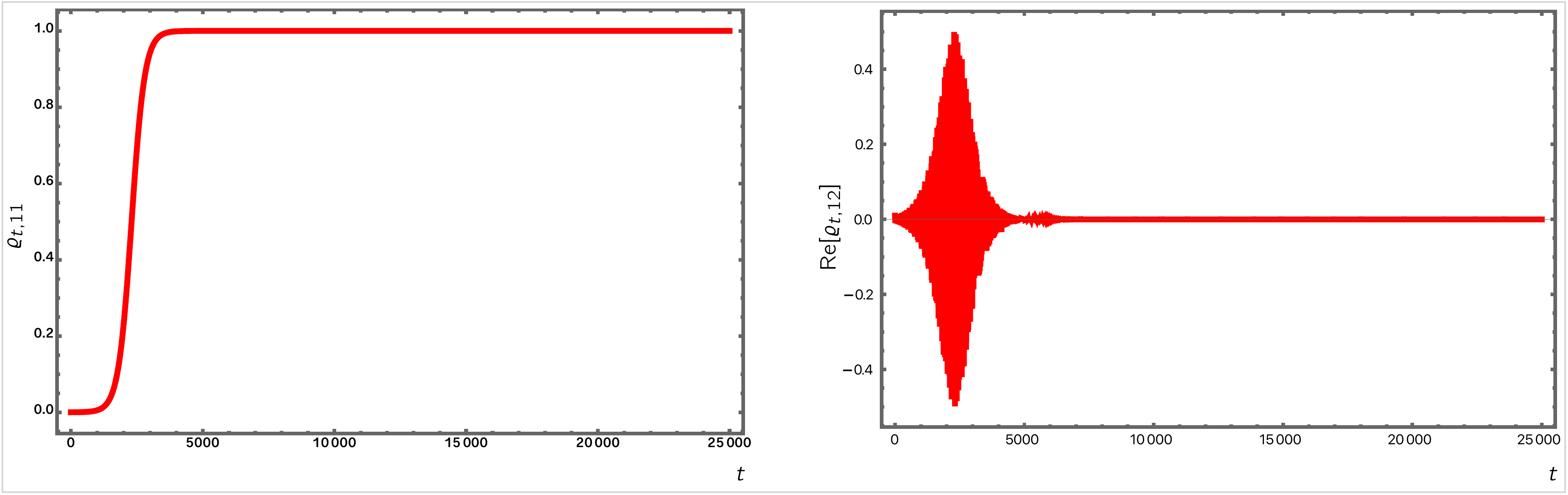}
\caption{EC monotonic localization of the one-qubit wavefunction to the ground state of $H=\sigma^{3}$ based on the additive EC deformation $-\xi\dot{\varrho}_{t}$, with $\xi=-0.001$ and the initial pure state $\varrho_{0}$ at an infinitesimal vicinity of $|2\rangle \langle 2|$, obtained by numerically solving the corresponding EC von Neumann equation (\ref{ecvovn}). We have used the eigenbasis of $\sigma^{3}$ to depict the plots, such that $\varrho_{t,11}$ is equal to the fidelity of the one-qubit wavefunction and the ground state of $H=\sigma^{3}$.}
\label{fig:slightpert}
\end{figure}
%%%%%%%%%%%%%%%%%%%%%%%%%%%%%%%%%%%%%%%%%%%%%%%%%%%%%%%%%%%%%%%%

As we see in Eq. \eqref{urfvo}, all physical phase contributions to the wavefunction, $e^{i (\varphi_{t,n > 1} - \varphi_{t,1})} $'s or $e^{i (\varphi_{t,d} - \varphi_{t,n < d} )}$'s, develop fixed-period oscillations asymptotically. However, the amplitude of these oscillations shrink monotonically following the exponentially fast decay of the non-attractor prefactors $\sqrt{p_{t,n}}s$. These points reflect the fidelity behavior of the state---Eq. \eqref{fdlw8}. \\

The result (\ref{fdlw8}) is a concrete manifestation of \textit{unitary monotonic self-focusing} behavior of the global wavefunction; \textit{a novel phenomenon} for closed quantum systems in the context of ECQT which, as will be shown in Secs. \ref{subsec:PRI-B} and \ref{subsec:PRI-C}, is robustly realizable upon EC fine-tuning-free deformations or even arbitrarily small perturbations of the SQT Hamiltonians. In Fig. \ref{fig:slightpert} we present the numerical solution to the one-qubit EC von Neumann equation (\ref{ecvovn}) based on the EC Hamiltonian (\ref{wow}) with $H=\sigma^{3}$ corresponding to the perturbation parameter $\xi=-0.001$. As these plots show in the $\sigma^{3}$ eigenbasis, the one-qubit wavefunction develops a monotonic localization to the ground state of $\sigma^{3}$, with the profile matching the analytic solution (\ref{1Ns}). \\

We now \textit{conclude} the above study as follows. Let the time-independent Hamiltonian $H$ of an arbitrary closed system get EC deformed ($\forall \xi  \in \mathrm{I\!R}$) or infinitesimally perturbed ($\xi  \to 0^{\pm}$) as in Eq. \eqref{wow}. For $\xi  > 0$ ($\xi  < 0$), the hybrid EC unitary evolution turns the excited state (the ground state) of $H$ into the asymptotic fixed-point attractor; that is, the global wavefunction \textit{self focuses} itself in a monotonic unitary manner and with an exponentially fast rate to \textit{localize} in the excited state (the ground state) of the SQT Hamiltonian $H$. 

%%%%%%%%%%%%%%%%%%%%%%%%%%%%%%%%%%%%%%%%%%%%%%%%%%%%%%%%%%%%%%%%
\subsection{Global Wavefunction Localization by Near-Markovian Hybrid EC Hamiltonians} 
\label{subsec:PRI-B}

Now we consider near-Markovian regimes. These regimes are when the chosen quantum memory distances $a_{t}$'s are sufficiently small at every present moment $t$ such that state-history monomials can be approximated by time-local state-dependent monomials upon replacing $\varrho_{t-b_t}$'s with their Taylor-series expansions,
\begin{equation*}
\varrho_{t - a_{t}} = \varrho_{t} - a_{t} \dot{\varrho_{t}} + \frac{1}{2} (a_{t})^{2} \ddot{\varrho}_{t} + \ldots,
\end{equation*}
perturbatively truncatable to any desired nontrivial order. We remind two observations from Ref. \cite{ECQT-1}. (i) In the absence of accompanying SQT Hamiltonians, all EC Hamiltonians which are entirely made of the system's past-to-present density operators have trivial near-Markovian regimes to all orders in their Taylor-series expansions. This is in the sense that these Hamiltonians generate nontrivial dynamics only when the largest quantum memory distances are not smaller than finite lower bounds set by the EC couplings. (ii) In contrast, hybrid $[\hskip-.8mm[N,L]\hskip-.8mm]$ EC Hamiltonians develop nontrivial near-Markovian regimes already in their first-order Taylor series truncations. The simplest example in category (ii) is a hybrid $[\hskip-.8mm[1,1]\hskip-.8mm]$ EC Hamiltonian 
\begin{equation} 
\label{woww11} 
\mathbbmss{H}_{t} = H + \lambda_{t-a} \varrho_{t-a}. 
\end{equation}

As in Sec. \ref{susbec:PRI-A}, we assume that the SQT Hamiltonian $H$ and the arbitrary coupling $\lambda_{t-a}$ are time independent. Moreover, we choose the quantum memory distance $a$ time independent and sufficiently small, $a \ll 1$. Then the first-order term in the Taylor series of $\varrho_{t-a}$ is sufficient. Hence in the near-Markovian regime, the hybrid $[\hskip-.8mm[1,1]\hskip-.8mm]$ EC Hamiltonian \eqref{woww11} reproduces the time-local EC Hamiltonian \eqref{wow} with the coupling
\begin{equation}
\xi  \equiv a  \lambda_{t-a}.
\end{equation}
Hence the results (\ref{ap8ipr}), (\ref{1Ns}), and (\ref{fdlw8}) are reproduced, developing a unitary monotonic self-focusing behavior. \\

We now consider a time-independent Hamiltonian deformed by the quadratic interactions of two states which are temporally separated by a sufficiently small distance $a$. Truncated to the first-order expansion in $a$, we have 
\begin{equation}
\label{woww22ip}
\begin{split} 
\mathbbmss{H}_{t} =& H + \lambda_{t-a} \varrho_{t-a} + i \lambda_{t-a,t}^{\mathrm{I}} [\varrho_{t-a},\varrho_{t}] + \lambda_{t-a,t}^{\mathrm{R}} \{ \varrho_{t-a}, \varrho_{t}\} \\ 
 =& H - a \lambda_{t-a} \dot{\varrho}_{t} -  a \lambda_{t-a,t}^{\mathrm{R}} \{\varrho_{t} , \dot{\varrho_{t}}\} + i a \lambda_{t-a,t}^{\mathrm{I}}  [\varrho_{t} , \dot{\varrho_{t}}] + O(a^{2}). 
\end{split} 
\end{equation}
By imposing state-history purity during its hybrid EC evolution, the differential relation $\{\varrho_{t} , \dot{\varrho}_{t}\} = \dot{ \varrho}_{t}$ is implied. Thus, the near-Markovian version of the $[\hskip-.8mm[2,2]\hskip-.8mm]$ EC deformation \eqref{woww22ip} becomes 
\begin{equation}
\label{woww22}
\mathbbmss{H}_{t} = H - \xi _{1} \dot{\varrho}_{t} + i \, \xi ^{\mathrm{I}}_{2}\,  [ \varrho_{t}  , \dot{\varrho_{t}} ], 
\end{equation}
where
\begin{equation}
\label{woww22-}
\begin{split}
\xi _{1} &\equiv a ( \lambda_{t-a} + \lambda_{t-a,t}^{\mathrm{R}}), \\
\xi ^{\mathrm{I}}_{2} &\equiv a \lambda_{t-a,t}^{\mathrm{I}}. 
\end{split} 
\end{equation}
This yields the corresponding hybrid EC von Neumann equation as
\begin{equation}
\label{woww22ase} 
\dot{\varrho}_{t} = i[ \varrho_{t},  \frac{H}{1 + \xi ^{\mathrm{I}}_{2} } ] - i  \frac{\xi _{1}}{1 + \xi ^{\mathrm{I}}_{2} } [ \varrho_{t}  , \dot{\varrho}_{t} ].  
\end{equation}
From this form, one can observe that the near-Markovian dynamics is equivalent to the deformed von Neumann Eq. \eqref{ecsovn} for a $[\hskip-.8mm[1,1]\hskip-.8mm]$ EC dynamics upon the mapping 
\begin{equation}
( \xi  , H ) \to \big(\xi _{1}/[1 + \xi ^{\mathrm{I}}_{2}] ,  H/[1 + \xi ^{\mathrm{I}}_{2}] \big).
\end{equation}
Hence one concludes that all near-Markovian deformations of SQT Hamiltonians of the $[\hskip-.8mm[2,2]\hskip-.8mm]$ type, or infinitesimal perturbations with $(\lambda_{t-a})^{2}  + |\lambda_{t-a,t}|^{2} \ll 1$, similarly develop unitary monotonic self-focusing behaviors. We highlight that the attractor of the hybrid EC evolution remains invariant under (sign or magnitude) variations of $\xi ^{\mathrm{I}}_{2}$ as it depends (similar to the  $[\hskip-.8mm[1,1]\hskip-.8mm]$ EC deformations) only on the sign of the coupling $\xi _{1}$.
\\

The above analytical result was based on the leading-order time-local truncation of the hybrid $[\hskip-.8mm[N \leqslant 2,L \leqslant 2]\hskip-.8mm]$ von Neumann equation which itself is time-nonlocal even in the near-Markovian regime $a \ll 1$. The full dynamics of the hybrid EC unitary evolution for $[\hskip-.8mm[2,2]\hskip-.8mm]$ in its near-Markovian regime is described by the time-nonlocal von Neumann equation, 
\begin{equation}
\label{ceb} 
\begin{split} 
\dot{\varrho}_{t} & = -i [\mathbbmss{H}_{t}  , \varrho_{t}],\\
\mathbbmss{H}_{t} & = H + \big( \lambda_{t-a} \varrho_{t-a} + i \lambda_{t-a,t}^{\mathrm{I}} [\varrho_{t-a},\varrho_{t}] + \lambda_{t-a,t}^{\mathrm{R}} \{ \varrho_{t-a}, \varrho_{t}\} \big) \bigl|_{a \ll 1}.
\end{split}
\end{equation}

\begin{figure}[tp]
\includegraphics[width=.47\linewidth]{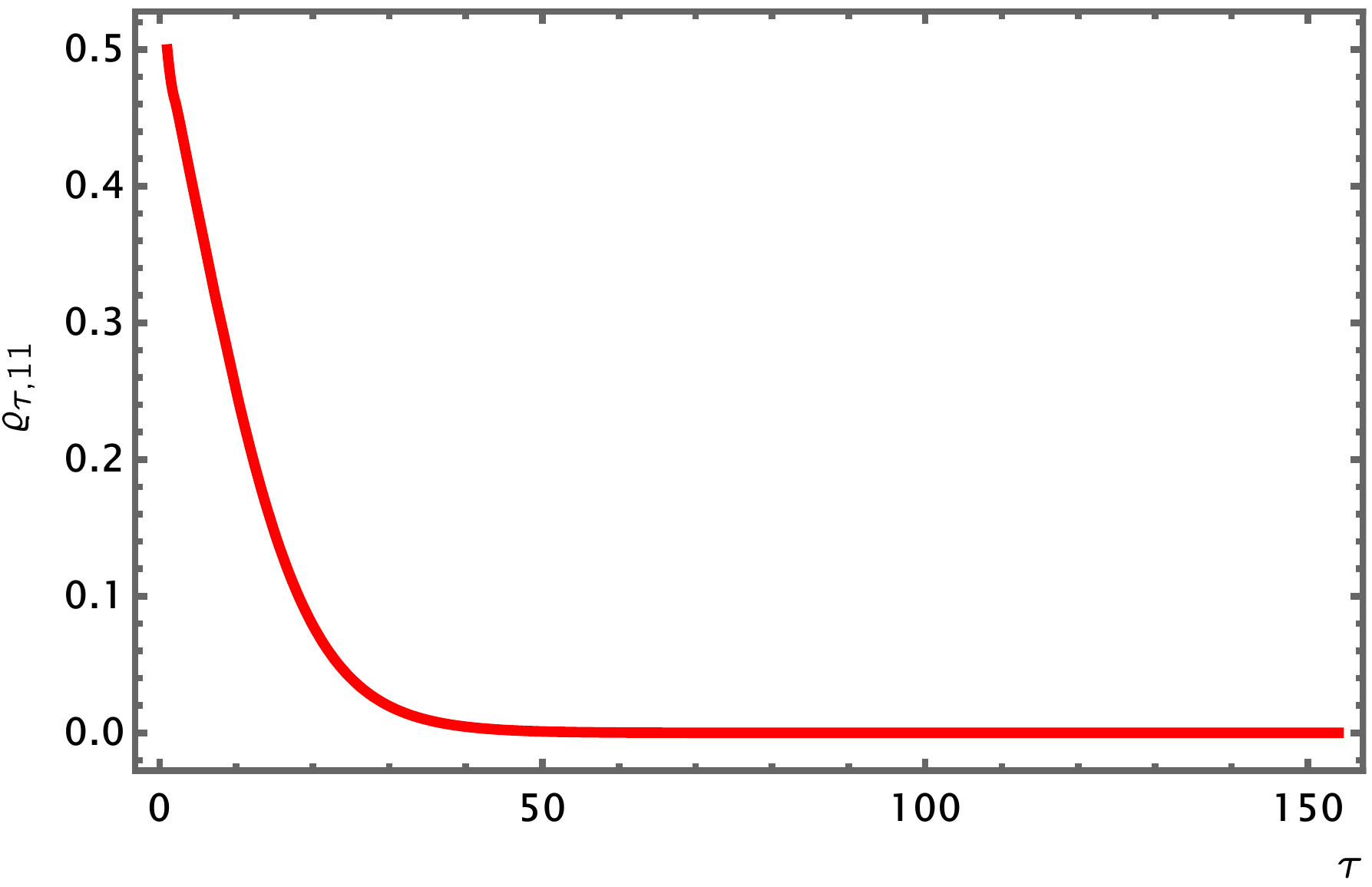} \hskip2mm \includegraphics[width=.47\linewidth]{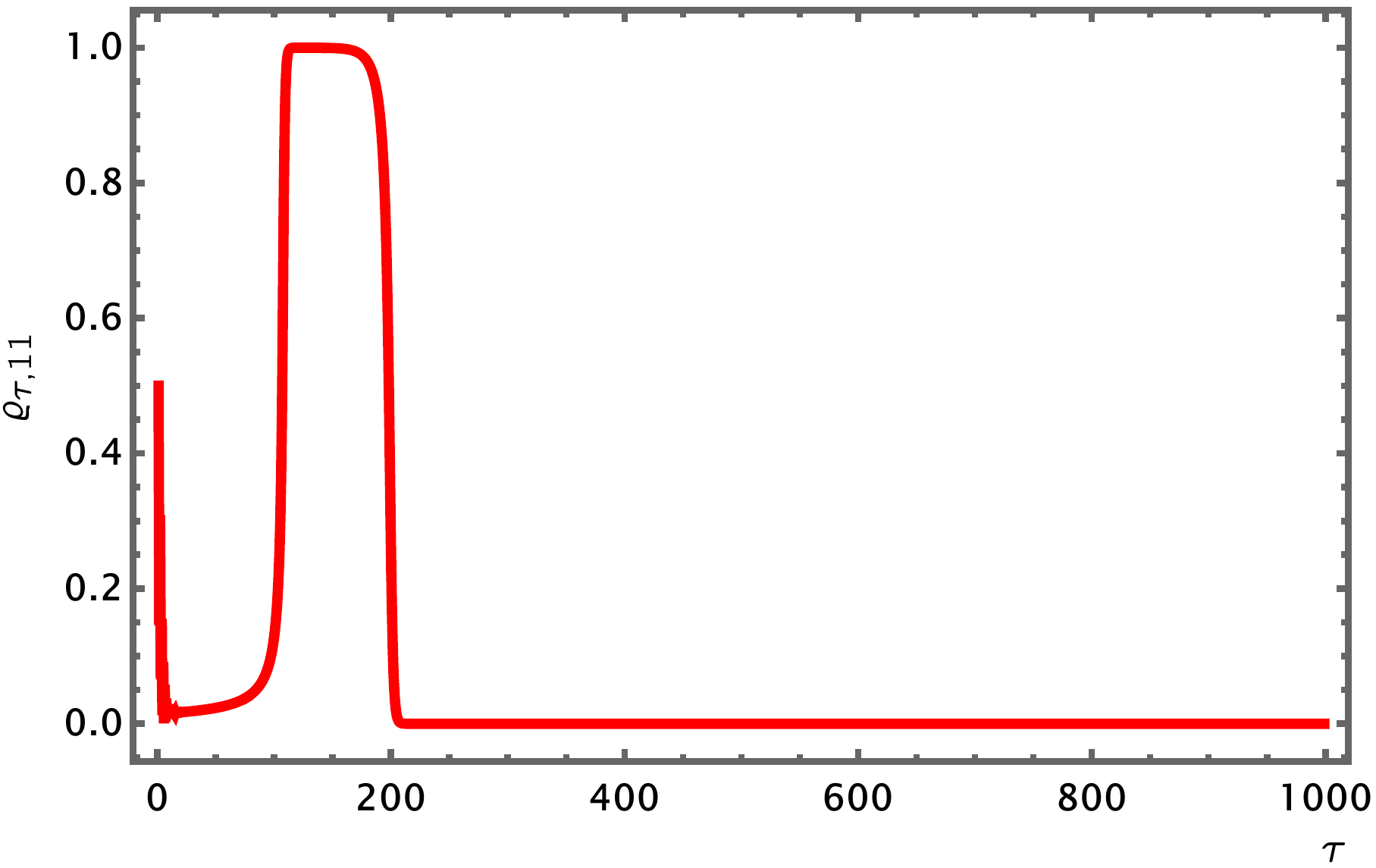}\\
\includegraphics[width=.47\linewidth]{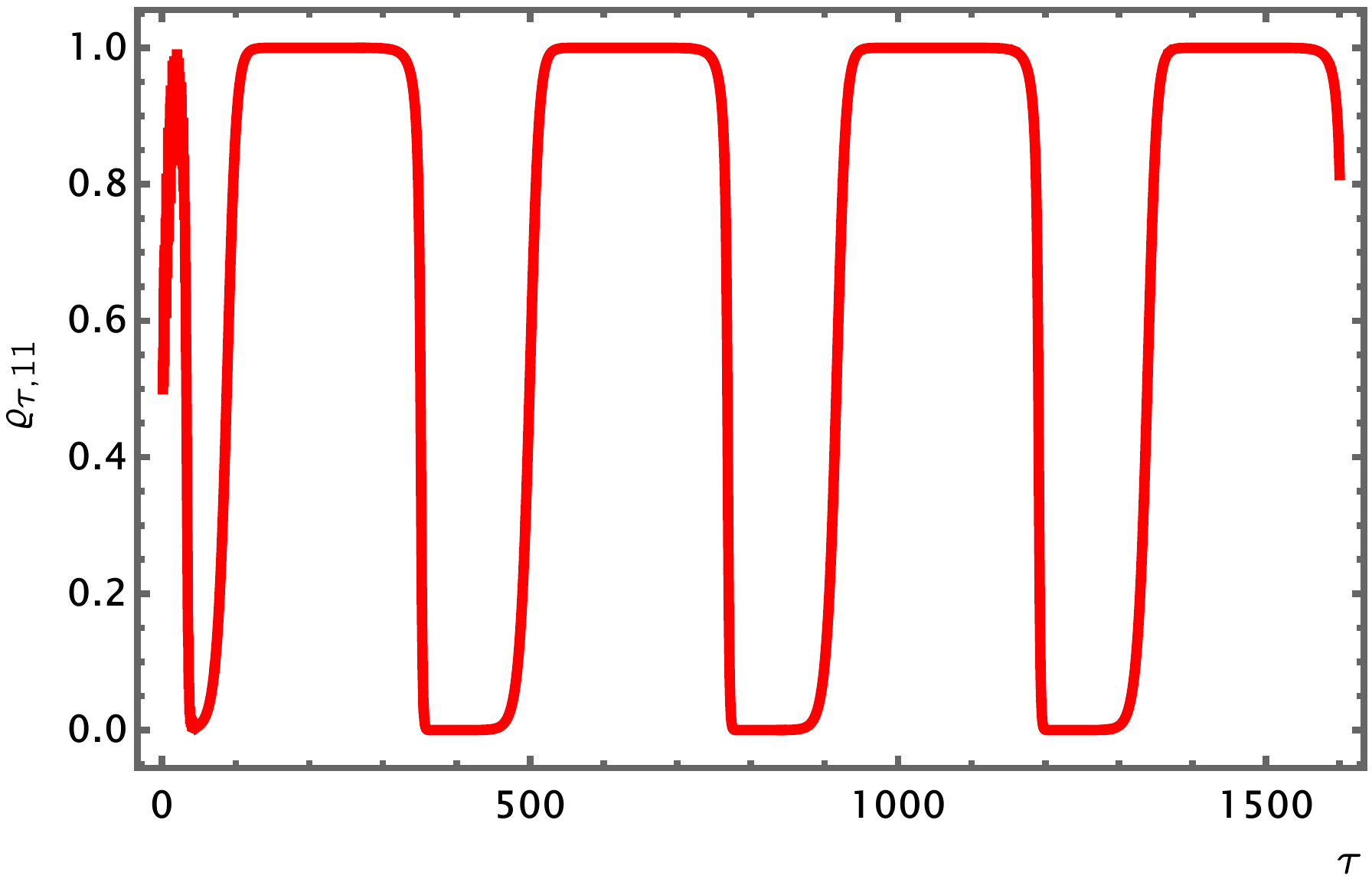} \hskip2mm \includegraphics[width=.47\linewidth]{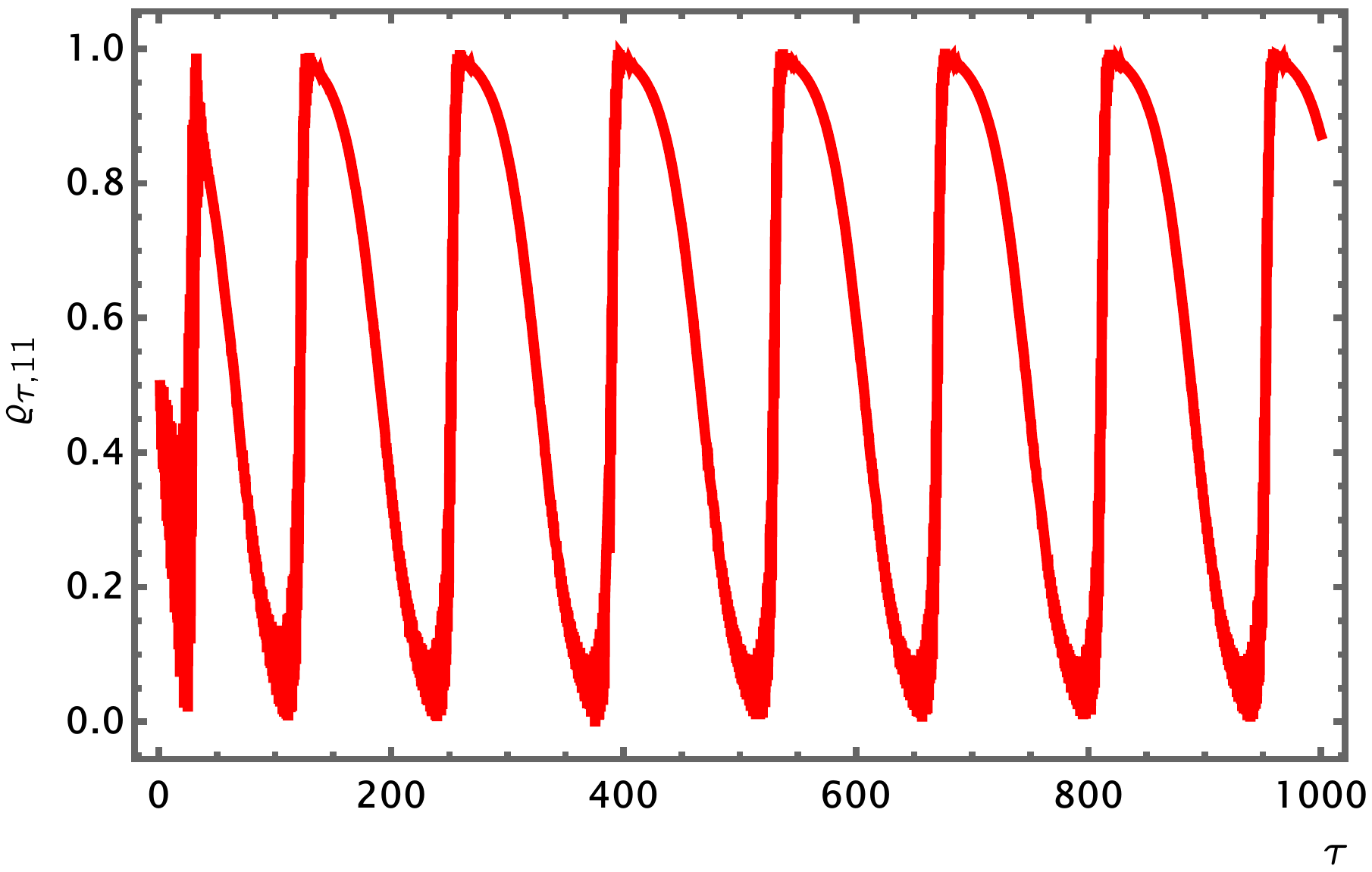}\\
\includegraphics[width=.47\linewidth]{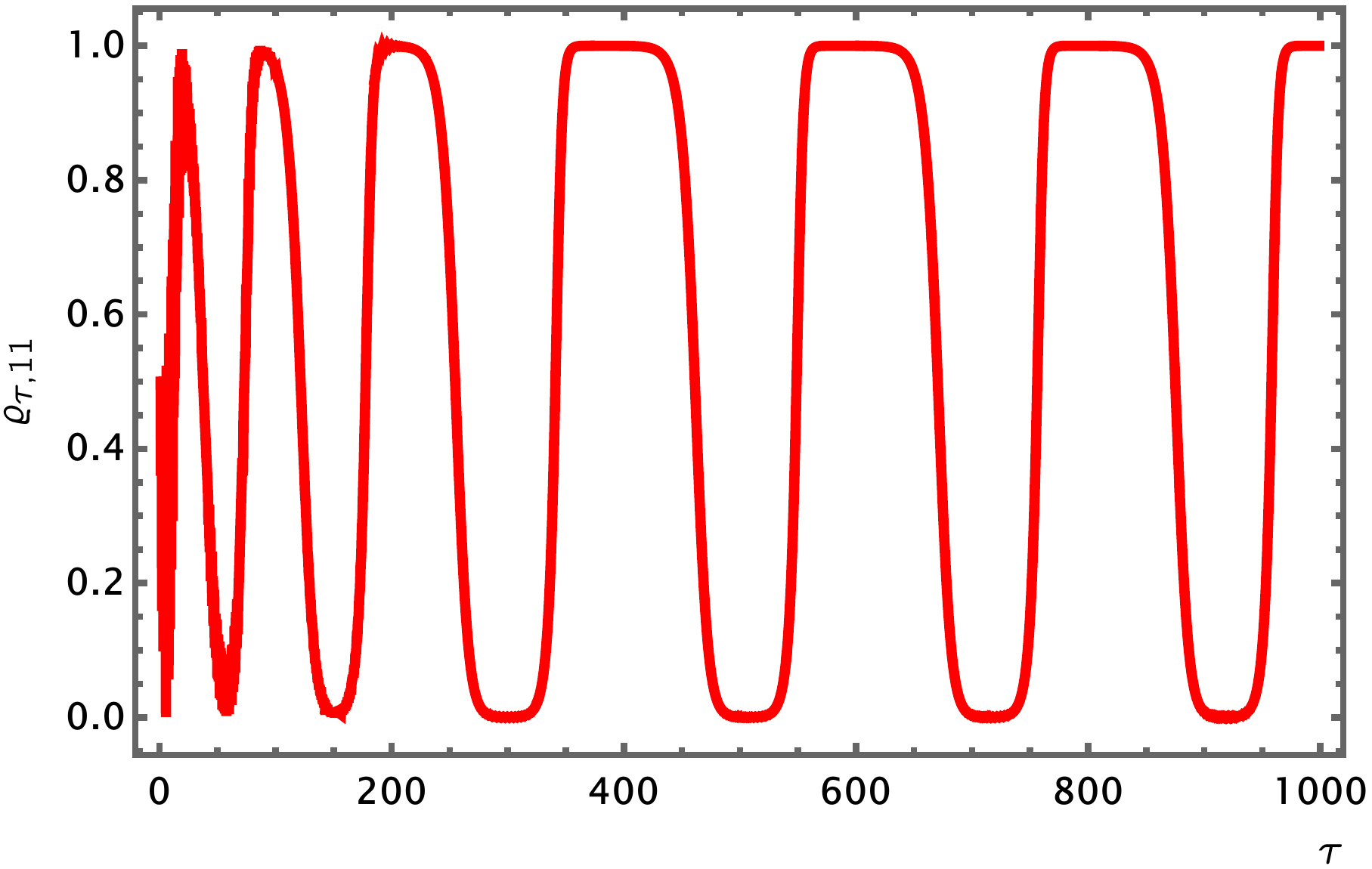} \hskip2mm \includegraphics[width=.47\linewidth]{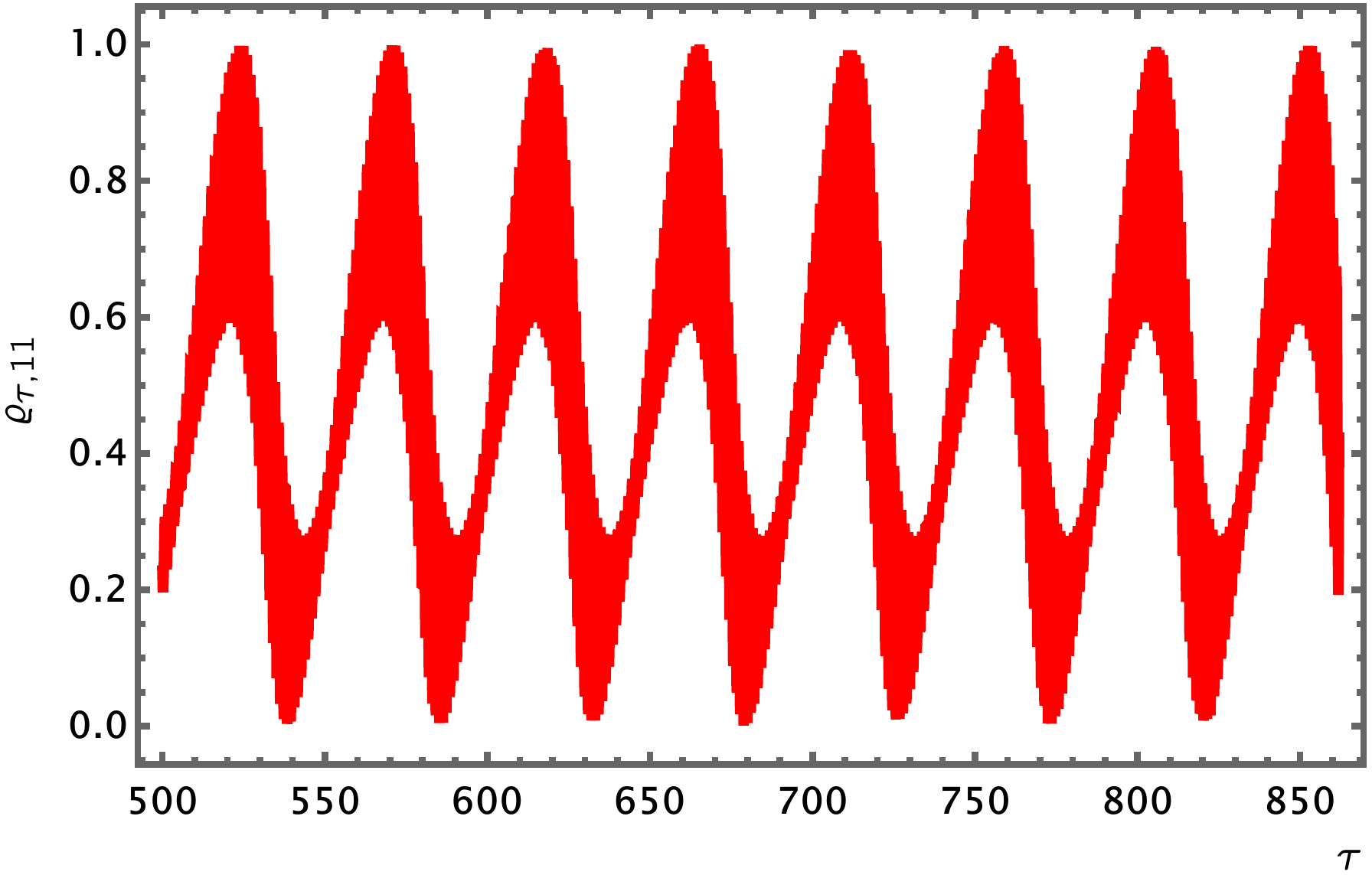}\\
\includegraphics[width=.47\linewidth]{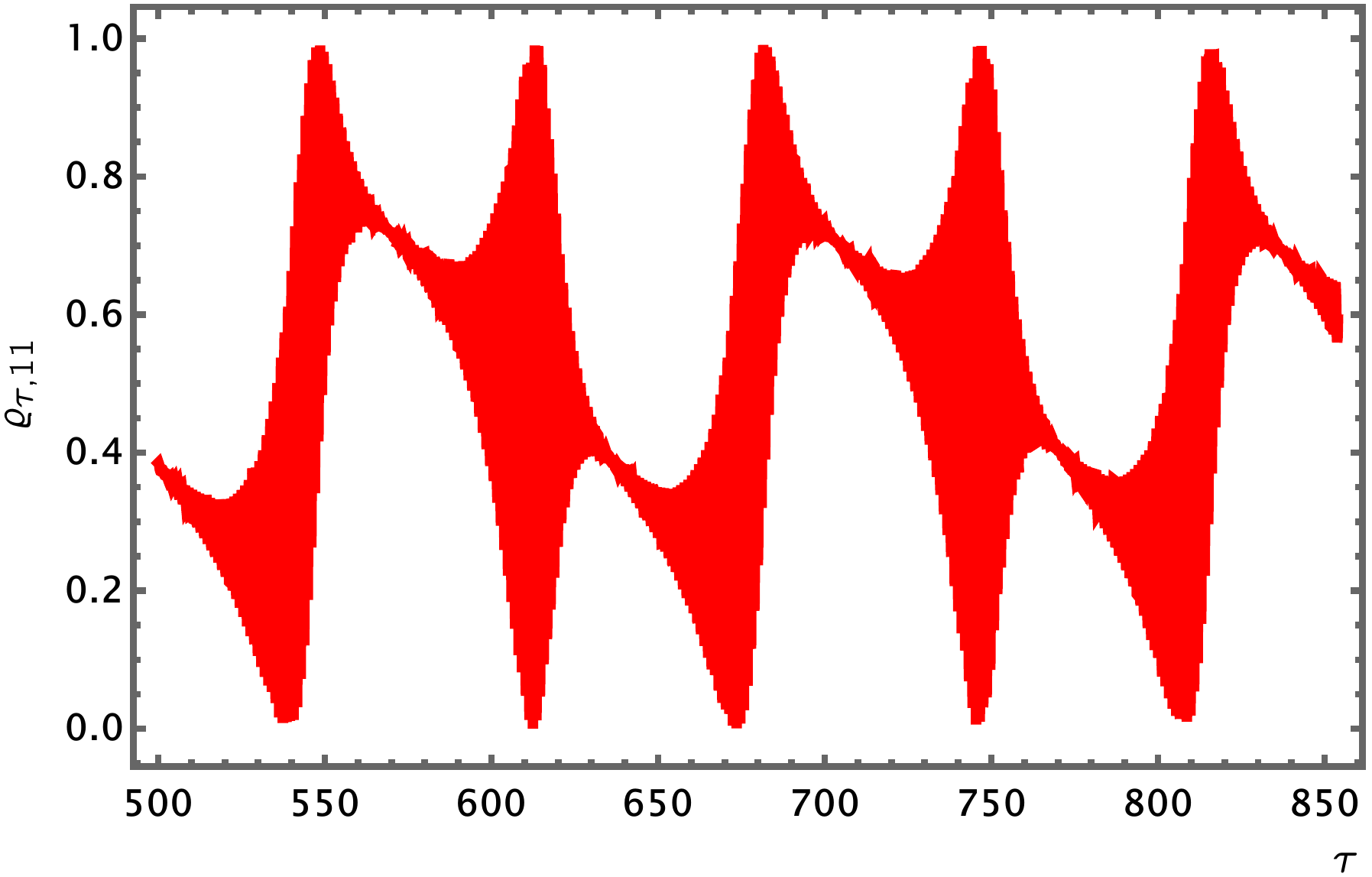} \hskip2mm \includegraphics[width=.47\linewidth]{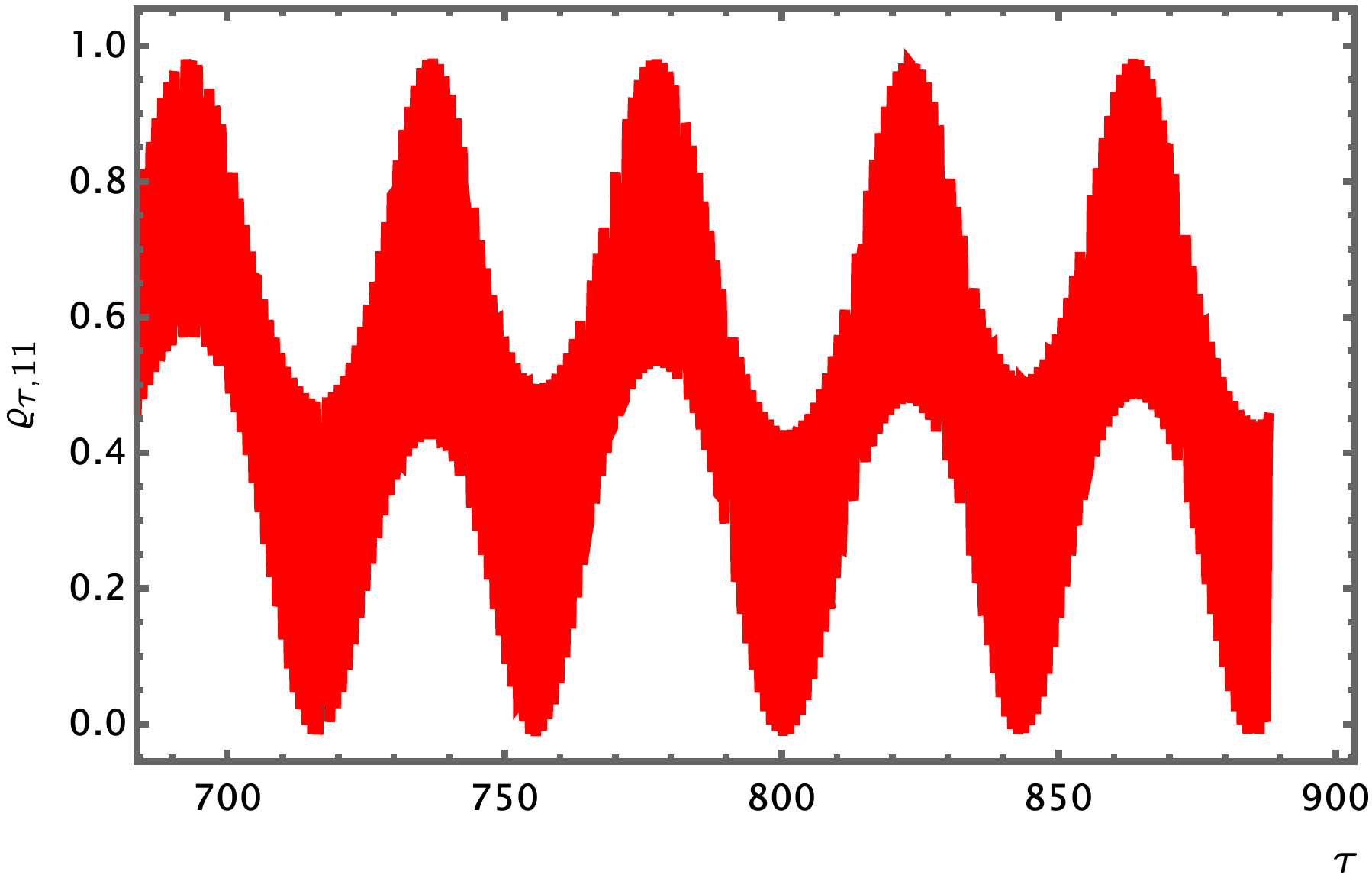}
\caption{Eight EC deformations of $H=\sigma^{3}$ by means of one-qubit $[\hskip-.8mm[2,2]\hskip-.8mm]$ Hamiltonians corresponding to $(\lambda_{t-a},\lambda^{\mathrm{R}}_{t-a,t},\lambda^{\mathrm{I}}_{t-a,t} )= (0,1,2)$ and $a$ being $0.25$ (first row left), $1.33$ (first row right), $2.5$ (second row left), $3.5$ (second row right), $4.25$ (third row left), $5.33$ (third row right), $6.5$ (fourth row left), $7.88$ (fourth row right). We have chosen $\varrho_{0}=(1/2)(\openone + \sigma^{1})=|+\rangle \langle +|$ as the initial state, and the initial state histories between $[0,a]$ using $H_{\mathrm{kicker}} = \sigma^{3}$. We have used the eigenbasis of $\sigma^{3}$ to depict the plots, such that $\varrho_{\tau,11}$ is equal to the fidelity of the one-qubit wavefunction and the ground state of $H=\sigma^{3}$. The plots show the EC evolution of $p_{\tau,11}$ with respect to $\tau\equiv t/a$. See the text in Sec. \ref{subsec:PRI-C} for detailed explanations of EC fidelity behaviors.}
\label{fig:deepnm2}
\end{figure}

We have numerically considered a one-qubit EC closed system \eqref{ceb} with $H = \sigma^{3}$ beyond the leading-order time-local truncation and obtained a representative solution to the full time-nonlocal EC dynamics \eqref{ceb}. The representative numerical solution has been depicted in the top left plot of Fig. \ref{fig:deepnm2}, which indeed validates monotonic unitary self-focusing attractor behavior of the system's wavefunction, as we have shown analytically in the present section. The relevant details of the solution shall be given in the next subsection along with the solutions for the deeply non-Markovian EC deformations of $H=\sigma^{3}$. 

%%%%%%%%%%%%%%%%%%%%%%%%%%%%%%%%%%%%%%%%%%%%%%%%%%%%%%%%%%%%%%%%
%%%%%%%%%%%%%%%%%%%%%%%%%%%%%%%%%%%%%%%%%%%%%%%%%%%%%%%%%%%%%%%%

%%%%%%%%%%%%%%%%%%%%%%%%%%%%%%%%%%%%%%%%%%%%%%%%%%%%%%%%%%%%%%%%
\subsection{Deforming SQT Evolutions by Deeply Non-Markovian EC Hamiltonians} 
\label{subsec:PRI-C}

Here we investigate nonlocal deformations of SQT Hamiltonians by EC Hamiltonians, which are by construction deeply non-Markovian. In such a deformation, one uses EC Hamiltonians whose largest quantum memory distances $\max_{t}a_{t} \equiv a$ are sufficiently large such that participating state-history monomials are not reachable by their time-local Taylor-series expansions within any finite or (even countably) infinite order of perturbation. Our aim here is to analyze whether and how these deeply non-Markovian deformations lead to unitary monotonic localizations of the system's wavefunction---as described by Eq. \eqref{fdlw8}. \\

Consider a closed system whose total Hamiltonian equals the sum of a time-independent SQT Hamiltonian $H$ and a $[\hskip-.8mm[1,1]\hskip-.8mm]$ EC Hamiltonian \eqref{11,22} whose quantum memory distance $a$ is chosen arbitrarily. We have shown in Sec. \ref{susbec:PRI-A} that for $a = \delta_{a} \ll 1$ the wavefunction in asymptotic times lands in a dynamical \textit{attractor}, which is one of the two extreme (or edge) eigenstates of $H$. The Taylor series of the only state-history monomial in Eq. \eqref{11,22}, $\varrho_{t-a}$, has a radius of convergence around the point $a = \delta_{a}$, which we denote by $D(\delta_{a})$. Now, consider shifting $\delta_{a}$ by some positive $\Delta_{a}$ constrained by $\Delta_{a} \leqslant  D(\delta_{a})$, which is to secure that the enlarged quantum memory distance $a = \delta_{a} + \Delta_{a}$ remains in the convergence domain. As we analytically show in what follows, the EC deformation based on such an enlarged $a$ should typically develop unitary monotonic self-focusing behavior. \\

It suffices to focus on $\lambda_{t-a} < 0$, as the case for $\lambda_{t-a}>0$ follows similarly. The evolution unitarity implies the dynamical constraint $\sum_{n=1}^{d} p_{t,n}  = 1$. Hence for validating the monotonic self-focusing behavior, it suffices to show that the ground-state fidelity $p_{t,1}$ develops (at least asymptotically) a monotonically increasing behavior as in Eq. \eqref{1Ns}. To perform the analytic derivation, we safely assume that the fidelity $p_{t,1}$ is an analytic function of $a$. Hence $p_{t,1} (a) \equiv p_{t|a,1}$ at any finite quantum memory distance $a \equiv \delta_{a} + \Delta_{a}$ can be recast via infinite-order Taylor series around the sufficiently small $\delta_{a} \ll 1$, 
\begin{equation}
\label{woww11adnm}
\begin{gathered} 
p_{t|a=\delta_{a}+\Delta_{a},1} = p_{t|\delta_{a},1} + \textstyle{\sum_{s = 1}^\infty}(\Delta^s/s!) \, R^{(s)}_{t|\delta_{a}},  
\end{gathered} 
\end{equation}
where $R^{(s)}_{t|\delta_{a}} \equiv  \Big( \frac{\partial^{s}}{\partial \tau^s} p_{t|\tau,1}  \Big)\Big|_{\tau = \delta_{a}}$. Recalling the formula for the near-Markovian ground-state fidelity \eqref{1Ns}, we now compute (up to behaviorally irrelevant subleading corrections) $R^{(s)}_{t|\delta_{a}}$'s and the remainder as the second term on the RHS of Eq. (\ref{woww11adnm}),  
\begin{equation}
\label{remainders1}
\begin{split}
R^{(s)}_{t|\delta_{a}} & = \frac{\partial^s}{\partial \delta_{a}^{s}}\Big[\Big( 1 + \textstyle{\sum_{l = 2}^{d}} (p_{0,l}/p_{0,1}) e^{ -  2 \delta_{a} | \lambda_{t-a} | (E_{l} - E_{1})t} \Big)^{-1} + O ( \mathrm{subleading\; in \;} \delta_{a} ) \Big]  \\
 &= - \textstyle{\sum_{l = 2}^{d}} (p_{0,l}/p_{0,1}) \frac{\partial^{s}}{\partial \delta_{a}^{s}} e^{- 2 \delta_{a} | \lambda_{t-a} | (E_{l} - E_{1}) t}  + O(\mathrm{asymp.\; irrelev.\; or\; subleading\; in\,} \delta_{a} )\\
 &= - \textstyle{\sum_{l = 2}^{d}} (p_{0,l}/p_{0,1})\, e^{- 2 \delta_{a} | \lambda_{t-a} | (E_{l} - E_{1}) t } \big( - 2 |\lambda_{t-a}| (E_{l} - E_{1}) t \big)^s + O(\mathrm{asymp.\; irrelev.\; or\; subleading\; in\,} \delta_{a} ) \\   
\,\,\,\Rightarrow\,\,\textstyle{\sum_{s = 1}^\infty}\Delta^s R^{(s)}_{t|\delta_{a}}/s! &= - \textstyle{\sum_{l = 2}^{d}} (p_{0,l}/p_{0,1}) \, e^{- 2 \delta_{a} | \lambda_{t-a} | (E_{l} - E_{1}) t } \big( e^{- 2 \Delta_{a} | \lambda_{t-a} | (E_{l} - E_{1}) t } - 1 \big) + O(\mathrm{irrelev.} ) \\ 
& \approx - \textstyle{\sum_{l = 2}^{d}} (p_{0,l}/p_{0,1}) \,e^{- 2 a | \lambda_{t-a} | (E_{l} - E_{1}) t } - p_{t|\delta_{a},1} + 1. 
\end{split}
\end{equation}

Finally, combining Eqs. (\ref{woww11adnm}) and (\ref{remainders1}) with (\ref{1Ns}), the dominant profile of the dynamical fidelity between the system's wavefunction and the ground state of the SQT Hamiltonian is obtained as
\begin{equation}
\label{woww11bhv} 
\begin{gathered}
F_{t|a}(\varrho_{t},|1\rangle\langle 1|) = p_{t|a,1} \approx 1 - \textstyle{\sum_{l = 2}^{d}} (p_{0,l}/p_{0,1}) e^{- 2 a | \lambda_{t-a} | (E_{l} - E_{1}) t } ~ \Rightarrow \\
F_{t|a}(\varrho_{t},|1\rangle\langle 1|) \approx \Big( 1 + \textstyle{\sum_{l = 2}^{d}} (p_{0,l}/p_{0,1}) e^{- 2 a | \lambda_{t-a} | (E_{l} - E_{1}) t } \Big)^{-1} = F_{t|\delta_{a}}(\varrho_{t},|1\rangle \langle 1|)_{|_{\delta_{a} \to a}}.
\end{gathered}
\end{equation}
As we observe, the late-time behavior of the ground-state fidelity scales up in a covariant fashion at any finite $a = \delta_{a} + \Delta_{a}$ (with $\Delta_{a} \leqslant D(\delta_{a})$); the infinite-order resummation corresponds to simply replacing $\delta_{a} \ll 1$ with the enlarged $a$. As such, for any $a \in [ \delta_{a} , \delta_{a} + D(\delta_{a}) ]$, depending only on the sign of the EC coupling $\lambda_{t-a}$, the EC deformation or even infinitesimal perturbation makes either the ground state or the excited state of the SQT Hamiltonians the asymptotic fixed-point attractor of the unitary hybrid EC evolution. \\

The above derivation extends under the same conditions to the $[\hskip-.8mm[2,2]\hskip-.8mm]$ EC deformations of SQT Hamiltonians. That is because, as shown in Sec. \ref{subsec:PRI-B}, the near-Markovian regime of the $[\hskip-.8mm[2,2]\hskip-.8mm]$ EC Hamiltonians \eqref{woww22ip} is behaviorally equivalent to their $[\hskip-.8mm[1,1]\hskip-.8mm]$ versions, upon a linear mapping. Hence the same rational goes through. The derivation itself, however, is not sufficiently inclusive, even though it undertakes the resummation of the infinitely many asymptotic perturbative terms. Its behavioral coverage is limited because (i) it is based on assuming the \textit{analyticity} of the attractor fidelity with respect to $a$, and (ii) the noncompact range of $a$ beyond the domain of near-Markovian Taylor-series \textit{convergence} is not covered by it. Indeed, the analysis of the behavioral phases of one-qubit $[\hskip-.8mm[N \leqslant 3, L \leqslant 3]\hskip-.8mm]$ EC unitary evolutions (\ref{aiuvb}), (\ref{ech:g}), and (\ref{monomials}) presented in Ref. \cite{ECQT-1} (and reviewed in Sec. \ref{sec:phase1}) strongly suggest that either or both of the assumptions (i) and (ii)---i.e., series convergence and fidelity analyticity---are violated when one goes sufficiently deep into the non-Markovian regime. Thus, deeply non-Markovian EC deformations of SQT Hamiltonians develop a highly nontrivial behavioral phase diagram with large qualitative diversity and physical richness. In this light, we present in what follows a numerical exploration of deeply non-Markovian deformations of one-qubit EC Hamiltonians. \\

Let us look into one-qubit pure-state-history EC deformations where $\mathbbmss{H}_{t}$ is the sum of the SQT Hamiltonian $H = \sigma^{3}$ and a $[\hskip-.8mm[2,2]\hskip-.8mm]$ EC Hamiltonian (\ref{11,22}), with representative fixed-value couplings $(\lambda_{t-a},\lambda^{\mathrm{R}}_{t-a,t},\lambda^{\mathrm{I}}_{t-a,t} )= (0,1,2)$, for $8$ choices of the quantum memory distances given by $a\in\{0.25, 1.33, 2.5, 3.5, 4.25, 5.33, 6.5, 7.88\}$. The numerical solutions to the corresponding EC von Neumann equations, or equivalently EC Schr\"{o}dinger equations, have been shown in the eight plots of Fig. \ref{fig:deepnm2}. Among these eight plots the top left one corresponds to the near-Markovian deformation discussed in the previous subsection, while the remaining plots beginning from the top right one correspond to seven examples of deeply non-Markovian deformations. These plots show the EC time evolution of $p_{t,1}$, which is the fidelity between the one-qubit wavefunction and the ground state of $H=\sigma^{3}$, being equal to $\varrho_{t,11}$ in the chosen basis. It is observed that $p_{t,1}$ develops diverse remarkable behaviors across the chosen spectrum of $a$. The numerical solution in the top-right plot, corresponding to $a=1.33$, which is close to the outset $a=1$ of deep non-Markovianity, still presents the near-Markovian localization on \textit{the excited states} of $H=\sigma^{3}$. In doing so, however, it first develops a sizable transient localization on the ground state of $H=\sigma^{3}$ preceded and followed by sharp transitions. The plots corresponding to $a\in\{3.5, 5.33, 6.5, 7.88\}$ present correlated oscillations of the fidelity $p_{t,1}$, which develop collectively at longer scales large structured regular oscillations between the ground state and the excited state of $H=\sigma^{3}$. It is worthwhile to highlight that the unitary formation of such longer scale structured large oscillations---which resemble time-like solitonic waves---is one landmark of (non-hybrid and hybrid) unitary EC evolutions. \\

\begin{figure}[tp]
\includegraphics[width=\linewidth]{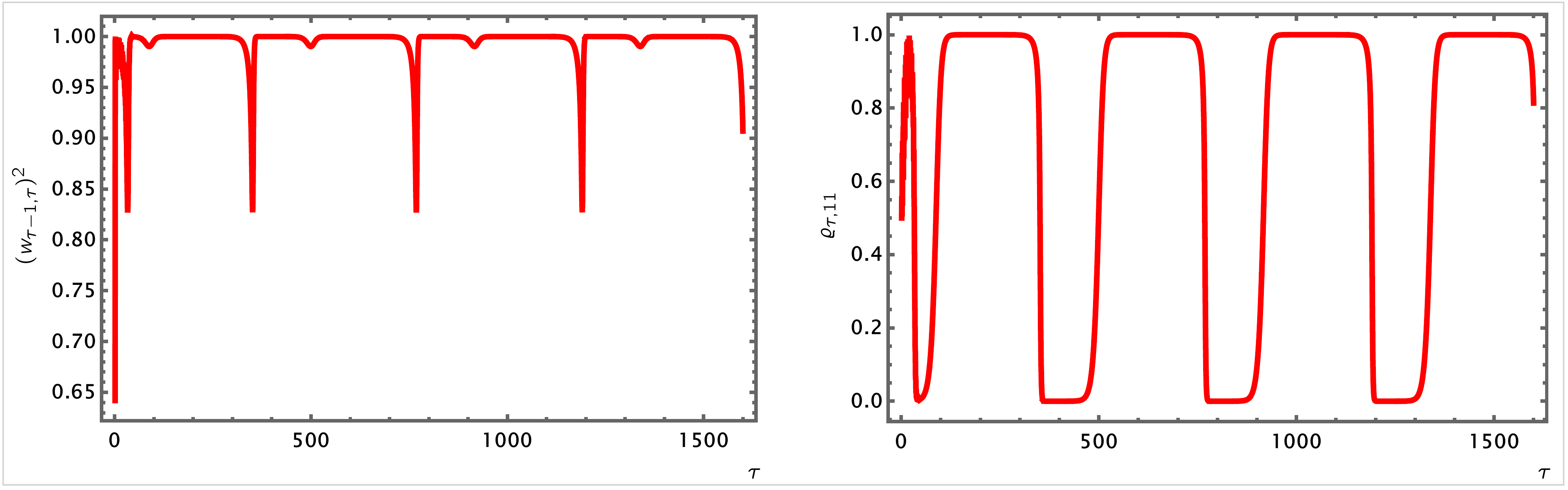}
\caption{The right plot shows the same EC deformation as detailed in the caption of Fig. \ref{fig:deepnm2}, for the case of $a=2.5$. The left plot shows the corresponding evolution of the one-qubit state-history fidelity, i.e., one-qubit autocorrelation with temporal distance $a$. See the text in Sec. \ref{subsec:PRI-C} for detailed explanations of the presented behaviors.}
\label{fig:deepnm1}
\end{figure}

A remarkable dynamic-attractor behavioral phase is presented by the solutions corresponding to $a\in\{2.5, 4.25\}$ in Fig. \ref{fig:deepnm2}. As we observe, these deeply non-Markovian $N=L=2$ deformations turn the two eigenstates of the SQT Hamiltonian $H=\sigma^{3}$ into metastable attractors between which the one-qubit wavefunction swaps sequentially in a structured manner with highly robust profile and quantitative regularity, involving sharp transitions between the two $\sigma^{3}$ eigenstates. In fact, Fig. \ref{fig:deepnm1} makes it clear that this bistable attractor behavior precisely corresponds to the behavioral Phase \textbf{\textsf{V}}, previously shown for non-hybrid $[\hskip-.8mm[2,2]\hskip-.8mm]$ EC Hamiltonians. As we see in the two plots of Fig. \ref{fig:deepnm1}, the sharp swaps between the ground state and the excited state of $\sigma^{3}$ are precisely synchronized and triggered by the one-qubit autocorrelation, i.e., its two-point state-history fidelity. \\

To sum up, it is clear from our analytic discussions and the presented numerical solutions that the deeply non-Markovian EC deformations of the SQT Hamiltonians constitute vastly diverse and qualitatively rich types of state and fidelity evolutions. 

%%%%%%%%%%%%%%%%%%%%%%%%%%%%%%%%%%%%%%%%%%%%%%%%%%%%%%%%%%%%%%%%
%%%%%%%%%%%%%%%%%%%%%%%%%%%%%%%%%%%%%%%%%%%%%%%%%%%%%%%%%%%%%%%%

%%%%%%%%%%%%%%%%%%%%%%%%%%%%%%%%%%%%%%%%%%%%%%%%%%%%%%%%%%%%%%%%
\subsection{EC Hamiltonians Endowed with State-History-Dependent Couplings}
\label{subsec:PRI-D}

We first remind that EC Hamiltonians by definition are Hermitian operators which selectively link and coalesce (the information inside) the past-to-present states of their closed systems. The essential substrate of EC Hamiltonians is the information content of the state history of the unitarily evolving quantum systems. For example, a linear combination of SQT Hamiltonians is itself an EC Hamiltonian once even a single coefficient is allowed to become sensitive to the unitary state history that the system has taken up to the present moment. \\

Indeed, typical Hamiltonians in ECQT not only employ products of the system's density operators but also contain scalar prefactors with diversely formed dependences on the state history. The defining couplings of typical $[\hskip-.8mm[N,L]\hskip-.8mm]$ Hamiltonians in ECQT can themselves be scalar quantities defined in terms of (say, expectation values, traces, or determinants of) the contextual combinations of past-to-present density operators of the closed system. The physical significance of state-history-dependent couplings in ECQT is already highlighted in Secs. \ref{sec:ECQT} and \ref{sec:sqt-vs-ecqt}. For example, we have shown in Sec. \ref{sec:sqt-vs-ecqt} that structurally fine-tuned $[\hskip-.8mm[N,L]\hskip-.8mm]$ Hamiltonians, whose defining couplings are tailored to be specific state-history-dependent scalars, can be equivalent to standard Hamiltonians. However, the role of state-history-dependent couplings in ECQT goes much beyond such equivalences; their significance is behavioral. Indeed, efficient generation of behaviorally desired trajectories of a closed system requires that the Hamiltonian couplings themselves be EC, namely history sensitive, in contextually distinctive ways. Let us highlight that it can become behaviorally relevant for a system developing an EC unitary evolution that even their quantum memory distances of its EC Hamiltonian become state-history dependent. We delegate a comprehensive study of this additional flexibility for a future publication. The central aim of this subsection is to use simple examples in the context of minimal EC deformations of standard Hamiltonians to show the behavioral significance of state-history-dependent couplings in EC unitary evolutions. \\

Let us now return to the ground-state localizations of the wavefunction developed by the $\dot{\varrho}_{t}$ EC deformation introduced in Sec. \ref{susbec:PRI-A}. As the solution of the fidelity dynamics \eqref{1Ns} manifests, EC wavefunction localizations under unitary evolution \eqref{wow} are characterized by the following three attributes:

\begin{itemize}

\item[(i)] The speed of wavefunction localization has a finite maximum value $v_{\max} =1$, for the whole spectrum of the deformation coupling. As one reads from the characteristic time scale of this EC evolution, $\tau(H,\xi ) = \frac{1}{2 \Delta E} \frac{1 + (\xi )^{2}}{|\xi |}$ [Eq. (\ref{time-tau})], the maximum speed is achieved at
\begin{equation}
\xi _{\mathrm{max~speed}} \equiv \arg\min_{\xi  \in \mathrm{I\!R}} \tau( H,\xi  ) = 1.
\end{equation}

\item[(ii)] The fixed-point attractor of the unitary evolution---the moment at which the ground-state fidelity reaches the value $1$---is infinitely distant. That is, the dynamical attractor is \textit{asymptotic} and the fidelity at any finite time has a value $<1$. 

\item[(iii)] The EC deformation in the minimal model \eqref{wow} does not dynamically rescale the reference Hamiltonian $H$ (see the first term on the RHS of Eq. (\ref{wow_optimal_AB}) throughout the unitary evolution.

\end{itemize}

We now  present a simple proof-of-principle toy model which demonstrates how bringing state-history dependence into the couplings of hybrid EC Hamiltonians, without any form of explicit time dependence, can be significant in generating desired behaviors. Specifically, we investigate how hybrid EC unitary evolutions generated by time-local couplings and operators depending only on ($\varrho_t$ or) $\dot{\varrho}_{t}$ meet the following localization optimality criteria: (a) by varying $\xi $ the speed defined by the time scale of wavefunction localization can increase arbitrarily; (b) for any choice of $\mu$, the time distance of the fixed-point attractor of the wavefunction localization is finite; that is, there is a finite time $\tau$ such that for all $t \geqslant \tau$, the ground-state fidelity is stably $1$. Based on the result in Eq. \eqref{1Ns}, the first criterion is satisfied by a slight variation of the profile \eqref{wow} as
\begin{equation}
\label{wow_optimal_a} 
\mathbbmss{H}_{t} = \big( 1 + (\xi )^{2}\big) H - \xi  \dot{\varrho}_{t}, 
\end{equation} 
involving a $\xi$-dependent rescaling of the standard Hamiltonian which, keeping the reference eigenstates and hence the attractor state intact, changes the time scale and the speed of the wavefunction localization to $\tau^{-1} = 2 \xi  \Delta E$ and $v_{\max} = 2 \xi $. \\

The consistent realization of the second criterion, however, involves a deformationally stronger and behaviorally richer EC modification of the profile \eqref{wow}. It is sufficient to work out the simplest model of an EC one-qubit closed system whose unitary evolution, based on $\dot{\varrho}_{t}$-dependent deformations of an arbitrary SQT Hamiltonian $H$ meets the two criteria mentioned above. Anticipating the necessity of state-history dependence as dynamical EC rescaling of $H$ and/or in the effective coupling of the additive EC operator, we introduce the following EC ansatz 
\begin{equation}
\label{wow_optimal_AB} 
\mathbbmss{H}_{t} = \eta_{t} H - \lambda_{t} \dot{\varrho}_{t},    
\end{equation} 
whose complete identification must be deduced as the simplest solution to the inverse problem. Here, $\eta_{t} = \eta (\xi , \dot{\varrho}_{t})$ and $\lambda_{t} = \lambda (\xi  , \dot{\varrho}_{t})$. Integrating the EC von Neumann equation \eqref{ecvovn} for the Hamiltonian \eqref{wow_optimal_AB} can be performed similarly to what we did in Sec. \ref{susbec:PRI-A} for the model of Eq. \eqref{wow}. In particular, considering the case of one-qubit wavefunction localization to the ground state of the SQT Hamiltonian $H$ (with the energy gap $\Delta E$), the equations read as 
\begin{equation}
\label{sol}
 \dot{p}_{t,1} = - 2 \Delta E \frac{\lambda_{t} \eta_{t}}{1 + (\lambda_{t})^{2}} p_{t,1} ( 1- p_{t,1} ) .
\end{equation}

If $\eta_{t} =|\eta|$ and $\lambda_{t} =-| \lambda|$ are state-history independent (hence taken constant upon excluding explicit time dependence), the solution of Eq. \eqref{sol} by choosing $p_{0,1}=1/2$ becomes   
\begin{equation}
 p_{t,1} = 1/(1+e^{-t/\tau}), 
\end{equation}
where $\tau = \frac{1}{2 |\eta| \Delta E} \frac{1 + (\lambda)^{2}}{|\lambda|}$. This gives $t = \tau \ln \big(p_{t,1}/[1- p_{t,1}] \big)$, from which the \textit{landing} time on the fixed-point attractor state is obtained as $t_{\mathrm{land}} \equiv t|_{p_{t,1} = 1} = \infty$. To obtain a finite landing time, at least one of the defining couplings in Eq. \eqref{wow_optimal_AB} needs to become state-history dependent. The characteristic solution which meets  criteria (a) and (b)---as stated earlier in this section---is the EC evolution marked with the linear growth of the ground-state fidelity, based on the monotonically increasing slope $\mathfrak{s}(\xi ) \equiv 2 v(\xi )/\Delta E$, with $v(\xi ) = |\xi |$, all the way from the initial moment $t=0$ up to $t=t_{\mathrm{land}}<\infty$. Hence the characteristic evolution is defined by 
\begin{equation}
\label{cue}
\dot{p}_{t,1} = 2\theta( t_{\mathrm{land}} - t) \,v(\xi )/\Delta E = 2\theta( t_{\mathrm{land}} - t) |\xi|/\Delta E.
\end{equation}
Combining Eqs. (\ref{sol}) and (\ref{cue}) implies that the state-history dependences of the defining couplings of Eq. \eqref{wow_optimal_AB} must be such that
\begin{equation}
\label{cuewsol}
\frac{1 + (\lambda_{t})^{2}}{\lambda_{t} \eta_{t}} \propto  p_{t,1} ( 1- p_{t,1} ) .
\end{equation}
One can observe that the fidelity quantity on the RHS of Eq. \eqref{cuewsol} is itself proportional to $\det[\dot{\varrho}_{t}]$ during the unitary evolution; an \textit{on shell} identity which every solution to the EC von Neumann equation \eqref{ecvovn} realizes. By using the matrix representation of the one-qubit $\varrho_t$ in the eigenbasis of $H$, recalling Eq. \eqref{wow_optimal_AB}, and the assumption of the state-history purity, we conclude 
\begin{equation}
\begin{split} 
\label{onshell}
\det[ \dot{\varrho}_{t}] = - \frac{(\Delta E \,\eta_{t})^{2}} {1 + (\lambda_{t})^{2}} \,p_{t,1} (1- p_{t,1}).
\end{split} 
\end{equation}
Thus, introducing the constant velocity control parameter $\xi $, the condition \eqref{cuewsol} is dynamically equivalent to the following constraint on the state-history dependence of the couplings of the EC Hamiltonian  \eqref{wow_optimal_AB}: 
\begin{equation}
\label{cuewsolerw}
\frac{\eta_{t}  }{\lambda_{t}} =- \frac{\det[\dot{\varrho}_{t}]}{\xi} .
\end{equation}

As a result, we observe the possibility of two complementary categories of characteristic solutions which feature linear wavefunction localization on the ground state---Eq. \eqref{cue} with arbitrarily large localization velocity $v(\xi)$. These categories are
\begin{equation}
\begin{split}
&\textsc{cat a}:\;\;\;  \eta_{t} = 1 ~\mathrm{and}~ \lambda_{t} = |\mu|/\det[ \dot{\varrho}_{t}],\\
&\textsc{cat b}:\;\;\;  \eta_{t} = - \det[ \dot{\varrho}_{t}]  ~\mathrm{and}~ \lambda_{t}  = - |\xi |.
\end{split}
\label{cuewsolerwi2cc}
\end{equation}
Interestingly, we find that \textsc{cat a} can only lead to unitary state histories with some finite temporal extents $t_{\max}$, which are not even sufficiently long for the complete localization of the wavefunction, $t_{\max} <  t_{\mathrm{land}}$. Combining Eq. \eqref{cuewsolerw} and the defining condition of \textsc{cat a} in Eq. \eqref{cuewsolerwi2cc}, we obtain the algebraic equation
\begin{equation}
\label{why}
(\lambda_{t})^{2} \xi  - ( \Delta E )^{2}  p_{t,1} ( 1- p_{t,1} ) \lambda_{t} +  \xi  = 0.
\end{equation}
Reality of the time-dependent roots of this equation, which correspond to the dynamical couplings $\lambda_{t}$, inevitably sets an upper bound $t_{\max}$ in the range $0 < t_{\max} <  t_{\mathrm{land}} $, on the time extent of the state history of any \textsc{cat a} solution. For example, considering the choices $H = \sigma^{3}$, $|\xi | < 1/2$, and the symmetric initial condition $ p_{0,1} = 1/2$, we obtain
\begin{equation}
\label{whyimplies}
|\xi | \leqslant 2  ( 1- p_{t,1} ) p_{t,1} = 2 \Big( \frac{1}{2} - |\xi | t \Big) \Big( \frac{1}{2} + |\xi |t \Big)  \;\;\;\Rightarrow\;\;\; t_{\max}=  \frac{\sqrt{1 - 2 |\xi | }}{2|\xi |} < \frac{1}{2 |\xi | } =  t_{\mathrm{land}}.
\end{equation}

The solutions to \textsc{cat b}---in contrast to \textsc{cat a}---always realize unitary evolutions of arbitrarily long age, which meet criteria (a) and (b); they attain complete stable localization at $t_{\mathrm{land}} = (1 - p_{0,1})/\mathfrak{s}(\xi ) = \Delta E ( 1 - p_{0,1})/(2 |\xi |)$ and realize $t_{\max} = \infty$. In particular, Eqs. (\ref{cuewsolerw}) and (\ref{cuewsolerwi2cc}) for \textsc{cat b} result in the following dynamics of the state-history-dependent scaling $\eta_{t}$:
\begin{equation}
\begin{split}
\eta_{t} =& \frac{1 + (\xi )^{2}}{ ( \Delta E )^{2} } \big[ p_{t,1} ( 1- p_{t,1} ) \big]^{-1} \\
=&  \frac{1 + (\xi )^{2}}{ ( \Delta E )^{2} } \Big[ \Big( p_{0,1} + \frac{2|\xi |}{\Delta E} t  \Big) \Big( 1 - p_{0,1}  - \frac{2|\xi |}{\Delta E} t  \Big)\Big]^{-1}.
\end{split}
\label{s}
\end{equation}
It is important to ensure the fulfillment of the fixed-point attractor condition at the first moment when \textsc{cat b} solutions arrive at complete wavefunction localization, that is, $p_{t_{\mathrm{land}},2} = 0$ and $p_{t_{\mathrm{land}},1} = 1$. To this aim, following the steps as in Sec. \ref{susbec:PRI-A} in solving the EC von Neumann equation \eqref{ecvovn} for the ansatz Hamiltonian \eqref{wow_optimal_AB}, the adjusted analog of the fidelity differential dynamics in Eqs. \eqref{lds} are found to be
\begin{equation} 
\begin{split}
\dot{p}_{t,1} = \frac{2 \lambda_{t} \eta_{t}}{1 + (\lambda_{t})^{2} } \big( E_{1} - \varepsilon_{t} \big) p_{t,1},\\
\dot{p}_{t,2} = \frac{2 \lambda_{t} \eta_{t}}{1 + (\lambda_{t})^{2} } \big( E_{2} - \varepsilon_{t} \big) p_{t,2}. 
\end{split}
\label{lds_318}
\end{equation}
Having $p_{t_{\mathrm{land}},2} =0$ and $\varepsilon_{t_{\mathrm{land}}} = E_{1}$, these equations show that $\dot{p}_{t_{\mathrm{land}},2} = \dot{p}_{t_{\mathrm{land}},1} = 0$. Likewise, Eqs. \eqref{lds_318} imply that all higher-order derivatives of the $p_{t,n}$ variables at $t_{\mathrm{land}}$ vanish, which realizes total termination of the dynamics. \\

%%%%%%%%%%%%%%%%%%%%%%%%%%%%%%%%%%%%%%%%%%%%%%%%%%%%%%%%%%%%%%%%
\begin{figure}[tp]
\includegraphics[width=\linewidth]{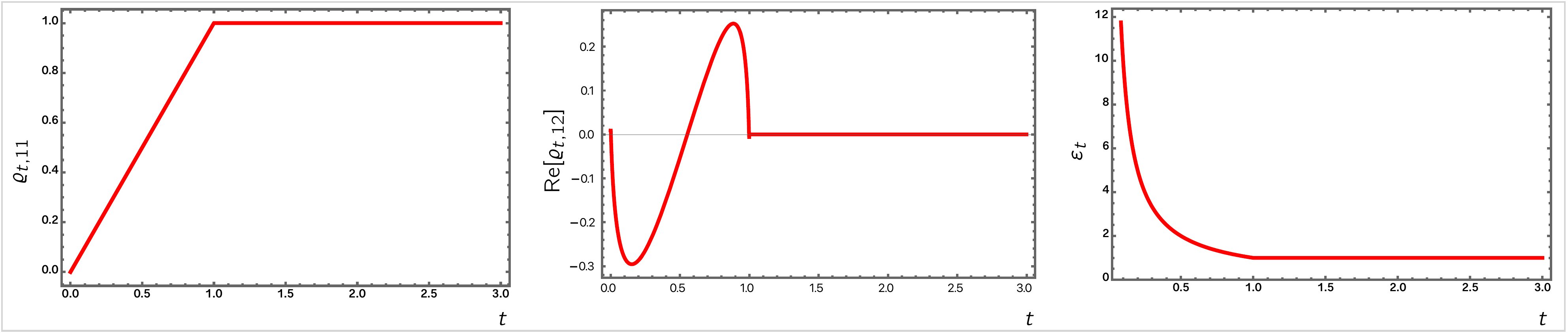}
\caption{Finite-time EC localization of the one-qubit wavefunction to the ground state of $H=\sigma^{3}$, starting from an infinitesimal vicinity of the $\sigma^{3}$ excited state $|2\rangle$, corresponding to the EC Hamiltonian (\ref{wow_optimalAB_renormalized}), with $\xi=-1$ , obtained by numerically solving the corresponding EC von Neumann equation (\ref{ecvovn}). We have used the eigenbasis of $\sigma^{3}$ to depict the plots, such that $\varrho_{t,11}$ is equal to the fidelity of the one-qubit wavefunction and the ground state of $H=\sigma^{3}$. The observed profile of localization is linear, matching Eq. (\ref{cue}), realizing $t_{\mathrm{land}}=1$ according to Eq. (\ref{t_land}), and the Lyapunov function following the analytic solution (\ref{wow_optimalAB_Lyapunov}).}
\label{fig:couplings}
\end{figure}
%%%%%%%%%%%%%%%%%%%%%%%%%%%%%%%%%%%%%%%%%%%%%%%%%%%%%%%%%%%%%%%%

The final point to address is a scaling feature of the characteristic solutions in \textsc{cat b} of the classification \eqref{cuewsolerwi2cc}. As we see in Eq. \eqref{s}, the state dependence of $\eta_{t}$ pushes it dynamically to increase unboundedly toward the landing moment, hence becoming singular at the finitely distant moment $t_{\mathrm{land}}$. This diverging behavior is not exclusive to the characteristic solution with the linear growth of ground-sate fidelity; based on the differential dynamics of the fidelity given in Eq. \eqref{sol}, we can easily prove that regardless of the ramping profile of $p_{t,1}$, any solution with finite landing moment must develop this scaling divergence. Fortunately, we can show that this diverging behavior can be trivially renormalized, hence securing the finiteness of all measurement outcomes. As there is no physical sense in the absolute values of the energy levels of a closed system (in the absence of a cosmological constant), all it takes to erase the formal divergence is to impose a physically trivial ($\varrho_t$ commuting operator) shift to the initial ansatz given in Eq. \eqref{wow_optimal_AB}, so that while the EC dynamics and the eigenstates of $H$ remain invariant, the ground-state energy becomes finite. The remedied version of the Hamiltonian \eqref{wow_optimal_AB} endowed with \textsc{cat b} of state-history-dependent couplings can be presented in the following two behaviorally equivalent forms:
\begin{equation}
\label{wow_optimalAB_renormalized} 
\mathbbmss{H}_{t} = -\det[ \dot{\varrho}_{t}] (H - E_{1} \openone)  +  |\xi | \dot{\varrho}_{t} \overset{\mathrm{behavioral}}{\cong}  -\det[ \dot{\varrho}_{t}] (H - E_{1} \varrho_{t}) +  |\xi | \dot{\varrho}_{t}.   
\end{equation}   
The unitary evolution corresponding to the above Hamiltonian yields an EC state history which lasts forever ($t_{\max} = \infty$), realizes sharp ground-state localization and fixed-point attractor at
\begin{equation}
t_{\mathrm{land}} = (1 - p_{0,1})/ \mathfrak{s}(\xi) = \frac{(1 - p_{0,1}) \Delta E}{2 v(\xi )} = \frac{\Delta E}{2|\xi |} (1 - p_{0,1}),
\label{t_land}
\end{equation}
with arbitrarily large ramping velocity $v(\xi ) = |\xi |$, and has a smooth \textit{Lyapunov function} with a finitely distant attractor of finite value---the energy function $\varepsilon_{t}$. Here $\varepsilon_{t}$ monotonically decreases to a stable plateau of finite value, which is the ground state of the one-qubit closed system; that is, $\lim_{t \to t_\mathrm{land}} \varepsilon_{t}  = \varepsilon_{\mathrm{land}}$. By using Eqs. (\ref{onshell}), (\ref{cuewsolerw}), (\ref{cuewsolerwi2cc}) [\textsc{cat b}], \eqref{cue}, and after some algebra, one can determine the Lyapunov function of the EC Hamiltonian \eqref{wow_optimalAB_renormalized}, and its attractor value $ \varepsilon_{\mathrm{land}}$ as follows: 
\begin{equation}
\label{wow_optimalAB_Lyapunov} 
\begin{split}
\varepsilon_{t} =& \mathrm{Tr}[ \varrho_{t} \mathbbmss{H}_{t}] = \frac{1 + (\xi )^{2}}{\Delta E}\;\frac{1}{p_{t,1}} = \frac{1 + (\xi )^{2}}{\Delta E}\; \Big[ p_{0,1} + \frac{2|\xi |} {\Delta E} \theta\left(\frac{\Delta E}{2|\xi |} ( 1 - p_{0,1} ) - t\right) t\Big]^{-1},\\
\varepsilon_{\mathrm{land}} =& \frac{1 + (\xi )^{2}}{\Delta E}.
\end{split}  
\end{equation}   
Following the above analytic analysis, in Fig. \ref{fig:couplings} we have depicted a representative unitary evolution corresponding to the refined hybrid EC Hamiltonian \eqref{wow_optimalAB_renormalized}, which is obtained by numerically solving the corresponding von Neumann equation \eqref{ecvovn}. As a measure of the effectiveness of the localization performance due to state-history-dependent couplings, we remark that the effective localization under the initial EC hybrid Hamiltonian \eqref{wow} with the same control parameters is about one order of magnitude slower than the one realized by Eq. \eqref{wow_optimalAB_renormalized}.

%%%%%%%%%%%%%%%%%%%%%%%%%%%%%%%%%%%%%%%%%%%%%%%%%%%%%%%%%%%%%%%%
\section{Characterization of General Quantum Behaviors and Phenomenological Significance of ECQT}
\label{sec:toward}

Quantum mechanics is a robust cornerstone of modern physics. It has been extremely successful in explaining as well as predicting diverse natural phenomena across many scales. What has been enabled by standard quantum theory thus far is already fundamentally and technologically vast and promising. Notwithstanding all achievements of this theory, it is open to fundamental (conceptual and structural) scrutiny, in particular in light of insights gained by long-standing attempts to understand quantum nature of spacetime as well as recent information-theoretic derivations of the theory in the new millennium. In fact, there are various natural motivations and observations which suggest deformations or generalizations of standard quantum theory. In particular, we discern at least three dimensions which constructively contribute to such a picture: (i) quantum gravity and emergent spacetime, especially formulated within the ``it-from-(qu)bit'' paradigm \cite{wheeler1, wheeler2, wheeler3}; (ii) information-theoretic reconstructions of SQT, especially within \textit{agent-based} frameworks \cite{qbism1, qbism2, qbism3}, as well as alternative models of emergent quantum theory; and (iii) conceptual and formal impacts from abstract theories of computational intelligence \cite{cibv8, mit}, which can formulate interactive dynamical networks of multiagent Bayesian inference and decision making.\\

Generalizing SQT is a grand project, which likely leads to significant outcomes and fundamental impacts. Here, we concentrate on one of these outcomes which is of particular interest to us: constructive development of an \textit{interactome of quantum theories}, or shortly, \textit{quantum interactome (QI)}, which we denote by $\mathpzc{M}_{Q}$. This interactive landscape is the theory land of all consistent, coexisting, and phenomenologically successful theories of \textit{context-based} (or \textit{situational}) \textit{quantum behaviors}. By ``context'' one means a sufficiently large domain of empirical phenomena and observables which share a \textit{common foundational theme}. We stress that contexts are neither exclusive to individual problems nor narrow and detail sensitive. One can naturally structure $\mathpzc{M}_{Q}$ as a (hypergraphic) \textit{network} wherein each \textit{node} is an independent theory of ``elementary quantum phenomenon'' \cite{wheeler1, wheeler2, wheeler3} within a specific context  (which can be different from the physical spacetime). In other words, each node of $\mathpzc{M}_{Q}$ is a theory of abstract manybody systems comprised of cooperative agents, their chosen queries in the context, and quantum information units of the answers to those queries. This abstract manybody system yields probabilistic \emph{(co)creative inferences} of configurated quantum entities in its own context. The \textit{links} of the network represent various \textit{structural interactions} between these quantum theories such as inclusions, reductions, hierarchies, or emergence relations. In the following, we shall elaborate the elements described in this paragraph and their rationale. In the present paper we have made a first foray into characterization of $\mathpzc{M}_{Q}$ by investigating various interrelations SQT and ECQT \cite{ECQT-1}. \\

Here we briefly outline the mindset behind developing the ECQT and QI. In order to develop generalized quantum theories, a proven methodology to start with, as done in SQT, is to give precedence to \textit{closed} systems rather than \textit{open} systems, and to obtain the latter as a subordinate of the former. In addition, we believe that the plausible approach is to start with purely \textit{dynamical} deformations of SQT, whereas other structural, including kinematical, deformations can be implemented (whenever necessary) through context-based requirements or internal consistency. This precedence is in particular natural when one begins with an abstract quantum system in $0+1$ dimensions out of which physical phenomena are going to emerge. In any such quantum system, the only primal observable at hand is the Hamiltonian. Accordingly, any deformation of the standard structure of the Hamiltonian gives rise to dynamical deformations of SQT. Assuming the nodes of $\mathpzc{M}_{Q}$ are theories of dynamical probability amplitudes in Hilbert spaces, their probability-theoretic consistency requires that the time evolution of quantum states, namely rays $|\Psi_{t}\rangle$ and density operators $\varrho_{t}$, are \textit{isometry}, or in a strong sense, \textit{unitarity}. This principle can be considered as the only fundamental postulate on dynamics in quantum theory. All other dynamical properties of quantum dynamics are context-based in that their rationalizations are conditioned on the associated context of the theory. \\

We remind that the basic context for which SQT, and local quantum field theories as its descendants, has been developed is description of particles which propagate and interact in spacetime, with the majority of relevant observables being accordingly spatiotemporal. Thus, it is important to discern that some of the incorporated principles in SQT have been imposed due to this very context. In particular, to our knowledge, the postulate of strict state-history independence of the Hamiltonians---hence linearity and time locality of the unitary dynamics---in quantum mechanics has been adopted for its simplicity or due to heuristic reasonings or analogies. This postulate turned out to be empirically successful all the way up to the standard model of elementary particles. \\

Naturally, however, theories of \textit{nonlinear} quantum mechanics have also been proposed as alternatives. But these theories have been mostly incoherent and scattered explorations or models. A common feature of the proposed models has been the possibility of anomalies and exotic features in them. Nevertheless, we want to point out that nonlinearity is not fundamentally forbidden and still remains a serious possibility. In particular, a deeper inspection reveals the following points: (i) Derivations of the anomalies in the proposed models, such as acausality \cite{Weinberg-testQM, Weinberg-testQM3, gisin1} or entropy-related concerns \cite{peres1}, for example, in the Weinberg model, or the attempts to prove a universal theorem for linearity of quantum dynamics \cite{jordan1} are not conclusive. Such derivations were intrinsically model dependent, or they had unnecessary underlying assumptions, or often contained hidden elements such as ignoring the possibility of necessary redefinitions, modifications, or less conventional scenarios. It is precisely in this regard that numerous independent constructive reconsiderations have been proposed, e.g., in Refs. \cite{Mielnik, Weinberg-testQM2, polchinski1, czachor1, jordan3, czachor2, bona1, refonlqm_a, fss1, fss2, kent1, jordan2, refonlqm_b, kent2, Beretta-ns-general, hc1, kentrelated, rc1, Beretta-ns}. All these show that there is no fundamental no-go theorem for nonlinearity of quantum theory. (ii) Various quantum theories in $\mathpzc{M}_{Q}$ can be typically immune by context against these anomalies. (iii) Interestingly within the very context of SQT, a natural framework for causal, unitary, measurement-wise consistent nonlinear quantum mechanics from \textit{state-dependent} deformations of quantum field theory has been recently proposed in Ref. \cite{kaplan1}, with experimental proposals and tests in Refs. \cite{kaplanrelated1, Kaplan-Berkeley}. Moreover, as we shall indicate below, recent developments especially in reconstructions of SQT and also quantum gravity encourage toward extensive explorations of physically relevant deformations of SQT. \\

A pervasive thesis to synthesize quantum theory and gravity has been the search for a framework which enables emergence of the physical \textit{spacetime}, including all associated geometrical features such as locality and causality, and spatiotemporal observables from within some fundamental \textit{spaceless} manybody theory in $0+1$ (or $0+0$) dimensions. Within this paradigm, several frameworks have been proposed such as formulations and models coming from the holographic principle (especially those within or inspired by string theory) \cite{AdS/CFT, bfss1, seiberg1, raamsdonk1, swingle1, mera1, syk1, sc1, lee1, lee2, sc2, Carroll-finite}, group field theory (and spin networks) \cite{oriti1, gftw1}, and various alternative pregeometric models of quantum gravity \cite{wheeler4, wheeler5, dp1, sorkin1, antonsen1, sa1, ch1, sethlloyd1, qgraphity1, qgraphity2, qgraphity3, crowther1, swiss1,  sc3}.\\

The basic assumption is that the fundamental manybody system is itself \textit{quantum}, in the sense that its defining degrees of freedom are \textit{prephysical} quantum information, which feature some version of \textit{abstract} quantum behavior. That is, the constituent quantum-information degrees of freedom are defined without any notion of physical spacetime, and hence they are spatially featureless. As such, although inherently no spatiotemporal observables, interactions, or communications exist \textit{a priori} in this theory, semiclassical spacetimes and spatiotemporal interactions and observables (within the standard model and general relativity as effective theories) are to emerge from the underlying abstract quantum behavior at some integrated higher levels. \\

It is evident that such a fundamental theory of abstract pregeometric quantum information does not need to be confined within the SQT conditioning. In fact, even in the contexts less ambitious than complete emergence of spacetime, various independent disciplines of quantum gravity (particularly, string theory and loop quantum gravity) suggest the relevance of nonlinear generalizations of SQT; see, e.g., Refs. \cite{Minic, minic2, minic3} and \cite{Ashtekar}. Moreover, nonlinear quantum mechanics can contribute to resolving deeper puzzles of black hole quantum information processing \cite{salaam}, especially within generalized frameworks of AdS/CFT to reconstruct the interior regions of black holes \cite{hm, s1, bousso1, mp1, mp2,pr1,lp1, susskind1, maldacena1, snow1}. We want to point out that the prerequisite of having an emergent local and causal spacetime out of some abstract quantum behaviors by no means imposes linearity or time locality on the unitary dynamics in the fundamental theory. Hence we propose that the fundamental quantum theory of emergent spacetime can be a distinct node in the QI $\mathpzc{M}_{Q}$. Moreover, we anticipate that in this theory, the unitary evolution operator of the original system can also be experience-centric, hence exploiting the state history as a source for updating manybody interactions. As we have observed in Sec. \ref{sec:phase2-perturbation} (and following Ref. \cite{ECQT-1}), having non-Markovianity and unitarity functionally united enables a broad set of phase diagrams for EC unitary evolutions, which can potentially serve as a promising resource for diverse features such as emergence of dynamical physical spacetimes and standard-model-like interactions and matter particles. \\

An alternative portal to $\mathpzc{M}_{Q}$ is spurred by various approaches, frameworks, or schemes through which SQT becomes a derivative rather than a fundamental theory. In fact, in light of numerous complementary reasons as well as variety of parent models, the idea of having SQT as an (exact or effective) emergent theory has gained significant momentum in recent years. There are two major categories where the frameworks or scenarios of this kind are classified. First, one can reconstruct SQT from first principles based on pure information theory and probabilistic reasoning: some abstract theory of probabilistic information acquisition and objective or subjective Bayesian inference; see the insightful reconstructions proposed recently, e.g., in Refs. \cite{lhardy1, lhardy2, goyal, cb1, cap, mm1, mm2, mm3, mm4, mm5, hohn1, hohn2, qbism4}---and also Ref. \cite{woottersqit1} as a relevant precursor. Alternatively, there are independent frameworks and specific models wherein SQT is recast as approximate or sometimes exact emergent descriptions of other abstract or physical theories or (even classical) manybody systems of various kinds, or as description of some particular phases of these systems or theories; see, e.g., Refs. \cite{adler1, thooft1, thooft2, thooft3, elze1, wetterich1, wetterich2, minic1, smolinweqm, vanchurin1, vanchurin2, QLike, sp1}. It is imperative to highlight the breadth and flexibility of these parent theories yielding approximate emergent quantum descriptions. In particular, SQT can emerge from manybody systems whose microscopic degrees of freedom and interactions are vastly different, encompassing manybody theories as diverse as matrix models, various distinctive classes of cellular automata, \textit{classical} neural networks or complex systems, and holographic bulk-boundary systems. \\

Along with technical evidence coming from all these diverse models, placing SQT and quantum behavior in its generality within the emergence framework also has purely conceptual roots. For example, in SQT the evolution of states or observables are parametrized based on physical, classically measurable time variables. On the ground that any such time is a derivative phenomenon, SQT as a theory for probabilities amplitudes should also be emergent. Now since some consequential properties of emergence can themselves be context dependent, mathematical and physical structures of the Schr\"{o}dinger and von Neumann equations of the alternative quantum theories in $\mathpzc{M}_{Q}$ can admit a wide variety. In particular, in the category of emergent quantum theories the Schr\"{o}dinger equation typically receives various nonlinear deformations. For a few relevant examples and remarks, see Refs. \cite{smolinweqm, minic1, vanchurin1, vanchurin2, sp1}. Moreover, we want to point out that even when restricting ourselves to the exact reconstructions of SQT (in the first category), the exact linearity and time locality of the unitary evolution are consequences of supplemental conditions or technical structures, which are external to the very abstract information theory in a general context. The principal moral of this discussion is that alternative quantum theories in $\mathpzc{M}_{Q}$ can have \textit{largely variant structures of unitary quantum dynamics}, manifesting the variety of their defining contexts. \\

The third dimension of the synthesis which can lead us to contextual quantum theories in $\mathpzc{M}_{Q}$ is deformational structural interactions and mutual impacts between the general disciplines of \textit{intelligent behavior} and \textit{quantum behavior}. It is noted that applications of SQT, especially quantum computation, have been vastly investigated in quantum machine learning and quantum artificial intelligence to devise advantageous algorithms for supervised, unsupervised, and reinforcement learning, and computational intelligence \cite{revoqai1, revoqai2, revoqai3, revoqai4, revoqai5, revoqai6}. There, basically SQT as a resource provides tools for remanufacturing learning systems in order to improve their performance. However, as a powerful forward step we propose that, principles and formalisms of computational intelligence and cognition (artificial \cite{cibv8, mit} or biological \cite{cmpcgn1, cmpcgn2}) can serve as guidelines for formulating contextual quantum behaviors beyond the reach of SQT. To motivate naturalness and utility of this reverse impact, in the following we discern three reasons which indicate how the abstract discipline of computational intelligence can project deep structural deformations for SQT. \\

\begin{itemize}

\item[(i)] Impacts of computational intelligence and cognition on formulating quantum theories in $\mathpzc{M}_{Q}$ can be linked to a fundamental origin, as follows. Information-theoretic constructions of quantum theories are ultimately based on observer-participatory networks (in the Wheelerian sense), where agentive decision making and inferential information processing are developed cooperatively. In effect, this ``interactome'' of agents function as \textit{abstract cognitive machines} performing inferential, probabilistic computations which in particular construct (experiential representations of) the defining contexts of the alternative quantum theories. The bottomline of this reasoning is that mutual structural interactions between quantum and intelligent behaviors can be natural, fundamental, and multipurpose.  

\item[(ii)] Behavioral states and functional phases of complex manybody systems and neural networks can give rise to alternative quantum theories in $\mathpzc{M}_{Q}$, as their \textit{emergent} descriptions. Thus, correspondingly intelligent behavior can be the dominantly determining factor on the defining structure of such quantum theories. This impact can manifest in either of the following forms: (a) the manybody complex system is itself a system with \textit{intelligent behavior} by architecture, e.g., a classical or quantum neural network, or (b) it is some complex manybody system whose sufficiently rich behavioral phase diagram hosts specific states and phases which can feature intelligent behavior. 

\item[(iii)] One can formulate abstract, contextual quantum theories in $\mathpzc{M}_{Q}$ which, by definition and independent of any emergence mechanism, model abstract computational systems which solve goal-driven complex tasks, such as complex decision making, inference, or meta-learning. 

\end{itemize}
To sum up, any of the above scenarios implies that the structural impacts of abstract computational intelligence and cognition on quantum theories can be direct and pivotal. Moreover, it can be argued based on thought experiments and general considerations that profound dynamical deformations of SQT are required for \textit{quantum general intelligence}, as proposed in Ref. \cite{ECQT-1}. The primary components of such deformations are a merge of experience centricity and unitarity. \\

Thus far we have substantiated why and how the notions of \textit{general quantum behavior} and QI are to be  conceived. In the light of these reasonings based on \textit{information theory}, \textit{fundamental or functional emergence}, and \textit{participatory agency\textit}, we establish an interactome of quantum theories $\mathpzc{M}_{Q}$ whose constituent theories all share the following basic elements and characteristics, as \textit{the prerequisites of general quantum behaviors}:   

\begin{itemize}

\item[(i)] \textit{elementary quantum phenomena in agent-participatory networks}: the elementary degrees of freedom, being the superposable ``\textsc{yes}'' or ``\textsc{no}'' (or even ``\textsc{no comment}'') responses to the context-based-chosen queries which are posed by the associated agents in their participatory inferential networks. These information degrees of freedom, which are meaning-based (e.g., association-specifying), interact, evolve, and form Hilbert-space-representable experiential systems.

\item[(ii)] \textit{complementarity}: existence of nonempty classes of queries whose answers are informationally incompatible/nonadditive, 

\item[(iii)] \textit{irreversibility of measurement actions}: registration of acquired information, meaning its retention in the absence of new informational incompatibilities, 

\item[(iv)] \textit{states}: momentary states identified with density operators, encoding the complete sets of updated Bayesian predictions corresponding to agent-system pairs,

\item[(v)] \textit{distinctive role of the pure states}: probability amplitudes encoded in pure states---being maximally informative density operators---with amplitudes identified by Born's rule or its possible consistent generalizations or appropriate context-based deformations,

\item[(vi)] \textit{evolution}: unitarity (or isometry) of time evolution of the states of closed systems. 

\end{itemize}

Upon this abstract characterization, we identify \textit{quantum theory} as the general theory of (fundamental or emergent) quantum behaviors. Beyond the above general quantum behavior prerequisites, any additional postulate or structure (whether dynamical or kinematical) is to be imposed as a \textit{context-defining constraint}, which singles out the specific context-based quantum theory, represented by a constituting node of the QI, $\mathpzc{M}_{Q}$. As we conjecture, many of these nodes should be formulated within ECQT---which is the quantum theory of (fundamental or emergent) experience-centric unitary evolutions as defined in Sec. \ref{sec:ECQT}. The hypergraphical essence of $\mathpzc{M}_{Q}$ manifests the essential fact that there are structural interactions (as explained earlier in this section) between these independent quantum theories, which are represented by different types of hyperlinks.\\

It is highlighted again that ``contexts'' are broad empirical categories of phenomenological observables organized by a central common theme. To bring an example, the organizational theme in quantum mechanics is spacetime, that is, spatiotemporal propagations, correlations, interactions, and observables. It is due to this specific context that SQT has been defined as the quantum theory formulated with the additional postulates of   time locality and linearity of the unitary dynamics. In a way, this important point was revealed when Feynman succeeded in reformulating quantum mechanics---without primarily imposing the standard properties of the Schr\"{o}dinger equation---by choosing a methodology where the very spacetime context plays \textit{a} direct determinant role in constructing SQT \cite{Feynman-I, Feynman-II}. For deeper investigations in recent times on the structural constraints imposed on SQT due to its spacetime context, see, e.g., Refs. \cite{Muller-youtube, Muller-new, Muller-new2}. Along this line, it should be reminded that even in the very spacetime context (including its relativistic extensions) neither dynamical linearity nor dynamical time locality are necessary conditions for the consistency of quantum physics; see, e.g., Refs. \cite{CorrPic, kaplan1, Beretta-ns} regarding nonlinearity and Refs. \cite{Smith-Ahmadi, Trinity, Oreshkov, Guerin, Brukner-, Paiva} regarding time nonlocality. Moving beyond SQT within its ECQT generalization, one becomes able to formulate numerous other contextual nodes of $\mathpzc{M}_{Q}$, some of which as we shall point out in Sec. \ref{sec:outlook}. \\

To sum up, in this section we have presented a general abstract, fundamental basis for the interactive quantum landscape $\mathpzc{M}_{Q}$ and ECQT (serving as its major component). However, we stress that ECQT can also be viewed independent of all the rationale and motivations presented in this section (entirely based on the formulation presented in Sec. \ref{sec:ECQT}---\textit{as is}.\\

Finally, we articulate the transformative insight that may conclusively arise from the complementary, abstract rationale, and general, independent considerations, presented above. In principle, one sees two broad domains of natural and artificial phenomena whose adequate theoretical formulations motivate and welcome larger quantum theories which by construction allow for strongly state-history-sourced quantum behaviors already at the core level of unitarily coherent processes---hence ECQT. Intriguingly, these domains represent the two extreme, opposite limits of the phenomenological spectrum in nature and universe.\\

In one end, we have utterly fundamental theories of the pregeometric phases of universe and multiverse from within which physical, geometric spacetime(s), and the entire standard theory of particle physics are to emerge, perhaps level by level, in their phase diagrams. In a theory of such self-evolving, purely-informational closed quantum systems there is nothing but the experienced information in the state-history of the closed system to update its manybody interactions and guide its evolution. Importantly a pregeometric and immaterial closed system as such cannot have the structural symmetries/invariances/constraints of the to-be-emergent physical spacetime and matter as built-in features imposed on it. Indeed, the principle of Experience Centricity implies that phenomenologically established features such as causal locality, diffeomorphism, CPT, quantum mechanical linearity, and gauge groups all ought to emerge experientially from internal ``experimentations'' of the closed system of a young universe, which follows its self-driven phases of evolution.\\

In the opposite end, we have sufficiently complex manybody systems in which EC quantum behaviors genuinely self-emerge on higher organizational levels. These complex manybody systems, being ubiquitous in nature and technology, can be material, purely symbolic, or hybrid. In particular, they can be widely diverse in their structural instantiations, and can be microscopically quantum mechanical or even classical. On general grounds, it is expected that generic higher-level processes in such complex manybody systems---that generically speaking, fall in our characterization of general quantum behaviors---can be influenced and led by the state-history of the system even in intervals in which the system is decoupled from its environment. Thus, ECQT serves as the emergent, larger quantum theory that formulates their higher-level behaviors.\\

Moreover, a variety of natural and technological domains residing in between the two extremes can substantially benefit from their (possibly effective) reformulations within ECQT. In the next section, we elaborate further the envisioned applications of ECQT across this broad spectrum end-to-end.

%%%%%%%%%%%%%%%%%%%%%%%%%%%%%%%%%%%%%%%%%%%%%%%%%%%%%%%%%%%%%%%%
\section{Foundations for Applications of ECQT}
\label{sec:outlook0}

We lay out in this section foundations for the broad applications of ECQT and QI, which span established areas of physics as well as fronts and domains which have remained thus far out of the direct reach of the traditional theories of physics. These foundations follow, in particular, the independent lines of reasoning put forward in both Sec. \ref{sec:toward} and Ref. \cite{ECQT-1}.  \\

First, we stress that there are various established domains of quantum physics which reasonably invite intriguing and promising applications of ECQT and QI. \\

\begin{enumerate}

\item ECQT can substantially impact the subfields of quantum information processing, including quantum computation, quantum communication, and quantum networks. In this respect, three points should be highlighted. \\

(i) The standard disciplines of quantum information theory and quantum computation have thus far offered an abstract language for SQT. In the past few decades, the significant productivity of this language has proven itself in diverse areas of physics such condensed matters, holographic spacetimes, etc. Yet beneath its abstractness, this language is bound to the contextual constraints of quantum mechanics, which renders it non-universal---``universality'' meant as faithful theoretical inclusivity, not simulational coverage. To bestow universality as such, one is required to revisit and generalize quantum information theory and quantum computation beyond this hidden \textit{mechanical} context. This that quantum information and quantum computation should transcend quantum mechanics to be universal becomes more conspicuous especially where the computation units and processes are \textit{functional} and \textit{emergent} at higher levels. The computational modules and processes of in such systems and states neither mirror the mechanisms which take place between the underlying elements at the anatomical level, nor do they bound themselves necessarily to the mechanistic constraints on the architectures. The QI and ECQT, by their nature and construction, do have the capacity and potency to develop a universal theory of fundamental and emergent information processing and computation. We shall return to this important point in the forthcoming applications of ECQT and $\mathpzc{M}_{Q}$.  \\

(ii) EC unitaries enable a larger space of observable quantities in which quantum information and computational elements can be implanted reliably. In particular, state-history fidelities, their corresponding phase factors, and their many-point descendants (as defined in Eq. \eqref{fam}) can naturally encode a variety of evolving data. Indeed, there are numerous phases of EC unitary evolutions in which such time-nonlocal information encodings are highly robust. Concrete examples of such behavioral phases include Phases \textbf{\textsf{IV}} and \textbf{\textsf{V}} of the $[\hskip-.8mm[2 , 2]\hskip-.8mm]$ EC Hamiltonians of the one-qubit closed systems---Fig. \ref{fig:phase1}---and their higher-order ramifications---Fig. \ref{fig:higherphase}. \\

(iii) It has been shown that nonlinear unitarities can significantly enhance quantum computation \cite{Abrams-Lloyd} and quantum search algorithms \cite{Childs-search, Meyer-search, Wong-thesis}. The time nonlocality, memory resource uses, and especially EC steerability of EC unitary evolutions allow for dissecting, composing, developing, and processing quantum information in novel manners which are more complex and highly orchestrated. These features endow ECQT with the potential to enrich and enhance quantum computation, e.g., by enabling novel and structurally unprecedented quantum circuits, quantum walks, and quantum algorithms, which surpass performance and limits of SQT.\\

(iv) In ECQT systems of information processing, the computational units and modules can be continually redefined and updated according to the experience of the system and its subsystems (Element \ref{def:2}, Sec. \ref{sec:ECQT}). One can anticipate that this characteristic can bring about substantial advantages in computational processes and tasks, which are highly complex. \\
 
Once these points are appreciated, it becomes transparent that one can launch drastically more powerful and versatile quantum information and quantum computation theories---which have the luxury of a universal, context-free framework---within ECQT. Moreover, the generalized families of EC Hamiltonians defined in Element \ref{def:7} of Sec. \ref{sec:ECQT} not only can further enrich these quantum computational systems, but also provide natural candidates for novel quantum-network architectures and quantum communication protocols, which are more intricate and resourceful. Finally we highlight that ECQT can naturally provide unique error correcting processes and protocols with which resilience of quantum computation becomes more easily achievable. Ultimately, these investigations should lead us to the computational landscape---e.g., in the sense of Ref. \cite{Barrett-npj}---of ECQT and QI, which are applicable in highly complex computations in broad disciplines and domains. \\

\item Systematic EC generalizations of the standard theory of open quantum systems---and its different contextual realizations within $\mathpzc{M}_{Q}$---sounds to be promising and can have potential profound impacts. In methodological similarity to quantum mechanics, ECQT has been initiated as a theory for \textit{closed} quantum systems with their \textit{unitary} evolutions. The next natural level shall be moving toward its \textit{open-system} descendent. The formulation begins with the EC unitary evolution of a closed \textit{composite} system (comprised of an open system interacting with an environment) and deducing the nonunitary dynamics the open quantum system of interest. The closed-to-open transition in ECQT, however, brings about the following fundamental differences compared to SQT: \\

(a) The innermost idea in ECQT is that nonunitarity (environmentally induced variations in the information content of open systems) and non-Markovianity (dynamical memory dependence) are essentially two independent aspects of general quantum behaviors in open quantum systems, including diverse contextual specializations. In fact, thinking reversely, ECQT of closed quantum systems can be thought of as the dissipationless limit of a generalized open-system theory where---unlike SQT---non-Markovianity and nonunitarity can have their presences as \textit{independent variable aspects}, while still impacting each other dynamically. \\

(b) Indeed, this decoupling can be explicitly observed upon tracing out environmental degrees of freedom of composite closed systems under EC unitary evolution. The EC open systems inherit their non-Markovianity from two independent sources; the first source being the intrinsic non-Markovianity of their total closed systems; the second source---the one inducing nonunitarity---being correlations and interactions with the environment. This EC trinity of the nonunitarity and the two different kinds of non-Markovianity, in turn, sources rich mutual interactions between these variable aspects (bipartite and tripartite) beyond open-system SQT. Moreover, the EC Hamiltonian of the mother system---as, e.g., in the resolution-refined families of $[\hskip-0.8mm[N,L]\hskip-0.8mm]$ EC Hamiltonians presented in Sec. \ref{sec:ECQT}---typically contains inter-subsystem and environmental interactions of the natures or forms, none of which exist in SQT. In particular, due to the experience centricity of all these novel interactions, the information exchange between the interplaying parties (even including the mother system herself as one of these parties) can be highly more constructive and orchestrated (resembling meaningful communications) than the information backflow in SQT. \\

(c) The contextual realizations of EC open systems, hinging on the above-mentioned characteristics, can naturally lead to diverse novel kinds of nonunitary quantum behaviors which are outstanding phenomenologically and empirically. As already witnessed in the smallest closed quantum system of one qubit, the experience centricity of the quantum dynamics leads to behavioral phases, transitions, and observable effects of diverse and unprecedented kinds. It is interesting to note that already in $4$-dimensional open ECQT, which describes two open-system qubits ``Alice'' and ``Bob'' arising experientially from within their mother systems ``Celine,'' (according to Element \ref{def:2}) become three-body interacting systems and generate highly rich behavioral phase diagrams. More generally, a plethora of even richer behavioral phases and observable effects show up in more complex EC open-system evolutions. Indeed, even in a single EC open system one has numerous phases only some of which would qualitatively show up only in a collection of distinct controlled SQT open systems. In particular, one can mention the following two natural features of EC open-system dynamics: recurrent phase revivals; complex cooperative learning on the basis of dynamically interactive exchanges of experiences among agents.  

\item In recent years an exciting progression of quantum computation theory has risen as quantum artificial intelligence and quantum machine learning. These disciplines apply the enriching techniques of quantum algorithms, quantum circuits, and open quantum dynamics---all---within SQT to advance artificial intelligent systems including machine learning and reinforcement learning \cite{revoqai4, revoqai3}. Following the reasonings presented in Ref. \cite{ECQT-1} and in the above items (1) and (2), combined with the ones presented in the forthcoming paragraphs about general cognitive processes, phenomena, and intelligent behaviors, we envision that ECQT can play a transformative role in all these disciplines and approaches. These ECQT applications include (structurally and functionally) novel generations of neural networks, deep-learning, convolutional neural networks, transformers and large-language models. In a broader and even independent perspective, ECQT can lead to novel intelligent systems featuring meta learning and meta cognition toward \textit{artificial general intelligence}. \\

\item Prototypical spin, fermionic, and bosonic condensed matter systems such as quantum spin models, Hubbard models, and Bose-Hubbard models can be drastically generalized by allowing their internal interactions to become EC. In fact, the simplest example of this type was formulated in Ref. \cite{ECQT-1}. It is exciting to obtain the enlarged phase diagram and novel phase transitions of these EC condensed matter systems, for example, EC Hubbard models, whose SQT versions already capture some phenomenological aspects of strange metals, high-temperature superconductors, and emergent gauge fields. Along these lines, it is intriguing to investigate novel aspects of nonequilibrium dynamics, effective thermalization, and especially scrambling of information in EC quantum manybody systems.  \\

\item It is necessary to point out the possible applications of ECQT in cosmology and ``beyond the standard model'' physics. First, the discernment of possible ECQT observable phenomena in particle physics and cosmology mainly boils down to addressing right, novel empirical questions which can be distinct from the established question sets traditionally addressed within the standard theories of particle physics and cosmology. Second, we highlight that the usual smallness of the empirical bounds on quantum nonlinearities do not necessarily render such ECQT observable phenomena insignificant. Indeed, SQT perturbations with infinitesimally coupled Hermitian EC operators can accumulatively give rise to sizable observable effects in sufficiently long course of time, which can be particularly relevant for cosmological scenarios. Moreover, as we have seen in Sec. \ref{sec:phase2-perturbation}, there are certain classes of \textit{slight} EC perturbations of SQT which can generate nonnegligible and phenomenologically interesting effects. One possible scenario for investigating such phenomena can be obtained via ECQT extensions of nonlinear quantum mechanics of Ref. \cite{kaplan1}, possibly along the lines of Refs. \cite{kaplanrelated1, Kaplan-Berkeley}.  \\

\item At the fundamental level, QI and ECQT may lead to unprecedented, more plausible theories and models of quantum gravity. In this direction, it is primarily interesting to examine different approaches to formulating holographic models within ECQT---in particular, EC generalizations of AdS/CFT and its de Sitter-type variants \cite{AdS/CFT}, matrix or tensor theories such as BFSS model \cite{bfss1}, SYK model \cite{SYK} and its proposed dS-dual variants \cite{SusDual} stand out. Moreover, various independent proposals of quantum gravity such as the ones introduced in Refs. \cite{minic2, Banks-old, Carroll-finite} can benefit from suitable EC generalizations. In addition, revisiting quantum-information theoretic aspects of causal horizons, including blackholes and observer-dependent horizons, within the framework of EC unitary processing and computing sounds promising in enhancement of existing models  or development of new models and formalisms. Finally, understanding that the ultimate and the only perfect closed system is the universe itself, it is natural to advance toward richer models of quantum cosmology using the resolution-refined versions of ECQT. Moving beyond, one can use QI and ECQT to revisit and reinvigorate formulations of grand unified theories particularly within the context of ``it-from-(qu)bit'' paradigms, for example in the spirit of Refs. \cite{book:Wen, Wen2}. \\

\textit{Thus far}, we have suggested plausible applications of ECQT in various disciplines of quantum physics, which however---as we envision---constitute \textit{a small subset of the phenomenological coverage of ECQT and QI}. One is aware that there have been several remarkable investigations about diverse ways to generalize quantum mechanics; see, e.g., Refs. \cite{Mielnik, Kibble-G, Sorkin-MPLA, Adler-etal, weakQT-1, weakQT-2, Hartle-Glafka, Giddings-, QLandscape, Kornyak2, Banks-21, Hartle-21, Muller-generalizing, Kent-new, Mullerresponse}. However, overwhelming majority of this volume is marked by the following three characteristics: (i) it has been basically presumed (sometimes implicitly) that quantum theory is---by context---quantum mechanics including its finite-dimensional reductions and quantum field-theoretical formulations; (ii) these generalizations are mostly motivated and hinted by the search for more fulfilling theories of quantum gravity with sound conceptual basis and sharp, falsifiable, and predictive phenomenological outreach. (iii) Typically they are marked by giving precedence to structural generalizations of SQT which are mainly mathematical or kinematical or on the formalism of measurement operators. \\

However, the standpoints of ECQT and QI is \textit{starkly different}. In this regard, there are several major points, which should be highlighted. \\

(a) Quantum theory \textit{is not} quantum mechanics. Quantum mechanics is quantum theory tailored for a specific context, that is, the organizational theme: spacetime and spatiotemporal propagations, correlations, interactions, and observables. Otherwise, quantum theory---the theory of general quantum behaviors---is a multi-context grand theory which is vastly larger, richer, and more flexible than quantum mechanics.\\

(b) Quantum gravity and more generally emergent spacetime have neither any privileged compass role nor any fundamental phenomenological precedence to guide generalizations of quantum mechanics. \\ 

(c) In transcending quantum mechanics within ECQT, the primary direction of generalizing SQT is dynamics rather than kinematics---merging the time-evolution unitarity with non-Markovian experience centricity. Any other possible kinematical generalizations (including observables and measurements) should follow this dynamical generalization and are based on the context. \\

(d) This that quantum mechanics has taken the role of identifying and representing quantum theory is mainly a historical (not principal) point. Quantum mechanics coexists and interacts with other structurally different context-based quantum theories which are independent and capable of predictive empirical success across domains and phenomena beyond the reach of traditional quantum physics. Indeed, the abstract fundamental nature of $\mathpzc{M}_{Q}$---participatory-agentive ``it-from-(qu)bit'' developing diverse contexts---together with associated flexibility and generality of ECQT formalism places alternative contextual specializations and phenomenological domains of quantum theory at \textit{an equal footing}. In this light broader applications of ECQT are outlined selectively as follows. \\

\item Formulating (``natural'' or artificial) \textit{cognitive science} \cite{book:Cog-1, book:Cog-2}---that is, theoretical and empirical studies of general mental, cognitive, and intelligent phenomena and behaviors---and predicting its empirical observables are among the natural applications of ECQT. As such, \textit{cognitive quantum theories}, as empirically predictive theories of general quantum intelligence and cognition, shall correspond to particular context-based nodes of $\mathpzc{M}_{Q}$. By \textit{general} agentive intelligence and cognition, we especially mean that the agent is able to: cognize, form abstract representations of, and anticipate diverse (internal or external) situations; have meaningful thoughts, emotions, goal orientations, and decision-for-action abilities; process and (in principle) solve general spectrum of problems of diverse kinds and levels, which in particular require general-scope learning, meta learning, abstraction, predictions, and continual self-updating; and meaningfully communicate and interact with other agents. As we see and now explain, ECQT naturally leads to \textit{a universal theory of (emergent) general intelligent and cognitive phenomena and behaviors}. One general prescription to formulate these phenomena and behaviors within ECQT is outlined in what follows. \\ 

Consider a sufficiently complex (``natural'' or artificial) \textit{classical} manybody system $\mathpzc{B}$, regardless of its microscopic units, interactions, and architecture. Moreover, suppose that (as typical) $\mathpzc{B}$ interacts with a dynamical, sufficiently complex environment $\mathpzc{W}$. To be general, it is assumed that  $\mathpzc{B}$-$\mathpzc{W}$ interactions can switch between ``off'' (passive) and ``on'' (active) states of arbitrary durations, and that there can be considerable strength modulations or type variations in the environmental interactions during the ``on'' states. In general, $\mathpzc{W}$ can also include subsystems similar to $\mathpzc{B}$. We stress that, as supposed, $\mathpzc{B}$ and $\mathpzc{W}$ form a classical closed system in total. \\

We now suppose that the ``mental'' states of an emergent cognizant agent $\mathpzc{A}$---whose cognitive behaviors fall into the broad definition of general quantum behaviors in $\mathpzc{M}_{Q}$ (Sec. \ref{sec:toward})---have (somehow) emerged from the internal mechanisms of $\mathpzc{B}$. As such, the (classical) $\mathpzc{BW}$ pair induces an \textit{interacting EC closed quantum system} whose \textit{abstract information-theoretic degrees of freedom} constitute the mental states of $\mathpzc{A}$ in the form of a complementary (and generally time-dependent) pair of \textit{intentional- extrospective} open quantum systems. The \textit{intentional} subsystem ``$\mathpzc{I}_{t}$'' is comprised of those quantum information degrees of freedom which constitute both introspective and intentional sub-states of $\mathpzc{A}$'s mind---those especially associated with $\mathpzc{A}$'s behavioral choices and their (re)organizations. The \textit{extrospective} subsystem ``$\mathpzc{E}_{t}$'' is comprised of the complementary subset of quantum information degrees of freedom constituting $\mathpzc{A}$'s mental states by which---based on both perceived sensory stimuli and behavioral outcomes---the \textit{inferred} internal representations, namely ``cognitive maps,'' together with \textit{predictive} formal models of $\mathpzc{W}$ are formed. As such, $\mathpzc{A} \approx (\mathpzc{I}_{t}, \mathpzc{E}_{t})$, and $\mathpzc{H}^{(\mathpzc{A})}=\mathpzc{H}^{(\mathpzc{I}_{t})} \otimes \mathpzc{H}^{(\mathpzc{E}_{t})}$, where $\mathpzc{H}^{(\mathpzc{I}_{t})}$ and $\mathpzc{H}^{(\mathpzc{E}_{t})}$ accommodate, respectively, the evolving quantum states $\varrho^{(\mathpzc{I}_{t})}_{t}$ and $\varrho^{(\mathpzc{E}_{t})}_{t}$, identified as follows. The intentional sub-states \textit{encode} $\mathpzc{A}$'s chosen probability distribution across its behavioral modes---$\mathpzc{A}$'s behavioral ``policy''---at every present moment $t$. These behavioral encodings can be \textit{time local}, namely inside $\varrho^{(\mathpzc{I}_{t})}_{t}$, or \textit{time nonlocal}, namely within $\{\varrho^{(\mathpzc{I}_{t})}_{t'\leqslant t} \}$, for example, encoded in the state-history $n$-point functions $m_{t_{1}, \ldots , t_{n\geqslant 2}}$. The extrospective sub-states $\varrho^{(\mathpzc{E}_{t})}_{t}$ \textit{encode} at every present moment $t$ the last updated cognitive map that $\mathpzc{A}$ has (passively or actively) inferred about $\mathpzc{W}$. \\

There are seven points in order about the characterization of quantum degrees of freedom and their state spaces as presented above. (i) The time evolutions of extrospective and intentional sub-states are both centered on the integrated \textit{experiences} of $\mathpzc{A}$ and follow Bayesian probabilistic reasoning under uncertainties, since we assume that at typical moments $t$, $\mathpzc{A}$ has only partial knowledge of $\mathpzc{W}$, due to complexity, practical, or epistemic restrictions. (ii) The extrospective Hilbert space $\mathpzc{H}^{(\mathpzc{E}_{t})}$ can correspond to an effective low-dimensional reduction of an otherwise large Hilbert space, capturing only those environmental aspects which $\mathpzc{A}$ cares about. (iii) $\mathpzc{A}$'s cognitive processes (even when their motivations or contents are arisen environmentally) can also go on during the ``off'' states, resembling dreaming states or seclusive contemplative states of $\mathpzc{A}$'s mind. (iv) As typical, there can be within the total $\mathpzc{A} \approx (\mathpzc{I}_{t},\mathpzc{E}_{t})$ closed system a set of noise variables which are particularly associated with the involved complexities. (v) As $\mathpzc{W}$ can include a number of subsystems $\mathpzc{B}$, the considered agent $\mathpzc{A}$ can know about and interact with other agents in its surrounding world. We, however, focus on the single-agent scenario, while we emphasize that the multi-agent scenarios have substantially enriching features. (vi) The time variable with respect to which the mental states and cognitive phenomena of the emergent $\mathpzc{A}$ evolve is typically distinct from the time variable with which the physical closed system $\mathpzc{BW}$ evolves. This time variable, which can be called \textit{cognitive time}, can be discrete or continuous and arises from the internal mechanisms of the physical system $\mathpzc{B}$. The point is that this cognitive time is emergent and arises across higher levels. (vii) The cognitive bipartitioning of the mental Hilbert space $\mathpzc{H}^{(\mathpzc{A})}$ into the subspace factors $\mathpzc{H}^{(\mathpzc{I}_{t})}$ and $\mathpzc{H}^{(\mathpzc{E}_{t})}$ can typically be time-dependent and \textit{experiential}, as explained in Sec. \ref{sec:ECQT}. Moreover, these two Hilbert spaces $\mathpzc{H}^{(\mathpzc{I}_{t})}$ and $\mathpzc{H}^{(\mathpzc{E}_{t})}$ can further take experiential resolutions into cognitively meaningful subsystem Hilbert spaces depending on the empirical specificities. \\ 

As we laid out, the cognizant agent $\mathpzc{A}$ \textit{decides} its behaviors by going through introspective-extrospective cognitive computational processes which are centered on its (sensory and action-outcome) experiences. The adequate formulation of such \textit{general cognitive processes} mirrors unitary or isometric time evolutions of $\mathpzc{A}$'s mental states $\varrho^{(\mathpzc{A})}_{t}$, generated by the instantaneous Hamiltonians $\mathbbmss{H}_{t}^{(\mathpzc{A})}$ which are EC, $(\mathpzc{I}_{t}, \mathpzc{E}_{t})$-resolution refined, and goal-oriented discriminative. That is, the quantum information degrees of freedom constituting $\mathpzc{A}$'s mental states in $\mathpzc{H}^{(\mathpzc{A})} = \mathpzc{H}^{(\mathpzc{I}_{t})} \otimes \mathpzc{H}^{(\mathpzc{E}_{t})}$ evolve based on instantaneous Hamiltonians $\mathbbmss{H}^{(\mathpzc{A})}_{t}$ which (at every present moment $t$) are directly (re)made from the triplet of the past-to-present quantum states $(\{\varrho_{t'}^{(\mathpzc{A})}\}, \{\varrho^{(\mathpzc{I}_{t})}_{t'}\},\{\varrho^{(\mathpzc{E}_{t})}_{t'}\})_{|t'\in[t_{0},t]}$, or the information contained therein. Thus, the structure of the \textit{cognitive EC Hamiltonians} of $\mathpzc{A}$'s mental states, at every present moment of the (continuous or discrete) time ``$t$'' reads as
\begin{equation}
\begin{gathered}
\mathbbmss{H}_{t}^{(\mathpzc{A})} = O_{t}\Big( \mathpzc{P}^{(\mathpzc{A})}_{[t_{0},t]} \textstyle{\bigcup} \mathpzc{P}^{(\mathpzc{A},2|\{\mathpzc{I}_{t} , \mathpzc{E}_{t}\})}_{[t_{0},t]}; \vec{A}^{(\mathpzc{A})}_{t}, \vec{\phi}_{t}; \vec{G}^{(\mathpzc{A}|\vec{\mu}_{t})}_{t} , \vec{P}^{+}_{\vec{G}^{(\mathpzc{A}|\vec{\mu}_{t})}_{t}}\Big), \\
\mathrm{continuous:}\,\,i\dot{\varrho}^{(\mathpzc{A})}_{t} = [\mathbbmss{H}_{t}^{(\mathpzc{A})}, \varrho^{(\mathpzc{A})}_{t}]; \hspace{5cm} \mathrm{discrete:}\,\,\varrho^{(\mathpzc{A})}_{t+1} = e^{-i\mathbbmss{H}^{(\mathpzc{A})}_{t}} \varrho^{(\mathpzc{A})}_{t+1} e^{i\mathbbmss{H}^{(\mathpzc{A})}_{t}}, \\
\\
\Upsilon = \mathbbmss{V}^{(\mathpzc{I}_{t})}_{t} \Big(\mathpzc{P}^{(\mathpzc{I}_{t})}_{[t_{0},t]}, \mathpzc{R}^{(\mathpzc{I}_{t})}_{\,[t_{0},t]}; \vec{A}^{(\mathpzc{I}_{t})}_{t}, \vec{\phi}_{t}; \vec{G}^{(\mathpzc{I}_{t}|\vec{\mu}_{t})}_{t} , \vec{P}^{+}_{\vec{G}^{(\mathpzc{I}_{t}|\vec{\mu}_{t})}_{t}}\Big);\,\,\,\,\,\Upsilon \in\{ \dot{\varrho}^{(\mathpzc{I}_{t})}_{t} \, \mathrm{for\,continuous};\, \varrho^{(\mathpzc{I}_{t})}_{t+1}\,\mathrm{for\, discrete}\}.
\end{gathered}
\label{eq:cogn}
\end{equation}
The character specifications on the first and second lines are as followed: $\mathpzc{P}^{(\mathpzc{A})}_{[t_{0},t]}$ and $\mathpzc{P}^{(\mathpzc{A},2|\{\mathpzc{I}_{t},\mathpzc{E}_{t}\})}_{[t_{0},t]}$ are defined as in Elements \ref{def:1} and \ref{def:2}; $\vec{\phi}_{t}$ and $\vec{A}_{t}$ are, respectively, sets of noise variables and state-history-dependent operators, some of which are induced by $\mathpzc{B}$-$\mathpzc{W}$ interactions; $\vec{G}^{(\mathpzc{A}|\vec{\mu}_{t})}_{t}$ are a set of \textit{goal operators} depending on some \textit{subjective value} parameters $\vec{\mu}_{t}$; and $\vec{P}^{+}_{\vec{G}^{(\mathpzc{A}|\vec{\mu}_{t})}_{t}}$ are projections operators corresponding to favorite eigenstates of the goal operators $\vec{G}^{(\mathpzc{A}|\vec{\mu}_{t})}_{t}$. On the third line of Eq. (\ref{eq:cogn}) we have the same operators but now reduced to the behavioral Hilbert space $\mathpzc{H}^{(\mathpzc{I}_{t})}$, while $\mathpzc{R}^{(\mathpzc{I}_{t})}$ denotes the history of $\mathpzc{A}$'s behavioral outcomes or ``rewards'' up to the present moment $t$. \\

The cognitive EC Hamiltonians (\ref{eq:cogn}) can particularly be in the form of the \textit{three major classes} introduced in Element \ref{def:3} of Sec. \ref{sec:ECQT}. Accordingly, the forms of $\mathpzc{A}$'s EC Hamiltonians can (in part) resemble the numerous explicit examples throughout Elements \ref{def:6}, \ref{def:7}, and \ref{def:xx} of Sec. \ref{sec:ECQT}. The central marker of the agent's intelligence---the \textit{recurrent organizations of its behavior and policy updates}---is mirrored in the experiential re-creations of the instantaneous operators $\mathbbmss{V}^{(\mathpzc{I}_{t})}_{t}$, which are conjunct with or obtainable from the EC Hamiltonian $\mathbbmss{H}^{(\mathpzc{I}_{t})}_{t}$ upon a relevant state-space reduction $\mathpzc{H}^{(\mathpzc{A})}  \to \mathpzc{H}^{(\mathpzc{I}_{t})}$. Finally we want to remark on cognitive quantum theories in $\mathpzc{M}_{Q}$ which are to be derived from ECQT. \\

(i) Thus far we have considered that the physical system $\mathpzc{B}$ and its environment $\mathpzc{W}$ are classical. Aside from concreteness, this assumption is motivated by the foremost examples of $\mathpzc{B}$ representing body and brain, or in the ``artificial'' realm, a sufficiently advanced classical neural network, and $\mathpzc{W}$ representing a large physical world surrounding it. This assumption is, however, by no means necessary. In fact, $\mathpzc{B}$ and $\mathpzc{W}$ can both or either be quantum mechanical systems, or any type of classical-quantum hybrid. One goes back in all these cases to the central idea of the emergence paradigm \cite{Anderson, morereallydiff, emerg2}, according to which a theory that describes observable properties of a system at higher levels of organizations in complex manybody systems can be a theory whose defining degrees of freedom, their interactions, and dynamics are by nature distinct from their macroscopic ones. Cognitive quantum theories are to formulate general cognitive processes and predict their phenomena and observable quantities, which are emergent at higher levels.  \\

(ii) We have stated that emergent cognitive processes respect the principles of general quantum behaviors (Sec. \ref{sec:toward}), such that the ECQT formalism in Sec. \ref{sec:ECQT} formulates them. We now highlight that these assumptions should not taken as being restrictive. Indeed, it is plausible to consider cognitive processes which across diverse internal-external situations can appropriately be quantum, classical, or simultaneous mixtures of them. Still, all these cognitive processes can be formulated, in appropriate ways, within ECQT. Even when cognitive processes are purely classical, they can be flexibly formulated as \textit{classical derivatives} of their general EC formulation (\ref{eq:cogn}). \\

(iii) In SQT treatments of agentive behavioral processes, individual decision-making events are typically incorporated as quantum measurement operations. We now highlight that the ECQT formalism provides various \textit{additional} ways to realize decisional action selections and events alike, which are (a) measurement free and decoherence-scenario independent, (b) naturally arise out of the experience centricity of quantum state evolutions without external controls and without fine tuning, and (c) can bring about substantial utilities in the formulation of cognitive processes outlined in Eq. (\ref{eq:cogn}). This leverage comes about because the total states, sub-states, and state-history $n$-point autocorrelations (in all of which cognitive and behavioral data can be encoded appropriately) can develop harmoniously throughout their EC evolutions recurrent patterns of (perfect or effective) localization-delocalization profiles. One observes this fact in the one-qubit behavioral analyses in Sec. \ref{sec:phase2-perturbation} throughout Secs. \ref{sec:phase1} --  \ref{subsec:PRI-D}, while appreciating that such behavioral features become more diverse and flexible in one-qubit scenarios with momentarily reconfigurable EC Hamiltonian operators ($\mathbbmss{H}_{t}=O_{t}(\mathpzc{P}^{\mathrm{chosen}}_{[t_{0,t}]})$) and (even substantially more) in EC multi-qubit scenarios. This point serves as a clear example of large technical possibilities that the ECQT formalism provides advantageously in modeling general and context-based quantum behaviors.  \\

Following the formulation of general cognitive processes within ECQT portrayed in Eq. (\ref{eq:cogn}), one can move on to extract \textit{cognitive quantum theories} as specific nodes of $\mathpzc{M}_{Q}$ which can predict cognitive observables and phenomena successfully. These extractions require discerning additional principles on top of those of general quantum behaviors (presented in Sec. \ref{sec:toward}), which are specific to the defining context of (general or special) cognition. As the results of these complementary principles, the cognitive EC Hamiltonian (\ref{eq:cogn}) admits additional constraints and structures. \\

In recent times, several streams of exciting investigations and proposals for quantum-theoretical description of cognitive processes and behaviors, mathematical psychology, human decision-making, and consciousness have been put forth---see, e.g., Refs. \cite{book:Qcog, Qwalk-, QQ, humanDM, Khrenn, QCogn, HumanQ, book:Khrenn, Khrenn-Z, Love, Busemeyer1, Aert, Brody}. As we envision, the contributions of this paper together with Ref. \cite{ECQT-1} to the natural evolution of this \textit{profound research field}---which develops constructive dialog between quantum physics and cognitive science---can be promising and substantial. In the perspective of ECQT and QI, the cognitive quantum theories, namely distinct nodes of $\mathpzc{M}_{Q}$ following the above general recipe (\ref{eq:cogn}), should not be taken merely as mathematical modeling. Rather, we highlight that (i) these cognitive quantum theories, being ECQT offsprings, do necessarily differ from SQT in several drastic ways; (ii) what these theories (based on ECQT) are to the higher-level emergent realm of cognitive science and phenomena is what standard theory of particle physics (based on quantum mechanics) is to the lower-level real of elementary particles and their spatiotemporal phenomena; and (iii) by conception and construction, cognitive quantum theories are genuine quantum theories in which general quantum behaviors (Sec. \ref{sec:toward}) specialize on the general context of mental, cognitive, and intelligent processes and phenomena, and (admitting all the involved complexities) shall be able to \textit{predict} their empirical observables.  \\     

\item Versatile and natural applications of ECQT---the major component of QI which formulates \textit{abstract} information-theoretic general and context-based quantum behaviors---can include quantum formulations of diverse formal and phenomenological areas in broader domains and disciplines such as linguistics, sociophysics, economy, and game theory. Earlier investigations have focused on mostly mathematical modeling of such phenomena within (closed- and open-system) SQT---see, e.g., Refs. \cite{book:Linguistic, Khrenn-5, book:KhrennS, book:Qfinance}. Advancing further, one can harness advantageous predictive formulations of these complex phenomena, processes, and their observables within the profound generalization of SQT offered by ECQT. Moreover, substantially novel ways of formulating, modeling, or behaviorally simulating and computing nonlinear classical dynamical systems featuring chaos or turbulence, such as weather forecast, can be investigated within ECQT. Finally we point to promising applications of ECQT for modeling classical probabilistic systems such as stochastic time-series and hidden Markov systems. \\

\end{enumerate}

Indeed, there are numerous questions in different directions and fronts which are pivotal to advance the understanding of ECQT and to extract context-based quantum theories within the QI. We now mention only a few of these exciting questions which are closer to the immediate interest of the present paper. \\

--An intriguing inter-theoretical inquiry is to address whether and how ECQT can emerge from SQT in novel Wilsonian frameworks as a larger quantum theory which stands on its own. \\

--There have been various epistemic derivations of quantum mechanics and particular nonlinear deformations of SQT, or at least some of their main features, from classical statistical theories---see, e.g., Refs. \cite{spekkens-toy, Budiyono1, Budiyono2}. Given the primacy of the Bayesian nature of EC quantum states, it is interesting to work out similar derivations of EC closed and open quantum systems. \\

--One immediate direction to investigate is experimental realizations of EC unitary evolutions and descendant open-system dynamics. Indeed, our proof-of-principle quantum simulation protocols presented in Sec. \ref{sec:SQTsim} already demonstrates that a considerable class of phenomenologically interesting EC unitary evolutions can in principle be simulated using novel SQT quantum circuits. Though, one can look for designing alternative, more resource-efficient quantum simulation networks with enhanced performance and larger coverage of ECQT processes. Independently, based on the understanding of EC quantum behaviors as higher-level functional emergences and their prevalence in natural systems, a promising alternative way to experimentally realize ECQT processes and phenomena is by synthesizing machines which---e.g., in the form of classical neural networks---serves as artificial cognitive connectomes. \\

--It is significant to explore novel features of entanglement dynamics under EC unitary evolutions and their open-system derivatives. For example, one can begin with two-qubit EC quantum systems or the ECQT versions of Heisenberg or Hubbard models.  \\

--It it natural to discern ECQT gates, circuits, and algorithms motivated as follows. In an abstract general perspective, information-theoretic computations can ultimately boil down to a finite number of elementary input-output operations, which upon composing developed modules of information generate new processed information modules. As such, one envisions abstract quantum circuits as formal dynamical systems of information-theoretic grammar and meaningful text generation, which function based on experience centricity. Along these lines, ECQT algorithms can be naturally developed, resembling transition from Turing machines to computational circuits in computer science. Moreover, in the context of quantum cognitive theories, mental and cognitive processes, and especially thought developments, should naturally be based on such abstract elementary gates and circuits. In the paradigm of computational cognition, discerning versatile, finite formal systems of cognitive computations are also practically relevant due to the finiteness of available physical resources for the underlying machines such as the brain. Consequently, this leads to a generalization and enhancement of standard theories of quantum logic and quantum computation within ECQT. Independently, from experimental points of view, it is interesting to design and investigate gates, circuits, and networks for ECQT.  \\

--ECQT can generate distinct spatiotemporal sub-theories in $\mathpzc{M}_{Q}$, which nevertheless are larger and more flexible that quantum mechanics. This requires sub-theories of ECQT with spatiotemporal causality, whose one example can be EC extensions of causal nonlinear quantum mechanics of Ref. \cite{kaplan1}. A fundamental way to develop such causal spatiotemporal sub-theories is to formulate abstract information-theoretic degrees of freedom whose EC interactions are constrained by appropriate adjacency matrices of an abstract incomplete graph. In such scenarios, there can be analog variants of Lieb-Robinson bounds---especially their abstract variants \cite{AdLRB}---from which spatiotemporal locality and causality can emerge. It is evident that from a technical point of view, some structural elements of the so-called ``quantum graphity'' models can be relevant \cite{qgraphity3}. \\

--It is evident that ECQT can generate various novel families of classical theories in its appropriate mechanisms or limits of classicalization. For example, unprecedented models of classical random systems can be deduced from the classical limits of the $[\hskip-.8mm[N,L]\hskip-.8mm]$ EC Hamiltonians and their generalizations and ramifications, presented in Sec. \ref{sec:phase2-perturbation} throughout Secs. \ref{sec:phase1} -- \ref{subsec:PRI-D}. Along these lines, it is interesting to explicitly derive various deformed versions of Fokker-Planck equations, Langevin equations, and classical neural networks, within ECQT classicalizations. Independently, EC classicalization procedures can lead to novel classical nonlinear differential equations---such as drastically deformed Navier-Stokes equations---especially for hydrodynamics which can help capture uncharted spectra of phenomena in fluid systems. Moreover, it is particularly relevant, to work out purely classical formulations of cognitive processes and intelligent behaviors obtainable from classical limits of general cognitive EC Hamiltonians (\ref{eq:cogn}). \\

--In the present paper, as a matter of convenience, we have focused on finite-dimensional quantum systems even in the very context of SQT. The reason has been three-fold. First, the message and main results of the paper are insensitive to the dimensionality of the Hilbert space of the system. Second, finite-dimensional quantum systems are more crucial in abstract information-theoretic computations and inferential versions of quantum theory. However, certainly it is interesting to directly investigate distinct properties of EC closed and open quantum systems with infinitely many degrees of freedom. Furthermore, formulating EC quantum field theories requires systems with continua of degrees of freedom. \\

--Some crystalized features of quantum mechanics such as no-cloning theorem \cite{cloning} and violation of Bell's inequalities \cite{Bell} should be revisited within the much broader scope of ECQT and QI. As we remind, by conception and construction, ECQT formulates emergent or fundamental general quantum behaviors alongside its various context-based specializations, as presented in Sec. \ref{sec:toward}. Such behaviors, being typically marked with experience centricity are described by generalized quantum dynamics in which inherent time-nonlocality and nonlinearity are merged with isometry and unitarity. Thus, since some of the assumptions or structural settings underlying quantum no-cloning and Bell's inequalities lose fundamental ground in ECQT, a major revisit of these quantum mechanical properties become relevant. Quantum no-cloning theorem has already been revisited in works such as Ref. \cite{Tomamichel}, while experience centricity provides a much larger space for scrutiny and generalizations. Furthermore, we highlight that the ECQT reconsiderations of these aspects can be double-faceted, bringing distinctive technicalities with conceptual foundations. Especially one notes in revisiting Bell's inequalities in cognitive quantum theories formulated within the general formulation (\ref{eq:cogn}) the following points. Within the totality of abstract information-theoretic degrees of freedom which comprise the emergent mental states in $\mathpzc{H}^{(\mathpzc{A})}$ (regardless of their partitioning into introspective-extrospective subsystems) the traditional boundaries between ontic-vs.-epistemic and local-vs.-nonlocal characteristics melt intrinsically.  \\

--All foundational landmarks and techniques of SQT can be naturally revisited in ECQT. This list especially includes EC generalization of: uncertainty relations \cite{UR-1,UR-2} for general and context-based quantum behaviors; quantum reference frames \cite{QRef-1, QRef-2, QRef-4, Fields-1, Fields-2} formulated within ECQT and QI; formulating second-quantized (and third-quantized) descendants of ECQT and developing EC quantum field theories.    \\ 

--Following the significance of the adiabatic evolutions, processes, and computations in quantum systems, it is relevant both phenomenologically and technologically to formulate and investigate EC adiabatic phenomena and investigate their properties. Moreover, in similar vein but independently, it is of immediate interest and relevance to formulate and investigate generalized quantum walks \cite{QRW} with EC unitarity and their open-system variants. \\

--Quantifying experience centricity (i.e., interfusion of non-Markovianity and unitarity) in generalized quantum dynamics proposed in Ref. \cite{ECQT-1} and in the present paper is of special interest. This requires formulating observables which can witness or measure locally or globally deviations of EC unitary evolutions away from SQT. \\

--To initiate broader applications of ECQT in domains beyond the traditional disciplines of SQT, as we have explained in this section, we suggest formulating and investigating sufficiently rich models of (a) human decision making within ECQT which advance earlier models such as those of Refs. \cite{Qwalk-, humanDM}, (b) quantum games \cite{QGames} enriched within ECQT, and (c) EC social phenomena, expanding on earlier models, e.g., as in Ref. \cite{JohnsonPRL}. Moreover, it is of special interest to formulate and investigate self-organized complex structures in multi-agent decision-making networks---such as Ref. \cite{Ziepke-etal}---within ECQT. \\

--In this work and in Ref. \cite{ECQT-1}, we have focused on investigating the behaviors of EC Hamiltonians whose chosen experience resources (chosen state histories) do not involve subsystem resolution refining. It is crucial to advance our understanding of distinctive behaviors of EC evolutions based on resolution-refined Hamiltonians for the closed quantum systems \cite{future}.  

%%%%%%%%%%%%%%%%%%%%%%%%%%%%%%%%%%%%%%%%%%%%%%%%%%%%%%%%%%%%%%%%
\section{Closing Remarks on a Portal to a Paradigm Shift} 
\label{sec:outlook}

The quantum interactome (QI) proposed in this work and its major and rigorous component experience-centric quantum theory (ECQT) are, respectively, a powerful framework and rigorous theory which transcend quantum mechanics by relaxing time-locality and linearity of the unitary dynamics and formulating a variety of consistent, independent context-based quantum theories based on structural experience centricity. The developments of ECQT and QI conceptually analogize the transition from Euclidean geometry to non-Euclidean ones, which are formulated in differential and algebraic geometries. In particular, for now two points should be highlighted along this line. First, extensive attempts of mathematicians across numerous centuries to prove the fifth postulate of Euclidian geometry and the impossibility of consistent generalizations of Euclidean geometry basically paved the way to the opposite accomplishment of formulating the early versions of concrete, consistent non-Euclidean geometries, where the fifth postulate transmuted to two distinct categories divided and directed by their foreseen natural contexts \cite{book:Geometry}. Second, it is evident that the elevation of Euclidean geometry to non-Euclidean ones has been not only far-reaching in pure mathematics but also inevitable and essential for applications to describe diverse natural phenomena. \\

In light of this comparison, which is meaningful even from the technical point of view, we point out an important remark as follows about inter-theoretical interactions between SQT and ECQT. (a) It is clear that SQT, being a measure-zero subset of ECQT (Sec. \ref{sec:sqt-vs-ecqt}), can be derived from EQCT in straightforward manners. (b) As we showed in Sec. \ref{sec:SQTsim}, exceedingly higher-dimensional SQT can in principle simulate some interesting classes of ECQT or even might realize ECQT in infinite dimensions. In the above geometrical perspective, this likens the hypersurface embedding of non-Euclidean geometries in higher-dimensional Euclidean spaces. Such embeddings are possible (globally or locally) and can be convenient in addressing certain problems in differential geometry. Nevertheless, one important point still stands. What is responsible for the mathematical richness and phenomenological relevance of non-Euclidean geometries is their curvature, which marks their deviation from the Euclidean postulates. Now, this that the curvature is considered to be embedding induced or intrinsic is not essential. Likewise, it is the very experience centricity of unitary evolutions beyond SQT which causes their behavioral richness and phenomenological versatilities. \\

The present work and Ref. \cite{ECQT-1} have initiated the ambitious and radical program of formulating (from first principles) quantum theory meant as the grand theory of general (elementary or high-level emergent) quantum behaviors which gives rise to a rich variety of context-based quantum theories. These context-based quantum theories include: quantum mechanics and possible quantum gravitational extensions of it; structurally and functionally novel computational systems; theories of emergent complex systems, in particular (but not limited to), cognitive quantum theories and the theory of general intelligence. At the theoretical level, the result shall be a vast and rich theory land comprised of interacting context-based quantum theories which stand at equal footing. At the phenomenological and experimental levels, the significant gain shall be expanding the predictive power of quantum physics especially to new domains of emergent highly complex phenomena. Future works are in progress \cite{future2} in the light of this wisdom.

%%%%%%%%%%%%%%%%%%%%%%%%%%%%%%%%%%%%%%%%%%%%%%%%%%%%%%%%%%%%%%%%
\newpage
\appendix
\section{A Useful Identity} 
\label{sec:identity}

A convenient way to present the reformulation algorithm of Sec. \ref{sec:sqt-reformulation} is based on a straightforward \textit{identity} which is satisfied by state-history couplings of any EC Hamiltonian (\ref{ech:g}). In other words, this identity mirrors Eq. (\ref{ech:g}) inversely. For simplicity and specificity of the discussion, we only focus on the \textit{primitive} EC Hamiltonians. We introduce three matrices denoted by $\boldsymbol{h}_{t}$, $\boldsymbol{\lambda}_{t}$, and $\mathbbmss{T}_{t}$ as follows:

(i) $\boldsymbol{h}_{t}$: The real-valued $(d^{-} + d^{+})$-dimensional vector defined as 
\begin{equation}
\label{vhb}
\boldsymbol{h}_{t} \equiv \big( h_{t}^{1-},\ldots,h_{t}^{d^{-}-},h_{t}^{1+},\ldots,h_{t}^{d^{+}+}\big)^{T},
\end{equation}
whose elements are 
\begin{equation}
\label{vhb-a}
\begin{split}
h_{t}^{j-} &\equiv \mathrm{Tr}\big[i\mathbbmss{h}_{t}^{j-} \mathbbmss{H}_{t}^{\Xi} \big], \\
h_{t}^{k+} &\equiv \mathrm{Tr} \big[\mathbbmss{h}_{t}^{k+} \mathbbmss{H}_{t}^{\Xi} \big].
\end{split}
\end{equation}
Here superscript $T$ denotes transposition. 

(ii) $\boldsymbol{\lambda}_{t}$: The $(d^{-}+d^{+})$-dimensional real-valued vector which consists of all the couplings of the EC Hamiltonian,
\begin{equation}
\boldsymbol{\lambda}_{t} \equiv \big(\lambda_{t}^{1-},\ldots,\lambda_{t}^{d^{-}-},\lambda_{t}^{1+},\ldots,\lambda_{t}^{d^{+}+}\big)^{T}.
\label{ech:cm}
\end{equation}

(iii) $\mathbbmss{T}_{t}$: The state-history-made real-valued $(d^{-} + d^{+})\times (d^{-} + d^{+})$ matrix which collects all the traces of the pairwise products of the state-history monomials, and whose elements are 
\begin{equation}
\begin{split}
\label{mshwt:8}
i T_{t}^{s\mp, r\pm} &\equiv i \mathrm{Tr} \big[\mathbbmss{h}^{s\mp}_{t} \mathbbmss{h}^{r\pm}_{t}\big],\\
i T_{t}^{s\pm, r\mp} &\equiv i \mathrm{Tr}\big[\mathbbmss{h}^{s\pm}_{t} \mathbbmss{h}^{r\mp}_{t}\big],\\
 T_{t}^{s\pm, r\pm} &\equiv  \mathrm{Tr} \big[ \mathbbmss{h}^{s\pm}_{t} \mathbbmss{h}^{r\pm}_{t}\big].
\end{split}
\end{equation}
This matrix provides a dynamical EC characterization of the closed quantum system. Hence its elements can be fully determined as specific time-dependent functions of the state-history two-point functions, $\mathbbmss{T}_{t} = \mathbbmss{T}_{t} \big( \{ m_{t,t''} \}_{|_{\mathpzc{P}^{\mathrm{chosen}}_{[t_{0} , t]}}} \big)$.

The identity in the form of a matrix equation relates the triplet of $(\boldsymbol{h}_{t},\boldsymbol{\lambda}_{t},\mathbbmss{T}_{t})$ for any EC Hamiltonian according Eq. (\ref{ech:g}) and reads as $\boldsymbol{h}_{t} = \mathbbmss{T}_{t} \boldsymbol{\lambda}_{t}$, or equivalently,
\begin{equation}
\label{wgi:ech}
\boldsymbol{\lambda}_{t} = (\mathbbmss{T}_{t})^{-1} \boldsymbol{h}_{t}. 
\end{equation}
Note that the invertibility condition of $\mathbbmss{T}_{t}$ is a mild constraint on the choice of the quantum memory distance $a_{t}$. Independently, in atypical cases in which $\mathbbmss{T}_{t}$ is singular, the invertibility can be retrieved by adding appropriate dynamically trivial terms to the EC Hamiltonian.

The state-history purity $\varrho_{t_{1}} = |\Psi_{t_{1}}\rangle \langle \Psi_{t_{1}}|$ for all moments $t_{1} \in [t_0,t]$ implies that any product of the density operators can be reduced to a sequential product of the state-history two-point functions $m_{t_{1} t_{2}} = w_{t_{1} t_{2}} \, e^{i \alpha_{t_{1} t_{2}}}$, multiplied by a single state-history two-point operator $M_{t' t''} \equiv |\Psi_{t_{1}}\rangle \langle \Psi_{t_{2}}|$. Hence introducing the shorthand $t_{l} \equiv t-a_{l}$ for each present moment $t$ and $a_{l}$ for each quantum memory time distance, one immediately observes that the state-history monomials \eqref{monomials} are reducible to
\begin{equation}
\begin{split}
\label{ksi:8}
\mathbbmss{h}_{t}^{j-} &= w [ j ] \big( e^{ i \alpha [ j ] } |\Psi_{j_{1}}\rangle \langle\Psi_{j_{L_{j}}}|  -  e^{- i \alpha [ j ] } |\Psi_{j_{L_{j}}}\rangle \langle\Psi_{ j_{1}}| \big), \\
\mathbbmss{h}_{t}^{k+} &= w [ k ]\big( e^{ i \alpha [ k ] } |\Psi_{k_{1}}\rangle \langle\Psi_{k_{L_{j}}}|  +  e^{- i \alpha [ k ] } |\Psi_{k_{L_{k}}} \rangle \langle \Psi_{ k_{1}}| \big),
\end{split}
\end{equation}
where 
\begin{equation}
\begin{split}
w [s] &\equiv \textstyle{ \prod_{r=1}^{L_s -1} w_{s_{r} s_{r+1} } },\\
\alpha [s] & \equiv \textstyle{ \sum_{r=1}^{L_s -1} \alpha_{s_{r} s_{r+1} }} .
\end{split}
\end{equation}
Having this result, we can now determine the elements of $\mathbbmss{T}_{t}$ via a straightforward computation,
\begin{equation*} 
\label{cti:8a8} 
\begin{split}
T_{t}^{s\pm, r\pm}  &= 2 w [ s ] \,w [ r ] \big[  w_{ s_{L_{s}} r_{1}} w_{ r_{L_{r}}  s_{1}} \cos\big( \alpha [ s ] + \alpha [ r ] + \alpha_{ s_{L_{s}}  r_{1}} + \alpha_{ r_{L_{r}} s_{1}} \big) \pm w_{ s_{1} r_{1}} w_{ r_{L_{r}}  s_{L_{s}}}  \cos \big( - \alpha [ s ] + \alpha [ r ] + \alpha_{ s_{1} r_{1}} + \alpha_{ r_{L_{r}}  s_{L_{s}}} \big) \big],\\
i T_{t}^{s\pm, r\mp} &= - 2 w [ s ] \,w [ r ] \big[  w_{ s_{L_{s}} r_{1}} w_{ r_{L_{r}}  s_{1}} \sin \big( \alpha [ s ] + \alpha [ r ] + \alpha_{ s_{L_{s}}  r_{1}} + \alpha_{ r_{L_{r}} s_{1}} \big) \pm w_{s_{1} r_{1}} w_{r_{L_{r}}  s }  \sin \big( - \alpha [ s ] + \alpha [ r ] + \alpha_{ s_{1} r_{1}} + \alpha_{ r_{L_{r}}  s_{L_{s}}} \big) \big],\\
i T_{t}^{s\mp, r\pm}  &= - 2 w [ r ] \, w [ s ] \big[  w_{ r_{L_{r}} s_{1}} w_{ s  r_{1}} \sin \big( \alpha [ r ] + \alpha [ s ] + \alpha_{ r_{L_{r}}  s_{1}} + \alpha_{ s r_{1}} \big) \pm w_{ r_{1} s_{1}} w_{s  r_{L_{r}}}  \sin \big( - \alpha [ r ] + \alpha [ s ] + \alpha_{ r_{1} s_{1}} + \alpha_{ s_{L_{r}}  r_{L_{r}}} \big) \big].
\end{split}
\end{equation*}
In addition, a straightforward calculation gives
\begin{equation} 
\label{scp:1}  
\begin{split}
& h_{t}^{j-} = - 2 w [ j] \sin(\alpha [ j])  \,\mathrm{Re} \big(\mathrm{Tr}\big[ |\Psi_{ j_{1}}\rangle \langle\Psi_{ j_{L_{j}}}|  \mathbbmss{H}_{t} \big] \big) - 2 w [ j] \cos(\alpha [ j]) \,\mathrm{Im}\big(\mathrm{Tr}[|\Psi_{ j_{1}}\rangle \langle\Psi_{ j_{L_{j}}}| \mathbbmss{H}_{t}]\big), \\ 
& h_{t}^{k+} =2 w [ k] \cos(\alpha [ k ]  ) \,\mathrm{Re} \big(\mathrm{Tr}\big[|\Psi_{ k_{1}}\rangle \langle\Psi_{ k_{L_{k}}}| \mathbbmss{H}_{t} \big]\big)- 2w [ k ] \sin(\alpha [ k]) \,\mathrm{Im} \big(\mathrm{Tr}[|\Psi_{ k_{1}}\rangle \langle\Psi_{ k_{L_{k}}}| \mathbbmss{H}_{t} ] \big). 
\end{split}  
\end{equation} 
Thus, we now have the RHS of Eq. (\ref{wgi:ech}) in terms of $\mathbbmss{H}^{\Xi}_{t}$ and the states $|\Psi_{t}\rangle$. 

%%%%%%%%%%%%%%%%%%%%%%%%%%%%%%%%%%%%%%%%%%%%%%%%%%%%%%%%%%%%%%%%
%%%%%%%%%%%%%%%%%%%%%%%%%%%%%%%%%%%%%%%%%%%%%%%%%%%%%%%%%%%%%%%%
%\bibliography{ecqt-ref}

\begin{thebibliography}{219}%
\makeatletter
\providecommand \@ifxundefined [1]{%
 \@ifx{#1\undefined}
}%
\providecommand \@ifnum [1]{%
 \ifnum #1\expandafter \@firstoftwo
 \else \expandafter \@secondoftwo
 \fi
}%
\providecommand \@ifx [1]{%
 \ifx #1\expandafter \@firstoftwo
 \else \expandafter \@secondoftwo
 \fi
}%
\providecommand \natexlab [1]{#1}%
\providecommand \enquote  [1]{``#1''}%
\providecommand \bibnamefont  [1]{#1}%
\providecommand \bibfnamefont [1]{#1}%
\providecommand \citenamefont [1]{#1}%
\providecommand \href@noop [0]{\@secondoftwo}%
\providecommand \href [0]{\begingroup \@sanitize@url \@href}%
\providecommand \@href[1]{\@@startlink{#1}\@@href}%
\providecommand \@@href[1]{\endgroup#1\@@endlink}%
\providecommand \@sanitize@url [0]{\catcode `\\12\catcode `\$12\catcode
  `\&12\catcode `\#12\catcode `\^12\catcode `\_12\catcode `\%12\relax}%
\providecommand \@@startlink[1]{}%
\providecommand \@@endlink[0]{}%
\providecommand \url  [0]{\begingroup\@sanitize@url \@url }%
\providecommand \@url [1]{\endgroup\@href {#1}{\urlprefix }}%
\providecommand \urlprefix  [0]{URL }%
\providecommand \Eprint [0]{\href }%
\providecommand \doibase [0]{https://doi.org/}%
\providecommand \selectlanguage [0]{\@gobble}%
\providecommand \bibinfo  [0]{\@secondoftwo}%
\providecommand \bibfield  [0]{\@secondoftwo}%
\providecommand \translation [1]{[#1]}%
\providecommand \BibitemOpen [0]{}%
\providecommand \bibitemStop [0]{}%
\providecommand \bibitemNoStop [0]{.\EOS\space}%
\providecommand \EOS [0]{\spacefactor3000\relax}%
\providecommand \BibitemShut  [1]{\csname bibitem#1\endcsname}%
\let\auto@bib@innerbib\@empty
%</preamble>
\bibitem [{\citenamefont {Tavanfar}\ \emph {et~al.}(2023)\citenamefont
  {Tavanfar}, \citenamefont {Parvizi},\ and\ \citenamefont
  {Pezzutto}}]{ECQT-1}%
  \BibitemOpen
  \bibfield  {author} {\bibinfo {author} {\bibfnamefont {A.}~\bibnamefont
  {Tavanfar}}, \bibinfo {author} {\bibfnamefont {A.}~\bibnamefont {Parvizi}},\
  and\ \bibinfo {author} {\bibfnamefont {M.}~\bibnamefont {Pezzutto}},\
  }\bibfield  {title} {\bibinfo {title} {Unitary evolutions sourced by
  interacting quantum memories: Closed quantum systems directing themselves
  using their state histories},\ }\href
  {https://doi.org/10.22331/q-2023-05-15-1007} {\bibfield  {journal} {\bibinfo
  {journal} {Quantum}\ }\textbf {\bibinfo {volume} {7}},\ \bibinfo {pages}
  {1007} (\bibinfo {year} {2023})}\BibitemShut {NoStop}%
\bibitem [{\citenamefont {Kadanoff}(2009)}]{MFT}%
  \BibitemOpen
  \bibfield  {author} {\bibinfo {author} {\bibfnamefont {L.~P.}\ \bibnamefont
  {Kadanoff}},\ }\bibfield  {title} {\bibinfo {title} {More is the same; phase
  transitions and mean field theories},\ }\href
  {https://doi.org/10.1007/s10955-009-9814-1} {\bibfield  {journal} {\bibinfo
  {journal} {J. Stat. Phys.}\ }\textbf {\bibinfo {volume} {137}},\ \bibinfo
  {pages} {777} (\bibinfo {year} {2009})}\BibitemShut {NoStop}%
\bibitem [{\citenamefont {Hohenberg}\ and\ \citenamefont {Kohn}(1964)}]{DFT}%
  \BibitemOpen
  \bibfield  {author} {\bibinfo {author} {\bibfnamefont {P.}~\bibnamefont
  {Hohenberg}}\ and\ \bibinfo {author} {\bibfnamefont {W.}~\bibnamefont
  {Kohn}},\ }\bibfield  {title} {\bibinfo {title} {Inhomogeneous electron
  gas},\ }\href {https://doi.org/10.1103/PhysRev.136.B864} {\bibfield
  {journal} {\bibinfo  {journal} {Phys. Rev.}\ }\textbf {\bibinfo {volume}
  {136}},\ \bibinfo {pages} {B864} (\bibinfo {year} {1964})}\BibitemShut
  {NoStop}%
\bibitem [{\citenamefont {White}(1992)}]{DMRG}%
  \BibitemOpen
  \bibfield  {author} {\bibinfo {author} {\bibfnamefont {S.~R.}\ \bibnamefont
  {White}},\ }\bibfield  {title} {\bibinfo {title} {{Density Matrix Formulation
  for Quantum Renormalization Groups}},\ }\href
  {https://doi.org/10.1103/PhysRevLett.69.2863} {\bibfield  {journal} {\bibinfo
   {journal} {Phys. Rev. Lett.}\ }\textbf {\bibinfo {volume} {69}},\ \bibinfo
  {pages} {2863} (\bibinfo {year} {1992})}\BibitemShut {NoStop}%
\bibitem [{\citenamefont {Rahimi-Keshari}\ \emph {et~al.}(2016)\citenamefont
  {Rahimi-Keshari}, \citenamefont {Ralph},\ and\ \citenamefont
  {Caves}}]{Keshari}%
  \BibitemOpen
  \bibfield  {author} {\bibinfo {author} {\bibfnamefont {S.}~\bibnamefont
  {Rahimi-Keshari}}, \bibinfo {author} {\bibfnamefont {T.~C.}\ \bibnamefont
  {Ralph}},\ and\ \bibinfo {author} {\bibfnamefont {C.~M.}\ \bibnamefont
  {Caves}},\ }\bibfield  {title} {\bibinfo {title} {Sufficient conditions for
  efficient classical simulation of quantum optics},\ }\href
  {https://doi.org/10.1103/PhysRevX.6.021039} {\bibfield  {journal} {\bibinfo
  {journal} {Phys. Rev. X}\ }\textbf {\bibinfo {volume} {6}},\ \bibinfo {pages}
  {021039} (\bibinfo {year} {2016})}\BibitemShut {NoStop}%
\bibitem [{\citenamefont {Feynman}(1982)}]{Feynman-qsim}%
  \BibitemOpen
  \bibfield  {author} {\bibinfo {author} {\bibfnamefont {R.~P.}\ \bibnamefont
  {Feynman}},\ }\bibfield  {title} {\bibinfo {title} {Simulating physics with
  computers},\ }\href {https://doi.org/10.1007/BF02650179} {\bibfield
  {journal} {\bibinfo  {journal} {Int. J. Theor. Phys.}\ }\textbf {\bibinfo
  {volume} {21}},\ \bibinfo {pages} {467} (\bibinfo {year} {1982})}\BibitemShut
  {NoStop}%
\bibitem [{\citenamefont {Lloyd}(1996)}]{Lloyd-qsim}%
  \BibitemOpen
  \bibfield  {author} {\bibinfo {author} {\bibfnamefont {S.}~\bibnamefont
  {Lloyd}},\ }\bibfield  {title} {\bibinfo {title} {Universal quantum
  simulators},\ }\href {https://doi.org/science.273.5278.107} {\bibfield
  {journal} {\bibinfo  {journal} {Science}\ }\textbf {\bibinfo {volume}
  {273}},\ \bibinfo {pages} {1073} (\bibinfo {year} {1996})}\BibitemShut
  {NoStop}%
\bibitem [{\citenamefont {Altman}\ \emph {et~al.}(2021)\citenamefont {Altman},
  \citenamefont {Brown}, \citenamefont {Carleo}, \citenamefont {Carr},
  \citenamefont {Demler}, \citenamefont {Chin}, \citenamefont {DeMarco},
  \citenamefont {Economou}, \citenamefont {Eriksson}, \citenamefont {Fu},
  \citenamefont {Greiner}, \citenamefont {Hazzard}, \citenamefont {Hulet},
  \citenamefont {Koll\'{a}r}, \citenamefont {Lev}, \citenamefont {Lukin},
  \citenamefont {Ma}, \citenamefont {Mi}, \citenamefont {Misra}, \citenamefont
  {Monroe}, \citenamefont {Murch}, \citenamefont {Nazario}, \citenamefont {Ni},
  \citenamefont {Potter}, \citenamefont {Roushan}, \citenamefont {Saffman},
  \citenamefont {Schleier-Smith}, \citenamefont {Siddiqi}, \citenamefont
  {Simmonds}, \citenamefont {Singh}, \citenamefont {Spielman}, \citenamefont
  {Temme}, \citenamefont {Weiss}, \citenamefont {Vu\v{c}kovi\'{c}},
  \citenamefont {Vuleti\'{c}}, \citenamefont {Ye},\ and\ \citenamefont
  {Zwierlein}}]{qsim-techniques:PRXQ}%
  \BibitemOpen
  \bibfield  {author} {\bibinfo {author} {\bibfnamefont {E.}~\bibnamefont
  {Altman}}, \bibinfo {author} {\bibfnamefont {K.~R.}\ \bibnamefont {Brown}},
  \bibinfo {author} {\bibfnamefont {G.}~\bibnamefont {Carleo}}, \bibinfo
  {author} {\bibfnamefont {L.~D.}\ \bibnamefont {Carr}}, \bibinfo {author}
  {\bibfnamefont {E.}~\bibnamefont {Demler}}, \bibinfo {author} {\bibfnamefont
  {C.}~\bibnamefont {Chin}}, \bibinfo {author} {\bibfnamefont {B.}~\bibnamefont
  {DeMarco}}, \bibinfo {author} {\bibfnamefont {S.~E.}\ \bibnamefont
  {Economou}}, \bibinfo {author} {\bibfnamefont {M.~A.}\ \bibnamefont
  {Eriksson}}, \bibinfo {author} {\bibfnamefont {K.-M.~C.}\ \bibnamefont {Fu}},
  \bibinfo {author} {\bibfnamefont {M.}~\bibnamefont {Greiner}}, \bibinfo
  {author} {\bibfnamefont {K.~R.~A.}\ \bibnamefont {Hazzard}}, \bibinfo
  {author} {\bibfnamefont {R.~G.}\ \bibnamefont {Hulet}}, \bibinfo {author}
  {\bibfnamefont {A.~J.}\ \bibnamefont {Koll\'{a}r}}, \bibinfo {author}
  {\bibfnamefont {B.~L.}\ \bibnamefont {Lev}}, \bibinfo {author} {\bibfnamefont
  {M.~D.}\ \bibnamefont {Lukin}}, \bibinfo {author} {\bibfnamefont
  {R.}~\bibnamefont {Ma}}, \bibinfo {author} {\bibfnamefont {X.}~\bibnamefont
  {Mi}}, \bibinfo {author} {\bibfnamefont {S.}~\bibnamefont {Misra}}, \bibinfo
  {author} {\bibfnamefont {C.}~\bibnamefont {Monroe}}, \bibinfo {author}
  {\bibfnamefont {K.}~\bibnamefont {Murch}}, \bibinfo {author} {\bibfnamefont
  {Z.}~\bibnamefont {Nazario}}, \bibinfo {author} {\bibfnamefont {K.-K.}\
  \bibnamefont {Ni}}, \bibinfo {author} {\bibfnamefont {A.~C.}\ \bibnamefont
  {Potter}}, \bibinfo {author} {\bibfnamefont {P.}~\bibnamefont {Roushan}},
  \bibinfo {author} {\bibfnamefont {M.}~\bibnamefont {Saffman}}, \bibinfo
  {author} {\bibfnamefont {M.}~\bibnamefont {Schleier-Smith}}, \bibinfo
  {author} {\bibfnamefont {I.}~\bibnamefont {Siddiqi}}, \bibinfo {author}
  {\bibfnamefont {R.}~\bibnamefont {Simmonds}}, \bibinfo {author}
  {\bibfnamefont {M.}~\bibnamefont {Singh}}, \bibinfo {author} {\bibfnamefont
  {I.~B.}\ \bibnamefont {Spielman}}, \bibinfo {author} {\bibfnamefont {K.~V.}\
  \bibnamefont {Temme}}, \bibinfo {author} {\bibfnamefont {D.~S.}\ \bibnamefont
  {Weiss}}, \bibinfo {author} {\bibfnamefont {J.}~\bibnamefont
  {Vu\v{c}kovi\'{c}}}, \bibinfo {author} {\bibfnamefont {V.}~\bibnamefont
  {Vuleti\'{c}}}, \bibinfo {author} {\bibfnamefont {J.}~\bibnamefont {Ye}},\
  and\ \bibinfo {author} {\bibfnamefont {M.}~\bibnamefont {Zwierlein}},\
  }\bibfield  {title} {\bibinfo {title} {Quantum simulators: Architectures and
  opportunities},\ }\href {https://doi.org/10.1103/PRXQuantum.2.017003}
  {\bibfield  {journal} {\bibinfo  {journal} {PRX Quantum}\ }\textbf {\bibinfo
  {volume} {2}},\ \bibinfo {pages} {017003} (\bibinfo {year}
  {2021})}\BibitemShut {NoStop}%
\bibitem [{\citenamefont {Ekert}\ \emph {et~al.}(2002)\citenamefont {Ekert},
  \citenamefont {Alves}, \citenamefont {Oi}, \citenamefont {Horodecki},
  \citenamefont {Horodecki},\ and\ \citenamefont {Kwek}}]{Ekert-etal}%
  \BibitemOpen
  \bibfield  {author} {\bibinfo {author} {\bibfnamefont {A.~K.}\ \bibnamefont
  {Ekert}}, \bibinfo {author} {\bibfnamefont {C.~M.}\ \bibnamefont {Alves}},
  \bibinfo {author} {\bibfnamefont {D.~K.~L.}\ \bibnamefont {Oi}}, \bibinfo
  {author} {\bibfnamefont {M.}~\bibnamefont {Horodecki}}, \bibinfo {author}
  {\bibfnamefont {P.}~\bibnamefont {Horodecki}},\ and\ \bibinfo {author}
  {\bibfnamefont {L.~C.}\ \bibnamefont {Kwek}},\ }\bibfield  {title} {\bibinfo
  {title} {{Direct Estimations of Linear and Nonlinear Functionals of a Quantum
  State}},\ }\href {https://doi.org/10.1103/PhysRevLett.88.217901} {\bibfield
  {journal} {\bibinfo  {journal} {Phys. Rev. Lett.}\ }\textbf {\bibinfo
  {volume} {88}},\ \bibinfo {pages} {217901} (\bibinfo {year}
  {2002})}\BibitemShut {NoStop}%
\bibitem [{\citenamefont {Hagan}\ and\ \citenamefont {Wiebe}()}]{qHsim}%
  \BibitemOpen
  \bibfield  {author} {\bibinfo {author} {\bibfnamefont {M.}~\bibnamefont
  {Hagan}}\ and\ \bibinfo {author} {\bibfnamefont {N.}~\bibnamefont {Wiebe}},\
  }\href@noop {} {\bibinfo {title} {Composite quantum simulations}},\ \Eprint
  {https://arxiv.org/abs/2206.06409} {arXiv:2206.06409} \BibitemShut {NoStop}%
\bibitem [{\citenamefont {Poulin}\ \emph {et~al.}(2011)\citenamefont {Poulin},
  \citenamefont {Qarry}, \citenamefont {Somma},\ and\ \citenamefont
  {Verstraete}}]{Poulin-qsim-tH}%
  \BibitemOpen
  \bibfield  {author} {\bibinfo {author} {\bibfnamefont {D.}~\bibnamefont
  {Poulin}}, \bibinfo {author} {\bibfnamefont {A.}~\bibnamefont {Qarry}},
  \bibinfo {author} {\bibfnamefont {R.}~\bibnamefont {Somma}},\ and\ \bibinfo
  {author} {\bibfnamefont {F.}~\bibnamefont {Verstraete}},\ }\bibfield  {title}
  {\bibinfo {title} {{Quantum Simulation of Time-Dependent Hamiltonians and the
  Convenient Illusion of Hilbert Space}},\ }\href
  {https://doi.org/10.1103/PhysRevLett.106.170501} {\bibfield  {journal}
  {\bibinfo  {journal} {Phys. Rev. Lett.}\ }\textbf {\bibinfo {volume} {106}},\
  \bibinfo {pages} {170501} (\bibinfo {year} {2011})}\BibitemShut {NoStop}%
\bibitem [{\citenamefont {Lloyd}\ \emph {et~al.}(2014)\citenamefont {Lloyd},
  \citenamefont {Mohseni},\ and\ \citenamefont {Rebentrost}}]{Lloyd-DME}%
  \BibitemOpen
  \bibfield  {author} {\bibinfo {author} {\bibfnamefont {S.}~\bibnamefont
  {Lloyd}}, \bibinfo {author} {\bibfnamefont {M.}~\bibnamefont {Mohseni}},\
  and\ \bibinfo {author} {\bibfnamefont {P.}~\bibnamefont {Rebentrost}},\
  }\bibfield  {title} {\bibinfo {title} {Quantum principal component
  analysis},\ }\href {https://doi.org/10.1038/nphys3029} {\bibfield  {journal}
  {\bibinfo  {journal} {Nat. Phys.}\ }\textbf {\bibinfo {volume} {10}},\
  \bibinfo {pages} {631} (\bibinfo {year} {2014})}\BibitemShut {NoStop}%
\bibitem [{\citenamefont {Kjaergaard}\ \emph {et~al.}(2022)\citenamefont
  {Kjaergaard}, \citenamefont {Schwartz}, \citenamefont {Greene}, \citenamefont
  {Samach}, \citenamefont {Bengtsson}, \citenamefont {O'Keeffe}, \citenamefont
  {McNally}, \citenamefont {Braum\"{u}ller}, \citenamefont {Kim}, \citenamefont
  {Krantz}, \citenamefont {Marvian}, \citenamefont {Melville}, \citenamefont
  {Niedzielski}, \citenamefont {Sung}, \citenamefont {Winik}, \citenamefont
  {Yoder}, \citenamefont {Rosenberg}, \citenamefont {Obenland}, \citenamefont
  {Lloyd}, \citenamefont {Orlando}, \citenamefont {Marvian}, \citenamefont
  {Gustavsson},\ and\ \citenamefont {Oliver}}]{DME-experimental}%
  \BibitemOpen
  \bibfield  {author} {\bibinfo {author} {\bibfnamefont {M.}~\bibnamefont
  {Kjaergaard}}, \bibinfo {author} {\bibfnamefont {M.~E.}\ \bibnamefont
  {Schwartz}}, \bibinfo {author} {\bibfnamefont {A.}~\bibnamefont {Greene}},
  \bibinfo {author} {\bibfnamefont {G.~O.}\ \bibnamefont {Samach}}, \bibinfo
  {author} {\bibfnamefont {A.}~\bibnamefont {Bengtsson}}, \bibinfo {author}
  {\bibfnamefont {M.}~\bibnamefont {O'Keeffe}}, \bibinfo {author}
  {\bibfnamefont {C.~M.}\ \bibnamefont {McNally}}, \bibinfo {author}
  {\bibfnamefont {J.}~\bibnamefont {Braum\"{u}ller}}, \bibinfo {author}
  {\bibfnamefont {D.~K.}\ \bibnamefont {Kim}}, \bibinfo {author} {\bibfnamefont
  {P.}~\bibnamefont {Krantz}}, \bibinfo {author} {\bibfnamefont
  {M.}~\bibnamefont {Marvian}}, \bibinfo {author} {\bibfnamefont
  {A.}~\bibnamefont {Melville}}, \bibinfo {author} {\bibfnamefont {B.~M.}\
  \bibnamefont {Niedzielski}}, \bibinfo {author} {\bibfnamefont
  {Y.}~\bibnamefont {Sung}}, \bibinfo {author} {\bibfnamefont {R.}~\bibnamefont
  {Winik}}, \bibinfo {author} {\bibfnamefont {J.}~\bibnamefont {Yoder}},
  \bibinfo {author} {\bibfnamefont {D.}~\bibnamefont {Rosenberg}}, \bibinfo
  {author} {\bibfnamefont {K.}~\bibnamefont {Obenland}}, \bibinfo {author}
  {\bibfnamefont {S.}~\bibnamefont {Lloyd}}, \bibinfo {author} {\bibfnamefont
  {T.~P.}\ \bibnamefont {Orlando}}, \bibinfo {author} {\bibfnamefont
  {I.}~\bibnamefont {Marvian}}, \bibinfo {author} {\bibfnamefont
  {S.}~\bibnamefont {Gustavsson}},\ and\ \bibinfo {author} {\bibfnamefont
  {W.~D.}\ \bibnamefont {Oliver}},\ }\bibfield  {title} {\bibinfo {title}
  {Demonstration of density matrix exponentiation using a superconducting
  quantum processor},\ }\href {https://doi.org/10.1103/PhysRevX.12.011005}
  {\bibfield  {journal} {\bibinfo  {journal} {Phys. Rev. X}\ }\textbf {\bibinfo
  {volume} {12}},\ \bibinfo {pages} {011005} (\bibinfo {year}
  {2022})}\BibitemShut {NoStop}%
\bibitem [{\citenamefont {Richard}(2003)}]{delay-DE}%
  \BibitemOpen
  \bibfield  {author} {\bibinfo {author} {\bibfnamefont {J.-P.}\ \bibnamefont
  {Richard}},\ }\bibfield  {title} {\bibinfo {title} {{Time delay systems: An
  overview of some recent advances and open problems}},\ }\href
  {https://doi.org/10.1016/S0005-1098(03)00167-5} {\bibfield  {journal}
  {\bibinfo  {journal} {Automatica}\ }\textbf {\bibinfo {volume} {39}},\
  \bibinfo {pages} {1667} (\bibinfo {year} {2003})}\BibitemShut {NoStop}%
\bibitem [{\citenamefont {Ericsson}(2002)}]{Ericsson-thesis}%
  \BibitemOpen
  \bibfield  {author} {\bibinfo {author} {\bibfnamefont {M.}~\bibnamefont
  {Ericsson}},\ }\emph {\bibinfo {title} {Geometric and topological phases with
  applications to quantum computation}},\ \href@noop {} {Ph.D. thesis},\
  \bibinfo  {school} {Uppsala University} (\bibinfo {year} {2002})\BibitemShut
  {NoStop}%
\bibitem [{\citenamefont {Oi}(2003)}]{Oi}%
  \BibitemOpen
  \bibfield  {author} {\bibinfo {author} {\bibfnamefont {D.~K.~L.}\
  \bibnamefont {Oi}},\ }\bibfield  {title} {\bibinfo {title} {{Interference of
  Quantum Channels}},\ }\href {https://doi.org/10.1103/PhysRevLett.91.067902}
  {\bibfield  {journal} {\bibinfo  {journal} {Phys. Rev. Lett.}\ }\textbf
  {\bibinfo {volume} {91}},\ \bibinfo {pages} {067902} (\bibinfo {year}
  {2003})}\BibitemShut {NoStop}%
\bibitem [{\citenamefont {Oi}\ and\ \citenamefont {{\AA}berg}(2006)}]{OiAb}%
  \BibitemOpen
  \bibfield  {author} {\bibinfo {author} {\bibfnamefont {D.~K.~L.}\
  \bibnamefont {Oi}}\ and\ \bibinfo {author} {\bibfnamefont {J.}~\bibnamefont
  {{\AA}berg}},\ }\bibfield  {title} {\bibinfo {title} {{Fidelity and Coherence
  Measures from Interference}},\ }\href
  {https://doi.org/10.1103/PhysRevLett.97.220404} {\bibfield  {journal}
  {\bibinfo  {journal} {Phys. Rev. Lett.}\ }\textbf {\bibinfo {volume} {97}},\
  \bibinfo {pages} {220404} (\bibinfo {year} {2006})}\BibitemShut {NoStop}%
\bibitem [{\citenamefont {Kanjilal}\ \emph {et~al.}(2023)\citenamefont
  {Kanjilal}, \citenamefont {Pandey},\ and\ \citenamefont {Pati}}]{Pati-etal}%
  \BibitemOpen
  \bibfield  {author} {\bibinfo {author} {\bibfnamefont {S.}~\bibnamefont
  {Kanjilal}}, \bibinfo {author} {\bibfnamefont {V.}~\bibnamefont {Pandey}},\
  and\ \bibinfo {author} {\bibfnamefont {A.~K.}\ \bibnamefont {Pati}},\
  }\bibfield  {title} {\bibinfo {title} {Entanglement meter: {E}stimation of
  entanglement with single copy in interferometer},\ }\href
  {https://doi.org/0.1088/1367-2630/accd8d} {\bibfield  {journal} {\bibinfo
  {journal} {New J. Phys.}\ }\textbf {\bibinfo {volume} {25}},\ \bibinfo
  {pages} {043026} (\bibinfo {year} {2023})}\BibitemShut {NoStop}%
\bibitem [{\citenamefont {Marvian}\ and\ \citenamefont
  {Lloyd}()}]{Marvian-Lloyd}%
  \BibitemOpen
  \bibfield  {author} {\bibinfo {author} {\bibfnamefont {I.}~\bibnamefont
  {Marvian}}\ and\ \bibinfo {author} {\bibfnamefont {S.}~\bibnamefont
  {Lloyd}},\ }\href@noop {} {\bibinfo {title} {Universal quantum emulator}},\
  \Eprint {https://arxiv.org/abs/1606.02734} {arXiv:1606.02734} \BibitemShut
  {NoStop}%
\bibitem [{\citenamefont {Alicki}\ \emph {et~al.}(2009)\citenamefont {Alicki},
  \citenamefont {Fannes},\ and\ \citenamefont {Pogorzelska}}]{Alicki}%
  \BibitemOpen
  \bibfield  {author} {\bibinfo {author} {\bibfnamefont {R.}~\bibnamefont
  {Alicki}}, \bibinfo {author} {\bibfnamefont {M.}~\bibnamefont {Fannes}},\
  and\ \bibinfo {author} {\bibfnamefont {M.}~\bibnamefont {Pogorzelska}},\
  }\bibfield  {title} {\bibinfo {title} {Quantum generalized subsystems},\
  }\href {https://doi.org/10.1103/PhysRevA.79.052111} {\bibfield  {journal}
  {\bibinfo  {journal} {Phys. Rev. A}\ }\textbf {\bibinfo {volume} {79}},\
  \bibinfo {pages} {052111} (\bibinfo {year} {2009})}\BibitemShut {NoStop}%
\bibitem [{\citenamefont {Duarte}\ \emph {et~al.}(2017)\citenamefont {Duarte},
  \citenamefont {Carvalho}, \citenamefont {Bernardes},\ and\ \citenamefont
  {de~Melo}}]{noTr1}%
  \BibitemOpen
  \bibfield  {author} {\bibinfo {author} {\bibfnamefont {C.}~\bibnamefont
  {Duarte}}, \bibinfo {author} {\bibfnamefont {G.~D.}\ \bibnamefont
  {Carvalho}}, \bibinfo {author} {\bibfnamefont {N.~K.}\ \bibnamefont
  {Bernardes}},\ and\ \bibinfo {author} {\bibfnamefont {F.}~\bibnamefont
  {de~Melo}},\ }\bibfield  {title} {\bibinfo {title} {Emerging dynamics arising
  from coarse-grained quantum systems},\ }\href
  {https://doi.org/10.1103/PhysRevA.96.032113} {\bibfield  {journal} {\bibinfo
  {journal} {Phys. Rev. A}\ }\textbf {\bibinfo {volume} {96}},\ \bibinfo
  {pages} {032113} (\bibinfo {year} {2017})}\BibitemShut {NoStop}%
\bibitem [{\citenamefont {Singh}\ and\ \citenamefont
  {Carroll}(2018)}]{Carroll-coarse}%
  \BibitemOpen
  \bibfield  {author} {\bibinfo {author} {\bibfnamefont {A.}~\bibnamefont
  {Singh}}\ and\ \bibinfo {author} {\bibfnamefont {S.~M.}\ \bibnamefont
  {Carroll}},\ }\bibfield  {title} {\bibinfo {title} {{Quantum decimation in
  Hilbert space: Coarse graining without structure}},\ }\href
  {https://doi.org/10.1103/PhysRevA.97.032111} {\bibfield  {journal} {\bibinfo
  {journal} {Phys. Rev. A}\ }\textbf {\bibinfo {volume} {97}},\ \bibinfo
  {pages} {032111} (\bibinfo {year} {2018})}\BibitemShut {NoStop}%
\bibitem [{\citenamefont {Agon}\ \emph {et~al.}(2018)\citenamefont {Agon},
  \citenamefont {Balasubramanian}, \citenamefont {Kasko},\ and\ \citenamefont
  {Lawrence}}]{Subra}%
  \BibitemOpen
  \bibfield  {author} {\bibinfo {author} {\bibfnamefont {C.}~\bibnamefont
  {Agon}}, \bibinfo {author} {\bibfnamefont {V.}~\bibnamefont
  {Balasubramanian}}, \bibinfo {author} {\bibfnamefont {S.}~\bibnamefont
  {Kasko}},\ and\ \bibinfo {author} {\bibfnamefont {A.}~\bibnamefont
  {Lawrence}},\ }\bibfield  {title} {\bibinfo {title} {Coarse grained quantum
  dynamics},\ }\href {https://doi.org/10.1103/PhysRevD.98.025019} {\bibfield
  {journal} {\bibinfo  {journal} {Phys. Rev. D}\ }\textbf {\bibinfo {volume}
  {98}},\ \bibinfo {pages} {025019} (\bibinfo {year} {2018})}\BibitemShut
  {NoStop}%
\bibitem [{\citenamefont {Kabernik}\ \emph {et~al.}(2020)\citenamefont
  {Kabernik}, \citenamefont {Pollack},\ and\ \citenamefont {Singh}}]{Kaber}%
  \BibitemOpen
  \bibfield  {author} {\bibinfo {author} {\bibfnamefont {O.}~\bibnamefont
  {Kabernik}}, \bibinfo {author} {\bibfnamefont {J.}~\bibnamefont {Pollack}},\
  and\ \bibinfo {author} {\bibfnamefont {A.}~\bibnamefont {Singh}},\ }\bibfield
   {title} {\bibinfo {title} {Quantum state reduction: Generalized bipartitions
  from algebras of observables},\ }\href
  {https://doi.org/10.1103/PhysRevA.101.032303} {\bibfield  {journal} {\bibinfo
   {journal} {Phys. Rev. A}\ }\textbf {\bibinfo {volume} {101}},\ \bibinfo
  {pages} {032303} (\bibinfo {year} {2020})}\BibitemShut {NoStop}%
\bibitem [{\citenamefont {Cotler}\ and\ \citenamefont {Strominger}()}]{cs1}%
  \BibitemOpen
  \bibfield  {author} {\bibinfo {author} {\bibfnamefont {J.}~\bibnamefont
  {Cotler}}\ and\ \bibinfo {author} {\bibfnamefont {A.}~\bibnamefont
  {Strominger}},\ }\href@noop {} {\bibinfo {title} {The universe as a quantum
  encoder}},\ \Eprint {https://arxiv.org/abs/2201.11658} {arXiv:2201.11658}
  \BibitemShut {NoStop}%
\bibitem [{\citenamefont {Kornyak}(2012)}]{Kornyak1}%
  \BibitemOpen
  \bibfield  {author} {\bibinfo {author} {\bibfnamefont {V.~V.}\ \bibnamefont
  {Kornyak}},\ }\bibfield  {title} {\bibinfo {title} {Mathematical modeling of
  finite quantum systems},\ }in\ \href@noop {} {\emph {\bibinfo {booktitle}
  {Mathematical Modeling and Computational Science, MMCP 2011}}},\ \bibinfo
  {editor} {edited by\ \bibinfo {editor} {\bibfnamefont {G.}~\bibnamefont
  {Adam}}, \bibinfo {editor} {\bibfnamefont {J.}~\bibnamefont {Bu{\v{s}}a}},\
  and\ \bibinfo {editor} {\bibfnamefont {M.}~\bibnamefont {Hnati{\v{c}}}}}\
  (\bibinfo  {publisher} {Springer},\ \bibinfo {address} {Berlin},\ \bibinfo
  {year} {2012})\ p.~\bibinfo {pages} {79}\BibitemShut {NoStop}%
\bibitem [{\citenamefont {Kornyak}(2018)}]{Kornyak2}%
  \BibitemOpen
  \bibfield  {author} {\bibinfo {author} {\bibfnamefont {V.}~\bibnamefont
  {Kornyak}},\ }\bibfield  {title} {\bibinfo {title} {Modeling quantum behavior
  in the framework of permutation groups},\ }\href
  {https://doi.org/10.1051/epjconf/201817301007} {\bibfield  {journal}
  {\bibinfo  {journal} {EPJ Web Conf.}\ }\textbf {\bibinfo {volume} {173}},\
  \bibinfo {pages} {01007} (\bibinfo {year} {2018})}\BibitemShut {NoStop}%
\bibitem [{\citenamefont {Banks}({\natexlab{a}})}]{Banks-21}%
  \BibitemOpen
  \bibfield  {author} {\bibinfo {author} {\bibfnamefont {T.}~\bibnamefont
  {Banks}},\ }\href@noop {} {\bibinfo {title} {Finite deformations of quantum
  mechanics}} ({\natexlab{a}}),\ \Eprint {https://arxiv.org/abs/2001.07662}
  {arXiv:2001.07662} \BibitemShut {NoStop}%
\bibitem [{\citenamefont {Pitaevskii}\ and\ \citenamefont
  {Stringari}(2003)}]{book:GP}%
  \BibitemOpen
  \bibfield  {author} {\bibinfo {author} {\bibfnamefont {L.}~\bibnamefont
  {Pitaevskii}}\ and\ \bibinfo {author} {\bibfnamefont {S.}~\bibnamefont
  {Stringari}},\ }\href@noop {} {\emph {\bibinfo {title} {Bose-Einstein
  Condensation}}}\ (\bibinfo  {publisher} {Oxford University Press},\ \bibinfo
  {address} {Oxford},\ \bibinfo {year} {2003})\BibitemShut {NoStop}%
\bibitem [{\citenamefont {Weinberg}(1989{\natexlab{a}})}]{Weinberg-testQM}%
  \BibitemOpen
  \bibfield  {author} {\bibinfo {author} {\bibfnamefont {S.}~\bibnamefont
  {Weinberg}},\ }\bibfield  {title} {\bibinfo {title} {{Precision Tests of
  Quantum Mechanics}},\ }\href {https://doi.org/10.1103/PhysRevLett.62.485}
  {\bibfield  {journal} {\bibinfo  {journal} {Phys. Rev. Lett.}\ }\textbf
  {\bibinfo {volume} {62}},\ \bibinfo {pages} {485} (\bibinfo {year}
  {1989}{\natexlab{a}})}\BibitemShut {NoStop}%
\bibitem [{\citenamefont {Weinberg}(1989{\natexlab{b}})}]{Weinberg-testQM3}%
  \BibitemOpen
  \bibfield  {author} {\bibinfo {author} {\bibfnamefont {S.}~\bibnamefont
  {Weinberg}},\ }\bibfield  {title} {\bibinfo {title} {Testing quantum
  mechanics},\ }\href {https://doi.org/10.1016/0003-4916(89)90276-5} {\bibfield
   {journal} {\bibinfo  {journal} {Ann. Phys. (N.Y.)}\ }\textbf {\bibinfo
  {volume} {194}},\ \bibinfo {pages} {336} (\bibinfo {year}
  {1989}{\natexlab{b}})}\BibitemShut {NoStop}%
\bibitem [{\citenamefont {Wheeler}(1982)}]{wheeler1}%
  \BibitemOpen
  \bibfield  {author} {\bibinfo {author} {\bibfnamefont {J.~A.}\ \bibnamefont
  {Wheeler}},\ }\bibfield  {title} {\bibinfo {title} {The computer and the
  universe},\ }\href {https://doi.org/10.1007/BF02650185} {\bibfield  {journal}
  {\bibinfo  {journal} {Int. J. Theor. Phys.}\ }\textbf {\bibinfo {volume}
  {21}},\ \bibinfo {pages} {557} (\bibinfo {year} {1982})}\BibitemShut
  {NoStop}%
\bibitem [{\citenamefont {Wheeler}(1988)}]{wheeler2}%
  \BibitemOpen
  \bibfield  {author} {\bibinfo {author} {\bibfnamefont {J.~A.}\ \bibnamefont
  {Wheeler}},\ }\bibfield  {title} {\bibinfo {title} {World as system
  self-synthesized by quantum networking},\ }\href
  {https://doi.org/10.1147/rd.321.0004} {\bibfield  {journal} {\bibinfo
  {journal} {IBM J. Res. Dev.}\ }\textbf {\bibinfo {volume} {32}},\ \bibinfo
  {pages} {4} (\bibinfo {year} {1988})}\BibitemShut {NoStop}%
\bibitem [{\citenamefont {Wheeler}(1990)}]{wheeler3}%
  \BibitemOpen
  \bibfield  {author} {\bibinfo {author} {\bibfnamefont {J.~A.}\ \bibnamefont
  {Wheeler}},\ }\bibinfo {title} {Information, physics, quantum: The search for
  links},\ in\ \href@noop {} {\emph {\bibinfo {booktitle} {Complexity, Entropy,
  and the Physics of Information}}},\ \bibinfo {editor} {edited by\ \bibinfo
  {editor} {\bibfnamefont {W.~H.}\ \bibnamefont {Zurek}}}\ (\bibinfo
  {publisher} {Addison-Wesley},\ \bibinfo {address} {Redwood, CA},\ \bibinfo
  {year} {1990})\ p.\ \bibinfo {pages} {354}\BibitemShut {NoStop}%
\bibitem [{\citenamefont {Fuchs}\ and\ \citenamefont {Stacey}(2019)}]{qbism1}%
  \BibitemOpen
  \bibfield  {author} {\bibinfo {author} {\bibfnamefont {C.~A.}\ \bibnamefont
  {Fuchs}}\ and\ \bibinfo {author} {\bibfnamefont {B.~C.}\ \bibnamefont
  {Stacey}},\ }\bibfield  {title} {\bibinfo {title} {Qbism: Quantum theory as a
  hero's handbook},\ }in\ \href@noop {} {\emph {\bibinfo {booktitle}
  {Proceedings of the International School of Physics ``Enrico Fermi''}}},\
  Vol.\ \bibinfo {volume} {197}\ (\bibinfo {year} {2019})\ p.\ \bibinfo {pages}
  {133}\BibitemShut {NoStop}%
\bibitem [{\citenamefont {Fuchs}(2017)}]{qbism2}%
  \BibitemOpen
  \bibfield  {author} {\bibinfo {author} {\bibfnamefont {C.~A.}\ \bibnamefont
  {Fuchs}},\ }\bibinfo {title} {On participatory realism},\ in\ \href@noop {}
  {\emph {\bibinfo {booktitle} {Information and Interaction}}},\ \bibinfo
  {editor} {edited by\ \bibinfo {editor} {\bibfnamefont {I.}~\bibnamefont
  {Durham}}\ and\ \bibinfo {editor} {\bibfnamefont {D.}~\bibnamefont
  {Rickles}}}\ (\bibinfo  {publisher} {Springer},\ \bibinfo {address} {Cham,
  Switzerland},\ \bibinfo {year} {2017})\BibitemShut {NoStop}%
\bibitem [{\citenamefont {Fuchs}(2001)}]{qbism3}%
  \BibitemOpen
  \bibfield  {author} {\bibinfo {author} {\bibfnamefont {C.~A.}\ \bibnamefont
  {Fuchs}},\ }\bibinfo {title} {Quantum foundations in the light of quantum
  information},\ in\ \href@noop {} {\emph {\bibinfo {booktitle} {Decoherence
  and its Implications in Quantum Computation and Information Transfer}}},\
  \bibinfo {editor} {edited by\ \bibinfo {editor} {\bibfnamefont
  {A.}~\bibnamefont {Gonis}}\ and\ \bibinfo {editor} {\bibfnamefont {P.~E.~A.}\
  \bibnamefont {Turchi}}}\ (\bibinfo  {publisher} {IOS Press},\ \bibinfo
  {address} {Amsterdam},\ \bibinfo {year} {2001})\BibitemShut {NoStop}%
\bibitem [{\citenamefont {Engelbrecht}(2007)}]{cibv8}%
  \BibitemOpen
  \bibfield  {author} {\bibinfo {author} {\bibfnamefont {A.~P.}\ \bibnamefont
  {Engelbrecht}},\ }\href@noop {} {\emph {\bibinfo {title} {Computational
  Intelligence: An Introduction}}}\ (\bibinfo  {publisher} {Wiley},\ \bibinfo
  {address} {Chichester, U.K.},\ \bibinfo {year} {2007})\BibitemShut {NoStop}%
\bibitem [{\citenamefont {Kochenderfer}\ \emph {et~al.}(2022)\citenamefont
  {Kochenderfer}, \citenamefont {Wheeler},\ and\ \citenamefont {Wray}}]{mit}%
  \BibitemOpen
  \bibfield  {author} {\bibinfo {author} {\bibfnamefont {M.~J.}\ \bibnamefont
  {Kochenderfer}}, \bibinfo {author} {\bibfnamefont {T.~A.}\ \bibnamefont
  {Wheeler}},\ and\ \bibinfo {author} {\bibfnamefont {K.~H.}\ \bibnamefont
  {Wray}},\ }\href@noop {} {\emph {\bibinfo {title} {Algorithms for Decision
  Making}}}\ (\bibinfo  {publisher} {MIT Press},\ \bibinfo {address} {London},\
  \bibinfo {year} {2022})\BibitemShut {NoStop}%
\bibitem [{\citenamefont {Gisin}(1990)}]{gisin1}%
  \BibitemOpen
  \bibfield  {author} {\bibinfo {author} {\bibfnamefont {N.}~\bibnamefont
  {Gisin}},\ }\bibfield  {title} {\bibinfo {title} {Weinberg's non-linear
  quantum mechanics and supraluminal communication},\ }\href
  {https://doi.org/10.1016/0375-9601(90)90786-N} {\bibfield  {journal}
  {\bibinfo  {journal} {Phys. Lett. A}\ }\textbf {\bibinfo {volume} {143}},\
  \bibinfo {pages} {1} (\bibinfo {year} {1990})}\BibitemShut {NoStop}%
\bibitem [{\citenamefont {Peres}(1989)}]{peres1}%
  \BibitemOpen
  \bibfield  {author} {\bibinfo {author} {\bibfnamefont {A.}~\bibnamefont
  {Peres}},\ }\bibfield  {title} {\bibinfo {title} {{Nonlinear Variants of
  Schr\"{o}dinger's Equation Violate the Second Law of Thermodynamics}},\
  }\href {https://doi.org/10.1103/PhysRevLett.63.1114} {\bibfield  {journal}
  {\bibinfo  {journal} {Phys. Rev. Lett.}\ }\textbf {\bibinfo {volume} {63}},\
  \bibinfo {pages} {1114} (\bibinfo {year} {1989})}\BibitemShut {NoStop}%
\bibitem [{\citenamefont {Jordan}(2009)}]{jordan1}%
  \BibitemOpen
  \bibfield  {author} {\bibinfo {author} {\bibfnamefont {T.~F.}\ \bibnamefont
  {Jordan}},\ }\bibfield  {title} {\bibinfo {title} {Why quantum dynamics is
  linear},\ }\href {https://doi.org/10.1088/1742-6596/196/1/012010} {\bibfield
  {journal} {\bibinfo  {journal} {J. Phys.: Conf. Ser.}\ }\textbf {\bibinfo
  {volume} {196}},\ \bibinfo {pages} {012010} (\bibinfo {year}
  {2009})}\BibitemShut {NoStop}%
\bibitem [{\citenamefont {Mielnik}(1974)}]{Mielnik}%
  \BibitemOpen
  \bibfield  {author} {\bibinfo {author} {\bibfnamefont {B.}~\bibnamefont
  {Mielnik}},\ }\bibfield  {title} {\bibinfo {title} {Generalized quantum
  mechanics},\ }\href {https://doi.org/10.1007/BF01646346} {\bibfield
  {journal} {\bibinfo  {journal} {Commun. Math. Phys.}\ }\textbf {\bibinfo
  {volume} {37}},\ \bibinfo {pages} {221} (\bibinfo {year} {1974})}\BibitemShut
  {NoStop}%
\bibitem [{\citenamefont {Weinberg}(1989{\natexlab{c}})}]{Weinberg-testQM2}%
  \BibitemOpen
  \bibfield  {author} {\bibinfo {author} {\bibfnamefont {S.}~\bibnamefont
  {Weinberg}},\ }\bibfield  {title} {\bibinfo {title} {Weinberg replies},\
  }\href {https://doi.org/10.1103/PhysRevLett.63.1115} {\bibfield  {journal}
  {\bibinfo  {journal} {Phys. Rev. Lett.}\ }\textbf {\bibinfo {volume} {63}},\
  \bibinfo {pages} {1115} (\bibinfo {year} {1989}{\natexlab{c}})}\BibitemShut
  {NoStop}%
\bibitem [{\citenamefont {Polchinski}(1991)}]{polchinski1}%
  \BibitemOpen
  \bibfield  {author} {\bibinfo {author} {\bibfnamefont {J.}~\bibnamefont
  {Polchinski}},\ }\bibfield  {title} {\bibinfo {title} {{Weinberg’s
  Nonlinear Quantum Mechanics and the Einstein-Podolsky-Rosen Paradox}},\
  }\href {https://doi.org/10.1103/PhysRevLett.66.397} {\bibfield  {journal}
  {\bibinfo  {journal} {Phys. Rev. Lett.}\ }\textbf {\bibinfo {volume} {66}},\
  \bibinfo {pages} {397} (\bibinfo {year} {1991})}\BibitemShut {NoStop}%
\bibitem [{\citenamefont {Czachor}(1991)}]{czachor1}%
  \BibitemOpen
  \bibfield  {author} {\bibinfo {author} {\bibfnamefont {M.}~\bibnamefont
  {Czachor}},\ }\bibfield  {title} {\bibinfo {title} {Mobility and
  non-separability},\ }\href {https://doi.org/10.1007/BF00665894} {\bibfield
  {journal} {\bibinfo  {journal} {Found. Phys. Lett.}\ }\textbf {\bibinfo
  {volume} {4}},\ \bibinfo {pages} {351} (\bibinfo {year} {1991})}\BibitemShut
  {NoStop}%
\bibitem [{\citenamefont {Jordan}(1993)}]{jordan3}%
  \BibitemOpen
  \bibfield  {author} {\bibinfo {author} {\bibfnamefont {T.~F.}\ \bibnamefont
  {Jordan}},\ }\bibfield  {title} {\bibinfo {title} {Reconstructing a nonlinear
  dynamical framework for testing quantum mechanics},\ }\href
  {https://doi.org/10.1006/aphy.1993.1053} {\bibfield  {journal} {\bibinfo
  {journal} {Ann. Phys. (N.Y.)}\ }\textbf {\bibinfo {volume} {225}},\ \bibinfo
  {pages} {83} (\bibinfo {year} {1993})}\BibitemShut {NoStop}%
\bibitem [{\citenamefont {Czachor}(1998)}]{czachor2}%
  \BibitemOpen
  \bibfield  {author} {\bibinfo {author} {\bibfnamefont {M.}~\bibnamefont
  {Czachor}},\ }\bibfield  {title} {\bibinfo {title} {Nonlocal-looking
  equations can make nonlinear quantum dynamics local},\ }\href
  {https://doi.org/10.1103/PhysRevA.57.4122} {\bibfield  {journal} {\bibinfo
  {journal} {Phys. Rev. A}\ }\textbf {\bibinfo {volume} {57}},\ \bibinfo
  {pages} {4122} (\bibinfo {year} {1998})}\BibitemShut {NoStop}%
\bibitem [{\citenamefont {B\'{o}na}(2000)}]{bona1}%
  \BibitemOpen
  \bibfield  {author} {\bibinfo {author} {\bibfnamefont {P.}~\bibnamefont
  {B\'{o}na}},\ }\bibfield  {title} {\bibinfo {title} {Extended quantum
  mechanics},\ }\href {http://www.physics.sk/aps/pubs/2000/aps-2000-50-1-1.pdf}
  {\bibfield  {journal} {\bibinfo  {journal} {Acta Phys. Slov.}\ }\textbf
  {\bibinfo {volume} {50}},\ \bibinfo {pages} {1} (\bibinfo {year}
  {2000})}\BibitemShut {NoStop}%
\bibitem [{\citenamefont {Aerts}\ \emph {et~al.}(2002)\citenamefont {Aerts},
  \citenamefont {Czachor},\ and\ \citenamefont {Durt}}]{refonlqm_a}%
  \BibitemOpen
  \bibinfo {editor} {\bibfnamefont {D.}~\bibnamefont {Aerts}}, \bibinfo
  {editor} {\bibfnamefont {M.}~\bibnamefont {Czachor}},\ and\ \bibinfo {editor}
  {\bibfnamefont {T.}~\bibnamefont {Durt}},\ eds.,\ \href@noop {} {\emph
  {\bibinfo {title} {Probing the Structure of Quantum Mechanics: Nonlinearity,
  Nonlocality, Computation and Axiomatics}}}\ (\bibinfo  {publisher} {World
  Scientific},\ \bibinfo {address} {Singapore},\ \bibinfo {year}
  {2002})\BibitemShut {NoStop}%
\bibitem [{\citenamefont {Ferrero}\ \emph {et~al.}(2004)\citenamefont
  {Ferrero}, \citenamefont {Salgado},\ and\ \citenamefont
  {S\'{a}nchez-G\'{o}mez}}]{fss1}%
  \BibitemOpen
  \bibfield  {author} {\bibinfo {author} {\bibfnamefont {M.}~\bibnamefont
  {Ferrero}}, \bibinfo {author} {\bibfnamefont {D.}~\bibnamefont {Salgado}},\
  and\ \bibinfo {author} {\bibfnamefont {J.~L.}\ \bibnamefont
  {S\'{a}nchez-G\'{o}mez}},\ }\bibfield  {title} {\bibinfo {title} {Nonlinear
  quantum evolution does not imply supraluminal communication},\ }\href
  {https://doi.org/10.1103/PhysRevA.70.014101} {\bibfield  {journal} {\bibinfo
  {journal} {Phys. Rev. A}\ }\textbf {\bibinfo {volume} {70}},\ \bibinfo
  {pages} {014101} (\bibinfo {year} {2004})}\BibitemShut {NoStop}%
\bibitem [{\citenamefont {Ferrero}\ \emph {et~al.}(2005)\citenamefont
  {Ferrero}, \citenamefont {Salgado},\ and\ \citenamefont
  {S\'{a}nchez-G\'{o}mez}}]{fss2}%
  \BibitemOpen
  \bibfield  {author} {\bibinfo {author} {\bibfnamefont {M.}~\bibnamefont
  {Ferrero}}, \bibinfo {author} {\bibfnamefont {D.}~\bibnamefont {Salgado}},\
  and\ \bibinfo {author} {\bibfnamefont {J.~L.}\ \bibnamefont
  {S\'{a}nchez-G\'{o}mez}},\ }\bibfield  {title} {\bibinfo {title} {On
  nonlinear evolution and supraluminal communication between finite quantum
  systems},\ }\href {https://doi.org/10.1142/S0219749905000840} {\bibfield
  {journal} {\bibinfo  {journal} {Int. J. Quantum Inf.}\ }\textbf {\bibinfo
  {volume} {3}},\ \bibinfo {pages} {257} (\bibinfo {year} {2005})}\BibitemShut
  {NoStop}%
\bibitem [{\citenamefont {Kent}(2005)}]{kent1}%
  \BibitemOpen
  \bibfield  {author} {\bibinfo {author} {\bibfnamefont {A.}~\bibnamefont
  {Kent}},\ }\bibfield  {title} {\bibinfo {title} {Nonlinearity without
  superluminality},\ }\href {https://doi.org/10.1103/PhysRevA.72.012108}
  {\bibfield  {journal} {\bibinfo  {journal} {Phys. Rev. A}\ }\textbf {\bibinfo
  {volume} {72}},\ \bibinfo {pages} {012108} (\bibinfo {year}
  {2005})}\BibitemShut {NoStop}%
\bibitem [{\citenamefont {Jordan}(2006)}]{jordan2}%
  \BibitemOpen
  \bibfield  {author} {\bibinfo {author} {\bibfnamefont {T.~F.}\ \bibnamefont
  {Jordan}},\ }\bibfield  {title} {\bibinfo {title} {Assumptions that imply
  quantum dynamics is linear},\ }\href
  {https://doi.org/10.1103/PhysRevA.73.022101} {\bibfield  {journal} {\bibinfo
  {journal} {Phys. Rev. A}\ }\textbf {\bibinfo {volume} {73}},\ \bibinfo
  {pages} {022101} (\bibinfo {year} {2006})}\BibitemShut {NoStop}%
\bibitem [{\citenamefont {Pienaar}(2013)}]{refonlqm_b}%
  \BibitemOpen
  \bibfield  {author} {\bibinfo {author} {\bibfnamefont {J.}~\bibnamefont
  {Pienaar}},\ }\emph {\bibinfo {title} {Causality violation and nonlinear
  quantum mechanics}},\ \href@noop {} {Ph.D. thesis},\ \bibinfo  {school} {The
  University of Queensland} (\bibinfo {year} {2013}),\ \Eprint
  {https://arxiv.org/abs/1401.0167} {arXiv:1401.0167} \BibitemShut {NoStop}%
\bibitem [{\citenamefont {Kent}(2021)}]{kent2}%
  \BibitemOpen
  \bibfield  {author} {\bibinfo {author} {\bibfnamefont {A.}~\bibnamefont
  {Kent}},\ }\bibfield  {title} {\bibinfo {title} {Testing quantum gravity near
  measurement events},\ }\href {https://doi.org/10.1103/PhysRevD.103.064038}
  {\bibfield  {journal} {\bibinfo  {journal} {Phys. Rev. D}\ }\textbf {\bibinfo
  {volume} {103}},\ \bibinfo {pages} {064038} (\bibinfo {year}
  {2021})}\BibitemShut {NoStop}%
\bibitem [{\citenamefont {Beretta}(2005)}]{Beretta-ns-general}%
  \BibitemOpen
  \bibfield  {author} {\bibinfo {author} {\bibfnamefont {G.~P.}\ \bibnamefont
  {Beretta}},\ }\bibfield  {title} {\bibinfo {title} {Nonlinear extensions of
  {S}chr\"{o}dinger-von {N}eumann quantum dynamics: a list of conditions for
  compatibility with thermodynamics},\ }\href
  {https://doi.org/10.1142/S0217732305017263} {\bibfield  {journal} {\bibinfo
  {journal} {Mod. Phys. Lett. A}\ }\textbf {\bibinfo {volume} {20}},\ \bibinfo
  {pages} {977} (\bibinfo {year} {2005})}\BibitemShut {NoStop}%
\bibitem [{\citenamefont {Helou}\ and\ \citenamefont {Chen}(2017)}]{hc1}%
  \BibitemOpen
  \bibfield  {author} {\bibinfo {author} {\bibfnamefont {B.}~\bibnamefont
  {Helou}}\ and\ \bibinfo {author} {\bibfnamefont {Y.}~\bibnamefont {Chen}},\
  }\bibfield  {title} {\bibinfo {title} {Different interpretations of quantum
  mechanics make different predictions in non-linear quantum mechanics, and
  some do not violate the no-signaling condition},\ }\href
  {https://doi.org/10.1088/1742-6596/880/1/012021} {\bibfield  {journal}
  {\bibinfo  {journal} {J. Phys.: Conf. Ser.}\ }\textbf {\bibinfo {volume}
  {880}},\ \bibinfo {pages} {012021} (\bibinfo {year} {2017})}\BibitemShut
  {NoStop}%
\bibitem [{\citenamefont {Giulini}\ \emph {et~al.}()\citenamefont {Giulini},
  \citenamefont {Gro{\ss}ardt},\ and\ \citenamefont {Schwartz}}]{kentrelated}%
  \BibitemOpen
  \bibfield  {author} {\bibinfo {author} {\bibfnamefont {D.}~\bibnamefont
  {Giulini}}, \bibinfo {author} {\bibfnamefont {A.}~\bibnamefont
  {Gro{\ss}ardt}},\ and\ \bibinfo {author} {\bibfnamefont {P.~K.}\ \bibnamefont
  {Schwartz}},\ }\href@noop {} {\bibinfo {title} {Coupling quantum matter and
  gravity}},\ \Eprint {https://arxiv.org/abs/2207.05029} {arXiv:2207.05029}
  \BibitemShut {NoStop}%
\bibitem [{\citenamefont {Rembieli\'{n}ski}\ and\ \citenamefont
  {Caban}(2020)}]{rc1}%
  \BibitemOpen
  \bibfield  {author} {\bibinfo {author} {\bibfnamefont {J.}~\bibnamefont
  {Rembieli\'{n}ski}}\ and\ \bibinfo {author} {\bibfnamefont {P.}~\bibnamefont
  {Caban}},\ }\bibfield  {title} {\bibinfo {title} {Nonlinear evolution and
  signaling},\ }\href {https://doi.org/10.1103/PhysRevResearch.2.012027}
  {\bibfield  {journal} {\bibinfo  {journal} {Phys. Rev. Res.}\ }\textbf
  {\bibinfo {volume} {2}},\ \bibinfo {pages} {012027} (\bibinfo {year}
  {2020})}\BibitemShut {NoStop}%
\bibitem [{\citenamefont {Ray}\ and\ \citenamefont {Beretta}()}]{Beretta-ns}%
  \BibitemOpen
  \bibfield  {author} {\bibinfo {author} {\bibfnamefont {R.~K.}\ \bibnamefont
  {Ray}}\ and\ \bibinfo {author} {\bibfnamefont {G.~P.}\ \bibnamefont
  {Beretta}},\ }\href@noop {} {\bibinfo {title} {No-signaling in nonlinear
  extensions of quantum mechanics}},\ \Eprint
  {https://arxiv.org/abs/2301.11548} {arXiv:2301.11548} \BibitemShut {NoStop}%
\bibitem [{\citenamefont {Kaplan}\ and\ \citenamefont
  {Rajendran}(2022)}]{kaplan1}%
  \BibitemOpen
  \bibfield  {author} {\bibinfo {author} {\bibfnamefont {D.~E.}\ \bibnamefont
  {Kaplan}}\ and\ \bibinfo {author} {\bibfnamefont {S.}~\bibnamefont
  {Rajendran}},\ }\bibfield  {title} {\bibinfo {title} {Causal framework for
  nonlinear quantum mechanics},\ }\href
  {https://doi.org/10.1103/PhysRevD.105.055002} {\bibfield  {journal} {\bibinfo
   {journal} {Phys. Rev. D}\ }\textbf {\bibinfo {volume} {105}},\ \bibinfo
  {pages} {055002} (\bibinfo {year} {2022})}\BibitemShut {NoStop}%
\bibitem [{\citenamefont {Polkovnikov}\ \emph {et~al.}(2023)\citenamefont
  {Polkovnikov}, \citenamefont {Gramolin}, \citenamefont {Kaplan},
  \citenamefont {Rajendran},\ and\ \citenamefont {Sushkov}}]{kaplanrelated1}%
  \BibitemOpen
  \bibfield  {author} {\bibinfo {author} {\bibfnamefont {M.}~\bibnamefont
  {Polkovnikov}}, \bibinfo {author} {\bibfnamefont {A.~V.}\ \bibnamefont
  {Gramolin}}, \bibinfo {author} {\bibfnamefont {D.~E.}\ \bibnamefont
  {Kaplan}}, \bibinfo {author} {\bibfnamefont {S.}~\bibnamefont {Rajendran}},\
  and\ \bibinfo {author} {\bibfnamefont {A.~O.}\ \bibnamefont {Sushkov}},\
  }\bibfield  {title} {\bibinfo {title} {{Experimental Limit on Non-Linear
  State-Dependent Terms in Quantum Theory}},\ }\href
  {https://doi.org/10.1103/PhysRevLett.130.040202} {\bibfield  {journal}
  {\bibinfo  {journal} {Phys. Rev. Lett.}\ }\textbf {\bibinfo {volume} {130}},\
  \bibinfo {pages} {040202} (\bibinfo {year} {2023})}\BibitemShut {NoStop}%
\bibitem [{\citenamefont {Broz}\ \emph {et~al.}(2023)\citenamefont {Broz},
  \citenamefont {You}, \citenamefont {Khan}, \citenamefont {H\"{a}ffner},
  \citenamefont {Kaplan},\ and\ \citenamefont {Rajendran}}]{Kaplan-Berkeley}%
  \BibitemOpen
  \bibfield  {author} {\bibinfo {author} {\bibfnamefont {J.}~\bibnamefont
  {Broz}}, \bibinfo {author} {\bibfnamefont {B.}~\bibnamefont {You}}, \bibinfo
  {author} {\bibfnamefont {S.}~\bibnamefont {Khan}}, \bibinfo {author}
  {\bibfnamefont {H.}~\bibnamefont {H\"{a}ffner}}, \bibinfo {author}
  {\bibfnamefont {D.~E.}\ \bibnamefont {Kaplan}},\ and\ \bibinfo {author}
  {\bibfnamefont {S.}~\bibnamefont {Rajendran}},\ }\bibfield  {title} {\bibinfo
  {title} {{Test of Causal Nonlinear Quantum Mechanics by Ramsey Interferometry
  with a Trapped Ion}},\ }\href
  {https://doi.org/10.1103/PhysRevLett.130.200201} {\bibfield  {journal}
  {\bibinfo  {journal} {Phys. Rev. Lett.}\ }\textbf {\bibinfo {volume} {130}},\
  \bibinfo {pages} {200201} (\bibinfo {year} {2023})}\BibitemShut {NoStop}%
\bibitem [{\citenamefont {Maldacena}(1999)}]{AdS/CFT}%
  \BibitemOpen
  \bibfield  {author} {\bibinfo {author} {\bibfnamefont {J.}~\bibnamefont
  {Maldacena}},\ }\bibfield  {title} {\bibinfo {title} {{The large-$N$ limit of
  superconformal field theories and supergravity}},\ }\href
  {https://doi.org/10.1023/A:1026654312961} {\bibfield  {journal} {\bibinfo
  {journal} {Int. J. Theor. Phys.}\ }\textbf {\bibinfo {volume} {38}},\
  \bibinfo {pages} {1113} (\bibinfo {year} {1999})}\BibitemShut {NoStop}%
\bibitem [{\citenamefont {Banks}\ \emph {et~al.}(1997)\citenamefont {Banks},
  \citenamefont {Fischler}, \citenamefont {Shenker},\ and\ \citenamefont
  {Susskind}}]{bfss1}%
  \BibitemOpen
  \bibfield  {author} {\bibinfo {author} {\bibfnamefont {T.}~\bibnamefont
  {Banks}}, \bibinfo {author} {\bibfnamefont {W.}~\bibnamefont {Fischler}},
  \bibinfo {author} {\bibfnamefont {S.~H.}\ \bibnamefont {Shenker}},\ and\
  \bibinfo {author} {\bibfnamefont {L.}~\bibnamefont {Susskind}},\ }\bibfield
  {title} {\bibinfo {title} {{$M$ theory as a matrix model: A conjecture}},\
  }\href {https://doi.org/10.1103/PhysRevD.55.5112} {\bibfield  {journal}
  {\bibinfo  {journal} {Phys. Rev. D}\ }\textbf {\bibinfo {volume} {55}},\
  \bibinfo {pages} {5112} (\bibinfo {year} {1997})}\BibitemShut {NoStop}%
\bibitem [{\citenamefont {Seiberg}(2007)}]{seiberg1}%
  \BibitemOpen
  \bibfield  {author} {\bibinfo {author} {\bibfnamefont {N.}~\bibnamefont
  {Seiberg}},\ }\bibfield  {title} {\bibinfo {title} {Emergent spacetime},\
  }in\ \href@noop {} {\emph {\bibinfo {booktitle} {The Quantum Structure of
  Space and Time -- Proceedings of the $23$rd Solvay Conference on Physics,
  Brussels, Belgium, 1-3 December 2005}}},\ \bibinfo {editor} {edited by\
  \bibinfo {editor} {\bibfnamefont {D.}~\bibnamefont {Gross}}, \bibinfo
  {editor} {\bibfnamefont {M.}~\bibnamefont {Henneaux}},\ and\ \bibinfo
  {editor} {\bibfnamefont {A.}~\bibnamefont {Sevrin}}}\ (\bibinfo  {publisher}
  {World Scientific},\ \bibinfo {address} {Singapore},\ \bibinfo {year}
  {2007})\ p.\ \bibinfo {pages} {163}\BibitemShut {NoStop}%
\bibitem [{\citenamefont {Van~Raamsdonk}(2010)}]{raamsdonk1}%
  \BibitemOpen
  \bibfield  {author} {\bibinfo {author} {\bibfnamefont {M.}~\bibnamefont
  {Van~Raamsdonk}},\ }\bibfield  {title} {\bibinfo {title} {Building up
  spacetime with quantum entanglement},\ }\href
  {https://doi.org/10.1007/s10714-010-1034-0} {\bibfield  {journal} {\bibinfo
  {journal} {Gen. Relativ. Gravit.}\ }\textbf {\bibinfo {volume} {42}},\
  \bibinfo {pages} {2323} (\bibinfo {year} {2010})}\BibitemShut {NoStop}%
\bibitem [{\citenamefont {Swingle}(2012)}]{swingle1}%
  \BibitemOpen
  \bibfield  {author} {\bibinfo {author} {\bibfnamefont {B.}~\bibnamefont
  {Swingle}},\ }\bibfield  {title} {\bibinfo {title} {Entanglement
  renormalization and holography},\ }\href
  {https://doi.org/10.1103/PhysRevD.86.065007} {\bibfield  {journal} {\bibinfo
  {journal} {Phys. Rev. D}\ }\textbf {\bibinfo {volume} {86}},\ \bibinfo
  {pages} {065007} (\bibinfo {year} {2012})}\BibitemShut {NoStop}%
\bibitem [{\citenamefont {Nozaki}\ \emph {et~al.}(2012)\citenamefont {Nozaki},
  \citenamefont {Ryu},\ and\ \citenamefont {Takayanagi}}]{mera1}%
  \BibitemOpen
  \bibfield  {author} {\bibinfo {author} {\bibfnamefont {M.}~\bibnamefont
  {Nozaki}}, \bibinfo {author} {\bibfnamefont {S.}~\bibnamefont {Ryu}},\ and\
  \bibinfo {author} {\bibfnamefont {T.}~\bibnamefont {Takayanagi}},\ }\bibfield
   {title} {\bibinfo {title} {Holographic geometry of entanglement
  renormalization in quantum field theories},\ }\href
  {https://doi.org/10.1007/JHEP10(2012)193} {\bibfield  {journal} {\bibinfo
  {journal} {J. High Energy Phys.}\ }\textbf {\bibinfo {volume} {2012}},\
  \bibinfo {pages} {193}}\BibitemShut {NoStop}%
\bibitem [{\citenamefont {Hartnoll}\ \emph {et~al.}(2016)\citenamefont
  {Hartnoll}, \citenamefont {Lucas},\ and\ \citenamefont {Sachdev}}]{syk1}%
  \BibitemOpen
  \bibfield  {author} {\bibinfo {author} {\bibfnamefont {S.~A.}\ \bibnamefont
  {Hartnoll}}, \bibinfo {author} {\bibfnamefont {A.}~\bibnamefont {Lucas}},\
  and\ \bibinfo {author} {\bibfnamefont {S.}~\bibnamefont {Sachdev}},\
  }\href@noop {} {\emph {\bibinfo {title} {Holographic Quantum Matter}}}\
  (\bibinfo  {publisher} {MIT Press},\ \bibinfo {address} {Cambridge},\
  \bibinfo {year} {2016})\BibitemShut {NoStop}%
\bibitem [{\citenamefont {Bao}\ \emph {et~al.}(2017)\citenamefont {Bao},
  \citenamefont {Cao}, \citenamefont {Carroll},\ and\ \citenamefont
  {Chatwin-Davies}}]{sc1}%
  \BibitemOpen
  \bibfield  {author} {\bibinfo {author} {\bibfnamefont {N.}~\bibnamefont
  {Bao}}, \bibinfo {author} {\bibfnamefont {C.}~\bibnamefont {Cao}}, \bibinfo
  {author} {\bibfnamefont {S.~M.}\ \bibnamefont {Carroll}},\ and\ \bibinfo
  {author} {\bibfnamefont {A.}~\bibnamefont {Chatwin-Davies}},\ }\bibfield
  {title} {\bibinfo {title} {{de Sitter space as a tensor network: Cosmic
  no-hair, complementarity, and complexity}},\ }\href
  {https://doi.org/10.1103/PhysRevD.96.123536} {\bibfield  {journal} {\bibinfo
  {journal} {Phys. Rev. D}\ }\textbf {\bibinfo {volume} {96}},\ \bibinfo
  {pages} {123536} (\bibinfo {year} {2017})}\BibitemShut {NoStop}%
\bibitem [{\citenamefont {Lee}(2018)}]{lee1}%
  \BibitemOpen
  \bibfield  {author} {\bibinfo {author} {\bibfnamefont {S.~S.}\ \bibnamefont
  {Lee}},\ }\bibfield  {title} {\bibinfo {title} {Emergent gravity from
  relatively local {Hamiltonians} and a possible resolution of the black hole
  information puzzle},\ }\href {https://doi.org/10.1007/JHEP10(2018)043}
  {\bibfield  {journal} {\bibinfo  {journal} {J. High Energy Phys.}\ }\textbf
  {\bibinfo {volume} {2018}},\ \bibinfo {pages} {43}}\BibitemShut {NoStop}%
\bibitem [{\citenamefont {Lee}(2020)}]{lee2}%
  \BibitemOpen
  \bibfield  {author} {\bibinfo {author} {\bibfnamefont {S.~S.}\ \bibnamefont
  {Lee}},\ }\bibfield  {title} {\bibinfo {title} {A model of quantum gravity
  with emergent spacetime},\ }\href {https://doi.org/10.1007/JHEP06(2020)070}
  {\bibfield  {journal} {\bibinfo  {journal} {J. High Energy Phys.}\ }\textbf
  {\bibinfo {volume} {2020}},\ \bibinfo {pages} {70}}\BibitemShut {NoStop}%
\bibitem [{\citenamefont {Cao}\ \emph {et~al.}(2017)\citenamefont {Cao},
  \citenamefont {Carroll},\ and\ \citenamefont {Michalakis}}]{sc2}%
  \BibitemOpen
  \bibfield  {author} {\bibinfo {author} {\bibfnamefont {C.}~\bibnamefont
  {Cao}}, \bibinfo {author} {\bibfnamefont {S.~M.}\ \bibnamefont {Carroll}},\
  and\ \bibinfo {author} {\bibfnamefont {S.}~\bibnamefont {Michalakis}},\
  }\bibfield  {title} {\bibinfo {title} {{Space from Hilbert space: Recovering
  geometry from bulk entanglement}},\ }\href
  {https://doi.org/10.1103/PhysRevD.95.024031} {\bibfield  {journal} {\bibinfo
  {journal} {Phys. Rev. D}\ }\textbf {\bibinfo {volume} {95}},\ \bibinfo
  {pages} {024031} (\bibinfo {year} {2017})}\BibitemShut {NoStop}%
\bibitem [{\citenamefont {Carroll}()}]{Carroll-finite}%
  \BibitemOpen
  \bibfield  {author} {\bibinfo {author} {\bibfnamefont {S.~M.}\ \bibnamefont
  {Carroll}},\ }\href@noop {} {\bibinfo {title} {Completely discretized, finite
  quantum mechanics}},\ \Eprint {https://arxiv.org/abs/2307.11927}
  {arXiv:2307.11927} \BibitemShut {NoStop}%
\bibitem [{\citenamefont {Oriti}(2009)}]{oriti1}%
  \BibitemOpen
  \bibfield  {author} {\bibinfo {author} {\bibfnamefont {D.}~\bibnamefont
  {Oriti}},\ }\href@noop {} {\emph {\bibinfo {title} {Approaches to Quantum
  Gravity: Toward a New Understanding of Space, Time and Matter}}}\ (\bibinfo
  {publisher} {Cambridge University Press},\ \bibinfo {address} {Cambridge},\
  \bibinfo {year} {2009})\BibitemShut {NoStop}%
\bibitem [{\citenamefont {Oriti}(2014)}]{gftw1}%
  \BibitemOpen
  \bibfield  {author} {\bibinfo {author} {\bibfnamefont {D.}~\bibnamefont
  {Oriti}},\ }\bibfield  {title} {\bibinfo {title} {Disappearance and emergence
  of space and time in quantum gravity},\ }\href
  {https://doi.org/10.1016/j.shpsb.2013.10.006} {\bibfield  {journal} {\bibinfo
   {journal} {Stud. Hist. Philos. Mod. Phys.}\ }\textbf {\bibinfo {volume}
  {46}},\ \bibinfo {pages} {186} (\bibinfo {year} {2014})}\BibitemShut
  {NoStop}%
\bibitem [{\citenamefont {Patton}\ and\ \citenamefont
  {Wheeler}(1975)}]{wheeler4}%
  \BibitemOpen
  \bibfield  {author} {\bibinfo {author} {\bibfnamefont {C.~M.}\ \bibnamefont
  {Patton}}\ and\ \bibinfo {author} {\bibfnamefont {J.~A.}\ \bibnamefont
  {Wheeler}},\ }\bibfield  {title} {\bibinfo {title} {Is physics legislated by
  cosmogony?},\ }in\ \href@noop {} {\emph {\bibinfo {booktitle} {Quantum
  Gravity: An Oxford Symposium}}},\ \bibinfo {editor} {edited by\ \bibinfo
  {editor} {\bibfnamefont {C.~J.}\ \bibnamefont {Isham}}, \bibinfo {editor}
  {\bibfnamefont {R.}~\bibnamefont {Penrose}},\ and\ \bibinfo {editor}
  {\bibfnamefont {D.~W.}\ \bibnamefont {Sciama}}}\ (\bibinfo  {publisher}
  {Clarendon Press},\ \bibinfo {address} {Oxford},\ \bibinfo {year} {1975})\
  p.\ \bibinfo {pages} {538}\BibitemShut {NoStop}%
\bibitem [{\citenamefont {Wheeler}(1980)}]{wheeler5}%
  \BibitemOpen
  \bibfield  {author} {\bibinfo {author} {\bibfnamefont {J.~A.}\ \bibnamefont
  {Wheeler}},\ }\bibinfo {title} {Pregeometry: Motivations and prospects},\ in\
  \href@noop {} {\emph {\bibinfo {booktitle} {Quantum Theory and
  Gravitation}}},\ \bibinfo {editor} {edited by\ \bibinfo {editor}
  {\bibfnamefont {A.~R.}\ \bibnamefont {Marlow}}}\ (\bibinfo  {publisher}
  {Academic Press},\ \bibinfo {address} {New York},\ \bibinfo {year}
  {1980})\BibitemShut {NoStop}%
\bibitem [{\citenamefont {Dadi\'{c}}\ and\ \citenamefont {Pisk}(1980)}]{dp1}%
  \BibitemOpen
  \bibfield  {author} {\bibinfo {author} {\bibfnamefont {I.}~\bibnamefont
  {Dadi\'{c}}}\ and\ \bibinfo {author} {\bibfnamefont {K.}~\bibnamefont
  {Pisk}},\ }\bibfield  {title} {\bibinfo {title} {Dynamics of discrete-space
  structure},\ }\href {https://doi.org/10.1007/BF00670430} {\bibfield
  {journal} {\bibinfo  {journal} {Int. J. Theor. Phys.}\ }\textbf {\bibinfo
  {volume} {18}},\ \bibinfo {pages} {345} (\bibinfo {year} {1980})}\BibitemShut
  {NoStop}%
\bibitem [{\citenamefont {Bombelli}\ \emph {et~al.}(1987)\citenamefont
  {Bombelli}, \citenamefont {Lee}, \citenamefont {Meyer},\ and\ \citenamefont
  {Sorkin}}]{sorkin1}%
  \BibitemOpen
  \bibfield  {author} {\bibinfo {author} {\bibfnamefont {L.}~\bibnamefont
  {Bombelli}}, \bibinfo {author} {\bibfnamefont {J.}~\bibnamefont {Lee}},
  \bibinfo {author} {\bibfnamefont {D.}~\bibnamefont {Meyer}},\ and\ \bibinfo
  {author} {\bibfnamefont {R.~D.}\ \bibnamefont {Sorkin}},\ }\bibfield  {title}
  {\bibinfo {title} {{Space-Time as a Causal Set}},\ }\href
  {https://doi.org/10.1103/PhysRevLett.59.521} {\bibfield  {journal} {\bibinfo
  {journal} {Phys. Rev. Lett.}\ }\textbf {\bibinfo {volume} {59}},\ \bibinfo
  {pages} {521} (\bibinfo {year} {1987})}\BibitemShut {NoStop}%
\bibitem [{\citenamefont {Antonsen}(1992)}]{antonsen1}%
  \BibitemOpen
  \bibfield  {author} {\bibinfo {author} {\bibfnamefont {F.}~\bibnamefont
  {Antonsen}},\ }\emph {\bibinfo {title} {Pregeometry}},\ \href@noop {}
  {Master's thesis},\ \bibinfo  {school} {University of Copenhagen} (\bibinfo
  {year} {1992})\BibitemShut {NoStop}%
\bibitem [{\citenamefont {Surya}(2019)}]{sa1}%
  \BibitemOpen
  \bibfield  {author} {\bibinfo {author} {\bibfnamefont {S.}~\bibnamefont
  {Surya}},\ }\bibfield  {title} {\bibinfo {title} {The causal set approach to
  quantum gravity},\ }\href {https://doi.org/10.1007/s41114-019-0023-1}
  {\bibfield  {journal} {\bibinfo  {journal} {Living Rev. Relativ.}\ }\textbf
  {\bibinfo {volume} {22}},\ \bibinfo {pages} {5} (\bibinfo {year}
  {2019})}\BibitemShut {NoStop}%
\bibitem [{\citenamefont {Hawkins}\ \emph {et~al.}(2003)\citenamefont
  {Hawkins}, \citenamefont {Markopoulou},\ and\ \citenamefont
  {Sahlmann}}]{ch1}%
  \BibitemOpen
  \bibfield  {author} {\bibinfo {author} {\bibfnamefont {E.}~\bibnamefont
  {Hawkins}}, \bibinfo {author} {\bibfnamefont {F.}~\bibnamefont
  {Markopoulou}},\ and\ \bibinfo {author} {\bibfnamefont {H.}~\bibnamefont
  {Sahlmann}},\ }\bibfield  {title} {\bibinfo {title} {Evolution in quantum
  causal histories},\ }\href {https://doi.org/10.1088/0264-9381/20/16/320}
  {\bibfield  {journal} {\bibinfo  {journal} {Class. Quantum Grav.}\ }\textbf
  {\bibinfo {volume} {20}},\ \bibinfo {pages} {16} (\bibinfo {year}
  {2003})}\BibitemShut {NoStop}%
\bibitem [{\citenamefont {Lloyd}()}]{sethlloyd1}%
  \BibitemOpen
  \bibfield  {author} {\bibinfo {author} {\bibfnamefont {S.}~\bibnamefont
  {Lloyd}},\ }\href@noop {} {\bibinfo {title} {A theory of quantum gravity
  based on quantum computation}},\ \Eprint
  {https://arxiv.org/abs/quant-ph/0501135} {arXiv:quant-ph/0501135}
  \BibitemShut {NoStop}%
\bibitem [{\citenamefont {Konopka}\ \emph {et~al.}(2008)\citenamefont
  {Konopka}, \citenamefont {Markopoulou},\ and\ \citenamefont
  {Severini}}]{qgraphity1}%
  \BibitemOpen
  \bibfield  {author} {\bibinfo {author} {\bibfnamefont {T.}~\bibnamefont
  {Konopka}}, \bibinfo {author} {\bibfnamefont {F.}~\bibnamefont
  {Markopoulou}},\ and\ \bibinfo {author} {\bibfnamefont {S.}~\bibnamefont
  {Severini}},\ }\bibfield  {title} {\bibinfo {title} {{Quantum graphity: A
  model of emergent locality}},\ }\href
  {https://doi.org/10.1103/PhysRevD.77.104029} {\bibfield  {journal} {\bibinfo
  {journal} {Phys. Rev. D}\ }\textbf {\bibinfo {volume} {77}},\ \bibinfo
  {pages} {104029} (\bibinfo {year} {2008})}\BibitemShut {NoStop}%
\bibitem [{\citenamefont {Hamma}\ \emph {et~al.}(2009)\citenamefont {Hamma},
  \citenamefont {Markopoulou}, \citenamefont {Premont-Schwarz},\ and\
  \citenamefont {Severini}}]{qgraphity2}%
  \BibitemOpen
  \bibfield  {author} {\bibinfo {author} {\bibfnamefont {A.}~\bibnamefont
  {Hamma}}, \bibinfo {author} {\bibfnamefont {F.}~\bibnamefont {Markopoulou}},
  \bibinfo {author} {\bibfnamefont {I.}~\bibnamefont {Premont-Schwarz}},\ and\
  \bibinfo {author} {\bibfnamefont {S.}~\bibnamefont {Severini}},\ }\bibfield
  {title} {\bibinfo {title} {{Lieb-Robinson Bounds and the Speed of Light from
  Topological Order}},\ }\href {https://doi.org/10.1103/PhysRevLett.102.017204}
  {\bibfield  {journal} {\bibinfo  {journal} {Phys. Rev. Lett.}\ }\textbf
  {\bibinfo {volume} {102}},\ \bibinfo {pages} {017204} (\bibinfo {year}
  {2009})}\BibitemShut {NoStop}%
\bibitem [{\citenamefont {Hamma}\ \emph {et~al.}(2010)\citenamefont {Hamma},
  \citenamefont {Markopoulou}, \citenamefont {Lloyd}, \citenamefont
  {Caravelli}, \citenamefont {Severini},\ and\ \citenamefont
  {Markstr\"{o}m}}]{qgraphity3}%
  \BibitemOpen
  \bibfield  {author} {\bibinfo {author} {\bibfnamefont {A.}~\bibnamefont
  {Hamma}}, \bibinfo {author} {\bibfnamefont {F.}~\bibnamefont {Markopoulou}},
  \bibinfo {author} {\bibfnamefont {S.}~\bibnamefont {Lloyd}}, \bibinfo
  {author} {\bibfnamefont {F.}~\bibnamefont {Caravelli}}, \bibinfo {author}
  {\bibfnamefont {S.}~\bibnamefont {Severini}},\ and\ \bibinfo {author}
  {\bibfnamefont {K.}~\bibnamefont {Markstr\"{o}m}},\ }\bibfield  {title}
  {\bibinfo {title} {{Quantum Bose-Hubbard model with an evolving graph as a
  toy model for emergent spacetime}},\ }\href
  {https://doi.org/10.1103/PhysRevD.81.104032} {\bibfield  {journal} {\bibinfo
  {journal} {Phys. Rev. D}\ }\textbf {\bibinfo {volume} {81}},\ \bibinfo
  {pages} {104032} (\bibinfo {year} {2010})}\BibitemShut {NoStop}%
\bibitem [{\citenamefont {Crowther}(2014)}]{crowther1}%
  \BibitemOpen
  \bibfield  {author} {\bibinfo {author} {\bibfnamefont {K.}~\bibnamefont
  {Crowther}},\ }\emph {\bibinfo {title} {Appearing out of nowhere: The
  emergence of spacetime in quantum gravity}},\ \href@noop {} {Ph.D. thesis},\
  \bibinfo  {school} {The University of Sydney} (\bibinfo {year} {2014}),\
  \Eprint {https://arxiv.org/abs/1410.0345} {arXiv:1410.0345} \BibitemShut
  {NoStop}%
\bibitem [{\citenamefont {Trugenberger}(2017)}]{swiss1}%
  \BibitemOpen
  \bibfield  {author} {\bibinfo {author} {\bibfnamefont {C.~A.}\ \bibnamefont
  {Trugenberger}},\ }\bibfield  {title} {\bibinfo {title} {Combinatorial
  quantum gravity: Geometry from random bits},\ }\href
  {https://doi.org/10.1007/JHEP09(2017)045} {\bibfield  {journal} {\bibinfo
  {journal} {J. High Energy Phys.}\ }\textbf {\bibinfo {volume} {2017}},\
  \bibinfo {pages} {45}}\BibitemShut {NoStop}%
\bibitem [{\citenamefont {Cao}\ and\ \citenamefont {Carroll}(2018)}]{sc3}%
  \BibitemOpen
  \bibfield  {author} {\bibinfo {author} {\bibfnamefont {C.}~\bibnamefont
  {Cao}}\ and\ \bibinfo {author} {\bibfnamefont {S.~M.}\ \bibnamefont
  {Carroll}},\ }\bibfield  {title} {\bibinfo {title} {{Bulk entanglement
  gravity without a boundary: Towards finding Einstein's equation in Hilbert
  space}},\ }\href {https://doi.org/10.1103/PhysRevD.97.086003} {\bibfield
  {journal} {\bibinfo  {journal} {Phys. Rev. D}\ }\textbf {\bibinfo {volume}
  {97}},\ \bibinfo {pages} {086003} (\bibinfo {year} {2018})}\BibitemShut
  {NoStop}%
\bibitem [{\citenamefont {Minic}\ and\ \citenamefont {Tze}(2003)}]{Minic}%
  \BibitemOpen
  \bibfield  {author} {\bibinfo {author} {\bibfnamefont {D.}~\bibnamefont
  {Minic}}\ and\ \bibinfo {author} {\bibfnamefont {C.-H.}\ \bibnamefont
  {Tze}},\ }\bibfield  {title} {\bibinfo {title} {Background independent
  quantum mechanics and gravity},\ }\href
  {https://doi.org/10.1103/PhysRevD.68.061501} {\bibfield  {journal} {\bibinfo
  {journal} {Phys. Rev. D}\ }\textbf {\bibinfo {volume} {68}},\ \bibinfo
  {pages} {061501(R)} (\bibinfo {year} {2003})}\BibitemShut {NoStop}%
\bibitem [{\citenamefont {Berglund}\ \emph {et~al.}(2022)\citenamefont
  {Berglund}, \citenamefont {H\"{u}bsch}, \citenamefont {Mattingly},\ and\
  \citenamefont {Minic}}]{minic2}%
  \BibitemOpen
  \bibfield  {author} {\bibinfo {author} {\bibfnamefont {P.}~\bibnamefont
  {Berglund}}, \bibinfo {author} {\bibfnamefont {T.}~\bibnamefont
  {H\"{u}bsch}}, \bibinfo {author} {\bibfnamefont {D.}~\bibnamefont
  {Mattingly}},\ and\ \bibinfo {author} {\bibfnamefont {D.}~\bibnamefont
  {Minic}},\ }\bibfield  {title} {\bibinfo {title} {Gravitizing the quantum},\
  }\href {https://doi.org/10.1142/S021827182242024X} {\bibfield  {journal}
  {\bibinfo  {journal} {Int. J. Mod. Phys. D}\ }\textbf {\bibinfo {volume}
  {31}},\ \bibinfo {pages} {2242024} (\bibinfo {year} {2022})}\BibitemShut
  {NoStop}%
\bibitem [{\citenamefont {Berglund}\ \emph {et~al.}()\citenamefont {Berglund},
  \citenamefont {Geraci}, \citenamefont {H\"{u}bsch}, \citenamefont
  {Mattingly},\ and\ \citenamefont {Minic}}]{minic3}%
  \BibitemOpen
  \bibfield  {author} {\bibinfo {author} {\bibfnamefont {P.}~\bibnamefont
  {Berglund}}, \bibinfo {author} {\bibfnamefont {A.}~\bibnamefont {Geraci}},
  \bibinfo {author} {\bibfnamefont {T.}~\bibnamefont {H\"{u}bsch}}, \bibinfo
  {author} {\bibfnamefont {D.}~\bibnamefont {Mattingly}},\ and\ \bibinfo
  {author} {\bibfnamefont {D.}~\bibnamefont {Minic}},\ }\href@noop {} {\bibinfo
  {title} {Triple interference, non-linear {T}albot effect and gravitization of
  the quantum}},\ \Eprint {https://arxiv.org/abs/2303.15645} {arXiv:2303.15645}
  \BibitemShut {NoStop}%
\bibitem [{\citenamefont {Ashtekar}\ and\ \citenamefont
  {Schilling}(1999)}]{Ashtekar}%
  \BibitemOpen
  \bibfield  {author} {\bibinfo {author} {\bibfnamefont {A.}~\bibnamefont
  {Ashtekar}}\ and\ \bibinfo {author} {\bibfnamefont {T.~A.}\ \bibnamefont
  {Schilling}},\ }\bibinfo {title} {Geometrical formulation of quantum
  mechanics},\ in\ \href@noop {} {\emph {\bibinfo {booktitle} {On Einstein’s
  Path}}},\ \bibinfo {editor} {edited by\ \bibinfo {editor} {\bibfnamefont
  {A.}~\bibnamefont {Harvey}}}\ (\bibinfo  {publisher} {Springer},\ \bibinfo
  {address} {New York},\ \bibinfo {year} {1999})\BibitemShut {NoStop}%
\bibitem [{\citenamefont {Harlow}(2016)}]{salaam}%
  \BibitemOpen
  \bibfield  {author} {\bibinfo {author} {\bibfnamefont {D.}~\bibnamefont
  {Harlow}},\ }\bibfield  {title} {\bibinfo {title} {Jerusalem lectures on
  black holes and quantum information},\ }\href
  {https://doi.org/10.1103/RevModPhys.88.015002} {\bibfield  {journal}
  {\bibinfo  {journal} {Rev. Mod. Phys.}\ }\textbf {\bibinfo {volume} {88}},\
  \bibinfo {pages} {15002} (\bibinfo {year} {2016})}\BibitemShut {NoStop}%
\bibitem [{\citenamefont {Horowitz}\ and\ \citenamefont
  {Maldacena}(2004)}]{hm}%
  \BibitemOpen
  \bibfield  {author} {\bibinfo {author} {\bibfnamefont {G.~T.}\ \bibnamefont
  {Horowitz}}\ and\ \bibinfo {author} {\bibfnamefont {J.}~\bibnamefont
  {Maldacena}},\ }\bibfield  {title} {\bibinfo {title} {The black hole final
  state},\ }\href {https://doi.org/10.1088/1126-6708/2004/02/008} {\bibfield
  {journal} {\bibinfo  {journal} {J. High Energy Phys.}\ }\textbf {\bibinfo
  {volume} {2004}},\ \bibinfo {pages} {008}}\BibitemShut {NoStop}%
\bibitem [{\citenamefont {Svetlichny}(2005)}]{s1}%
  \BibitemOpen
  \bibfield  {author} {\bibinfo {author} {\bibfnamefont {G.}~\bibnamefont
  {Svetlichny}},\ }\bibfield  {title} {\bibinfo {title} {Nonlinear quantum
  mechanics at the planck scale},\ }\href
  {https://doi.org/10.1007/s10773-005-8983-1} {\bibfield  {journal} {\bibinfo
  {journal} {Int. J. Theor. Phys.}\ }\textbf {\bibinfo {volume} {44}},\
  \bibinfo {pages} {2051} (\bibinfo {year} {2005})}\BibitemShut {NoStop}%
\bibitem [{\citenamefont {Bousso}(2013)}]{bousso1}%
  \BibitemOpen
  \bibfield  {author} {\bibinfo {author} {\bibfnamefont {R.}~\bibnamefont
  {Bousso}},\ }\bibfield  {title} {\bibinfo {title} {Complementarity is not
  enough},\ }\href {https://doi.org/10.1103/PhysRevD.87.124023} {\bibfield
  {journal} {\bibinfo  {journal} {Phys. Rev. D}\ }\textbf {\bibinfo {volume}
  {87}},\ \bibinfo {pages} {124023} (\bibinfo {year} {2013})}\BibitemShut
  {NoStop}%
\bibitem [{\citenamefont {Marolf}\ and\ \citenamefont
  {Polchinski}(2013)}]{mp1}%
  \BibitemOpen
  \bibfield  {author} {\bibinfo {author} {\bibfnamefont {D.}~\bibnamefont
  {Marolf}}\ and\ \bibinfo {author} {\bibfnamefont {J.}~\bibnamefont
  {Polchinski}},\ }\bibfield  {title} {\bibinfo {title} {{Gauge/Gravity Duality
  and the Black Hole Interior}},\ }\href
  {https://doi.org/10.1103/PhysRevLett.111.171301} {\bibfield  {journal}
  {\bibinfo  {journal} {Phys. Rev. Lett.}\ }\textbf {\bibinfo {volume} {111}},\
  \bibinfo {pages} {171301} (\bibinfo {year} {2013})}\BibitemShut {NoStop}%
\bibitem [{\citenamefont {Marolf}\ and\ \citenamefont
  {Polchinski}(2016)}]{mp2}%
  \BibitemOpen
  \bibfield  {author} {\bibinfo {author} {\bibfnamefont {D.}~\bibnamefont
  {Marolf}}\ and\ \bibinfo {author} {\bibfnamefont {J.}~\bibnamefont
  {Polchinski}},\ }\bibfield  {title} {\bibinfo {title} {{Violations of the
  Born rule in cool state-dependent horizons}},\ }\href
  {https://doi.org/10.1007/JHEP01(2016)008} {\bibfield  {journal} {\bibinfo
  {journal} {J. High Energy Phys.}\ }\textbf {\bibinfo {volume} {2016}},\
  \bibinfo {pages} {8}}\BibitemShut {NoStop}%
\bibitem [{\citenamefont {Papadodimas}\ and\ \citenamefont {Raju}(2014)}]{pr1}%
  \BibitemOpen
  \bibfield  {author} {\bibinfo {author} {\bibfnamefont {K.}~\bibnamefont
  {Papadodimas}}\ and\ \bibinfo {author} {\bibfnamefont {S.}~\bibnamefont
  {Raju}},\ }\bibfield  {title} {\bibinfo {title} {State-dependent
  bulk-boundary maps and black hole complementarity},\ }\href
  {https://doi.org/10.1103/PhysRevD.89.086010} {\bibfield  {journal} {\bibinfo
  {journal} {Phys. Rev. D}\ }\textbf {\bibinfo {volume} {89}},\ \bibinfo
  {pages} {086010} (\bibinfo {year} {2014})}\BibitemShut {NoStop}%
\bibitem [{\citenamefont {Lloyd}\ and\ \citenamefont {Preskill}(2014)}]{lp1}%
  \BibitemOpen
  \bibfield  {author} {\bibinfo {author} {\bibfnamefont {S.}~\bibnamefont
  {Lloyd}}\ and\ \bibinfo {author} {\bibfnamefont {J.}~\bibnamefont
  {Preskill}},\ }\bibfield  {title} {\bibinfo {title} {Unitarity of black hole
  evaporation in final-state projection models},\ }\href
  {https://doi.org/10.1007/JHEP08(2014)126} {\bibfield  {journal} {\bibinfo
  {journal} {J. High Energy Phys.}\ }\textbf {\bibinfo {volume} {2014}},\
  \bibinfo {pages} {126}}\BibitemShut {NoStop}%
\bibitem [{\citenamefont {Susskind}(2016)}]{susskind1}%
  \BibitemOpen
  \bibfield  {author} {\bibinfo {author} {\bibfnamefont {L.}~\bibnamefont
  {Susskind}},\ }\bibfield  {title} {\bibinfo {title} {Entanglement is not
  enough},\ }\href {https://doi.org/10.1002/prop.201500095} {\bibfield
  {journal} {\bibinfo  {journal} {Fortschr. Phys.}\ }\textbf {\bibinfo {volume}
  {64}},\ \bibinfo {pages} {49} (\bibinfo {year} {2016})}\BibitemShut {NoStop}%
\bibitem [{\citenamefont {Maldacena}()}]{maldacena1}%
  \BibitemOpen
  \bibfield  {author} {\bibinfo {author} {\bibfnamefont {J.}~\bibnamefont
  {Maldacena}},\ }\href@noop {} {\bibinfo {title} {Quantum mechanics and the
  geometry of spacetime}},\ \bibinfo {note} {{Strings} 2015 Talk, 100th
  anniversary of general relativity session, Bangalore, India}\BibitemShut
  {NoStop}%
\bibitem [{\citenamefont {Bousso}\ \emph {et~al.}()\citenamefont {Bousso},
  \citenamefont {Dong}, \citenamefont {Engelhardt}, \citenamefont {Faulkner},
  \citenamefont {Hartman}, \citenamefont {Shenker},\ and\ \citenamefont
  {Stanford}}]{snow1}%
  \BibitemOpen
  \bibfield  {author} {\bibinfo {author} {\bibfnamefont {R.}~\bibnamefont
  {Bousso}}, \bibinfo {author} {\bibfnamefont {X.}~\bibnamefont {Dong}},
  \bibinfo {author} {\bibfnamefont {N.}~\bibnamefont {Engelhardt}}, \bibinfo
  {author} {\bibfnamefont {T.}~\bibnamefont {Faulkner}}, \bibinfo {author}
  {\bibfnamefont {T.}~\bibnamefont {Hartman}}, \bibinfo {author} {\bibfnamefont
  {S.~H.}\ \bibnamefont {Shenker}},\ and\ \bibinfo {author} {\bibfnamefont
  {D.}~\bibnamefont {Stanford}},\ }\href@noop {} {\bibinfo {title} {Snowmass
  white paper: Quantum aspects of black holes and the emergence of
  spacetime}},\ \Eprint {https://arxiv.org/abs/2201.03096} {arXiv:2201.03096}
  \BibitemShut {NoStop}%
\bibitem [{\citenamefont {Hardy}()}]{lhardy1}%
  \BibitemOpen
  \bibfield  {author} {\bibinfo {author} {\bibfnamefont {L.}~\bibnamefont
  {Hardy}},\ }\href@noop {} {\bibinfo {title} {Quantum theory from five
  reasonable axioms}},\ \Eprint {https://arxiv.org/abs/quant-ph/0101012}
  {arXiv:quant-ph/0101012} \BibitemShut {NoStop}%
\bibitem [{\citenamefont {Hardy}(2016)}]{lhardy2}%
  \BibitemOpen
  \bibfield  {author} {\bibinfo {author} {\bibfnamefont {L.}~\bibnamefont
  {Hardy}},\ }\bibinfo {title} {Reconstructing quantum theory},\ in\ \href@noop
  {} {\emph {\bibinfo {booktitle} {Quantum Theory: Informational Foundations
  and Foils}}},\ \bibinfo {editor} {edited by\ \bibinfo {editor} {\bibfnamefont
  {G.}~\bibnamefont {Chiribella}}\ and\ \bibinfo {editor} {\bibfnamefont
  {R.}~\bibnamefont {Spekkens}}}\ (\bibinfo  {publisher} {Springer},\ \bibinfo
  {address} {Dordrecht, The Netherlands},\ \bibinfo {year} {2016})\BibitemShut
  {NoStop}%
\bibitem [{\citenamefont {Goyal}(2010)}]{goyal}%
  \BibitemOpen
  \bibfield  {author} {\bibinfo {author} {\bibfnamefont {P.}~\bibnamefont
  {Goyal}},\ }\bibfield  {title} {\bibinfo {title} {From information geometry
  to quantum theory},\ }\href {https://doi.org/10.1088/1367-2630/12/2/023012}
  {\bibfield  {journal} {\bibinfo  {journal} {New J. Phys.}\ }\textbf {\bibinfo
  {volume} {12}},\ \bibinfo {pages} {023012} (\bibinfo {year}
  {2010})}\BibitemShut {NoStop}%
\bibitem [{\citenamefont {Daki\'{c}}\ and\ \citenamefont
  {Brukner}(2011)}]{cb1}%
  \BibitemOpen
  \bibfield  {author} {\bibinfo {author} {\bibfnamefont {B.}~\bibnamefont
  {Daki\'{c}}}\ and\ \bibinfo {author} {\bibfnamefont {{\v{C}}.}~\bibnamefont
  {Brukner}},\ }\bibinfo {title} {Quantum theory and beyond: Is entanglement
  special?},\ in\ \href@noop {} {\emph {\bibinfo {booktitle} {Deep Beauty:
  Understanding the Quantum World through Mathematical Innovation}}},\ \bibinfo
  {editor} {edited by\ \bibinfo {editor} {\bibfnamefont {H.}~\bibnamefont
  {Halvorson}}}\ (\bibinfo  {publisher} {Cambridge University Press},\ \bibinfo
  {address} {New York},\ \bibinfo {year} {2011})\ p.\ \bibinfo {pages}
  {365}\BibitemShut {NoStop}%
\bibitem [{\citenamefont {Chiribella}\ \emph {et~al.}(2011)\citenamefont
  {Chiribella}, \citenamefont {D'Ariano},\ and\ \citenamefont
  {Perinotti}}]{cap}%
  \BibitemOpen
  \bibfield  {author} {\bibinfo {author} {\bibfnamefont {G.}~\bibnamefont
  {Chiribella}}, \bibinfo {author} {\bibfnamefont {G.}~\bibnamefont
  {D'Ariano}},\ and\ \bibinfo {author} {\bibfnamefont {P.}~\bibnamefont
  {Perinotti}},\ }\bibfield  {title} {\bibinfo {title} {Informational
  derivation of quantum theory},\ }\href
  {https://doi.org/10.1103/PhysRevA.84.012311} {\bibfield  {journal} {\bibinfo
  {journal} {Phys. Rev. A}\ }\textbf {\bibinfo {volume} {84}},\ \bibinfo
  {pages} {012311} (\bibinfo {year} {2011})}\BibitemShut {NoStop}%
\bibitem [{\citenamefont {Masanes}\ and\ \citenamefont
  {M\"{u}ller}(2011)}]{mm1}%
  \BibitemOpen
  \bibfield  {author} {\bibinfo {author} {\bibfnamefont {L.}~\bibnamefont
  {Masanes}}\ and\ \bibinfo {author} {\bibfnamefont {M.~P.}\ \bibnamefont
  {M\"{u}ller}},\ }\bibfield  {title} {\bibinfo {title} {A derivation of
  quantum theory from physical requirements},\ }\href
  {https://doi.org/10.1088/1367-2630/13/6/063001} {\bibfield  {journal}
  {\bibinfo  {journal} {New J. Phys.}\ }\textbf {\bibinfo {volume} {13}},\
  \bibinfo {pages} {063001} (\bibinfo {year} {2011})}\BibitemShut {NoStop}%
\bibitem [{\citenamefont {de~la Torre}\ \emph {et~al.}(2012)\citenamefont
  {de~la Torre}, \citenamefont {Masanes}, \citenamefont {Short},\ and\
  \citenamefont {M\"{u}ller}}]{mm2}%
  \BibitemOpen
  \bibfield  {author} {\bibinfo {author} {\bibfnamefont {G.}~\bibnamefont
  {de~la Torre}}, \bibinfo {author} {\bibfnamefont {L.}~\bibnamefont
  {Masanes}}, \bibinfo {author} {\bibfnamefont {A.~J.}\ \bibnamefont {Short}},\
  and\ \bibinfo {author} {\bibfnamefont {M.~P.}\ \bibnamefont {M\"{u}ller}},\
  }\bibfield  {title} {\bibinfo {title} {Deriving quantum theory from its local
  structure and reversibility},\ }\href
  {https://doi.org/10.1103/PhysRevLett.109.090403} {\bibfield  {journal}
  {\bibinfo  {journal} {Phys. Rev. Lett.}\ }\textbf {\bibinfo {volume} {109}},\
  \bibinfo {pages} {090403} (\bibinfo {year} {2012})}\BibitemShut {NoStop}%
\bibitem [{\citenamefont {Masanes}\ \emph {et~al.}(2013)\citenamefont
  {Masanes}, \citenamefont {M\"{u}ller}, \citenamefont {Augusiak},\ and\
  \citenamefont {P\'{e}rez-Garc\'{i}a}}]{mm3}%
  \BibitemOpen
  \bibfield  {author} {\bibinfo {author} {\bibfnamefont {L.}~\bibnamefont
  {Masanes}}, \bibinfo {author} {\bibfnamefont {M.~P.}\ \bibnamefont
  {M\"{u}ller}}, \bibinfo {author} {\bibfnamefont {R.}~\bibnamefont
  {Augusiak}},\ and\ \bibinfo {author} {\bibfnamefont {D.}~\bibnamefont
  {P\'{e}rez-Garc\'{i}a}},\ }\bibfield  {title} {\bibinfo {title} {Existence of
  an information unit as a postulate of quantum theory},\ }\href
  {https://doi.org/10.1073/pnas.1304884110} {\bibfield  {journal} {\bibinfo
  {journal} {Proc. Natl. Acad. Sci. U.S.A.}\ }\textbf {\bibinfo {volume}
  {110}},\ \bibinfo {pages} {16373} (\bibinfo {year} {2013})}\BibitemShut
  {NoStop}%
\bibitem [{\citenamefont {Barnum}\ \emph {et~al.}(2014)\citenamefont {Barnum},
  \citenamefont {M\"{u}ller},\ and\ \citenamefont {Ududec}}]{mm4}%
  \BibitemOpen
  \bibfield  {author} {\bibinfo {author} {\bibfnamefont {H.}~\bibnamefont
  {Barnum}}, \bibinfo {author} {\bibfnamefont {M.~P.}\ \bibnamefont
  {M\"{u}ller}},\ and\ \bibinfo {author} {\bibfnamefont {C.}~\bibnamefont
  {Ududec}},\ }\bibfield  {title} {\bibinfo {title} {Higher-order interference
  and single-system postulates characterizing quantum theory},\ }\href
  {https://doi.org/10.1088/1367-2630/16/12/123029} {\bibfield  {journal}
  {\bibinfo  {journal} {New J. Phys.}\ }\textbf {\bibinfo {volume} {16}},\
  \bibinfo {pages} {123029} (\bibinfo {year} {2014})}\BibitemShut {NoStop}%
\bibitem [{\citenamefont {Krumm}\ \emph {et~al.}(2017)\citenamefont {Krumm},
  \citenamefont {Barnum}, \citenamefont {Barrett},\ and\ \citenamefont
  {M\"{u}ller}}]{mm5}%
  \BibitemOpen
  \bibfield  {author} {\bibinfo {author} {\bibfnamefont {M.}~\bibnamefont
  {Krumm}}, \bibinfo {author} {\bibfnamefont {H.}~\bibnamefont {Barnum}},
  \bibinfo {author} {\bibfnamefont {J.}~\bibnamefont {Barrett}},\ and\ \bibinfo
  {author} {\bibfnamefont {M.~P.}\ \bibnamefont {M\"{u}ller}},\ }\bibfield
  {title} {\bibinfo {title} {Thermodynamics and the structure of quantum
  theory},\ }\href {https://doi.org/10.1088/1367-2630/aa68ef} {\bibfield
  {journal} {\bibinfo  {journal} {New J. Phys.}\ }\textbf {\bibinfo {volume}
  {19}},\ \bibinfo {pages} {043025} (\bibinfo {year} {2017})}\BibitemShut
  {NoStop}%
\bibitem [{\citenamefont {H\"{o}hn}(2017)}]{hohn1}%
  \BibitemOpen
  \bibfield  {author} {\bibinfo {author} {\bibfnamefont {P.~A.}\ \bibnamefont
  {H\"{o}hn}},\ }\bibfield  {title} {\bibinfo {title} {Toolbox for
  reconstructing quantum theory from rules on information acquisition},\ }\href
  {https://doi.org/10.22331/q-2017-12-14-38} {\bibfield  {journal} {\bibinfo
  {journal} {Quantum}\ }\textbf {\bibinfo {volume} {1}},\ \bibinfo {pages} {38}
  (\bibinfo {year} {2017})}\BibitemShut {NoStop}%
\bibitem [{\citenamefont {H\"{o}hn}\ and\ \citenamefont {Wever}(2017)}]{hohn2}%
  \BibitemOpen
  \bibfield  {author} {\bibinfo {author} {\bibfnamefont {P.~A.}\ \bibnamefont
  {H\"{o}hn}}\ and\ \bibinfo {author} {\bibfnamefont {C.~S.~P.}\ \bibnamefont
  {Wever}},\ }\bibfield  {title} {\bibinfo {title} {Quantum theory from
  questions},\ }\href {https://doi.org/10.1103/PhysRevA.95.012102} {\bibfield
  {journal} {\bibinfo  {journal} {Phys. Rev. A}\ }\textbf {\bibinfo {volume}
  {95}},\ \bibinfo {pages} {012102} (\bibinfo {year} {2017})}\BibitemShut
  {NoStop}%
\bibitem [{\citenamefont {Appleby}\ \emph {et~al.}(2017)\citenamefont
  {Appleby}, \citenamefont {Fuchs}, \citenamefont {Stacey},\ and\ \citenamefont
  {Zhu}}]{qbism4}%
  \BibitemOpen
  \bibfield  {author} {\bibinfo {author} {\bibfnamefont {M.}~\bibnamefont
  {Appleby}}, \bibinfo {author} {\bibfnamefont {C.~A.}\ \bibnamefont {Fuchs}},
  \bibinfo {author} {\bibfnamefont {B.~C.}\ \bibnamefont {Stacey}},\ and\
  \bibinfo {author} {\bibfnamefont {H.}~\bibnamefont {Zhu}},\ }\bibfield
  {title} {\bibinfo {title} {Introducing the qplex: A novel arena for quantum
  theory},\ }\href {https://doi.org/10.1140/epjd/e2017-80024-y} {\bibfield
  {journal} {\bibinfo  {journal} {Eur. Phys. J. D}\ }\textbf {\bibinfo {volume}
  {71}},\ \bibinfo {pages} {197} (\bibinfo {year} {2017})}\BibitemShut
  {NoStop}%
\bibitem [{\citenamefont {Wootters}(1980)}]{woottersqit1}%
  \BibitemOpen
  \bibfield  {author} {\bibinfo {author} {\bibfnamefont {W.~K.}\ \bibnamefont
  {Wootters}},\ }\emph {\bibinfo {title} {The acquisition of information from
  quantum measurements}},\ \href@noop {} {Ph.D. thesis},\ \bibinfo  {school}
  {University of Texas at Austin} (\bibinfo {year} {1980})\BibitemShut
  {NoStop}%
\bibitem [{\citenamefont {Adler}(2004)}]{adler1}%
  \BibitemOpen
  \bibfield  {author} {\bibinfo {author} {\bibfnamefont {S.~L.}\ \bibnamefont
  {Adler}},\ }\href@noop {} {\emph {\bibinfo {title} {Quantum Theory as an
  Emergent Phenomenon: The Statistical Mechanics of Matrix Models as the
  Precursor of Quantum Field Theory}}}\ (\bibinfo  {publisher} {Cambridge
  University Press},\ \bibinfo {address} {Cambridge},\ \bibinfo {year}
  {2004})\BibitemShut {NoStop}%
\bibitem [{\citenamefont {'t~Hooft}(2007)}]{thooft1}%
  \BibitemOpen
  \bibfield  {author} {\bibinfo {author} {\bibfnamefont {G.}~\bibnamefont
  {'t~Hooft}},\ }\bibfield  {title} {\bibinfo {title} {Emergent quantum
  mechanics and emergent symmetries},\ }\href
  {https://doi.org/10.1063/1.2823751} {\bibfield  {journal} {\bibinfo
  {journal} {AIP Conf. Proc.}\ }\textbf {\bibinfo {volume} {957}},\ \bibinfo
  {pages} {154} (\bibinfo {year} {2007})}\BibitemShut {NoStop}%
\bibitem [{\citenamefont {'t~Hooft}(2016)}]{thooft2}%
  \BibitemOpen
  \bibfield  {author} {\bibinfo {author} {\bibfnamefont {G.}~\bibnamefont
  {'t~Hooft}},\ }\href@noop {} {\emph {\bibinfo {title} {The Cellular Automaton
  Interpretation of Quantum Mechanics}}}\ (\bibinfo  {publisher} {Springer},\
  \bibinfo {address} {Cham, Switzerland},\ \bibinfo {year} {2016})\BibitemShut
  {NoStop}%
\bibitem [{\citenamefont {'t~Hooft}(2021)}]{thooft3}%
  \BibitemOpen
  \bibfield  {author} {\bibinfo {author} {\bibfnamefont {G.}~\bibnamefont
  {'t~Hooft}},\ }\bibfield  {title} {\bibinfo {title} {Fast vacuum fluctuations
  and the emergence of quantum mechanics},\ }\href
  {https://doi.org/10.1007/s10701-021-00464-7} {\bibfield  {journal} {\bibinfo
  {journal} {Found. Phys.}\ }\textbf {\bibinfo {volume} {51}},\ \bibinfo
  {pages} {63} (\bibinfo {year} {2021})}\BibitemShut {NoStop}%
\bibitem [{\citenamefont {Elze}(2015)}]{elze1}%
  \BibitemOpen
  \bibfield  {author} {\bibinfo {author} {\bibfnamefont {H.-T.}\ \bibnamefont
  {Elze}},\ }\bibfield  {title} {\bibinfo {title} {Qubit exchange interactions
  from permutations of classical bits},\ }\href
  {https://doi.org/10.1142/S021974991941003X} {\bibfield  {journal} {\bibinfo
  {journal} {Int. J. Quantum Info.}\ }\textbf {\bibinfo {volume} {17}},\
  \bibinfo {pages} {1941003} (\bibinfo {year} {2015})}\BibitemShut {NoStop}%
\bibitem [{\citenamefont {Wetterich}(2021)}]{wetterich1}%
  \BibitemOpen
  \bibfield  {author} {\bibinfo {author} {\bibfnamefont {C.}~\bibnamefont
  {Wetterich}},\ }\bibfield  {title} {\bibinfo {title} {Probabilistic cellular
  automata for interacting fermionic quantum field theories},\ }\href
  {https://doi.org/10.1016/j.nuclphysb.2020.11529} {\bibfield  {journal}
  {\bibinfo  {journal} {Nucl. Phys. B}\ }\textbf {\bibinfo {volume} {963}},\
  \bibinfo {pages} {115296} (\bibinfo {year} {2021})}\BibitemShut {NoStop}%
\bibitem [{\citenamefont {Wetterich}(2022)}]{wetterich2}%
  \BibitemOpen
  \bibfield  {author} {\bibinfo {author} {\bibfnamefont {C.}~\bibnamefont
  {Wetterich}},\ }\bibfield  {title} {\bibinfo {title} {Fermionic quantum field
  theories as probabilistic cellular automata},\ }\href
  {https://doi.org/10.1103/PhysRevD.105.074502} {\bibfield  {journal} {\bibinfo
   {journal} {Phys. Rev. D}\ }\textbf {\bibinfo {volume} {105}},\ \bibinfo
  {pages} {074502} (\bibinfo {year} {2022})}\BibitemShut {NoStop}%
\bibitem [{\citenamefont {Minic}\ and\ \citenamefont {Pajevic}(2016)}]{minic1}%
  \BibitemOpen
  \bibfield  {author} {\bibinfo {author} {\bibfnamefont {D.}~\bibnamefont
  {Minic}}\ and\ \bibinfo {author} {\bibfnamefont {S.}~\bibnamefont
  {Pajevic}},\ }\bibfield  {title} {\bibinfo {title} {Emergent ``quantum''
  theory in complex adaptive systems},\ }\href
  {https://doi.org/10.1142/S0217984916502018} {\bibfield  {journal} {\bibinfo
  {journal} {Mod. Phys. Lett. B}\ }\textbf {\bibinfo {volume} {30}},\ \bibinfo
  {pages} {1650201} (\bibinfo {year} {2016})}\BibitemShut {NoStop}%
\bibitem [{\citenamefont {Smolin}(2016)}]{smolinweqm}%
  \BibitemOpen
  \bibfield  {author} {\bibinfo {author} {\bibfnamefont {L.}~\bibnamefont
  {Smolin}},\ }\bibfield  {title} {\bibinfo {title} {Quantum mechanics and the
  principle of maximal variety},\ }\href
  {https://doi.org/10.1007/s10701-016-9994-x} {\bibfield  {journal} {\bibinfo
  {journal} {Found. Phys.}\ }\textbf {\bibinfo {volume} {46}},\ \bibinfo
  {pages} {736} (\bibinfo {year} {2016})}\BibitemShut {NoStop}%
\bibitem [{\citenamefont {Vanchurin}(2020)}]{vanchurin1}%
  \BibitemOpen
  \bibfield  {author} {\bibinfo {author} {\bibfnamefont {V.}~\bibnamefont
  {Vanchurin}},\ }\bibfield  {title} {\bibinfo {title} {Entropic mechanics:
  Towards a stochastic description of quantum mechanics},\ }\href
  {https://doi.org/10.1007/s10701-019-00315-6} {\bibfield  {journal} {\bibinfo
  {journal} {Found. Phys.}\ }\textbf {\bibinfo {volume} {50}},\ \bibinfo
  {pages} {40} (\bibinfo {year} {2020})}\BibitemShut {NoStop}%
\bibitem [{\citenamefont {Katsnelson}\ and\ \citenamefont
  {Vanchurin}(2021)}]{vanchurin2}%
  \BibitemOpen
  \bibfield  {author} {\bibinfo {author} {\bibfnamefont {M.~I.}\ \bibnamefont
  {Katsnelson}}\ and\ \bibinfo {author} {\bibfnamefont {V.}~\bibnamefont
  {Vanchurin}},\ }\bibfield  {title} {\bibinfo {title} {Emergent quantumness in
  neural networks},\ }\href {https://doi.org/10.1007/s10701-021-00503-3}
  {\bibfield  {journal} {\bibinfo  {journal} {Found. Phys.}\ }\textbf {\bibinfo
  {volume} {51}},\ \bibinfo {pages} {94} (\bibinfo {year} {2021})}\BibitemShut
  {NoStop}%
\bibitem [{\citenamefont {Filk}(2022)}]{QLike}%
  \BibitemOpen
  \bibfield  {author} {\bibinfo {author} {\bibfnamefont {T.}~\bibnamefont
  {Filk}},\ }\bibinfo {title} {The quantum-like behavior of neural networks},\
  in\ \href@noop {} {\emph {\bibinfo {booktitle} {From Electrons to Elephants
  and Elections}}},\ \bibinfo {editor} {edited by\ \bibinfo {editor}
  {\bibfnamefont {S.}~\bibnamefont {Wuppuluri}}\ and\ \bibinfo {editor}
  {\bibfnamefont {I.}~\bibnamefont {Stewart}}}\ (\bibinfo  {publisher}
  {Springer},\ \bibinfo {address} {Cham, Switzerland},\ \bibinfo {year}
  {2022})\BibitemShut {NoStop}%
\bibitem [{\citenamefont {Slagle}\ and\ \citenamefont {Preskill}(2023)}]{sp1}%
  \BibitemOpen
  \bibfield  {author} {\bibinfo {author} {\bibfnamefont {K.}~\bibnamefont
  {Slagle}}\ and\ \bibinfo {author} {\bibfnamefont {J.}~\bibnamefont
  {Preskill}},\ }\bibfield  {title} {\bibinfo {title} {Emergent quantum
  mechanics at the boundary of a local classical lattice model},\ }\href
  {https://doi.org/10.1103/PhysRevA.108.012217} {\bibfield  {journal} {\bibinfo
   {journal} {Phys. Rev. A}\ }\textbf {\bibinfo {volume} {108}},\ \bibinfo
  {pages} {012217} (\bibinfo {year} {2023})}\BibitemShut {NoStop}%
\bibitem [{\citenamefont {Wittek}(2014)}]{revoqai1}%
  \BibitemOpen
  \bibfield  {author} {\bibinfo {author} {\bibfnamefont {P.}~\bibnamefont
  {Wittek}},\ }\href@noop {} {\emph {\bibinfo {title} {Quantum Machine
  Learning: What Quantum Computing Means to Data Mining}}}\ (\bibinfo
  {publisher} {Academic Press},\ \bibinfo {address} {San Diego, CA},\ \bibinfo
  {year} {2014})\BibitemShut {NoStop}%
\bibitem [{\citenamefont {Schuld}\ \emph {et~al.}(2014)\citenamefont {Schuld},
  \citenamefont {Sinayskiy},\ and\ \citenamefont {Petruccione}}]{revoqai2}%
  \BibitemOpen
  \bibfield  {author} {\bibinfo {author} {\bibfnamefont {M.}~\bibnamefont
  {Schuld}}, \bibinfo {author} {\bibfnamefont {I.}~\bibnamefont {Sinayskiy}},\
  and\ \bibinfo {author} {\bibfnamefont {F.}~\bibnamefont {Petruccione}},\
  }\bibfield  {title} {\bibinfo {title} {The quest for a quantum neural
  network},\ }\href {https://doi.org/10.1007/s11128-014-0809-8} {\bibfield
  {journal} {\bibinfo  {journal} {Quantum Inf. Process.}\ }\textbf {\bibinfo
  {volume} {13}},\ \bibinfo {pages} {2567} (\bibinfo {year}
  {2014})}\BibitemShut {NoStop}%
\bibitem [{\citenamefont {Biamonte}\ \emph {et~al.}(2017)\citenamefont
  {Biamonte}, \citenamefont {Wittek}, \citenamefont {Pancotti}, \citenamefont
  {Rebentrost}, \citenamefont {Wiebe},\ and\ \citenamefont {Lloyd}}]{revoqai3}%
  \BibitemOpen
  \bibfield  {author} {\bibinfo {author} {\bibfnamefont {J.}~\bibnamefont
  {Biamonte}}, \bibinfo {author} {\bibfnamefont {P.}~\bibnamefont {Wittek}},
  \bibinfo {author} {\bibfnamefont {N.}~\bibnamefont {Pancotti}}, \bibinfo
  {author} {\bibfnamefont {P.}~\bibnamefont {Rebentrost}}, \bibinfo {author}
  {\bibfnamefont {N.}~\bibnamefont {Wiebe}},\ and\ \bibinfo {author}
  {\bibfnamefont {S.}~\bibnamefont {Lloyd}},\ }\bibfield  {title} {\bibinfo
  {title} {Quantum machine learning},\ }\href
  {https://doi.org/10.1038/nature23474} {\bibfield  {journal} {\bibinfo
  {journal} {Nature (London)}\ }\textbf {\bibinfo {volume} {549}},\ \bibinfo
  {pages} {195} (\bibinfo {year} {2017})}\BibitemShut {NoStop}%
\bibitem [{\citenamefont {Dunjko}\ and\ \citenamefont
  {Briegel}(2018)}]{revoqai4}%
  \BibitemOpen
  \bibfield  {author} {\bibinfo {author} {\bibfnamefont {V.}~\bibnamefont
  {Dunjko}}\ and\ \bibinfo {author} {\bibfnamefont {H.~J.}\ \bibnamefont
  {Briegel}},\ }\bibfield  {title} {\bibinfo {title} {Machine learning and
  artificial intelligence in the quantum domain: a review of recent progress},\
  }\href {https://doi.org/10.1088/1361-6633/aab406} {\bibfield  {journal}
  {\bibinfo  {journal} {Rep. Prog. Phys.}\ }\textbf {\bibinfo {volume} {81}},\
  \bibinfo {pages} {074001} (\bibinfo {year} {2018})}\BibitemShut {NoStop}%
\bibitem [{\citenamefont {Garg}\ and\ \citenamefont
  {Ramakrishnan}()}]{revoqai5}%
  \BibitemOpen
  \bibfield  {author} {\bibinfo {author} {\bibfnamefont {S.}~\bibnamefont
  {Garg}}\ and\ \bibinfo {author} {\bibfnamefont {G.}~\bibnamefont
  {Ramakrishnan}},\ }\href@noop {} {\bibinfo {title} {{Advances in quantum deep
  learning: An overview}}},\ \Eprint {https://arxiv.org/abs/2005.04316}
  {arXiv:2005.04316} \BibitemShut {NoStop}%
\bibitem [{\citenamefont {Cerezo}\ \emph {et~al.}(2022)\citenamefont {Cerezo},
  \citenamefont {Verdon}, \citenamefont {Huang}, \citenamefont {Cincio},\ and\
  \citenamefont {Coles}}]{revoqai6}%
  \BibitemOpen
  \bibfield  {author} {\bibinfo {author} {\bibfnamefont {M.}~\bibnamefont
  {Cerezo}}, \bibinfo {author} {\bibfnamefont {G.}~\bibnamefont {Verdon}},
  \bibinfo {author} {\bibfnamefont {H.-Y.}\ \bibnamefont {Huang}}, \bibinfo
  {author} {\bibfnamefont {L.}~\bibnamefont {Cincio}},\ and\ \bibinfo {author}
  {\bibfnamefont {P.~J.}\ \bibnamefont {Coles}},\ }\bibfield  {title} {\bibinfo
  {title} {Challenges and opportunities in quantum machine learning},\ }\href
  {https://doi.org/10.1038/s43588-022-00311-3} {\bibfield  {journal} {\bibinfo
  {journal} {Nat. Comput. Sci.}\ }\textbf {\bibinfo {volume} {2}},\ \bibinfo
  {pages} {567} (\bibinfo {year} {2022})}\BibitemShut {NoStop}%
\bibitem [{\citenamefont {O'Reilly}\ \emph {et~al.}(2020)\citenamefont
  {O'Reilly}, \citenamefont {Munakata}, \citenamefont {Frank}, \citenamefont
  {Hazy},\ and\ \citenamefont {Contributors}}]{cmpcgn1}%
  \BibitemOpen
  \bibfield  {author} {\bibinfo {author} {\bibfnamefont {R.~C.}\ \bibnamefont
  {O'Reilly}}, \bibinfo {author} {\bibfnamefont {Y.}~\bibnamefont {Munakata}},
  \bibinfo {author} {\bibfnamefont {M.~J.}\ \bibnamefont {Frank}}, \bibinfo
  {author} {\bibfnamefont {T.~E.}\ \bibnamefont {Hazy}},\ and\ \bibinfo
  {author} {\bibnamefont {Contributors}},\ }\href@noop {} {\emph {\bibinfo
  {title} {Computational Cognitive Neuroscience}}}\ (\bibinfo  {publisher}
  {Online Books},\ \bibinfo {year} {2020})\BibitemShut {NoStop}%
\bibitem [{\citenamefont {van Gerven}(2017)}]{cmpcgn2}%
  \BibitemOpen
  \bibfield  {author} {\bibinfo {author} {\bibfnamefont {M.~A.~J.}\
  \bibnamefont {van Gerven}},\ }\bibfield  {title} {\bibinfo {title}
  {Computational foundations of natural intelligence},\ }\href
  {https://doi.org/10.3389/fncom.2017.00112} {\bibfield  {journal} {\bibinfo
  {journal} {Front. Comput. Neurosci.}\ }\textbf {\bibinfo {volume} {11}},\
  \bibinfo {pages} {1} (\bibinfo {year} {2017})}\BibitemShut {NoStop}%
\bibitem [{\citenamefont {Feynman}(1948)}]{Feynman-I}%
  \BibitemOpen
  \bibfield  {author} {\bibinfo {author} {\bibfnamefont {R.~P.}\ \bibnamefont
  {Feynman}},\ }\bibfield  {title} {\bibinfo {title} {Space-time approach to
  non-relativistic quantum mechanics},\ }\href
  {https://doi.org/10.1103/RevModPhys.20.367} {\bibfield  {journal} {\bibinfo
  {journal} {Rev. Mod. Phys.}\ }\textbf {\bibinfo {volume} {20}},\ \bibinfo
  {pages} {367} (\bibinfo {year} {1948})}\BibitemShut {NoStop}%
\bibitem [{\citenamefont {Feynman}(1949)}]{Feynman-II}%
  \BibitemOpen
  \bibfield  {author} {\bibinfo {author} {\bibfnamefont {R.~P.}\ \bibnamefont
  {Feynman}},\ }\bibfield  {title} {\bibinfo {title} {Space-time approach to
  quantum electrodynamics},\ }\href {https://doi.org/10.1103/PhysRev.76.769}
  {\bibfield  {journal} {\bibinfo  {journal} {Phys. Rev.}\ }\textbf {\bibinfo
  {volume} {76}},\ \bibinfo {pages} {769} (\bibinfo {year} {1949})}\BibitemShut
  {NoStop}%
\bibitem [{\citenamefont {M{\"{u}}ller}()}]{Muller-youtube}%
  \BibitemOpen
  \bibfield  {author} {\bibinfo {author} {\bibfnamefont {M.~P.}\ \bibnamefont
  {M{\"{u}}ller}},\ }\href@noop {} {\bibinfo {title} {How spacetime constrains
  the structure of quantum theory}},\ \bibinfo {note} {virtual seminar, March
  17, 2023; \url{https://www.youtube.com/watch?v=NSZ7gtMhnSI}}\BibitemShut
  {NoStop}%
\bibitem [{\citenamefont {Garner}\ \emph {et~al.}(2017)\citenamefont {Garner},
  \citenamefont {M{\"{u}}ller},\ and\ \citenamefont {Dahlsten}}]{Muller-new}%
  \BibitemOpen
  \bibfield  {author} {\bibinfo {author} {\bibfnamefont {A.~J.~P.}\
  \bibnamefont {Garner}}, \bibinfo {author} {\bibfnamefont {M.~P.}\
  \bibnamefont {M{\"{u}}ller}},\ and\ \bibinfo {author} {\bibfnamefont
  {O.~C.~O.}\ \bibnamefont {Dahlsten}},\ }\bibfield  {title} {\bibinfo {title}
  {The complex and quaternionic quantum bit from relativity of simultaneity on
  aninterferometer},\ }\href {https://doi.org/10.1098/rspa.2017.0596}
  {\bibfield  {journal} {\bibinfo  {journal} {Proc. R. Soc. A}\ }\textbf
  {\bibinfo {volume} {473}},\ \bibinfo {pages} {20170596} (\bibinfo {year}
  {2017})}\BibitemShut {NoStop}%
\bibitem [{\citenamefont {Jones}\ \emph {et~al.}()\citenamefont {Jones},
  \citenamefont {Ludescher}, \citenamefont {Aloy},\ and\ \citenamefont
  {M{\"{u}}ller}}]{Muller-new2}%
  \BibitemOpen
  \bibfield  {author} {\bibinfo {author} {\bibfnamefont {C.~L.}\ \bibnamefont
  {Jones}}, \bibinfo {author} {\bibfnamefont {S.~L.}\ \bibnamefont
  {Ludescher}}, \bibinfo {author} {\bibfnamefont {A.}~\bibnamefont {Aloy}},\
  and\ \bibinfo {author} {\bibfnamefont {M.~P.}\ \bibnamefont {M{\"{u}}ller}},\
  }\href@noop {} {\bibinfo {title} {Theory-independent randomness generation
  with spacetime symmetries}},\ \Eprint {https://arxiv.org/abs/2210.14811}
  {arXiv:2210.14811} \BibitemShut {NoStop}%
\bibitem [{\citenamefont {Alipour}\ \emph {et~al.}(2020)\citenamefont
  {Alipour}, \citenamefont {Rezakhani}, \citenamefont {Babu}, \citenamefont
  {M{\o}lmer}, \citenamefont {M\"{o}tt\"{o}nen},\ and\ \citenamefont
  {Ala-Nissila}}]{CorrPic}%
  \BibitemOpen
  \bibfield  {author} {\bibinfo {author} {\bibfnamefont {S.}~\bibnamefont
  {Alipour}}, \bibinfo {author} {\bibfnamefont {A.~T.}\ \bibnamefont
  {Rezakhani}}, \bibinfo {author} {\bibfnamefont {A.~P.}\ \bibnamefont {Babu}},
  \bibinfo {author} {\bibfnamefont {K.}~\bibnamefont {M{\o}lmer}}, \bibinfo
  {author} {\bibfnamefont {M.}~\bibnamefont {M\"{o}tt\"{o}nen}},\ and\ \bibinfo
  {author} {\bibfnamefont {T.}~\bibnamefont {Ala-Nissila}},\ }\bibfield
  {title} {\bibinfo {title} {{Correlation-Picture Approach to
  Open-Quantum-System Dynamics}},\ }\href
  {https://doi.org/10.1103/PhysRevX.10.041024} {\bibfield  {journal} {\bibinfo
  {journal} {Phys. Rev. X}\ }\textbf {\bibinfo {volume} {10}},\ \bibinfo
  {pages} {041024} (\bibinfo {year} {2020})}\BibitemShut {NoStop}%
\bibitem [{\citenamefont {Smith}\ and\ \citenamefont
  {Ahmadi}(2019)}]{Smith-Ahmadi}%
  \BibitemOpen
  \bibfield  {author} {\bibinfo {author} {\bibfnamefont {A.~R.~H.}\
  \bibnamefont {Smith}}\ and\ \bibinfo {author} {\bibfnamefont
  {M.}~\bibnamefont {Ahmadi}},\ }\bibfield  {title} {\bibinfo {title}
  {Quantizing time: {I}nteracting clocks and systems},\ }\href
  {https://doi.org/10.22331/q-2019-07-08-160} {\bibfield  {journal} {\bibinfo
  {journal} {Quantum}\ }\textbf {\bibinfo {volume} {3}},\ \bibinfo {pages}
  {160} (\bibinfo {year} {2019})}\BibitemShut {NoStop}%
\bibitem [{\citenamefont {H\"{o}hn}\ \emph {et~al.}(2021)\citenamefont
  {H\"{o}hn}, \citenamefont {Smith},\ and\ \citenamefont {Lock}}]{Trinity}%
  \BibitemOpen
  \bibfield  {author} {\bibinfo {author} {\bibfnamefont {P.~A.}\ \bibnamefont
  {H\"{o}hn}}, \bibinfo {author} {\bibfnamefont {A.~R.~H.}\ \bibnamefont
  {Smith}},\ and\ \bibinfo {author} {\bibfnamefont {M.~P.~E.}\ \bibnamefont
  {Lock}},\ }\bibfield  {title} {\bibinfo {title} {Trinity of relational
  quantum dynamics},\ }\href {https://doi.org/10.1103/PhysRevD.104.066001}
  {\bibfield  {journal} {\bibinfo  {journal} {Phys. Rev. D}\ }\textbf {\bibinfo
  {volume} {104}},\ \bibinfo {pages} {066001} (\bibinfo {year}
  {2021})}\BibitemShut {NoStop}%
\bibitem [{\citenamefont {Oreshkov}(2019)}]{Oreshkov}%
  \BibitemOpen
  \bibfield  {author} {\bibinfo {author} {\bibfnamefont {O.}~\bibnamefont
  {Oreshkov}},\ }\bibfield  {title} {\bibinfo {title} {Time-delocalized quantum
  subsystems and operations: on the existence of processes with indefinite
  causal structure in quantum mechanics},\ }\href
  {https://doi.org/10.22331/q-2019-12-02-206} {\bibfield  {journal} {\bibinfo
  {journal} {Quantum}\ }\textbf {\bibinfo {volume} {3}},\ \bibinfo {pages}
  {206} (\bibinfo {year} {2019})}\BibitemShut {NoStop}%
\bibitem [{\citenamefont {Gu\'{r}rin}\ and\ \citenamefont
  {Brukner}(2018)}]{Guerin}%
  \BibitemOpen
  \bibfield  {author} {\bibinfo {author} {\bibfnamefont {P.~A.}\ \bibnamefont
  {Gu\'{r}rin}}\ and\ \bibinfo {author} {\bibfnamefont {{\v{C}}.}~\bibnamefont
  {Brukner}},\ }\bibfield  {title} {\bibinfo {title} {Observer-dependent
  locality of quantum events},\ }\href
  {https://doi.org/10.1088/1367-2630/aae742} {\bibfield  {journal} {\bibinfo
  {journal} {New J. Phys.}\ }\textbf {\bibinfo {volume} {20}},\ \bibinfo
  {pages} {103031} (\bibinfo {year} {2018})}\BibitemShut {NoStop}%
\bibitem [{\citenamefont {Castro-Ruiz}\ \emph {et~al.}(2020)\citenamefont
  {Castro-Ruiz}, \citenamefont {Giacomini}, \citenamefont {Belenchia},\ and\
  \citenamefont {Brukner}}]{Brukner-}%
  \BibitemOpen
  \bibfield  {author} {\bibinfo {author} {\bibfnamefont {E.}~\bibnamefont
  {Castro-Ruiz}}, \bibinfo {author} {\bibfnamefont {F.}~\bibnamefont
  {Giacomini}}, \bibinfo {author} {\bibfnamefont {A.}~\bibnamefont
  {Belenchia}},\ and\ \bibinfo {author} {\bibfnamefont {{\v{C}}.}~\bibnamefont
  {Brukner}},\ }\bibfield  {title} {\bibinfo {title} {Quantum clocks and the
  temporal localisability of events in the presence of gravitating quantum
  systems},\ }\href {https://doi.org/10.1038/s41467-020-16013-1} {\bibfield
  {journal} {\bibinfo  {journal} {Nat. Commun.}\ }\textbf {\bibinfo {volume}
  {11}},\ \bibinfo {pages} {2672} (\bibinfo {year} {2020})}\BibitemShut
  {NoStop}%
\bibitem [{\citenamefont {Paiva}\ \emph {et~al.}(2022)\citenamefont {Paiva},
  \citenamefont {Nowakowski},\ and\ \citenamefont {Cohen}}]{Paiva}%
  \BibitemOpen
  \bibfield  {author} {\bibinfo {author} {\bibfnamefont {I.~L.}\ \bibnamefont
  {Paiva}}, \bibinfo {author} {\bibfnamefont {M.}~\bibnamefont {Nowakowski}},\
  and\ \bibinfo {author} {\bibfnamefont {E.}~\bibnamefont {Cohen}},\ }\bibfield
   {title} {\bibinfo {title} {Dynamical nonlocality in quantum time via modular
  operators},\ }\href {https://doi.org/10.1103/PhysRevA.105.042207} {\bibfield
  {journal} {\bibinfo  {journal} {Phys. Rev. A}\ }\textbf {\bibinfo {volume}
  {105}},\ \bibinfo {pages} {042207} (\bibinfo {year} {2022})}\BibitemShut
  {NoStop}%
\bibitem [{\citenamefont {Abrams}\ and\ \citenamefont
  {Lloyd}(1998)}]{Abrams-Lloyd}%
  \BibitemOpen
  \bibfield  {author} {\bibinfo {author} {\bibfnamefont {D.~S.}\ \bibnamefont
  {Abrams}}\ and\ \bibinfo {author} {\bibfnamefont {S.}~\bibnamefont {Lloyd}},\
  }\bibfield  {title} {\bibinfo {title} {{Nonlinear Quantum Mechanics Implies
  Polynomial-Time Solution for $NP$-Complete and $\#P$ Problems}},\ }\href
  {https://doi.org/10.1103/PhysRevLett.81.3992} {\bibfield  {journal} {\bibinfo
   {journal} {Phys. Rev. Lett.}\ }\textbf {\bibinfo {volume} {81}},\ \bibinfo
  {pages} {3992} (\bibinfo {year} {1998})}\BibitemShut {NoStop}%
\bibitem [{\citenamefont {Childs}\ and\ \citenamefont
  {Young}(2016)}]{Childs-search}%
  \BibitemOpen
  \bibfield  {author} {\bibinfo {author} {\bibfnamefont {A.~M.}\ \bibnamefont
  {Childs}}\ and\ \bibinfo {author} {\bibfnamefont {J.}~\bibnamefont {Young}},\
  }\bibfield  {title} {\bibinfo {title} {Optimal state discrimination and
  unstructured search in nonlinear quantum mechanics},\ }\href
  {https://doi.org/10.1103/PhysRevA.93.022314} {\bibfield  {journal} {\bibinfo
  {journal} {Phys. Rev. A}\ }\textbf {\bibinfo {volume} {93}},\ \bibinfo
  {pages} {022314} (\bibinfo {year} {2016})}\BibitemShut {NoStop}%
\bibitem [{\citenamefont {Meyer}\ and\ \citenamefont
  {Wong}(2013)}]{Meyer-search}%
  \BibitemOpen
  \bibfield  {author} {\bibinfo {author} {\bibfnamefont {D.~A.}\ \bibnamefont
  {Meyer}}\ and\ \bibinfo {author} {\bibfnamefont {T.~G.}\ \bibnamefont
  {Wong}},\ }\bibfield  {title} {\bibinfo {title} {{Nonlinear quantum search
  using the Gross–Pitaevskii equation}},\ }\href
  {https://doi.org/10.1088/1367-2630/15/6/063014} {\bibfield  {journal}
  {\bibinfo  {journal} {New J. Phys.}\ }\textbf {\bibinfo {volume} {15}},\
  \bibinfo {pages} {063014} (\bibinfo {year} {2013})}\BibitemShut {NoStop}%
\bibitem [{\citenamefont {Wong}(2014)}]{Wong-thesis}%
  \BibitemOpen
  \bibfield  {author} {\bibinfo {author} {\bibfnamefont {T.~G.}\ \bibnamefont
  {Wong}},\ }\emph {\bibinfo {title} {Nonlinear quantum search}},\ \href@noop
  {} {Ph.D. thesis},\ \bibinfo  {school} {University of California San Diego}
  (\bibinfo {year} {2014}),\ \Eprint {https://arxiv.org/abs/1506.04388}
  {arXiv:1506.04388} \BibitemShut {NoStop}%
\bibitem [{\citenamefont {Barrett}\ \emph {et~al.}(2019)\citenamefont
  {Barrett}, \citenamefont {de~Beaudrap}, \citenamefont {Hoban},\ and\
  \citenamefont {Lee}}]{Barrett-npj}%
  \BibitemOpen
  \bibfield  {author} {\bibinfo {author} {\bibfnamefont {J.}~\bibnamefont
  {Barrett}}, \bibinfo {author} {\bibfnamefont {N.}~\bibnamefont
  {de~Beaudrap}}, \bibinfo {author} {\bibfnamefont {M.~J.}\ \bibnamefont
  {Hoban}},\ and\ \bibinfo {author} {\bibfnamefont {C.~M.}\ \bibnamefont
  {Lee}},\ }\bibfield  {title} {\bibinfo {title} {The computational landscape
  of general physical theories},\ }\href
  {https://doi.org/10.1038/s41534-019-0156-9} {\bibfield  {journal} {\bibinfo
  {journal} {npj Quantum Inf.}\ }\textbf {\bibinfo {volume} {5}},\ \bibinfo
  {pages} {41} (\bibinfo {year} {2019})}\BibitemShut {NoStop}%
\bibitem [{\citenamefont {Chowdhury}\ \emph {et~al.}(2022)\citenamefont
  {Chowdhury}, \citenamefont {Georges}, \citenamefont {Parcollet},\ and\
  \citenamefont {Sachdev}}]{SYK}%
  \BibitemOpen
  \bibfield  {author} {\bibinfo {author} {\bibfnamefont {D.}~\bibnamefont
  {Chowdhury}}, \bibinfo {author} {\bibfnamefont {A.}~\bibnamefont {Georges}},
  \bibinfo {author} {\bibfnamefont {O.}~\bibnamefont {Parcollet}},\ and\
  \bibinfo {author} {\bibfnamefont {S.}~\bibnamefont {Sachdev}},\ }\bibfield
  {title} {\bibinfo {title} {{Sachdev-Ye-Kitaev models and beyond: Window into
  non-Fermi liquids}},\ }\href {https://doi.org/10.1103/RevModPhys.94.035004}
  {\bibfield  {journal} {\bibinfo  {journal} {Rev. Mod. Phys.}\ }\textbf
  {\bibinfo {volume} {94}},\ \bibinfo {pages} {035004} (\bibinfo {year}
  {2022})}\BibitemShut {NoStop}%
\bibitem [{\citenamefont {Susskind}()}]{SusDual}%
  \BibitemOpen
  \bibfield  {author} {\bibinfo {author} {\bibfnamefont {L.}~\bibnamefont
  {Susskind}},\ }\href@noop {} {\bibinfo {title} {{De Sitter space,
  double-scaled SYK, and the separation of scales in the semiclassical
  limit}}},\ \Eprint {https://arxiv.org/abs/2209.09999} {arXiv:2209.09999}
  \BibitemShut {NoStop}%
\bibitem [{\citenamefont {Banks}({\natexlab{b}})}]{Banks-old}%
  \BibitemOpen
  \bibfield  {author} {\bibinfo {author} {\bibfnamefont {T.}~\bibnamefont
  {Banks}},\ }\href@noop {} {\bibinfo {title} {Old ideas for new physicists:1}},\ \Eprint {https://arxiv.org/abs/2208.08959}
  {arXiv:2208.08959} \BibitemShut {NoStop}%
\bibitem [{\citenamefont {Wen}(2007)}]{book:Wen}%
  \BibitemOpen
  \bibfield  {author} {\bibinfo {author} {\bibfnamefont {X.-G.}\ \bibnamefont
  {Wen}},\ }\href@noop {} {\emph {\bibinfo {title} {Quantum Field Theory of
  Many-body Systems: From the Origin of Sound to an Origin of Light and
  Electrons}}}\ (\bibinfo  {publisher} {Oxford University Press},\ \bibinfo
  {address} {Oxford},\ \bibinfo {year} {2007})\BibitemShut {NoStop}%
\bibitem [{\citenamefont {Wen}(2018)}]{Wen2}%
  \BibitemOpen
  \bibfield  {author} {\bibinfo {author} {\bibfnamefont {X.-G.}\ \bibnamefont
  {Wen}},\ }\bibfield  {title} {\bibinfo {title} {{Four revolutions in physics
  and the second quantum revolution -- A unification of force and matter by
  quantum information}},\ }\href {https://doi.org/10.1142/S0217979218300104}
  {\bibfield  {journal} {\bibinfo  {journal} {Int. J. Mod. Phys. B}\ }\textbf
  {\bibinfo {volume} {32}},\ \bibinfo {pages} {1830010} (\bibinfo {year}
  {2018})}\BibitemShut {NoStop}%
\bibitem [{\citenamefont {Kibble}(1979)}]{Kibble-G}%
  \BibitemOpen
  \bibfield  {author} {\bibinfo {author} {\bibfnamefont {T.~W.~B.}\
  \bibnamefont {Kibble}},\ }\bibfield  {title} {\bibinfo {title}
  {Geometrization of quantum mechanics},\ }\href
  {https://doi.org/10.1007/BF01225149} {\bibfield  {journal} {\bibinfo
  {journal} {Commun. Math. Phys.}\ }\textbf {\bibinfo {volume} {65}},\ \bibinfo
  {pages} {189} (\bibinfo {year} {1979})}\BibitemShut {NoStop}%
\bibitem [{\citenamefont {Sorkin}(1994)}]{Sorkin-MPLA}%
  \BibitemOpen
  \bibfield  {author} {\bibinfo {author} {\bibfnamefont {R.~D.}\ \bibnamefont
  {Sorkin}},\ }\bibfield  {title} {\bibinfo {title} {Quantum mechanics as
  quantum measure theory},\ }\href {https://doi.org/10.1142/S021773239400294X}
  {\bibfield  {journal} {\bibinfo  {journal} {Mod. Phys. Lett. A}\ }\textbf
  {\bibinfo {volume} {09}},\ \bibinfo {pages} {3119} (\bibinfo {year}
  {1994})}\BibitemShut {NoStop}%
\bibitem [{\citenamefont {Adler}\ and\ \citenamefont
  {Millard}(1996)}]{Adler-etal}%
  \BibitemOpen
  \bibfield  {author} {\bibinfo {author} {\bibfnamefont {S.~L.}\ \bibnamefont
  {Adler}}\ and\ \bibinfo {author} {\bibfnamefont {A.~C.}\ \bibnamefont
  {Millard}},\ }\bibfield  {title} {\bibinfo {title} {Generalized quantum
  dynamics as pre-quantum mechanics},\ }\href
  {https://doi.org/10.1016/0550-3213(96)00253-2} {\bibfield  {journal}
  {\bibinfo  {journal} {Nucl. Phys. B}\ }\textbf {\bibinfo {volume} {473}},\
  \bibinfo {pages} {199} (\bibinfo {year} {1996})}\BibitemShut {NoStop}%
\bibitem [{\citenamefont {Atmanspacher}\ \emph {et~al.}(2002)\citenamefont
  {Atmanspacher}, \citenamefont {R\"{o}mer},\ and\ \citenamefont
  {Walach}}]{weakQT-1}%
  \BibitemOpen
  \bibfield  {author} {\bibinfo {author} {\bibfnamefont {H.}~\bibnamefont
  {Atmanspacher}}, \bibinfo {author} {\bibfnamefont {H.}~\bibnamefont
  {R\"{o}mer}},\ and\ \bibinfo {author} {\bibfnamefont {H.}~\bibnamefont
  {Walach}},\ }\bibfield  {title} {\bibinfo {title} {Weak quantum theory:
  {C}omplementarity and entanglement in physics and beyond},\ }\href
  {https://doi.org/10.1023/A:1014809312397} {\bibfield  {journal} {\bibinfo
  {journal} {Found. Phys.}\ }\textbf {\bibinfo {volume} {32}},\ \bibinfo
  {pages} {379} (\bibinfo {year} {2002})}\BibitemShut {NoStop}%
\bibitem [{\citenamefont {Atmanspacher}\ \emph {et~al.}(2006)\citenamefont
  {Atmanspacher}, \citenamefont {Filk},\ and\ \citenamefont
  {R\"{o}mer}}]{weakQT-2}%
  \BibitemOpen
  \bibfield  {author} {\bibinfo {author} {\bibfnamefont {H.}~\bibnamefont
  {Atmanspacher}}, \bibinfo {author} {\bibfnamefont {T.}~\bibnamefont {Filk}},\
  and\ \bibinfo {author} {\bibfnamefont {H.}~\bibnamefont {R\"{o}mer}},\
  }\bibfield  {title} {\bibinfo {title} {Weak quantum theory: {F}ormal
  framework and selected applications},\ }\href
  {https://doi.org/10.1063/1.2158709} {\bibfield  {journal} {\bibinfo
  {journal} {AIP Conf. Proc.}\ }\textbf {\bibinfo {volume} {810}},\ \bibinfo
  {pages} {34} (\bibinfo {year} {2006})}\BibitemShut {NoStop}%
\bibitem [{\citenamefont {Hartle}(2006)}]{Hartle-Glafka}%
  \BibitemOpen
  \bibfield  {author} {\bibinfo {author} {\bibfnamefont {J.~B.}\ \bibnamefont
  {Hartle}},\ }\bibfield  {title} {\bibinfo {title} {Glafka 2004:
  {G}eneralizing quantum mechanics for quantum gravity},\ }\href
  {https://doi.org/10.1007/s10773-006-9134-z} {\bibfield  {journal} {\bibinfo
  {journal} {Int. J. Theo. Phys.}\ }\textbf {\bibinfo {volume} {45}},\ \bibinfo
  {pages} {1390} (\bibinfo {year} {2006})}\BibitemShut {NoStop}%
\bibitem [{\citenamefont {Giddings}(2008)}]{Giddings-}%
  \BibitemOpen
  \bibfield  {author} {\bibinfo {author} {\bibfnamefont {S.~B.}\ \bibnamefont
  {Giddings}},\ }\bibfield  {title} {\bibinfo {title} {Universal quantum
  mechanics},\ }\href {https://doi.org/10.1103/PhysRevD.78.084004} {\bibfield
  {journal} {\bibinfo  {journal} {Phys. Rev. D}\ }\textbf {\bibinfo {volume}
  {78}},\ \bibinfo {pages} {084004} (\bibinfo {year} {2008})}\BibitemShut
  {NoStop}%
\bibitem [{QLa()}]{QLandscape}%
  \BibitemOpen
  \href@noop {} {\bibinfo {title} {{The Quantum Landscape 2013 Conference}}},\
  \bibinfo {note} {{Perimeter Institute}, May 27-31, 2013, Waterloo, Canada,
  \url{https://www2.perimeterinstitute.ca/conferences/quantum-landscape}}\BibitemShut
  {NoStop}%
\bibitem [{\citenamefont {Hartle}()}]{Hartle-21}%
  \BibitemOpen
  \bibfield  {author} {\bibinfo {author} {\bibfnamefont {J.}~\bibnamefont
  {Hartle}},\ }\href@noop {} {\bibinfo {title} {Generalized quantum
  mechanics}},\ \Eprint {https://arxiv.org/abs/2110.11268} {arXiv:2110.11268}
  \BibitemShut {NoStop}%
\bibitem [{\citenamefont {M{\"{u}}ller}(2021)}]{Muller-generalizing}%
  \BibitemOpen
  \bibfield  {author} {\bibinfo {author} {\bibfnamefont {M.~P.}\ \bibnamefont
  {M{\"{u}}ller}},\ }\bibfield  {title} {\bibinfo {title} {Probabilistic
  theories and reconstructions of quantum theory},\ }\href
  {https://doi.org/10.21468/SciPostPhysLectNotes.28} {\bibfield  {journal}
  {\bibinfo  {journal} {SciPost Phys. Lect. Notes}\ ,\ \bibinfo {pages} {28}}
  (\bibinfo {year} {2021})}\BibitemShut {NoStop}%
\bibitem [{\citenamefont {Kent}()}]{Kent-new}%
  \BibitemOpen
  \bibfield  {author} {\bibinfo {author} {\bibfnamefont {A.}~\bibnamefont
  {Kent}},\ }\href@noop {} {\bibinfo {title} {The measurement postulates of
  quantum mechanics are not redundant}},\ \Eprint
  {https://arxiv.org/abs/2307.06191} {arXiv:2307.06191} \BibitemShut {NoStop}%
\bibitem [{\citenamefont {Masanes}\ \emph {et~al.}()\citenamefont {Masanes},
  \citenamefont {Galley},\ and\ \citenamefont {M{\"{u}}ller}}]{Mullerresponse}%
  \BibitemOpen
  \bibfield  {author} {\bibinfo {author} {\bibfnamefont {L.}~\bibnamefont
  {Masanes}}, \bibinfo {author} {\bibfnamefont {T.~D.}\ \bibnamefont
  {Galley}},\ and\ \bibinfo {author} {\bibfnamefont {M.~P.}\ \bibnamefont
  {M{\"{u}}ller}},\ }\href@noop {} {\bibinfo {title} {{Response to ``The
  measurement postulates of quantum mechanics are not redundant''}}},\ \Eprint
  {https://arxiv.org/abs/2309.01650} {arXiv:2309.01650} \BibitemShut {NoStop}%
\bibitem [{\citenamefont {Friedenberg}\ and\ \citenamefont
  {Silverman}(2015)}]{book:Cog-1}%
  \BibitemOpen
  \bibfield  {author} {\bibinfo {author} {\bibfnamefont {J.~D.}\ \bibnamefont
  {Friedenberg}}\ and\ \bibinfo {author} {\bibfnamefont {G.~W.}\ \bibnamefont
  {Silverman}},\ }\href@noop {} {\emph {\bibinfo {title} {Cognitive Science: An
  Introduction to the Study of Mind}}}\ (\bibinfo  {publisher} {Sage},\
  \bibinfo {address} {Thousand Oaks, CA},\ \bibinfo {year} {2015})\BibitemShut
  {NoStop}%
\bibitem [{\citenamefont {Poeppel}\ \emph {et~al.}(2020)\citenamefont
  {Poeppel}, \citenamefont {Mangun},\ and\ \citenamefont
  {Gazzaniga}}]{book:Cog-2}%
  \BibitemOpen
  \bibinfo {editor} {\bibfnamefont {D.}~\bibnamefont {Poeppel}}, \bibinfo
  {editor} {\bibfnamefont {G.~R.}\ \bibnamefont {Mangun}},\ and\ \bibinfo
  {editor} {\bibfnamefont {M.~S.}\ \bibnamefont {Gazzaniga}},\ eds.,\
  \href@noop {} {\emph {\bibinfo {title} {The Cognitive Neurosciences}}}\
  (\bibinfo  {publisher} {MIT Press},\ \bibinfo {address} {Cambridge, MA},\
  \bibinfo {year} {2020})\BibitemShut {NoStop}%
\bibitem [{\citenamefont {Anderson}(1972)}]{Anderson}%
  \BibitemOpen
  \bibfield  {author} {\bibinfo {author} {\bibfnamefont {P.~W.}\ \bibnamefont
  {Anderson}},\ }\bibfield  {title} {\bibinfo {title} {More is different},\
  }\href {https://doi.org/10.1126/science.177.4047.393} {\bibfield  {journal}
  {\bibinfo  {journal} {Science}\ }\textbf {\bibinfo {volume} {177}},\ \bibinfo
  {pages} {393} (\bibinfo {year} {1972})}\BibitemShut {NoStop}%
\bibitem [{\citenamefont {Gu}\ \emph {et~al.}(2009)\citenamefont {Gu},
  \citenamefont {Weedbrook}, \citenamefont {Perales},\ and\ \citenamefont
  {Nielsen}}]{morereallydiff}%
  \BibitemOpen
  \bibfield  {author} {\bibinfo {author} {\bibfnamefont {M.}~\bibnamefont
  {Gu}}, \bibinfo {author} {\bibfnamefont {C.}~\bibnamefont {Weedbrook}},
  \bibinfo {author} {\bibfnamefont {{\'{A}}.}~\bibnamefont {Perales}},\ and\
  \bibinfo {author} {\bibfnamefont {M.~A.}\ \bibnamefont {Nielsen}},\
  }\bibfield  {title} {\bibinfo {title} {{More really is different}},\ }\href
  {https://doi.org/10.1016/j.physd.2008.12.016} {\bibfield  {journal} {\bibinfo
   {journal} {Physica D}\ }\textbf {\bibinfo {volume} {238}},\ \bibinfo {pages}
  {835} (\bibinfo {year} {2009})}\BibitemShut {NoStop}%
\bibitem [{\citenamefont {Ellis}(2020)}]{emerg2}%
  \BibitemOpen
  \bibfield  {author} {\bibinfo {author} {\bibfnamefont {G.~F.~R.}\
  \bibnamefont {Ellis}},\ }\bibfield  {title} {\bibinfo {title} {Emergence in
  solid state physics and biology},\ }\href
  {https://doi.org/10.1007/s10701-020-00367-z} {\bibfield  {journal} {\bibinfo
  {journal} {Found. Phys.}\ }\textbf {\bibinfo {volume} {50}},\ \bibinfo
  {pages} {1098} (\bibinfo {year} {2020})}\BibitemShut {NoStop}%
\bibitem [{\citenamefont {Busemeyer}\ and\ \citenamefont
  {Bruza}(2012)}]{book:Qcog}%
  \BibitemOpen
  \bibfield  {author} {\bibinfo {author} {\bibfnamefont {J.~R.}\ \bibnamefont
  {Busemeyer}}\ and\ \bibinfo {author} {\bibfnamefont {P.~D.}\ \bibnamefont
  {Bruza}},\ }\href@noop {} {\emph {\bibinfo {title} {Quantum Models of
  Cognition and Decision}}}\ (\bibinfo  {publisher} {Cambridge University
  Press},\ \bibinfo {address} {New York},\ \bibinfo {year} {2012})\BibitemShut
  {NoStop}%
\bibitem [{\citenamefont {Mart\'{i}nez-Mart\'{i}nez}\ and\ \citenamefont
  {S\'{a}nchez-Burillo}(2016)}]{Qwalk-}%
  \BibitemOpen
  \bibfield  {author} {\bibinfo {author} {\bibfnamefont {I.}~\bibnamefont
  {Mart\'{i}nez-Mart\'{i}nez}}\ and\ \bibinfo {author} {\bibfnamefont
  {E.}~\bibnamefont {S\'{a}nchez-Burillo}},\ }\bibfield  {title} {\bibinfo
  {title} {Quantum stochastic walks on networks for decision-making},\ }\href
  {https://doi.org/10.1038/srep23812} {\bibfield  {journal} {\bibinfo
  {journal} {Sci. Rep.}\ }\textbf {\bibinfo {volume} {6}},\ \bibinfo {pages}
  {23812} (\bibinfo {year} {2016})}\BibitemShut {NoStop}%
\bibitem [{\citenamefont {Kent}(2018)}]{QQ}%
  \BibitemOpen
  \bibfield  {author} {\bibinfo {author} {\bibfnamefont {A.}~\bibnamefont
  {Kent}},\ }\bibfield  {title} {\bibinfo {title} {Quanta and qualia},\ }\href
  {https://doi.org/10.1007/s10701-018-0193-9} {\bibfield  {journal} {\bibinfo
  {journal} {Found. Phys.}\ }\textbf {\bibinfo {volume} {48}},\ \bibinfo
  {pages} {1021} (\bibinfo {year} {2018})}\BibitemShut {NoStop}%
\bibitem [{\citenamefont {Li}\ \emph {et~al.}(2020)\citenamefont {Li},
  \citenamefont {Dong}, \citenamefont {Wei}, \citenamefont {Liu}, \citenamefont
  {Pan}, \citenamefont {Nori},\ and\ \citenamefont {Zhang}}]{humanDM}%
  \BibitemOpen
  \bibfield  {author} {\bibinfo {author} {\bibfnamefont {J.-A.}\ \bibnamefont
  {Li}}, \bibinfo {author} {\bibfnamefont {D.}~\bibnamefont {Dong}}, \bibinfo
  {author} {\bibfnamefont {Z.}~\bibnamefont {Wei}}, \bibinfo {author}
  {\bibfnamefont {Y.}~\bibnamefont {Liu}}, \bibinfo {author} {\bibfnamefont
  {Y.}~\bibnamefont {Pan}}, \bibinfo {author} {\bibfnamefont {F.}~\bibnamefont
  {Nori}},\ and\ \bibinfo {author} {\bibfnamefont {X.}~\bibnamefont {Zhang}},\
  }\bibfield  {title} {\bibinfo {title} {Quantum reinforcement learning during
  human decision-making},\ }\href {https://doi.org/10.1038/s41562-019-0804-2}
  {\bibfield  {journal} {\bibinfo  {journal} {Nat. Hum. Behav.}\ }\textbf
  {\bibinfo {volume} {4}},\ \bibinfo {pages} {294} (\bibinfo {year}
  {2020})}\BibitemShut {NoStop}%
\bibitem [{\citenamefont {Khrennikov}(2021)}]{Khrenn}%
  \BibitemOpen
  \bibfield  {author} {\bibinfo {author} {\bibfnamefont {A.}~\bibnamefont
  {Khrennikov}},\ }\bibfield  {title} {\bibinfo {title} {Quantum-like model for
  unconscious–conscious interaction and emotional coloring of perceptions and
  other conscious experiences},\ }\href
  {https://doi.org/10.1016/j.biosystems.2021.104471} {\bibfield  {journal}
  {\bibinfo  {journal} {Biosys.}\ }\textbf {\bibinfo {volume} {208}},\ \bibinfo
  {pages} {104471} (\bibinfo {year} {2021})}\BibitemShut {NoStop}%
\bibitem [{\citenamefont {Pothos}\ and\ \citenamefont
  {Busemeyer}(2022)}]{QCogn}%
  \BibitemOpen
  \bibfield  {author} {\bibinfo {author} {\bibfnamefont {E.~M.}\ \bibnamefont
  {Pothos}}\ and\ \bibinfo {author} {\bibfnamefont {J.~R.}\ \bibnamefont
  {Busemeyer}},\ }\bibfield  {title} {\bibinfo {title} {Quantum cognition},\
  }\href {https://doi.org/10.1146/annurev-psych-033020-123501} {\bibfield
  {journal} {\bibinfo  {journal} {Annu. Rev. Psychol.}\ }\textbf {\bibinfo
  {volume} {73}},\ \bibinfo {pages} {749} (\bibinfo {year} {2022})}\BibitemShut
  {NoStop}%
\bibitem [{\citenamefont {Aerts}\ and\ \citenamefont
  {Argu\"{e}lles}(2022)}]{HumanQ}%
  \BibitemOpen
  \bibfield  {author} {\bibinfo {author} {\bibfnamefont {D.}~\bibnamefont
  {Aerts}}\ and\ \bibinfo {author} {\bibfnamefont {J.~A.}\ \bibnamefont
  {Argu\"{e}lles}},\ }\bibfield  {title} {\bibinfo {title} {Human perception as
  a phenomenon of quantization},\ }\href {https://doi.org/10.3390/e24091207}
  {\bibfield  {journal} {\bibinfo  {journal} {Entropy}\ }\textbf {\bibinfo
  {volume} {24}},\ \bibinfo {pages} {1207} (\bibinfo {year}
  {2022})}\BibitemShut {NoStop}%
\bibitem [{\citenamefont {Khrennikov}(2023)}]{book:Khrenn}%
  \BibitemOpen
  \bibfield  {author} {\bibinfo {author} {\bibfnamefont {A.~Y.}\ \bibnamefont
  {Khrennikov}},\ }\href@noop {} {\emph {\bibinfo {title} {Open Quantum Systems
  in Biology, Cognitive and Social Sciences}}}\ (\bibinfo  {publisher}
  {Springer},\ \bibinfo {address} {Cham, Switzerland},\ \bibinfo {year}
  {2023})\BibitemShut {NoStop}%
\bibitem [{\citenamefont {Conte}\ \emph {et~al.}(2009)\citenamefont {Conte},
  \citenamefont {Khrennikov}, \citenamefont {Todarello}, \citenamefont
  {Federici}, \citenamefont {Mendolicchio},\ and\ \citenamefont
  {Zbilut}}]{Khrenn-Z}%
  \BibitemOpen
  \bibfield  {author} {\bibinfo {author} {\bibfnamefont {E.}~\bibnamefont
  {Conte}}, \bibinfo {author} {\bibfnamefont {A.~Y.}\ \bibnamefont
  {Khrennikov}}, \bibinfo {author} {\bibfnamefont {O.}~\bibnamefont
  {Todarello}}, \bibinfo {author} {\bibfnamefont {A.}~\bibnamefont {Federici}},
  \bibinfo {author} {\bibfnamefont {L.}~\bibnamefont {Mendolicchio}},\ and\
  \bibinfo {author} {\bibfnamefont {J.~P.}\ \bibnamefont {Zbilut}},\ }\bibfield
   {title} {\bibinfo {title} {Mental states follow quantum mechanics during
  perception and cognition of ambiguous figures},\ }\href
  {https://doi.org/10.1142/S1230161209000074} {\bibfield  {journal} {\bibinfo
  {journal} {Open Sys. Info. Dyn.}\ }\textbf {\bibinfo {volume} {16}},\
  \bibinfo {pages} {85} (\bibinfo {year} {2009})}\BibitemShut {NoStop}%
\bibitem [{\citenamefont {DeBrota}\ and\ \citenamefont {Love}(2022)}]{Love}%
  \BibitemOpen
  \bibfield  {author} {\bibinfo {author} {\bibfnamefont {J.~B.}\ \bibnamefont
  {DeBrota}}\ and\ \bibinfo {author} {\bibfnamefont {P.~J.}\ \bibnamefont
  {Love}},\ }\bibfield  {title} {\bibinfo {title} {Quantum and classical
  {B}ayesian agents},\ }\href {https://doi.org/10.22331/q-2022-05-16-713}
  {\bibfield  {journal} {\bibinfo  {journal} {Quantum}\ }\textbf {\bibinfo
  {volume} {6}},\ \bibinfo {pages} {713} (\bibinfo {year} {2022})}\BibitemShut
  {NoStop}%
\bibitem [{\citenamefont {Busemeyer}\ \emph {et~al.}(2020)\citenamefont
  {Busemeyer}, \citenamefont {Zhang}, \citenamefont {Balakrishnan},\ and\
  \citenamefont {Wang}}]{Busemeyer1}%
  \BibitemOpen
  \bibfield  {author} {\bibinfo {author} {\bibfnamefont {J.}~\bibnamefont
  {Busemeyer}}, \bibinfo {author} {\bibfnamefont {Q.}~\bibnamefont {Zhang}},
  \bibinfo {author} {\bibfnamefont {S.}~\bibnamefont {Balakrishnan}},\ and\
  \bibinfo {author} {\bibfnamefont {Z.}~\bibnamefont {Wang}},\ }\bibfield
  {title} {\bibinfo {title} {Application of quantum—{M}arkov open system
  models to human cognition and decision},\ }\href
  {https://doi.org/10.3390/e22090990} {\bibfield  {journal} {\bibinfo
  {journal} {Entropy}\ }\textbf {\bibinfo {volume} {22}},\ \bibinfo {pages}
  {990} (\bibinfo {year} {2020})}\BibitemShut {NoStop}%
\bibitem [{\citenamefont {Aerts}\ \emph {et~al.}(2019)\citenamefont {Aerts},
  \citenamefont {Aerts~Argu\"{e}lles}, \citenamefont {Beltran}, \citenamefont
  {Geriente}, \citenamefont {Sassoli~de Bianchi}, \citenamefont {Sozzo},\ and\
  \citenamefont {Veloz}}]{Aert}%
  \BibitemOpen
  \bibfield  {author} {\bibinfo {author} {\bibfnamefont {D.}~\bibnamefont
  {Aerts}}, \bibinfo {author} {\bibfnamefont {J.}~\bibnamefont
  {Aerts~Argu\"{e}lles}}, \bibinfo {author} {\bibfnamefont {L.}~\bibnamefont
  {Beltran}}, \bibinfo {author} {\bibfnamefont {V.}~\bibnamefont {Geriente}},
  \bibinfo {author} {\bibfnamefont {M.}~\bibnamefont {Sassoli~de Bianchi}},
  \bibinfo {author} {\bibfnamefont {S.}~\bibnamefont {Sozzo}},\ and\ \bibinfo
  {author} {\bibfnamefont {T.}~\bibnamefont {Veloz}},\ }\bibfield  {title}
  {\bibinfo {title} {Quantum entanglement in physical and cognitive systems: A
  conceptual analysis and a general representation},\ }\href
  {https://doi.org/10.1140/epjp/i2019-12987-0} {\bibfield  {journal} {\bibinfo
  {journal} {Eur. Phys. J. Plus}\ }\textbf {\bibinfo {volume} {134}},\ \bibinfo
  {pages} {493} (\bibinfo {year} {2019})}\BibitemShut {NoStop}%
\bibitem [{\citenamefont {Brody}()}]{Brody}%
  \BibitemOpen
  \bibfield  {author} {\bibinfo {author} {\bibfnamefont {D.~C.}\ \bibnamefont
  {Brody}},\ }\href@noop {} {\bibinfo {title} {Quantum formalism for cognitive
  psychology}},\ \Eprint {https://arxiv.org/abs/2303.06055} {arXiv:2303.06055}
  \BibitemShut {NoStop}%
\bibitem [{\citenamefont {Heunen}\ \emph {et~al.}(2013)\citenamefont {Heunen},
  \citenamefont {Sadrzadeh},\ and\ \citenamefont
  {Grefenstette}}]{book:Linguistic}%
  \BibitemOpen
  \bibinfo {editor} {\bibfnamefont {C.}~\bibnamefont {Heunen}}, \bibinfo
  {editor} {\bibfnamefont {M.}~\bibnamefont {Sadrzadeh}},\ and\ \bibinfo
  {editor} {\bibfnamefont {E.}~\bibnamefont {Grefenstette}},\ eds.,\ \href@noop
  {} {\emph {\bibinfo {title} {Quantum Physics and Linguistics: A
  Compositional, Diagrammatic Discourse}}}\ (\bibinfo  {publisher} {Oxford
  University Press},\ \bibinfo {address} {Oxford},\ \bibinfo {year}
  {2013})\BibitemShut {NoStop}%
\bibitem [{\citenamefont {Surov}\ \emph {et~al.}(2021)\citenamefont {Surov},
  \citenamefont {Semenenko}, \citenamefont {Platonov}, \citenamefont
  {Bessmertny}, \citenamefont {Galofaro}, \citenamefont {Toffano},
  \citenamefont {Khrennikov},\ and\ \citenamefont {Alodjants}}]{Khrenn-5}%
  \BibitemOpen
  \bibfield  {author} {\bibinfo {author} {\bibfnamefont {I.~A.}\ \bibnamefont
  {Surov}}, \bibinfo {author} {\bibfnamefont {E.}~\bibnamefont {Semenenko}},
  \bibinfo {author} {\bibfnamefont {A.~V.}\ \bibnamefont {Platonov}}, \bibinfo
  {author} {\bibfnamefont {I.~A.}\ \bibnamefont {Bessmertny}}, \bibinfo
  {author} {\bibfnamefont {F.}~\bibnamefont {Galofaro}}, \bibinfo {author}
  {\bibfnamefont {Z.}~\bibnamefont {Toffano}}, \bibinfo {author} {\bibfnamefont
  {A.~Y.}\ \bibnamefont {Khrennikov}},\ and\ \bibinfo {author} {\bibfnamefont
  {A.~P.}\ \bibnamefont {Alodjants}},\ }\bibfield  {title} {\bibinfo {title}
  {Quantum semantics of text perception},\ }\href
  {https://doi.org/10.1038/s41598-021-83490-9} {\bibfield  {journal} {\bibinfo
  {journal} {Sci. Rep.}\ }\textbf {\bibinfo {volume} {11}},\ \bibinfo {pages}
  {4193} (\bibinfo {year} {2021})}\BibitemShut {NoStop}%
\bibitem [{\citenamefont {Haven}\ and\ \citenamefont
  {Khrennikov}(2013)}]{book:KhrennS}%
  \BibitemOpen
  \bibfield  {author} {\bibinfo {author} {\bibfnamefont {E.}~\bibnamefont
  {Haven}}\ and\ \bibinfo {author} {\bibfnamefont {A.}~\bibnamefont
  {Khrennikov}},\ }\href@noop {} {\emph {\bibinfo {title} {Quantum Social
  Science}}}\ (\bibinfo  {publisher} {Cambridge University Press},\ \bibinfo
  {address} {New York},\ \bibinfo {year} {2013})\BibitemShut {NoStop}%
\bibitem [{\citenamefont {Orrell}(2020)}]{book:Qfinance}%
  \BibitemOpen
  \bibfield  {author} {\bibinfo {author} {\bibfnamefont {D.}~\bibnamefont
  {Orrell}},\ }\href@noop {} {\emph {\bibinfo {title} {Quantum Economics and
  Finance: An Applied Mathematics Introduction}}}\ (\bibinfo  {publisher}
  {Panda Ohana},\ \bibinfo {address} {New York},\ \bibinfo {year}
  {2020})\BibitemShut {NoStop}%
\bibitem [{\citenamefont {Hausmann}\ \emph {et~al.}()\citenamefont {Hausmann},
  \citenamefont {Nurgalieva},\ and\ \citenamefont {del Rio}}]{spekkens-toy}%
  \BibitemOpen
  \bibfield  {author} {\bibinfo {author} {\bibfnamefont {L.}~\bibnamefont
  {Hausmann}}, \bibinfo {author} {\bibfnamefont {N.}~\bibnamefont
  {Nurgalieva}},\ and\ \bibinfo {author} {\bibfnamefont {L.}~\bibnamefont {del
  Rio}},\ }\href@noop {} {\bibinfo {title} {{A consolidating review of
  Spekkens' toy theory}}},\ \Eprint {https://arxiv.org/abs/2105.03277}
  {arXiv:2105.03277} \BibitemShut {NoStop}%
\bibitem [{\citenamefont {Budiyono}\ and\ \citenamefont
  {Dipojono}(2020)}]{Budiyono1}%
  \BibitemOpen
  \bibfield  {author} {\bibinfo {author} {\bibfnamefont {A.}~\bibnamefont
  {Budiyono}}\ and\ \bibinfo {author} {\bibfnamefont {H.~K.}\ \bibnamefont
  {Dipojono}},\ }\bibfield  {title} {\bibinfo {title} {Nonlinear
  {S}chr\"{o}dinger equations and generalized {H}eisenberg uncertainty
  principle from estimation schemes violating the principle of estimation
  independence},\ }\href {https://doi.org/10.1103/PhysRevA.102.012205}
  {\bibfield  {journal} {\bibinfo  {journal} {Phys. Rev. A}\ }\textbf {\bibinfo
  {volume} {102}},\ \bibinfo {pages} {012205} (\bibinfo {year}
  {2020})}\BibitemShut {NoStop}%
\bibitem [{\citenamefont {Budiyono}\ and\ \citenamefont
  {Rohrlich}(2017)}]{Budiyono2}%
  \BibitemOpen
  \bibfield  {author} {\bibinfo {author} {\bibfnamefont {A.}~\bibnamefont
  {Budiyono}}\ and\ \bibinfo {author} {\bibfnamefont {D.}~\bibnamefont
  {Rohrlich}},\ }\bibfield  {title} {\bibinfo {title} {Quantum mechanics as
  classical statistical mechanics with an ontic extension and an epistemic
  restriction},\ }\href {https://doi.org/10.1038/s41467-017-01375-w} {\bibfield
   {journal} {\bibinfo  {journal} {Nat. Commun.}\ }\textbf {\bibinfo {volume}
  {8}},\ \bibinfo {pages} {1306} (\bibinfo {year} {2017})}\BibitemShut
  {NoStop}%
\bibitem [{\citenamefont {Koochakie}\ \emph {et~al.}()\citenamefont
  {Koochakie}, \citenamefont {Alipour},\ and\ \citenamefont
  {Rezakhani}}]{AdLRB}%
  \BibitemOpen
  \bibfield  {author} {\bibinfo {author} {\bibfnamefont {M.~M.~R.}\
  \bibnamefont {Koochakie}}, \bibinfo {author} {\bibfnamefont {S.}~\bibnamefont
  {Alipour}},\ and\ \bibinfo {author} {\bibfnamefont {A.~T.}\ \bibnamefont
  {Rezakhani}},\ }\href@noop {} {\bibinfo {title} {{Lieb-Robinson bound and
  adiabatic evolution}}},\ \Eprint {https://arxiv.org/abs/1307.3726}
  {arXiv:1307.3726} \BibitemShut {NoStop}%
\bibitem [{\citenamefont {Scarani}\ \emph {et~al.}(2005)\citenamefont
  {Scarani}, \citenamefont {Iblisdir}, \citenamefont {Gisin},\ and\
  \citenamefont {Ac\'{\i}n}}]{cloning}%
  \BibitemOpen
  \bibfield  {author} {\bibinfo {author} {\bibfnamefont {V.}~\bibnamefont
  {Scarani}}, \bibinfo {author} {\bibfnamefont {S.}~\bibnamefont {Iblisdir}},
  \bibinfo {author} {\bibfnamefont {N.}~\bibnamefont {Gisin}},\ and\ \bibinfo
  {author} {\bibfnamefont {A.}~\bibnamefont {Ac\'{\i}n}},\ }\bibfield  {title}
  {\bibinfo {title} {Quantum cloning},\ }\href
  {https://doi.org/10.1103/RevModPhys.77.1225} {\bibfield  {journal} {\bibinfo
  {journal} {Rev. Mod. Phys.}\ }\textbf {\bibinfo {volume} {77}},\ \bibinfo
  {pages} {1225} (\bibinfo {year} {2005})}\BibitemShut {NoStop}%
\bibitem [{\citenamefont {Brunner}\ \emph {et~al.}(2014)\citenamefont
  {Brunner}, \citenamefont {Cavalcanti}, \citenamefont {Pironio}, \citenamefont
  {Scarani},\ and\ \citenamefont {Wehner}}]{Bell}%
  \BibitemOpen
  \bibfield  {author} {\bibinfo {author} {\bibfnamefont {N.}~\bibnamefont
  {Brunner}}, \bibinfo {author} {\bibfnamefont {D.}~\bibnamefont {Cavalcanti}},
  \bibinfo {author} {\bibfnamefont {S.}~\bibnamefont {Pironio}}, \bibinfo
  {author} {\bibfnamefont {V.}~\bibnamefont {Scarani}},\ and\ \bibinfo {author}
  {\bibfnamefont {S.}~\bibnamefont {Wehner}},\ }\bibfield  {title} {\bibinfo
  {title} {Bell nonlocality},\ }\href
  {https://doi.org/10.1103/RevModPhys.86.419} {\bibfield  {journal} {\bibinfo
  {journal} {Rev. Mod. Phys.}\ }\textbf {\bibinfo {volume} {86}},\ \bibinfo
  {pages} {419} (\bibinfo {year} {2014})}\BibitemShut {NoStop}%
\bibitem [{\citenamefont {Hu}\ and\ \citenamefont {Tomamichel}()}]{Tomamichel}%
  \BibitemOpen
  \bibfield  {author} {\bibinfo {author} {\bibfnamefont {Y.}~\bibnamefont
  {Hu}}\ and\ \bibinfo {author} {\bibfnamefont {M.}~\bibnamefont
  {Tomamichel}},\ }\href@noop {} {\bibinfo {title} {Fundamental limits on
  quantum cloning from the no-signalling principle}},\ \Eprint
  {https://arxiv.org/abs/2305.02002} {arXiv:2305.02002} \BibitemShut {NoStop}%
\bibitem [{\citenamefont {Busch}\ \emph {et~al.}(2007)\citenamefont {Busch},
  \citenamefont {Heinonen},\ and\ \citenamefont {Lahti}}]{UR-1}%
  \BibitemOpen
  \bibfield  {author} {\bibinfo {author} {\bibfnamefont {P.}~\bibnamefont
  {Busch}}, \bibinfo {author} {\bibfnamefont {T.}~\bibnamefont {Heinonen}},\
  and\ \bibinfo {author} {\bibfnamefont {P.}~\bibnamefont {Lahti}},\ }\bibfield
   {title} {\bibinfo {title} {Heisenberg's uncertainty principle},\ }\href
  {https://doi.org/10.1016/j.physrep.2007.05.006} {\bibfield  {journal}
  {\bibinfo  {journal} {Phys. Rep.}\ }\textbf {\bibinfo {volume} {452}},\
  \bibinfo {pages} {155} (\bibinfo {year} {2007})}\BibitemShut {NoStop}%
\bibitem [{\citenamefont {Coles}\ \emph {et~al.}(2017)\citenamefont {Coles},
  \citenamefont {Berta}, \citenamefont {Tomamichel},\ and\ \citenamefont
  {Wehner}}]{UR-2}%
  \BibitemOpen
  \bibfield  {author} {\bibinfo {author} {\bibfnamefont {P.~J.}\ \bibnamefont
  {Coles}}, \bibinfo {author} {\bibfnamefont {M.}~\bibnamefont {Berta}},
  \bibinfo {author} {\bibfnamefont {M.}~\bibnamefont {Tomamichel}},\ and\
  \bibinfo {author} {\bibfnamefont {S.}~\bibnamefont {Wehner}},\ }\bibfield
  {title} {\bibinfo {title} {Entropic uncertainty relations and their
  applications},\ }\href {https://doi.org/10.1103/RevModPhys.89.015002}
  {\bibfield  {journal} {\bibinfo  {journal} {Rev. Mod. Phys.}\ }\textbf
  {\bibinfo {volume} {89}},\ \bibinfo {pages} {015002} (\bibinfo {year}
  {2017})}\BibitemShut {NoStop}%
\bibitem [{\citenamefont {Bartlett}\ \emph {et~al.}(2007)\citenamefont
  {Bartlett}, \citenamefont {Rudolph},\ and\ \citenamefont
  {Spekkens}}]{QRef-1}%
  \BibitemOpen
  \bibfield  {author} {\bibinfo {author} {\bibfnamefont {S.~D.}\ \bibnamefont
  {Bartlett}}, \bibinfo {author} {\bibfnamefont {T.}~\bibnamefont {Rudolph}},\
  and\ \bibinfo {author} {\bibfnamefont {R.~W.}\ \bibnamefont {Spekkens}},\
  }\bibfield  {title} {\bibinfo {title} {Reference frames, superselection
  rules, and quantum information},\ }\href
  {https://doi.org/10.1103/RevModPhys.79.555} {\bibfield  {journal} {\bibinfo
  {journal} {Rev. Mod. Phys.}\ }\textbf {\bibinfo {volume} {79}},\ \bibinfo
  {pages} {555} (\bibinfo {year} {2007})}\BibitemShut {NoStop}%
\bibitem [{\citenamefont {Giacomini}\ \emph {et~al.}(2019)\citenamefont
  {Giacomini}, \citenamefont {Castro-Ruiz},\ and\ \citenamefont
  {Brukner}}]{QRef-2}%
  \BibitemOpen
  \bibfield  {author} {\bibinfo {author} {\bibfnamefont {F.}~\bibnamefont
  {Giacomini}}, \bibinfo {author} {\bibfnamefont {E.}~\bibnamefont
  {Castro-Ruiz}},\ and\ \bibinfo {author} {\bibfnamefont
  {{\v{C}}.}~\bibnamefont {Brukner}},\ }\bibfield  {title} {\bibinfo {title}
  {Quantum mechanics and the covariance of physical laws in quantum reference
  frames},\ }\href {https://doi.org/10.1038/s41467-018-08155-0} {\bibfield
  {journal} {\bibinfo  {journal} {Nat. Commun.}\ }\textbf {\bibinfo {volume}
  {10}},\ \bibinfo {pages} {494} (\bibinfo {year} {2019})}\BibitemShut
  {NoStop}%
\bibitem [{\citenamefont {Vanrietvelde}\ \emph {et~al.}(2020)\citenamefont
  {Vanrietvelde}, \citenamefont {H{\"{o}}hn}, \citenamefont {Giacomini},\ and\
  \citenamefont {Castro-Ruiz}}]{QRef-4}%
  \BibitemOpen
  \bibfield  {author} {\bibinfo {author} {\bibfnamefont {A.}~\bibnamefont
  {Vanrietvelde}}, \bibinfo {author} {\bibfnamefont {P.~A.}\ \bibnamefont
  {H{\"{o}}hn}}, \bibinfo {author} {\bibfnamefont {F.}~\bibnamefont
  {Giacomini}},\ and\ \bibinfo {author} {\bibfnamefont {E.}~\bibnamefont
  {Castro-Ruiz}},\ }\bibfield  {title} {\bibinfo {title} {A change of
  perspective: switching quantum reference frames via a perspective-neutral
  framework},\ }\href {https://doi.org/10.22331/q-2020-01-27-225} {\bibfield
  {journal} {\bibinfo  {journal} {{Quantum}}\ }\textbf {\bibinfo {volume}
  {4}},\ \bibinfo {pages} {225} (\bibinfo {year} {2020})}\BibitemShut {NoStop}%
\bibitem [{\citenamefont {Fields}\ \emph
  {et~al.}(2022{\natexlab{a}})\citenamefont {Fields}, \citenamefont
  {Glazebrook},\ and\ \citenamefont {Levin}}]{Fields-1}%
  \BibitemOpen
  \bibfield  {author} {\bibinfo {author} {\bibfnamefont {C.}~\bibnamefont
  {Fields}}, \bibinfo {author} {\bibfnamefont {G.~F.}\ \bibnamefont
  {Glazebrook}},\ and\ \bibinfo {author} {\bibfnamefont {M.}~\bibnamefont
  {Levin}},\ }\bibfield  {title} {\bibinfo {title} {Neurons as hierarchies of
  quantum reference frames},\ }\href
  {https://doi.org/10.1016/j.biosystems.2022.104714} {\bibfield  {journal}
  {\bibinfo  {journal} {Biosys.}\ }\textbf {\bibinfo {volume} {219}},\ \bibinfo
  {pages} {104714} (\bibinfo {year} {2022}{\natexlab{a}})}\BibitemShut
  {NoStop}%
\bibitem [{\citenamefont {Fields}\ \emph
  {et~al.}(2022{\natexlab{b}})\citenamefont {Fields}, \citenamefont {Friston},
  \citenamefont {Glazebrook},\ and\ \citenamefont {Levin}}]{Fields-2}%
  \BibitemOpen
  \bibfield  {author} {\bibinfo {author} {\bibfnamefont {C.}~\bibnamefont
  {Fields}}, \bibinfo {author} {\bibfnamefont {K.}~\bibnamefont {Friston}},
  \bibinfo {author} {\bibfnamefont {J.~F.}\ \bibnamefont {Glazebrook}},\ and\
  \bibinfo {author} {\bibfnamefont {M.}~\bibnamefont {Levin}},\ }\bibfield
  {title} {\bibinfo {title} {A free energy principle for generic quantum
  systems},\ }\href {https://doi.org/10.1016/j.pbiomolbio.2022.05.006}
  {\bibfield  {journal} {\bibinfo  {journal} {Prog. Biophys. Mol. Biol.}\
  }\textbf {\bibinfo {volume} {173}},\ \bibinfo {pages} {36} (\bibinfo {year}
  {2022}{\natexlab{b}})}\BibitemShut {NoStop}%
\bibitem [{\citenamefont {Venegas-Andraca}(2012)}]{QRW}%
  \BibitemOpen
  \bibfield  {author} {\bibinfo {author} {\bibfnamefont {S.~E.}\ \bibnamefont
  {Venegas-Andraca}},\ }\bibfield  {title} {\bibinfo {title} {Quantum walks: a
  comprehensive review},\ }\href {https://doi.org/10.1007/s11128-012-0432-5}
  {\bibfield  {journal} {\bibinfo  {journal} {Quantum Inf. Process.}\ }\textbf
  {\bibinfo {volume} {11}},\ \bibinfo {pages} {1015} (\bibinfo {year}
  {2012})}\BibitemShut {NoStop}%
\bibitem [{\citenamefont {Shah~Khan}\ \emph {et~al.}(2018)\citenamefont
  {Shah~Khan}, \citenamefont {Solmeyer}, \citenamefont {Balu},\ and\
  \citenamefont {Humble}}]{QGames}%
  \BibitemOpen
  \bibfield  {author} {\bibinfo {author} {\bibfnamefont {F.}~\bibnamefont
  {Shah~Khan}}, \bibinfo {author} {\bibfnamefont {N.}~\bibnamefont {Solmeyer}},
  \bibinfo {author} {\bibfnamefont {R.}~\bibnamefont {Balu}},\ and\ \bibinfo
  {author} {\bibfnamefont {T.~S.}\ \bibnamefont {Humble}},\ }\bibfield  {title}
  {\bibinfo {title} {Quantum games: a review of the history, current state, and
  interpretation},\ }\href {https://doi.org/10.1007/s11128-018-2082-8}
  {\bibfield  {journal} {\bibinfo  {journal} {Quantum Inf. Process.}\ }\textbf
  {\bibinfo {volume} {309}},\ \bibinfo {pages} {17} (\bibinfo {year}
  {2018})}\BibitemShut {NoStop}%
\bibitem [{\citenamefont {Manrique}\ \emph {et~al.}(2023)\citenamefont
  {Manrique}, \citenamefont {Huo}, \citenamefont {El~Oud}, \citenamefont
  {Zheng}, \citenamefont {Illari},\ and\ \citenamefont {Johnson}}]{JohnsonPRL}%
  \BibitemOpen
  \bibfield  {author} {\bibinfo {author} {\bibfnamefont {P.~D.}\ \bibnamefont
  {Manrique}}, \bibinfo {author} {\bibfnamefont {F.~Y.}\ \bibnamefont {Huo}},
  \bibinfo {author} {\bibfnamefont {S.}~\bibnamefont {El~Oud}}, \bibinfo
  {author} {\bibfnamefont {M.}~\bibnamefont {Zheng}}, \bibinfo {author}
  {\bibfnamefont {L.}~\bibnamefont {Illari}},\ and\ \bibinfo {author}
  {\bibfnamefont {N.~F.}\ \bibnamefont {Johnson}},\ }\bibfield  {title}
  {\bibinfo {title} {{Shockwavelike Behavior across Social Media}},\ }\href
  {https://doi.org/10.1103/PhysRevLett.130.237401} {\bibfield  {journal}
  {\bibinfo  {journal} {Phys. Rev. Lett.}\ }\textbf {\bibinfo {volume} {130}},\
  \bibinfo {pages} {237401} (\bibinfo {year} {2023})}\BibitemShut {NoStop}%
\bibitem [{\citenamefont {Ziepke}\ \emph {et~al.}(2022)\citenamefont {Ziepke},
  \citenamefont {Maryshev}, \citenamefont {Aranson},\ and\ \citenamefont
  {Frey}}]{Ziepke-etal}%
  \BibitemOpen
  \bibfield  {author} {\bibinfo {author} {\bibfnamefont {A.}~\bibnamefont
  {Ziepke}}, \bibinfo {author} {\bibfnamefont {I.}~\bibnamefont {Maryshev}},
  \bibinfo {author} {\bibfnamefont {I.~S.}\ \bibnamefont {Aranson}},\ and\
  \bibinfo {author} {\bibfnamefont {E.}~\bibnamefont {Frey}},\ }\bibfield
  {title} {\bibinfo {title} {Multi-scale organization in communicating active
  matter},\ }\href {https://doi.org/10.1038/s41467-022-34484-2} {\bibfield
  {journal} {\bibinfo  {journal} {Nat. Commun.}\ }\textbf {\bibinfo {volume}
  {13}},\ \bibinfo {pages} {6727} (\bibinfo {year} {2022})}\BibitemShut
  {NoStop}%
\bibitem [{fut({\natexlab{a}})}]{future}%
  \BibitemOpen
  \href@noop \ \bibinfo {note} {{Alireza Tavanfar
  \textit{et al.,} Experience-Centric, Multi-Level Quantum Systems (in preparation)}}\BibitemShut {NoStop}%
\bibitem [{\citenamefont {Trudeau}(2008)}]{book:Geometry}%
  \BibitemOpen
  \bibfield  {author} {\bibinfo {author} {\bibfnamefont {R.~J.}\ \bibnamefont
  {Trudeau}},\ }\href@noop {} {\emph {\bibinfo {title} {The Non-Euclidean
  Revolution}}}\ (\bibinfo  {publisher} {Birkh\"{a}user},\ \bibinfo {address}
  {Boston},\ \bibinfo {year} {2008})\BibitemShut {NoStop}%
\bibitem [{fut({\natexlab{b}})}]{future2}%
  \BibitemOpen
  \href@noop  \ \bibinfo {note} {{Alireza Tavanfar
  \textit{et al.,} Understanding the Principle of Inclusive Experience Centricity in Nature (in preparation)}}\BibitemShut {NoStop}%
\end{thebibliography}


%%%%%%%%%%%%%%%%%%%%%%%%%%%%%%%%%%%%%%%%%%%%%%%%%%%%%%%%%%%%%%%%
\end{document}